\documentclass[prd,twocolumn,nofootinbib]{revtex4}
\usepackage{bm}
\usepackage{graphicx}
\usepackage{multirow}
\usepackage{amsmath}
\usepackage{amssymb}
\usepackage{tabulary}

\begin{document}

\newcommand{\mwimp}{$m_\chi$}
\newcommand{\vmin}{$v_{\mathrm{min}}$}
\newcommand{\vlag}{$v_{\mathrm{lag}}$}
\newcommand{\vrms}{$v_{\mathrm{rms}}$}
\newcommand{\vesc}{$v_{\mathrm{esc}}$}
\newcommand{\sigmapsi}{$\sigma_{\mathrm{p}}^{\mathrm{SI}}$}
\newcommand{\sigmapsie}{\sigma_{\mathrm{p}}^{\mathrm{SI}}}
\newcommand{\sigmaNsi}{$\sigma_{\mathrm{N}}^{\mathrm{SI}}$}
\newcommand{\sigmapsd}{$\sigma_{\mathrm{p}}^{\mathrm{SD}}$}
\newcommand{\sigmapsde}{\sigma_{\mathrm{p}}^{\mathrm{SD}}}
\newcommand{\sigmansd}{$\sigma_{\mathrm{n}}^{\mathrm{SD}}$}
\newcommand{\jcap}{{JCAP}}
\newcommand{\mnras}{{Mon. Not. R. Astron. Soc.}}
\newcommand{\bain}{{Bulletin of the Astronomical Institutes of the Netherlands}}
\newcommand{\aap}{{Astron. Astrophys.}}
\newcommand{\aj}{{Astron. J.}}

\title{WIMP physics with ensembles of direct-detection experiments}

\author{Annika H. G. Peter}
\email{apeter@physics.osu.edu, +1 (614) 688-3373}
\affiliation{\mbox{CCAPP and Department of Physics, The Ohio State University, 191 W. Woodruff Ave., Columbus, OH 43210, USA}}
\affiliation{\mbox{Department of Astronomy, The Ohio State University, 140 W. 18th Ave., Columbus, OH 43210, USA}}

\author{Vera Gluscevic}
\email{verag@ias.edu}
\affiliation{Institute for Advanced Study, Einstein Drive, Princeton, NJ 08540, USA}

\author{Anne M. Green}
\email{anne.green@nottingham.ac.uk}
\author{Bradley J. Kavanagh}
\email{ppxbk2@nottingham.ac.uk}
\affiliation{School of Physics and Astronomy, University of Nottingham, University Park, Nottingham, NG7 2RD, UK}

\author{Samuel K. Lee}
\email{samuelkl@princeton.edu}
\affiliation{Princeton Center for Theoretical Science, Princeton University, Princeton, NJ 08544, USA}

\begin{abstract}
The search for weakly-interacting massive particle (WIMP) dark matter is multi-pronged.  Ultimately, the WIMP-dark-matter picture will only be confirmed if different classes of experiments see consistent signals and infer the same WIMP properties. In this work, we review the ideas, methods, and status of direct-detection searches.  We focus in particular on extracting WIMP physics (WIMP interactions and phase-space distribution) from direct-detection data in the early discovery days when multiple experiments see of order dozens to hundreds of events. To demonstrate the essential complementarity of different direct-detection experiments in this context, we create mock data intended to represent the data from the near-future Generation 2 experiments. We consider both conventional supersymmetry-inspired benchmark points (with spin-independent and -dependent elastic cross sections just below current limits), as well as benchmark points for other classes of models (inelastic and effective-operator paradigms).  We also investigate the effect on parameter estimation of loosening or dropping the assumptions about the local WIMP phase-space distribution. We arrive at two main conclusions.  Firstly, teasing out WIMP physics with experiments depends critically on having a wide set of detector target materials, spanning a large range of target nuclear masses and spin-dependent sensitivity.  It is also highly desirable to obtain data from low-threshold experiments.  Secondly, a general reconstruction of the local WIMP velocity distribution, which will only be achieved if there are multiple experiments using different target materials, is critical to obtaining a robust and unbiased estimate of the WIMP mass.
\end{abstract}

\maketitle

\section{Introduction}\label{sec:introduction}
Since the 1980's, the dominant paradigm for the nature of dark matter has been that of the weakly interacting massive particle (WIMP) \cite{steigman1985}.  This particle class has the virtue of being cold dark matter (CDM) \cite{blumenthal1982,white1983,frenk1985,davis1985}, consistent with observations of the cosmic microwave background and cosmological observations of the growth and distribution of structure \cite{ho2012,parkinson2012,planck2013xvi}.  By essentially dimensional analysis and an order-of-magnitude calculation, one may show that the WIMP thermal relic abundance can ``naturally'' match the measured dark-matter abundance \cite{gunn1978,scherrer1986}.  WIMPs also come ``for free'' in minimal extensions to the standard model beyond the electroweak scale, the most famous particle candidate being the supersymmetric (SUSY) neutralino \citep{ellis1984,griest1988}.  These desirable properties are responsible for making the WIMP the most experimentally sought-after dark-matter particle candidate.

There are three key ways in which WIMPs may be hunted.  First, WIMPs may be created in colliders.  Based on ensembles of kinematic cuts for modes of particle creation, the WIMP mass and quantum numbers may be revealed \cite{baltz2006,asano2011,conley2011,baer2013}.  Second, WIMPs may annihilate with each other in dense cosmic dark-matter structures, producing showers of standard-model particles.  At present, the most stringent constraints on WIMP annihilation come from gamma-ray observations (specifically, with Fermi/LAT, H.E.S.S., and VERITAS) of nearby galaxies \cite{abazajian2010,acciari2010,abramowski2011,geringer-sameth2011,ackermann2011a,abazajian2012,abramowski2012,ackermann2012,arlen2012,doro2013,wood2013}.  While there is an intriguing excess of GeV-energy photons at the Galactic Center, and possible hints of a line, its origin is still unknown.  If the excess results from dark matter, the WIMP mass can be determined from the sharp break in the energy spectrum, suggesting a WIMP with mass in the range 10-30 GeV \cite{hooper2011,abazajian2012b,abazajian2013,hooper2013}.  If the line has a dark-matter origin, it points to a WIMP mass of $\sim 130$ GeV \cite{weniger2012,finkbeiner2013}.  While the WIMP mass is relatively straightforward to infer from the annihilation energy spectrum, the annihilation cross section is degenerate with the WIMP mass density.

Finally, WIMPs may be detected via the low energy recoils ($\sim$ 10-100 keV) they impart to nuclei in terrestrial experiments \cite{goodman1985,wasserman1986}.  These ``direct-detection'' experiments operate in the extremely low-background regime, with current limits on WIMP recoils being at the level of $\lesssim 1$ event/kilogram/year.  More than a dozen experiments are being built, are running, or are planned for the near future (see Ref. \cite{Baudis:2012ig} for a review).  The goal for the next decade is to reach the sensitivity level at which solar, atmospheric, and supernova neutrinos are expected to become an irreducible background for non-directional experiments \cite{billard2013}.  Generation-2 experiments, which should be producing data within the next five years, will come within approximately one or two orders of magnitude of this goal (spin-independent WIMP cross sections of \sigmapsi$\sim 10^{-47}-10^{-46}\hbox{ cm}^2$ with standard Galactic-halo assumptions).  This is the near-term future set of experiments.

In this work, we review and further explore the prospects for direct-detection experiments to constrain the WIMP mass and nuclear scattering cross sections.  Accurate inference of these properties is critical for cross-checking with collider and indirect-detection experiments to confirm the WIMP identity of dark matter. Moreover, it is critically important to have multiple direct-detection experiments, with different target nuclei, in order to characterize WIMP physics with direct detection.  There are three key reasons why having multiple target nuclei is important.  First, even if one knew the WIMP phase-space density and type of scattering interaction (spin-dependent, spin-independent, elastic vs. inelastic, etc.), there are strong degeneracies in the WIMP mass--cross-section plane \cite{Drees:2008bv,Peter:2009ak,Pato:2010yq}.  The degeneracy direction depends on the target particle mass because of the scattering kinematics.  By having several target nuclei, we can break this degeneracy.  Second, uncertainties in the WIMP velocity distribution translate directly into uncertainties on the WIMP mass, for the simple reason that kinetic energy depends both on the particle's mass and  velocity.  Again, the degeneracy direction depends on target nuclear masses \cite{Drees:2007hr,Drees:2008bv,green2008,Pato:2010yq,Green:2010gw,McCabe:2010zh,Peter:2011eu} and therefore a combination of experiments yields better constraints on the WIMP mass.  Finally, different nuclei have different sensitivities to the types of possible interactions with WIMPs.  For example, $^{19}$F, with an unpaired proton, is far more sensitive to the WIMP-proton spin-dependent cross section \sigmapsd~than $^{73}$Ge, but the latter is far more sensitive to the WIMP-neutron spin-dependent cross section \sigmansd~and the spin-independent WIMP-nucleon cross section \sigmaNsi \cite{Pato:2011de,Shan:2011ka}.

To demonstrate the capabilities of direct-detection experiments to elucidate WIMP physics, we create mock data sets for idealized models of ``Generation 2'' direct-detection experiments, as well as experiments with directional sensitivity.  We use Bayesian inference to estimate the WIMP properties from these mock data.  This is intended to be an exploration of the power of experiments in the next few years, if the cross-section lies just below current limits, so that they discover WIMPs soon after they turn on.  Note, however, that our results are significantly broader in scope---they demonstrate the capabilities of experiments to characterize WIMPs when the total number of events in each experiment is in the neighborhood of tens or hundreds.  We consider not only benchmark points for traditional supersymmetry-inspired scattering models (spin-independent and -dependent elastic scattering), but also broader models for WIMP interactions with nuclei.  In addition, since the local WIMP phase-space density has not been experimentally probed, we discuss how its uncertainties can be addressed experimentally.  

This work is part literature review and part new calculations in order to highlight how combinations of direct-detection experiments can unveil WIMP properties.  The paper is organized as follows:  Sections~\ref{sec:theory} to \ref{sec:gvmin} are mostly reviews of the literature, focused on the theory of direct detection, current experimental status, and methods to compare experimental results.  In Sec.~\ref{sec:theory}, we show how the event rate in direct-detection experiments depends both on fundamental WIMP properties as well as its local phase-space density.  In Sec.~\ref{sec:experiment}, we briefly present a short summary of the current state of, and future plans for, direct-detection experiments.  We summarize one particular method of comparing experimental results, the ``halo-independent modeling'' originally proposed by Ref. \cite{Fox:2010bz}, in Sec. \ref{sec:gvmin}. The second part of the paper shows how ensembles of direct-detection experiments can constrain WIMP physics.  This second part, Sections \ref{sec:simulation} to \ref{sec:directional}, contains both literature reviews and new calculations.  In Sec.~\ref{sec:simulation}, we explain our method of creating mock experimental data for specific benchmark parameters, and inference of WIMP properties from them using a Bayesian approach.  
In Sec.~\ref{sec:WIMPphys}, we show the prospects for recovering the type and strength of WIMP-nucleus scattering with more restrictive assumptions about the WIMP velocity distribution than we consider in Sec.~\ref{sec:astrophysics}.  Sec.~\ref{sec:astrophysics} summarizes the prospects for reconstruction of the WIMP velocity distribution in the context of spin-independent elastic scattering.  We consider the power of directional detection to unveil the WIMP physics in Sec.~\ref{sec:directional}.  We conclude by highlighting the key points of this work in Sec.~\ref{sec:conclusion}.

\section{Theoretical considerations for direct-detection experiments}\label{sec:theory}
Direct-detection experiments consist of ensembles of nuclear targets $T$.  To first order, the event rate as a function of nuclear recoil energy $Q$ in an experiment is:
\begin{multline}\label{eq:energyspectrum}
  \frac{dR(Q,t)}{dQ} = \left( \frac{\rho_\chi}{m_\chi} \right) \epsilon(Q) \\
  \times \sum_T N_T \int \limits_{v_{\mathrm{min}(m_\chi,m_T,Q)}} \, d^3 v \frac{d\sigma_T(v)}{dQ} |\mathbf{v}| f(\mathbf{v},t).
\end{multline}
This energy spectrum is the primary data product for most direct-detection experiments, although several experiments have no energy sensitivity, and others have angular sensitivity.   

Breaking down Eq.~(\ref{eq:energyspectrum}), $N_T$ is the number of target nuclei in the experiment with isotope $T$.  All else being equal, experiments with a larger target mass (i.e. with larger $N_T$) should see more events than smaller experiments.  $\rho_\chi$ is the local WIMP density, which we discuss further in Sec. \ref{sec:theory:astrophysics}, and \mwimp~is the WIMP mass.  Thus, $\rho_\chi/m_\chi$ is the local WIMP number density---the density of potential scatterers in the experiment.  The physics of scattering between WIMPs and nuclei (including both particle/nuclear considerations as well as kinematics) are encompassed in the velocity integral.  $f(\mathbf{v},t)$ is the distribution of WIMPs as a function of their velocity $\mathbf{v}$ with respect to the experiment which, in general, varies along the Earth's path through the Galaxy.  $(\rho_\chi/m_\chi)|\mathbf{v}| f(\mathbf{v},t)$ is the velocity-weighted flux of WIMPs passing through the experiment.  The probability of a WIMP-nuclear scatter that imparts a target nucleus with energy $Q$ is the product of the WIMP-nuclear cross section $d\sigma_T/dQ$ and the velocity-weighted WIMP flux.  This is integrated over all WIMP velocities $\mathbf{v}$ that are kinematically allowed to induce a nuclear recoil of energy $Q$.  The minimum speed for this recoil, $v_{\mathrm{min}}$, depends on the WIMP mass, the target nucleus mass $m_T$, the nuclear recoil energy $Q$, and the nature of the scatter (elastic vs. inelastic).  Note that most experiments have more than one isotope, so we sum over WIMP interactions with each isotope.  Finally, $\epsilon(Q)$ is the experimental efficiency, or response function. It is the probability that a nuclear scatter with energy $Q$ is detected and survives all data cuts.  This is a rich subject in and of itself, and we point the interested reader to Refs. \cite{lindhard1963,Bozorgnia:2010xy,sorensen2011,baudis2013,collar2013} for several key physical considerations, in addition to papers specific to each experiment.  

\begin{figure}
  \includegraphics[width=0.49\textwidth]{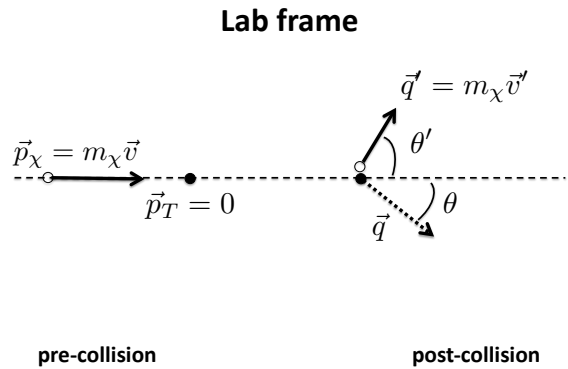}
  \caption{\label{fig:dd_angle}\textbf{Kinematics of WIMP-nuclear elastic scattering.} The open circle represents the WIMP, and the dark circle represents the target nucleus. The $\vec{p}$'s and $\vec{q}$'s represent momenta in the lab frame.}
\end{figure}

For directional detection, the event rate as a function of the lab-frame solid angle $\Omega_q$ looks similar to Eq. (\ref{eq:energyspectrum}),  
\begin{multline}\label{eq:energythetaspectrum}
  \frac{d^2R(Q,\Omega_q,t)}{dQd\Omega_q} = \left( \frac{\rho_\chi}{m_\chi} \right) \epsilon(Q) \\
  \times \sum_T N_T \int \limits_{v_{\mathrm{min}(m_\chi,T,Q)}} \, d^3 v \frac{d^2\sigma_T(\mathbf{v})}{dQd\Omega_q} |\mathbf{v}| f(\mathbf{v},t).
\end{multline}
The general geometry of scattering is illustrated in Fig. \ref{fig:dd_angle}, which shows a lab-frame view of the interaction.  A WIMP with momentum $\mathbf{p} = m_\chi \mathbf{v}$ interacts with a nucleus at rest, inducing a nuclear recoil $\mathbf{q}$ with an angle $\theta$ with respect to the incoming WIMP direction.  Thus, $\Omega_q \equiv \hat{\mathbf{q}} = (\theta, \phi)$, where the interactions are azimuthally symmetric around the WIMP incoming direction.  The specific relationship between $\mathbf{q}$ and the incoming WIMP momentum depends on whether the WIMP-nuclear recoil is elastic or inelastic (Sec. \ref{sec:theory:inelastic}).

Finally, the motion of the Earth about the Galaxy induces a signature in the WIMP recoil spectrum.  Annual modulation in the WIMP recoil spectrum occurs because the Earth's motion with respect to an inertial Galactocentric frame changes throughout the year.  Because of the Sun's motion with respect to the Galactocentric frame (and hence, with respect to the bulk of the WIMP population), the Sun's relative velocity induces a preferred direction for WIMP arrivals, as well as a characteristic energy scale for the WIMP-nuclear interactions.  Because of the Earth's motion around the Sun, during some times of the year the Earth is moving into the ``WIMP wind'' (increasing the relative kinetic energy of WIMPs with respect to experiments), or moving with it (reducing the relative WIMP kinetic energy) \cite{Drukier:1986tm,Griest:1987vc}.  In addition, gravitational focusing by the Sun means that the density of low-speed WIMPs is higher when we sit ``behind'' the Sun relative to the WIMP wind rather than in front of it \cite{Lee:2013wza}.  Both of these effects cause the recoil spectrum to change on an annual basis.  Since Lee et al. \cite{Lee:2013xxa} showed that to detect annual modulation, one needs an exposure $\sim 10^3$ times larger than that required to discover WIMPs (except under special conditions), we do not include annual modulation in the analysis section of this paper.  However, it will be an exciting probe of WIMP physics in the era in which thousands of WIMP events are detected in experiments.  We will discuss it briefly in the context of current experimental results in Sections \ref{sec:experiment} and \ref{sec:gvmin}.  We point the interested reader to Freese et al. \cite{Freese:2012xd} for a review of annual modulation.

In the next few subsections, we parse Eq. (\ref{eq:energyspectrum}) in terms of the WIMP cross section (Sec. \ref{sec:theory:crosssections}), the type of scattering (Sec. \ref{sec:theory:inelastic}), and the astrophysical dark matter distribution (Sec. \ref{sec:theory:astrophysics}).  Our goal is to highlight standard assumptions about WIMP-nuclear scattering within the context of each of these topics, and present a broad range of considerations for each of these pieces of WIMP scattering physics.  We will also highlight which pieces of physics we include in our mock data analysis in Sec. \ref{sec:simulation}-\ref{sec:astrophysics}.

\subsection{Cross sections}\label{sec:theory:crosssections}
In this section, we describe the cross section in phenomenological terms, considering coupling to nucleons instead of coupling to quarks.  The latter is the more fundamental description for specific particle models of WIMPs, but introduces complications in relating observations to fundamental theories because of hadronic uncertainties.  \citeauthor{Ellis:2008hf} explore the effects of the uncertainties in hadronic properties (quark masses and hadronic matrix elements), and find that measurement uncertainties in the $\pi$-nucleon $\sigma$ term can lead to order-of-magnitude uncertainties in WIMP-nucleon cross sections for a fixed point in supersymmetric parameter space \cite{Ellis:2008hf}.  See also Refs. \cite{Cheng:1988cz,Cheng:1988im,Iachello:1990ut,Kamionkowski:1994rm,Bottino:2001dj} for discussion of hadronic uncertainties in WIMP-nucleon cross sections.  By considering WIMP-nucleon cross sections instead of WIMP-quark couplings in this discussion, we can evade the biggest hadronic uncertainties at the moment.

However, we must consider the momentum-dependent response of nuclei (collections of nucleons) to their interactions with WIMPs, typically parametrized in terms of a ``form factor'' $F^2(Q)$.  This calculation is still plagued by hadronic and nuclear physics uncertainties, even for the standard minimal supersymmetric standard model (MSSM)-type WIMP interactions that dominate the direct-detection world.  There are a number of different ways to estimate the nuclear response functions---see Refs. \cite{engel1991,Engel:1992bf,ressell1993,dimitrov1995,HjorthJensen:1995ap,Ressell:1997kx,Toivanen:2009zza,Menendez:2012tm} for several calculations that are commonly used in the direct-detection literature.  (See Refs. \cite{Bednyakov:2004xq,Bednyakov:2006ux} for a compilation of calculations.  Note that these reviews precede new chiral effective field theory calculations and consideration of other WIMP-nuclear operators---see Refs. \cite{Menendez:2012tm,Klos:2013rwa,Fitzpatrick:2012ix,Fitzpatrick:2012ib,Anand:2013yka}.) For an example of how the uncertainty in the nuclear response functions affects limits on the spin-dependent coupling of WIMPs and nucleons in the XENON100 experiment, see Ref. \cite{Aprile:2013doa}.  The uncertainties are typically worse for spin-dependent than spin-independent interactions \cite{Co':2012ht}.

In the following subsections, we present our choices for the nuclear response functions in our simulations of experiments. In the analysis in the following sections we assume that they are known exactly.  Note, however, that this assumption is far from experimental reality.  In the future, it would be useful to explore how hadronic uncertainties affect our ability to infer fundamental properties of WIMPs from direct-detection experiments.

\subsubsection{Standard, MSSM-inspired WIMP scenario}
In standard MSSM scenarios in which the lightest neutralino is WIMP dark matter, there are generally two relevant types of interactions between WIMPs and nuclei: spin-independent (SI) and spin-dependent (SD) scattering.  The differential cross section is
\begin{eqnarray}\label{eq:sisdsigma}
  \frac{d\sigma_{\mathrm{T}}}{dQ} = \frac{m_T}{2\mu_T^2 v^2}\left[ \sigma_{\mathrm{T}}^{\mathrm{SI}} F^2_{\mathrm{SI}} (Q) + \sigma_{\mathrm{T}}^{\mathrm{SD}}F^2_{\mathrm{SD}}(Q) \right],
\end{eqnarray}
where $\mu_T \equiv $\, \mwimp\,$m_T/($\mwimp$ + m_T)$ is the WIMP-nuclear reduced mass.  The $\sigma_{\mathrm{T}}$'s are the interaction cross sections in the limit of no momentum transfer.  

Let us consider the spin-independent terms first.  In supersymmetric models, scalar spin-independent scattering can arise from WIMP couplings either to gluons or quarks (see Figs. 20 and 21 in Ref. \cite{jungman1996} for Feynman diagrams), with the quark couplings mediated through Higgs or squark exchange.  In Eq. (\ref{eq:sisdsigma}),
\begin{eqnarray}\label{eq:sisigma}
  \sigma_{\mathrm{T}}^{\mathrm{SI}} = \frac{4}{\pi} \mu_T^2 \left[ Zf_p + (A-Z)f_n\right]^2,
\end{eqnarray}
where $f_p$ is the coupling between protons and WIMPs, $f_n$ the coupling between neutrons and WIMPs, and $Z$ and $A$ are the nuclear electric charge and atomic mass, respectively.  For the rest of this work, we will assume equal WIMP couplings to nucleons, $f_p = f_n$.  While this relation holds well in the MSSM in general, model-builders have recently introduced strongly isospin-violating dark matter to reconcile seemingly discrepant direct-detection results \cite{Feng:2011vu}.  However, Pato \cite{Pato:2011de} showed that in general $f_p/f_n$ is difficult to constrain from direct-detection data alone if this variable is a free parameter.

Once we restrict ourselves to $f_p = f_n$, we make the usual choice of writing $\sigma_{\mathrm{T}}^{\mathrm{SI}}$ in terms of \sigmapsi,
\begin{eqnarray}
  \sigmapsie &= & \frac{4}{\pi}\mu_p^2 f_p^2, \\
  \sigma_{\mathrm{T}}^{\mathrm{SI}} &=& \left( \frac{\mu_T}{\mu_p} \right)^2 A^2 \sigmapsie,
\end{eqnarray}
where $\mu_p$ is the reduced mass of the WIMP-proton system.

We use the common assumption that the mass distribution in the nucleus follows the charge distribution, so that the charge distribution can be used to calculate the form factor \cite{Lewin:1995rx}.  We use the Helm form factor, with parameters in the form factor fit according to Engel \cite{engel1991},
\begin{eqnarray}
  F^2(Q) = \left( \frac{3j_1(qR_1)}{qR_1}\right)^2 e^{-s^2q^2},  
\end{eqnarray}
where $j_1$ is a spherical Bessel function, $R_1 = (R^2 -5s^2)^{1/2}$ is an effective nuclear radius, $s \approx 1$~fm is a skin thickness, and $R \approx 1.2 A^{1/3}$~fm.  (Other form factors are also used variously in the literature, but they do not affect our conclusions \cite{duda2007}.)

\begin{figure}
  \includegraphics[width=0.48\textwidth]{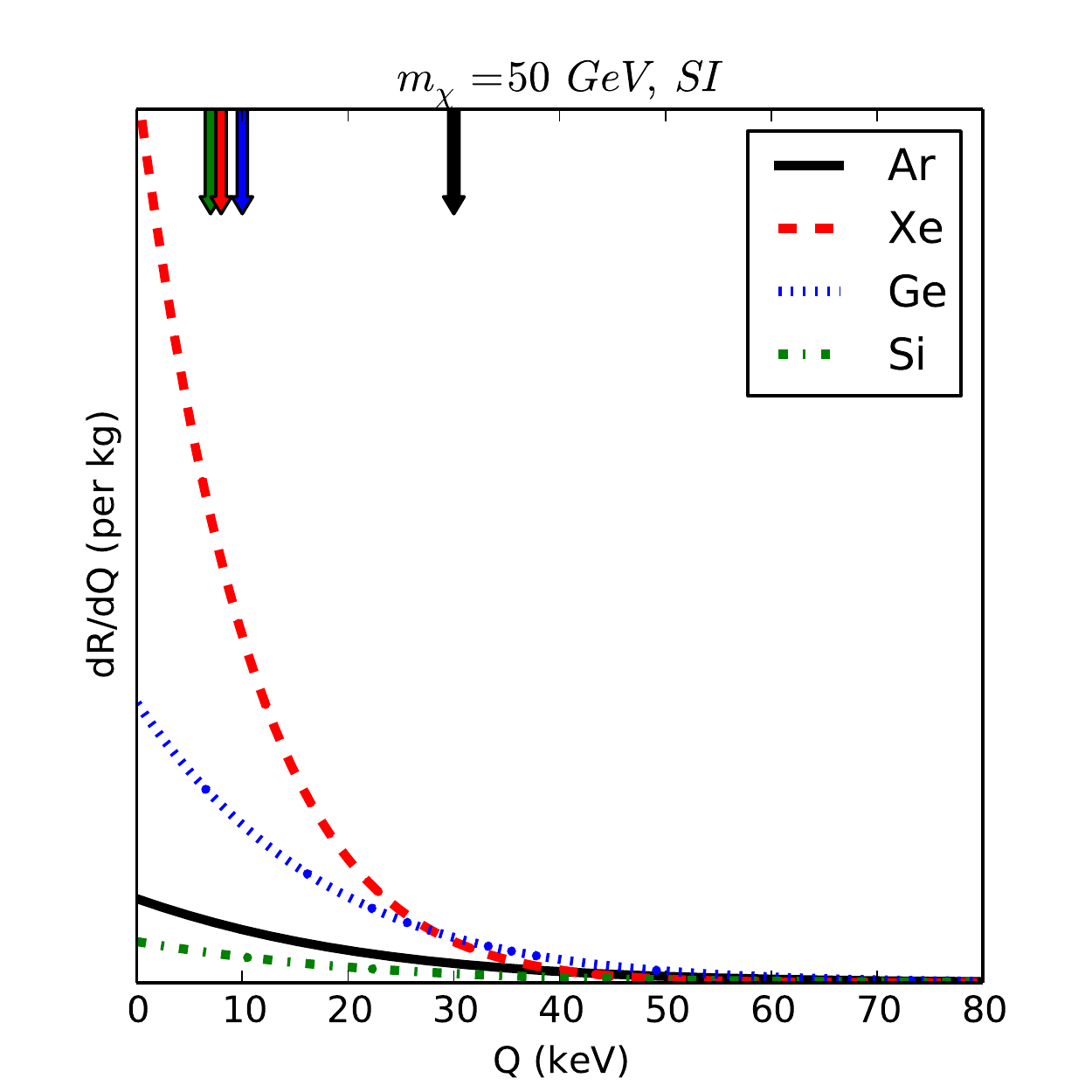}
  \caption{\label{fig:sispectrum}{\bf Recoil spectrum for several choices of target nuclei.}  These recoil spectra (per kilogram of target nuclei) show purely spin-independent scattering of a \mwimp$=50$ GeV WIMP on various target nuclei.  We use the fiducial SHM (Sec. \ref{sec:theory:astrophysics}) for the WIMP velocity distribution.  The strong $A-$dependence is apparent near zero momentum transfer.  The arrows show the energy thresholds for each target material assumed for the mock experiments used in this work.}
\end{figure}

We show examples of the recoil spectrum per kilogram of target material in Fig. \ref{fig:sispectrum} using the standard halo model (SHM; Sec. \ref{sec:theory:astrophysics}) for the WIMP velocities.  For low energy nuclear recoils, heavy target nuclei (e.g., Xenon) have the highest event rate at fixed \mwimp.  However, for larger energy nuclear recoils, lighter nuclei have larger event rates at fixed \mwimp.  The former effect is a result of the strong $A$-dependence of the spin-independent elastic scattering cross section.  The latter effect is dominated by a combination of the rapidly declining speed distribution function near the Galactic escape velocity and form-factor suppression.

For spin-dependent scattering,
\begin{eqnarray}
  \sigma_{\mathrm{T}}^{\mathrm{SD}} = \frac{32}{\pi} \mu_T^2 G_F^2 \frac{J+1}{J} \left[ a_p \langle S^T_p \rangle + a_n \langle S^T_n \rangle \right]^2.
\end{eqnarray}
Here, $G_F$ is the Fermi coupling constant, and $J$ is the nuclear angular momentum.  The $\langle S^T_i \rangle$ are the expectation values for the spin content of the nucleons.  There are minor differences in the set of $\langle S^T_i \rangle$ calculations \cite{Bednyakov:2004xq}.  Generally, $\langle S^T_n \rangle$ is two to three orders of magnitude higher than $\langle S^T_p \rangle$ for odd-neutron isotopes, and vice versa.  The $a_i$ are nuclear matrix elements (summed over the light quark contributions).  We follow Pato \cite{Pato:2011de} and parametrize $\sigma_{\mathrm{T}}^{\mathrm{SD}}$ by \sigmapsd~ and $a_n/a_p$.  This leads to 
\begin{eqnarray}
  \sigma_{\mathrm{T}}^{\mathrm{SD}} = \frac{4}{3}\left( \frac{\mu_T}{\mu_p}\right)^2 \frac{J+1}{J} \sigmapsde \left[ \langle S^T_p \rangle + \frac{a_n}{a_p} \langle S^T_n \rangle \right]^2.
\end{eqnarray}
Spin-dependent scattering only involves coupling to quarks, at least at tree level.

The spin-dependent nuclear form factor is given by
\begin{eqnarray}
F_{\mathrm{SD}}^2 (Q) &=& \frac{S(q)}{S(0)} \,, \\
S(q) &=& a_0^2 S_{00}(q) + a_0a_1S_{01}(q) + a_1^2S_{11}(q),
\end{eqnarray}
where $a_0 = a_p + a_n$ and $a_1 = a_p - a_n$.  These form factors emerge out of nuclear models for the distributions of nucleon spins within the nucleus in response to the interaction.  The $S_{ij}$ form factors can vary significantly among calculations.  For this work, we follow Pato's \cite{Pato:2011de} choices of form factor and $\langle S_i^T \rangle$.  For Silicon, which Pato does not consider, we use the zero-momentum spin expectation values and the form factors of Divari et al. \cite{Divari:2000dc}.

We now consider the angular dependence of the WIMP-nuclear cross section, which enters the directionally-dependent event rate in Eq. (\ref{eq:energythetaspectrum}).  The angular dependence is simply determined by the non-relativistic kinematics of the scattering, and is given by
\begin{equation}\label{eq:directionalcrosssection}
\frac{d^2\sigma_T(\mathbf{v})}{dQd\Omega_L} = \frac{d\sigma_T}{dQ} \frac{|\mathbf{v}|}{2\pi} \delta\left(\mathbf{v}\cdot\hat{\mathbf{q}} - v_{\mathrm{min}}\right).
\end{equation}
The delta function ensures that the incoming WIMP momentum $\mathbf{v}$ and the resulting nuclear-recoil direction $\hat{\mathbf{q}}$ are consistent with the kinematics encoded in the function $v_{\mathrm{min}}$. This function will depend on whether the scattering is elastic or inelastic (Sec. \ref{sec:theory:inelastic}).

\subsubsection{Non-relativistic effective operators}\label{sec:theory:effective}

In the midst of the discovery stage of dark-matter direct-detection experiments, there will be two complementary approaches: to constrain or identify MSSM WIMP-type cross sections (e.g., the spin-dependent and -independent interactions described in the last section); and to consider the most general set of interactions that dark-matter particles are allowed to have with nuclei.  The latter is motivated by our ignorance of dark-matter physics.  In the past few years, several authors have suggested using a non-relativistic effective-operator approach to categorize dark-matter physics in direct-detection experiments \cite{Fan:2010gt,Fitzpatrick:2012ix,Fitzpatrick:2012ib}.  This is part of a broader effort to decouple experimental constraints from specific microphysical dark-matter models  \cite{Feldstein:2009tr,Goodman:2010ku,Beltran:2010ww,Alves:2011wf}.  In this work, we focus on the effective-operator approach of Fan et al. \cite{Fan:2010gt}.  However, we note that Fitzpatrick et al. \cite{Fitzpatrick:2012ix,Fitzpatrick:2012ib} consider a much broader range of operators, including those with higher-order dependence on small parameters (WIMP-nuclear relative velocity $\mathbf{v}$ and momentum transfer $\mathbf{q}$).\footnote{They also calculate nuclear form factors for those interactions.  That group has also created a publicly available Mathematica package to estimate event rates for this broader range of interactions \cite{Anand:2013yka}.}

\begin{figure}
\begin{center}
\includegraphics[height=7cm,keepaspectratio=true]{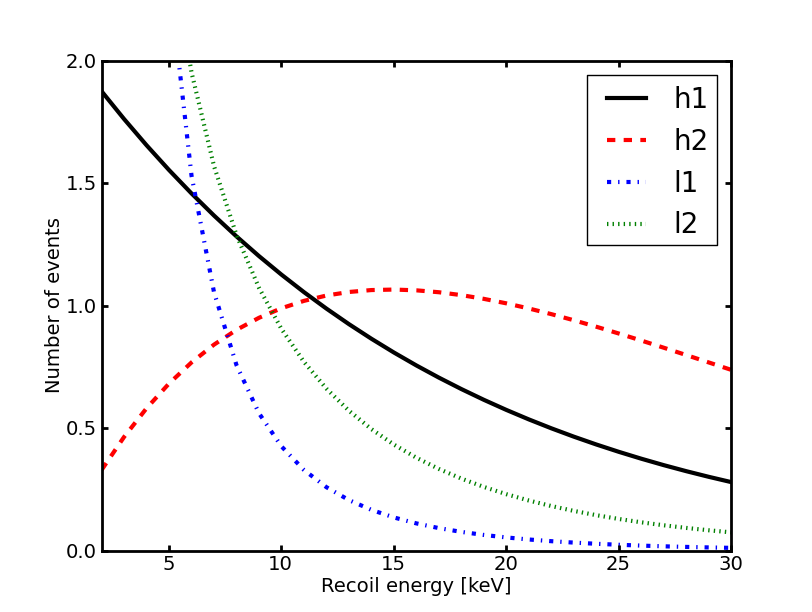}
\caption{{\bf Recoil spectra for spin-independent effective operators.}  The number of events per recoil energy, for the four different operators of Eq.~(\ref{eq:dRdQ_operators}), for the Xenon experiment (Sec.~\ref{sec:simulation}). The curves are normalized such that each operator produces about 100 nuclear-recoil events, with 1 ton-year of exposure.\label{fig:operators}}%
\end{center}
\end{figure}

Instead of categorizing a large number of proposed microphysical models for dark matter, we summarize the experiment-driven approach of parametrizing possible signals with Wilson coefficients of an effective theory. Since the speed of dark matter particles in the Earth's frame is highly non-relativistic, the energy scale for the scattering is low, and the interactions can be described with a non-relativistic effective potential.  To illustrate the diversity of effective-potential operators probed by direct detection, we focus on spin-independent interactions. Following Ref.~\cite{Fan:2010gt}, let us consider two scenarios---the one in which the mediator mass is much larger than the momentum transferred, and one in which it is much smaller. If we take only the operators suppressed by at most one power of the recoil energy and assume static potentials, the non-relativistic effective potential that is a minimal extension of the usual MSSM-inspired spin-independent scenario reads
\begin{equation}
V_{eff} = h_1\delta^3(\vec{r}) - h_2\vec{s}_\chi  \cdot \vec\nabla\delta^3(\vec{r}) + \ell_1\frac{1}{4\pi r} + \ell_2 \frac{\vec s_\chi \cdot \vec r}{4\pi r^3},
\label{eq:Veff}
\end{equation}
where $s_\chi$ is the spin of the dark matter particle, and the $\ell$'s and $h$'s are Wilson coefficients corresponding to the light and heavy mediator case, respectively. The first term represents the canonical case of contact interactions, and is directly related to the cross section of Eq.~(\ref{eq:sisigma}). The term containing $\vec{s}_\chi\cdot  \vec\nabla\delta^3(\vec{r})$ can arise from several scenarios, such as coupling of the dark electric dipole moment to a new gauge boson. The $1/r$ term, or the Coulomb potential, can arise through exchange of a new light boson with a mass smaller than the recoil energy. Finally, the term proportional to ${\vec s_\chi\cdot \vec r}/{r^3}$ can be due to dark-matter dipole coupling to the nucleus monopole. The differential spin-independent cross section corresponding to these four operators is \cite{Fan:2010gt}
\begin{eqnarray}\label{eq:dRdQ_operators}
\frac{d\sigma^{SI}}{dQ}& =& \frac{A^2F^2(Q)m_T}{2\pi v^2} \left( \left|h_1+\frac{\ell_1}{2m_TQ}\right|^2 \right.  \\ \nonumber 
&& \left. +\frac{1}{4} \left| h_2\sqrt{2m_TQ}+\frac{\ell_2}{\sqrt{2m_TQ}} \right|^2\right) \,.
\end{eqnarray}
See also the illustration of the shapes of the corresponding recoil energy spectra in Fig.~\ref{fig:operators}.

These four terms by no means exhaust the possible variety of effective-theory operators for dark matter-nucleus interactions at low energies, but they do represent some of the simplest and best-motivated extensions of the standard scenario and help illustrate key points about detectability of underlying physics, discussed in Section \ref{sec:WIMPphys}. Other possibilities, for example, include form-factor dark matter of Ref.~\cite{Feldstein:2009tr} with quadratic dependence on the recoil energy, and anapole moment and magnetic dipole moment couplings discussed in Ref.~\cite{McDermott:2011hx}.

Since Wilson coefficients link back to couplings in the underlying theories (i.e. they are determined by matching the operators in the effective theory to the operators in a UV complete theory at hand), their measurement  can probe different classes of theoretical models. In Section \ref{sec:WIMPphys}, we discuss the potential of the upcoming Generation 2 experiments to detect and discern between the effective operators using measurements of the recoil-energy spectra.

\subsection{Inelastic dark matter}\label{sec:theory:inelastic}
There are two primary ways in which dark matter and nuclei may experience inelastic interactions.  First, nuclei may transition to an excited state during interactions with dark-matter particles.  This type of inelastic scatter was first considered approximately twenty years ago, and has recently made a revival in the context of Xenon experiments \cite{ellis1988,belli1996,baudis2013b}.  Second, dark matter may either be a composite particle (like the proton) or exist as part of a multiplet of dark states in a hidden-sector theory \cite{TuckerSmith:2001hy,Feldstein:2009tr,Lisanti:2009am,Graham:2010ca}.  

\begin{figure}
  \includegraphics[width=0.48\textwidth]{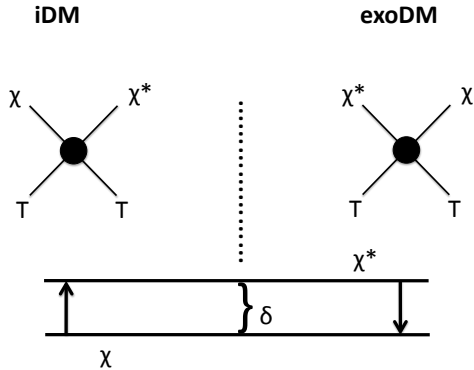}
  \caption{\label{fig:idm_cartoon}{\bf Inelastic dark matter.}  There are at least two dark-matter energy states, $\chi$ and $\chi^*$, separated by energy $\delta$.  iDM models have $\chi$ as the dark matter, which may transition to $\chi^*$ through interactions with nuclei.  exoDM models have $\chi^*$ as the dark matter, and may transition downward to $\chi$ by nuclear interactions.}
\end{figure}

In this work, we consider the case that dark matter belongs to a multiplet of dark states.  In particular, we consider two dark-matter states $\chi$ and $\chi^*$, where the former is the lowest energy state, and the latter is a higher energy state.  This is illustrated in Fig. \ref{fig:idm_cartoon}.  For ``inelastic dark matter'' (iDM) models, the dark matter is the lowest energy state $\chi$.  Interactions with nuclei cause a transition to the $\chi^*$ state, which is an energy $\delta$ higher than the ground state \cite{TuckerSmith:2001hy}. In elastic scattering the recoil energy depends smoothly on the center-of-mass frame scattering angle $\theta_{COM}$, 
\begin{eqnarray}\label{eq:Q_elastic}
  Q = \frac{\mu^2_T}{m_T} v^2 (1 - \cos\theta_{COM}), \hbox{ elastic}
\end{eqnarray}
and 
\begin{eqnarray}
  v_{\mathrm{min}} = \sqrt{m_TQ/2\mu^2_T},\hbox{ elastic}
\end{eqnarray}
However in iDM transitions are forbidden for small initial WIMP kinetic energies and instead,
\begin{eqnarray}\label{eq:vmin-inelastic}
  v_{\mathrm{min}} = \frac{ \left| m_T Q + \mu_T \delta \right| }{\mu \sqrt{2m_TQ}}, \hbox{ inelastic}.
\end{eqnarray}

\begin{figure*}
  \begin{center}
    \includegraphics[width=0.48\textwidth]{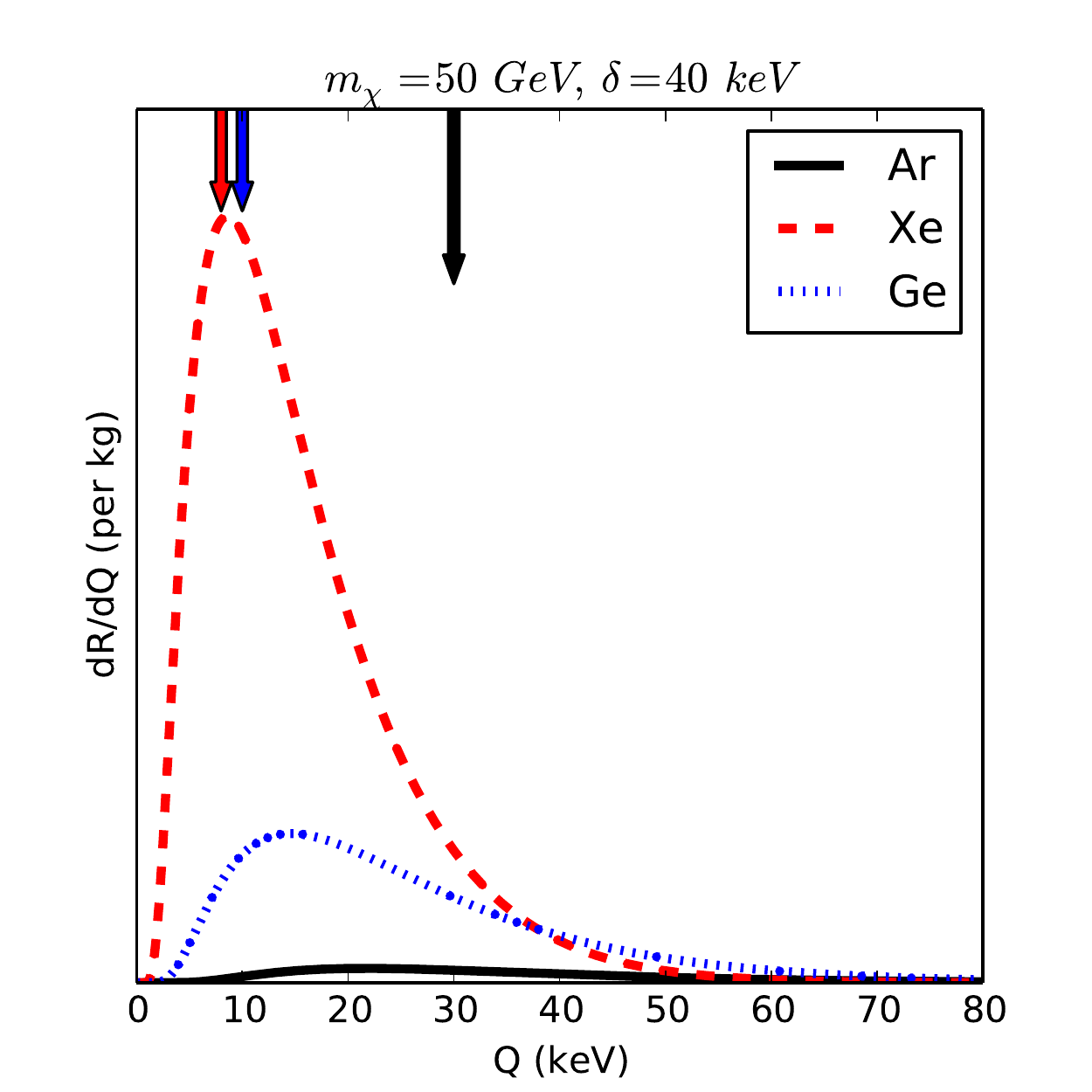} \includegraphics[width=0.48\textwidth]{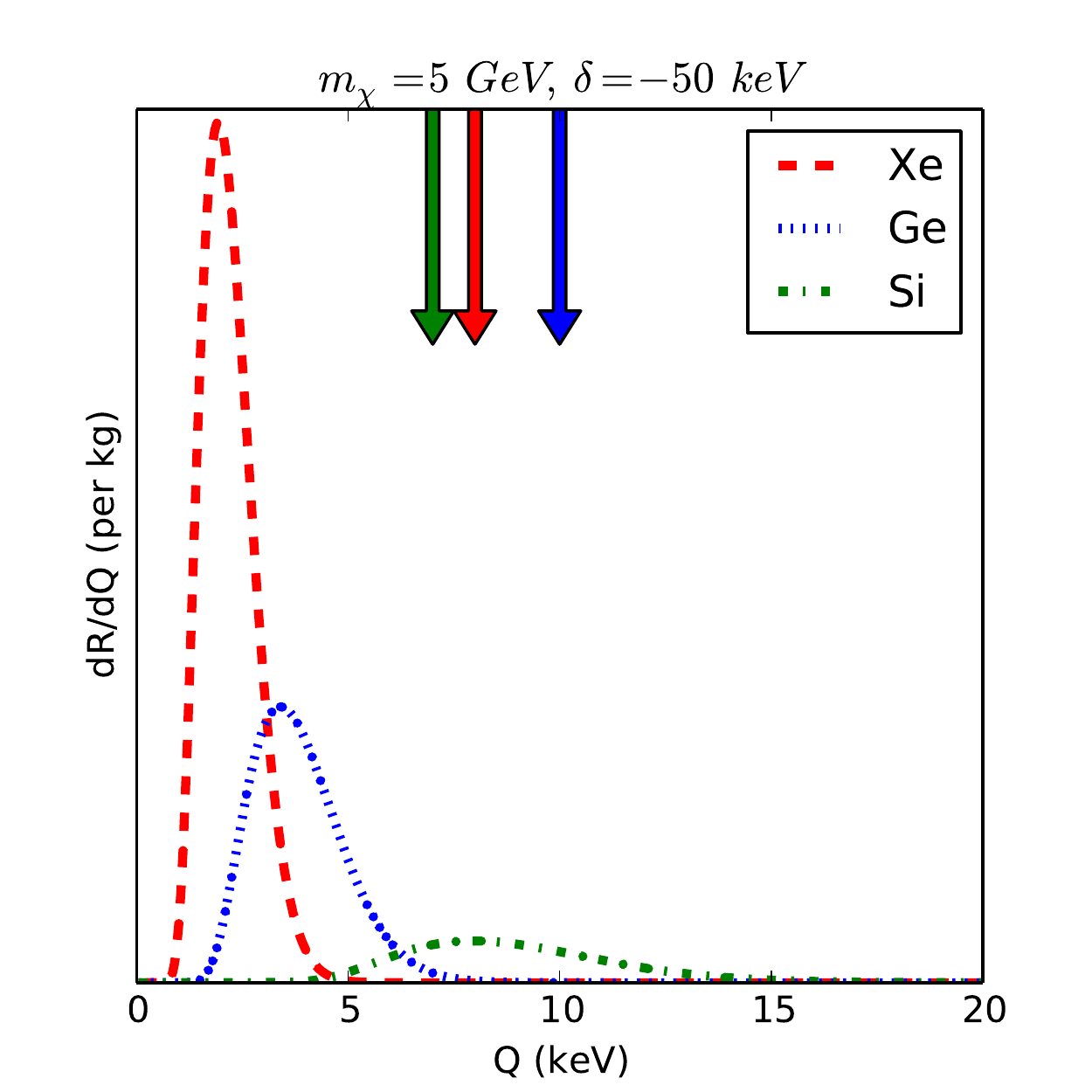}
    \caption{\label{fig:idmspectrum}{\bf Recoil spectra for inelastic and exothermic dark matter models.} The recoil rates are scaled per kilogram of target material.  We used the SHM (Sec. \ref{sec:theory:astrophysics}) for the WIMP velocity distribution.  To the left is an iDM model; to the right, an exoDM model.  Arrows mark the energy thresholds used for the mock experiments used in our analysis.}
  \end{center}
\end{figure*}

We can parse the physics of iDM using the recoil spectra shown in the left-hand side of Fig. \ref{fig:idmspectrum}, where we show the effects with several different target nuclei.  In this case, we assume purely spin-independent scattering.  The recoil spectrum vanishes for small $Q$, where the cut-off value of $Q$ increases as $A$ decreases.  This is what we expect, since WIMPs must have a minimum kinetic energy $E_\chi = \delta$ in order to kick $\chi$ into a higher energy state.  The overall number of events also decreases with decreasing $A$ because, for fixed $E_\chi$, we probe higher WIMP speeds.  Thus, a smaller fraction of WIMPs are allowed to scatter for nuclei with lower atomic number than higher atomic number.  This was in fact the motivation for iDM theories---the original goal was to reconcile DAMA data (assuming scattering off $^{127}$I) with CDMS null detections ($^{73}$Ge) \cite{TuckerSmith:2001hy,Chang:2008gd}.  While iDM cannot self-consistently reconcile DAMA with other experiments today \cite{Aprile:2011ts}, iDM models are still interesting from the perspective of hidden-sector theories \cite{ArkaniHamed:2008qn}.

The inelastic interaction can happen the other way around: dark matter can be produced in an excited state in the dark multiplet $\chi^*$ and be downscattered to the ground state $\chi$ through nuclear interactions.  This kind of model is called ``exothermic dark matter'', or exoDM \cite{Graham:2010ca}.  We choose to use the convention that $\delta > 0$ denotes iDM, and $\delta < 0$ denotes exoDM.  This model was initially proposed in order to reconcile DAMA (assuming low-mass WIMPs scattering off Na) with the XENON10 and CDMS experiments.  In this case, the low-energy part of the nuclear recoil spectrum is more sensitive to low-speed WIMPs than the elastic case, and scattering above experimental thresholds off low-$A$ targets is enhanced relative to high-$A$ targets.  This is shown in the right-hand panel of Fig. \ref{fig:idmspectrum}.  This model has been recently revived to reconcile the three events in the CDMS-Si data with null detections in other experiments \cite{Agnese:2013cvt,Agnese:2013rvf,Frandsen:2013cna}.  In Fig. \ref{fig:idmspectrum}, we see that for a model with \mwimp$=5$ GeV and $\delta = -50$ keV (the negative sign denotes downscattering), there can be a number of Silicon events above the experimental threshold, but essentially none in Xenon- and Germanium-based experiments.

\subsection{WIMP Astrophysics}\label{sec:theory:astrophysics}
The nuclear recoil spectrum depends on the local phase-space density of WIMPs.  Traditionally, the phase-space density is split into the configuration-space part (the WIMP number density $n_\chi = \rho_\chi$/\mwimp) and the velocity distribution $f(\mathbf{v},t)$ (normalized such that $\int d^3v f(\mathbf{v},t) = 1$).  We split our discussion along these lines.

\subsubsection{$\rho_\chi$}
The canonical assumption in WIMP searches is that the local WIMP mass density is $\rho_\chi = 0.3\hbox{ GeV cm}^{-3}$ ($=0.008 M_\odot \hbox{ pc}^{-3}$).   By fixing the value of the WIMP mass density, one can infer a WIMP-nucleon cross section from the event rate or recoil spectrum in experiments, since the normalization of the event rate (see Eq. (\ref{eq:energyspectrum})) depends on $\rho_{\chi} \sigma$.  Any uncertainty in $\rho_\chi$ bleeds directly into an uncertainty in the WIMP cross sections.  Note, however, that the ratio of cross sections (e.g., the ratio of the spin-independent to spin-dependent WIMP-proton cross sections, \sigmapsi/\sigmapsd) is completely independent of the local WIMP density.

Where does this number for the WIMP density come from?  There are traditionally two methods for estimating the local WIMP density.  The first method uses the vertical motion of stars through the Galactic plane to estimate the force induced on the stars from material (including dark matter) in the disk of the Milky Way.  The second method uses an ensemble of data sets throughout the virial radius of the Milky Way to constrain a parametrized model of the gravitational potential of the Milky Way.  The second method relies on a global, simplified equilibrium model of the Galaxy, whereas the first method typically depends primarily on local planar symmetry to interpret the stellar kinematic data.  In principle, the first method relies on fewer more-or-less motivated ansatzes for the Milky Way than the second method does, although a few simplifying assumptions (e.g., no $R-z$ cross term in the velocity dispersion tensor) sometimes crop up.  Both methods hinge on the assumption of dynamical equilibrium.

Estimates of the local density using the first method, using vertical motions of disk stars, have varied somewhat over the past three decades, but the mean value has always been near the canonical value of $\rho_\chi = 0.3\hbox{ GeV cm}^{-3}$.  The first estimate of the local dark-matter density was made by Jan Oort in 1932, in which he found that half the local matter density should consist of invisible material \cite{oort1932}.  The topic was revived in 1984 by Bahcall, who came to similar conclusions using a sample of $F$ stars, and later a sample of $K$ giants \cite{bahcall1984a,bahcall1984c,bahcall1992}.  Bahcall and collaborators found that near the Sun the density is $\rho_\chi = 0.1 M_\odot \hbox{ pc}^{-3}$ ($\sim 4$ GeV cm$^{-3}$), and that the dark matter should be disky in structure.  Around this time, Bienayme et al. \cite{bienayme1987}  and Kuijken \& Gilmore \cite{kuijken1989a,kuijken1989b,kuijken1989c,kuijken1991,kuijken1991b} disputed that result, finding no evidence for a thin disk of dark matter but allowing a local dark-matter halo density $\rho_\chi \sim 0.008 M_\odot \hbox{ pc}^{-3}$ (the canonical value; see also Ref. \cite{gould1990}).  Many other authors have found that there is no need for a thin (i.e., similar in scale height to the stellar thin disk) disk of dark matter in the Milky Way, and that even a local halo density of 0 is allowed \cite{flynn1994,creze1998,holmberg2000,holmberg2004,siebert2003}.  Bienaym{\'e} et al. \cite{bienayme2006} finds $\rho_\chi \lesssim 0.014 M_\odot \hbox{ pc}^{-3}$ ($0.53 \hbox{ GeV cm}^{-3}$) and Garbari et al. \cite{Garbari:2011dh,Garbari:2012ff} find $\rho_\chi = 0.003^{+0.0009}_{-0.007}M_\odot \hbox{ pc}^{-3}$ ($0.11^{+0.34}_{-0.27}\hbox{ GeV cm}^{-3}$) and $\rho_\chi = 0.025^{+0.014}_{-0.013} M_\odot\hbox{ pc}^{-3}$ ($0.95^{+0.57}_{-0.50}\hbox{ GeV cm}^{-3}$) depending on the stellar sample.  Salucci et al. \cite{salucci2010} find $\rho_\chi = 0.43^{+0.11}_{-0.10}$ GeV cm$^{-3}$.  Moni Bidin et al. \cite{Bidin:2012vt} initially found no dark matter locally.  Once Bovy \& Tremaine \cite{Bovy:2012tw} corrected the misinterpretation of asymmetric drift in Ref. \cite{Bidin:2012vt} they found that the data were consistent with $\rho_\chi = 0.3\pm 0.1$ GeV cm$^{-3}$.  Using a set of mono-abundance populations defined in the SEGUE survey data set, Bovy \& Rix \cite{Bovy:2013raa} also find the same central value and uncertainty in $\rho_\chi$ as Bovy \& Tremaine.  Zhang et al. \cite{Zhang:2012rsb} find, using the kinematics of K dwarfs,  $\rho_\chi = 0.25\pm 0.09$ GeV cm$^{-3}$.

Using the second method, an early estimate of the local density was $\rho_\chi = 0.2-0.8\hbox{ GeV cm}^{-3} $ depending on the form of the density profile of the Milky Way halo \cite{bergstrom1998}.  Using Bayesian inference, and including a wide range of dynamical data sets, Catena \& Ullio \cite{catena2010} find $\rho_\chi = 0.389\pm0.025\hbox{ GeV cm}^{-3}$ for a Navarro-Frenk-White (NFW) dark-matter halo density profile \cite{navarro1996,navarro1997}.  Using similar methods but a different data set, McMillan \cite{mcmillan2011} finds $\rho_\chi = 0.40 \pm 0.04$ GeV cm$^{-3}$ for an NFW halo.  Using frequentist statistics, Weber \& de Boer \cite{weber2010} find $\rho_\chi = (0.2-0.4)$ GeV cm$^{-3}$, and Nesti \& Salucci \cite{nesti2013} find $\rho_\chi = (0.4-0.55)$ GeV cm$^{-3}$ depending on the functional form of the halo profile. Iocco et al. include microlensing data in their analysis; this leads to a density estimate of $\rho_\chi = (0.2 - 0.56)$ GeV cm$^{-3}$ for a generalized NFW halo \cite{iocco2011}.

However, there are caveats to using these estimates for $\rho_\chi$ to infer cross sections from direct-detection data.  First, and perhaps most importantly, use of this value implies that we are assuming that all the dark matter in the Galaxy consists of a single species of WIMP.  This is a strong assumption.  Second, limits using the second method are often inferred using a fixed functional form for the density profile, one which is found in simulations without baryons. However the dark matter density profile of the real Milky Way in which baryons, and in particular the stellar disk, play an important role, may be quite different.  Third, these estimates are locally averaged or globally determined densities.  Several calculations show that there should be only small fluctuations, of order 15\%, in the density of the smooth dark-matter component  at fixed Galactocentric radius, which is good news \cite{Kamionkowski:2008vw,vogelsberger2009}.  On the other hand, the presence of a disk can compress dark matter in the disk plane, an effect not currently taken into account in the second method \cite{Pato:2010yq}.  Moreover, for direct detection, what matters is the dark-matter density on the $\sim$ milliparsec scales the Sun sweeps out through the Milky Way during the course of an experimental run.  While it is unlikely that we will encounter a distinct subhalo in the near future \cite{diemand2005,Schneider:2010jr}, potentially interesting microstructure may appear in the form of streams.  It is not clear what the filling factor or typical number density of streams is at any given location in the halo.  It is probably unlikely that the streams will have an enormous impact on the highly local WIMP mass density, but they could have more of an effect on the velocity distribution \cite{Helmi:1999uj,Vogelsberger:2007ny,vogelsberger2009}.   

Finally, this discussion rests on the assumption of the collisionless nature of dark matter.  If dark matter has strong interactions in a hidden sector, it may be possible to form a low-mass, high-density dark-matter disk thin enough to evade both vertical stellar velocity limits or global potential modeling \cite{Fan:2013tia,Fan:2013yva}.  Note, however, that the stability of such thin disks has not been ascertained.

\subsubsection{$f(\mathbf{v},t)$}
The velocity distribution most often used in the literature to predict or interpret direct-detection signals is the {\bf Standard Halo Model} (SHM).  This distribution emerges from isothermal (constant velocity-dispersion) models of dark-matter halos for which the circular velocity curve is flat, and the density profile goes as $r^{-2}$ (see Appendix A of Ref. \cite{peter2008}).  The SHM velocity distribution is
\begin{eqnarray}\label{eq:SHM}
  f_g(\mathbf{v}_g) = \begin{cases} \frac{N_{SHM}}{(2\pi \sigma_v^2)^{3/2}}e^{-v_g^2/2\sigma_v^2}, & \hbox{ if } v_g < v_{\mathrm{esc}}\\
    0 & \hbox{ if } v_g > v_{\mathrm{esc}} \end{cases}
\end{eqnarray}
where $\mathbf{v}_g$ is the WIMP velocity in the inertial Galactocentric frame, and $\sigma_v$ is the one-dimensional WIMP velocity dispersion.  This means that the three-dimensional rms speed of WIMPs is \vrms$=\sqrt{3}\sigma_v$.  
We assume that WIMPs with speeds above the local escape speed from the Milky Way halo, $v_{\mathrm{esc}}$, contribute only negligibly to the signal.  This is likely a reasonable assumption under most circumstances \cite{Freese:2001hk,Behroozi:2012cd}.  $N_{SHM}$ is a factor we include in order to normalize the integrated velocity distribution to unity.

This velocity distribution needs to be translated to the lab frame for recoil calculations.  Typically, the effect of the gravitational potential of the Sun and Earth on the velocity distribution (i.e., an accurate application of Liouville theory) is ignored.  This approximation is applied to the SHM because only a small population of low-speed WIMPs is affected significantly, and most of those WIMPs scatter well below the thresholds of current and near-future experiments.  Thus, typically only the Galilean translation to the Earth frame is included, such that
\begin{eqnarray}
  f(\mathbf{v},t) = f_g(\mathbf{v} + \mathbf{v}_E(t), t),
\end{eqnarray}
where $\mathbf{v}_E(t)$ is the velocity of the Earth with respect to the Galactocentric rest frame.  $\mathbf{v}_E$ includes contributions from the speed of the Local Standard of Rest $v_{LSR}$, the peculiar velocity of the Sun with respect to $v_{LSR}$, and the Earth's velocity around the Sun.  The standard value of $v_{LSR}$ is $220 \hbox{ km s}^{-1}$, with about a 10\% uncertainty \cite{kerr1986}.  Recent measurements skew about 10\% higher \cite{reid2009,bovy2009,mcmillan2010} (but see \cite{bovy2012}).

The typical value of $\sigma_v$ used in calculations is $v_{LSR}/\sqrt{2}$.  Note that this is \emph{not} an empirically determined quantity, but is inspired by the assumptions of the SHM (see again Appendix A of Ref. \cite{peter2008}).  
The escape velocity has recently been estimated using the RAVE survey to be 
$537^{+59}_{-43} \hbox{ km s}^{-1}$ \cite{smith2007,Piffl:2013mla}.

There is no reason that the SHM should be a good description of the local velocity distribution in detail.  High-resolution dark-matter-only simulations indicate that there is diversity in the local velocity distribution, both between halos and between patches at fixed galactocentric radius within the same halo, although the former is of bigger concern than the latter
\cite{vogelsberger2009,Kuhlen:2009vh}.  This is particularly significant in cases in which experiments are only sensitive to the tail of the distribution (e.g., if the WIMP mass is low).  Lisanti et al.  \cite{lisanti2010} and Mao et al. \cite{Mao:2013nda} find functional forms for the velocity distribution that are based on global fits to simulations (see Eq. (\ref{eq:lisanti})).  The high-velocity tail is imprinted by the halo's accretion history, for which there is significant halo-to-halo scatter \cite{vogelsberger2009,Lisanti:2011as,Kuhlen:2012fz}.  These halos also have some velocity anisotropy, which can also affect the recoil energy spectrum \cite{Green:2002hk,vogelsberger2009}.  In general, all of these effects are strongest for signals that are predominantly sensitive to the high-speed tail of the WIMP speed distribution.  

More serious is that most simulations do not actually have galaxies in them---only dark-matter halos.  The presence of a baryonic disk can change the halo WIMP velocity distribution locally from cold-dark-matter-only expectations \cite{Kuhlen:2013tra}, and can also allow another dark-matter macro structure to form even without strong dark-matter self-interactions: a dark disk.  Dark disks form out of the debris of subhalos that are dragged into and disrupted near the baryonic disk plane \cite{lake1989,Read:2008fh}.  The physical shape, mass, and velocity distribution of dark disks depend sensitively on the accretion history of the main halo, but they are generally fluffy compared even to the stellar thick disk \cite{Read:2008fh,read2009,purcell2009,ling2010}.  Moderate-resolution simulations suggest a wide possible range of dark-disk morphologies and properties.  Early simulations suggested that the local dark-disk-to-halo density ratio could be up to unity.  Recently, an analysis of the high-resolution Eris simulation of a spiral galaxy, which is $\sim 30\%$ less massive than the Milky Way and is in a halo up to a factor of two less massive than ours, indicated a fairly modest dark disk.  Such a weak dark disk would be unlikely to affect the nuclear recoil spectrum significantly \cite{Kuhlen:2013tra}.  However, a statistical ensemble of assembly histories of Milky Way analogs is required to make a statistical statement about the possible properties of a Milky Way dark disk.

Finally, streams of material from disrupting substructure may make the velocity distribution look choppy.  If there are many streams passing through the solar neighborhood, each containing only a small fraction of the local mass density, the distribution would be hard to distinguish from a smooth one.  Only if a handful of streams contribute significantly to the local density will the lack of smoothness cause strong deviations in recoil spectra from smooth halo-based models.  Several authors have discussed the possibility that the Sagittarius dwarf galaxy, which is reaching the end point of disruption, may have left some of its dark matter in the solar neighborhood \cite{Freese:2003na,Freese:2003tt}.  The Sagittarius stream is expected to leave a noticable signature in the nuclear recoil spectrum only if the WIMP mass is low \cite{Purcell:2012sh}.

Ultimately, the best estimate of the local velocity distribution will come from direct-detection experiments themselves.  This is the subject of Sec. \ref{sec:astrophysics}.

\section{WIMP direct-detection experiments}\label{sec:experiment}
WIMPs with mass in the ${\rm GeV}$ to ${\rm TeV}$ range and speed $ \sim100 \, {\rm km \, s}^{-1}$ produce nuclear recoils with energy of order tens of keV. The expected event rates are also small, less than one event per kilogram target mass per year. Therefore, to detect WIMPs a detector with low threshold energy, a large target mass and low backgrounds is required. The kinetic energy of the nuclear recoil can manifest itself as ionization, scintillation or phonons (leading to a rise in temperature). Electron and nuclear recoils deposit their energy between these channels differently. Therefore detectors which measure two channels can reject electron recoils efficiently.

From a theorist's perspective, ongoing and near-future experiments can be classified by function instead of technology.  The classification is as follows:
\begin{enumerate}
        \item Experiments with little discriminating power on an event-by-event basis.  Most of these experiments depend on time dependence in the WIMP interaction rate and recoil spectrum, such as annual modulation.
        \item Experiments with discriminating power but no energy sensitivity above threshold.
        \item Experiments with discriminating power and energy sensitivity.
        \item Experiments with directional resolution for the nuclear recoil.
\end{enumerate}

In this work, we primarily consider the latter three classes of experiment, but we say a few words about the first type of experiment for completeness.  Here we review the status and characteristics of direct-detection experiments, grouped by our theorist's classification.

\subsection{No event-by-event discrimination}
There exist two types of experiments in this class: scintillating crystal targets, and germanium-diode experiments.

With scintillating crystals, such as ${\rm Na I}$ and ${\rm Cs I}$, large detector arrays can be built. 
Pulse shape discrimination can be used to allow discrimination between electron and nuclear recoils on a statistical, rather than event by event, basis. Current experiments, DAMA/LIBRA~\cite{Bernabei:2013xsa} and KIMS~\cite{Kim:2012rza}, have $\sim 100 \, {\rm kg}$ target masses. The main focus of near-future scintillating crystal experiments, such as ANAIS~\cite{Amare:2012ex}, SABRE~\cite{shields2013} and DM-Ice \cite{Cherwinka:2011ij}, is the confirmation or refutation of the DAMA/LIBRA annual modulation signal. 

There are currently three experiments using p-type point-contact germanium detectors: TEXONO, CoGeNT and MALBEK \cite{Aalseth:2012if,Giovanetti:2012ek,Li:2013fla}.  The latter two experiments grew out of the \textsc{Majorana Demonstrator} $0\nu\beta\beta$ experiment.  They can distinguish events that occur in the bulk from those on the surface (a background) using pulse shapes.  Currently, CoGeNT sees an excess of events, as well as tentative evidence for annual modulation \cite{collar2013b}.  MALBEK results are consistent with backgrounds \cite{giovanetti2013}, as are those from TEXONO \cite{Li:2013fla}.

In addition, most experiments in the third category (discriminating power and energy resolution) can be operated in a mode in which one cannot distinguish electronic and nuclear recoils on an event-by-event basis.  The motivation for operating experiments in this mode is that one may achieve far lower energy thresholds \cite{Ahmed:2010wy,Angle:2011th,Agnese:2013lua}.

\subsection{Discriminating power but no energy resolution}

In superheated droplet detectors the energy deposited by a nuclear recoil can trigger a transition to the gas phase, leading to the formation of bubbles which can be detected acoustically and optically. The operating conditions (temperature and pressure) can be set so that only nuclear recoils, and not electron recoils, lead to bubble formation.
Fluorine is a common target, and is particularly sensitive to spin-dependent interactions since most of its spin is carried by the unpaired proton. Current detectors, COUPP~\cite{Behnke:2012ys},  PICASSO~\cite{Piro:2010qd} and SIMPLE~\cite{Felizardo:2010mi}, have $\sim 0.1-1 \, {\rm kg}$ target masses, with scale up to  
$\sim 10-100 \, {\rm kg}$ underway \cite{Girard:2012ep,neilson2013,lippincott2013}.  We consider a Generation-2 1-ton-year exposure of a COUPP-500-like experiment in our mock data sets in Sec.~\ref{sec:simulation}

\subsection{Discriminating power and energy resolution}
The strong temperature dependence of the heat-capacity of dielectric crystals at low temperatures means that a relativity small energy deposit leads to a measurable rise in temperature and therefore cryogenic detectors operated at sub-Kelvin temperatures can measure the total recoil energy. Combined with a measurement of, depending on the target material, either ionisation or scintillation this allows
discrimination between nuclear and electron recoils on an event by event basis. Cryogenic detectors also have low energy thresholds and excellent energy resolution.

CDMS~\cite{Ahmed:2009zw}  and Edelweiss~\cite{Armengaud:2011cy} both use Germanium (and in the case of CDMS, Silicon as well~\cite{Agnese:2013rvf}) and measure phonons and ionisation, while CRESST~\cite{Angloher:2011uu} uses Calcium Tungstate crystals and measures the phonons and scintillation. The next stage (toward Generation 2) is Edelweiss III~\cite{V.Y.KozlovfortheEDELWEISS:2013mba}  and SuperCDMS~\cite{Baudis:2012ig,Brink:2012zza} with $\sim100 \, {\rm kg}$ mass targets.  In the longer term (post 2020) GEODM is a proposed ton-scale experiment involving collaboration between CDMS and EURECA (itself a collaboration between CRESST and EDELWEISS~\cite{Kraus:2007zz})~\cite{Baudis:2012ig,Brink:2012zza}.  We consider 200 kg-year exposures of both Silicon- and Germanium-based Generation-2 experiments in our mock data sets in Secs.~\ref{sec:simulation}-\ref{sec:directional}.

There is also a new CCD-based experiment (DAMIC; \cite{barreto2012}), which we will not consider in our mock data analysis.

Large self-shielding detectors can be built using liquid noble elements. With liquid Argon (LAr) and Xenon (LXe) the simultaneous detection of scintillation and ionisation (via proportional scintillation) allows event by event electron recoil discrimination.
Current Xenon detectors have $\sim100 \, {\rm kg}$ mass targets. Xenon100 has reported results~\cite{Aprile:2012nq} while results from LUX~\cite{Akerib} and XMASS~\cite{xmass} are imminent. The Panda-X experiment in China is developing rapidly \cite{gong2013}.  The next stage is experiments with ton-scale targets,
LZ (a collaboration between LUX and Zeplin) and Xenon1T~\cite{Baudis:2012ig}. With liquid Argon $\sim100 - 1000 \, {\rm kg}$ detectors ArDM~\cite{ArDM}, DarkSide~\cite{DarkSide} and DEAP/CLEAN~\cite{Baudis:2012ig} are under development.
In the longer term, MAX~\cite{Baudis:2012ig} and DARWIN~\cite{Baudis:2012bc} are proposals for ten ton scale liquid noble detectors.  We consider 2-ton-year exposures of Argon- and Xenon-based Generation 2 experiments in our mock data sets.

\subsection{Directional detection}
Constructing a detector capable of measuring the directions of sub-100 keV nuclear recoils is a difficult challenge. Only in a gaseous detector are the recoil tracks long enough, $>{\cal O} ({\rm mm})$, for their directions to be measurable.  Low-pressure gas time projection chambers (TPCs) offer the best prospects for directional detection, and a number of prototype detectors are under development: DMTPC~\cite{Monroe:2011er},  DRIFT~\cite{Daw:2011wq}, MIMAC~\cite{Riffard:2013psa} and NEWAGE~\cite{Miuchi:2010hn} (see Refs.~\cite{Sciolla:2009ps,Ahlen:2009ev} for overviews).  Target gases under study include 
${\rm CF}_4$, ${\rm C S}_{2}$ and ${}^3{\rm He}$.  The use of gases means that such detectors will require scaling-up strategies different from those of non-directional detectors~\cite{Ahlen:2009ev}.  We consider the capabilities of a MIMAC-like m$^3$ experiment in Sec. \ref{sec:directional}.

There is also another type of proposed experiment using DNA to measure WIMP-induced recoil directions \cite{drukier2012}.

\vspace{1.0cm}

\begin{figure*}
        \includegraphics[trim=150 20 150 50,clip,width=0.48\textwidth]{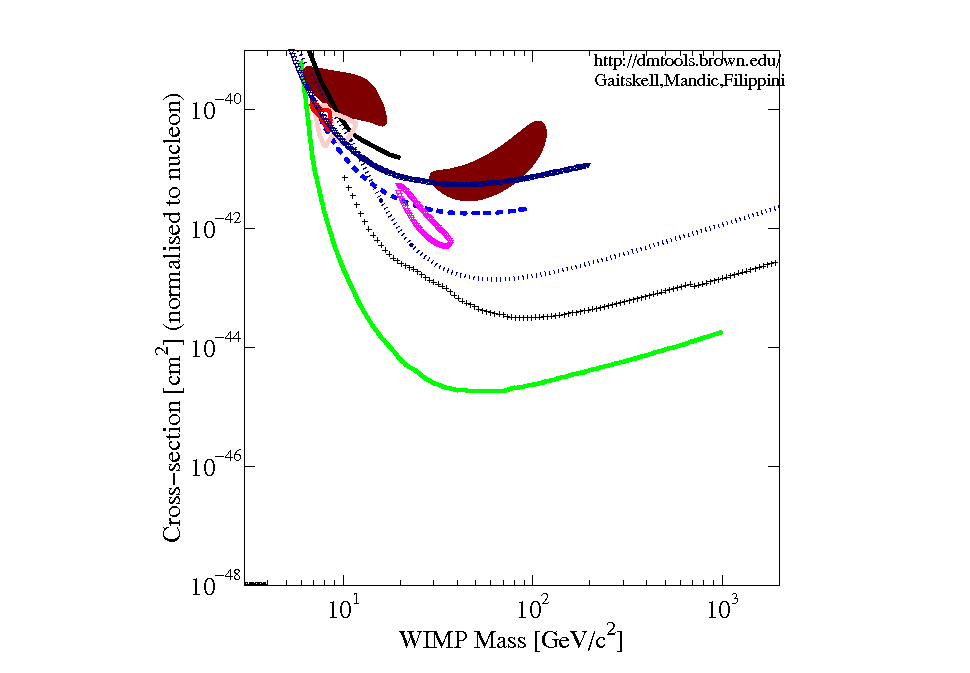} \includegraphics[trim=150 20 150 50,clip,width=0.48\textwidth]{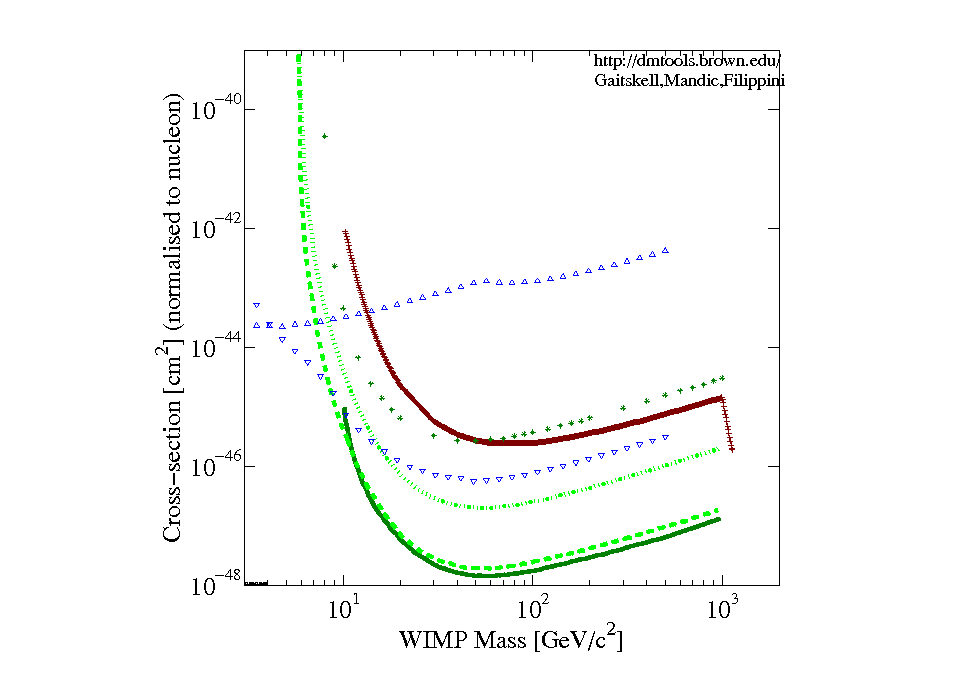}
        \caption{\label{fig:sicurrent} {\bf Spin-independent cross sections with current and future experiments.} This plot shows constraints and limits on the spin-independent WIMP-nucleon cross section as a function of WIMP mass for current (left panel) and future (right panel) experiments.  The constraints have been calculated assuming the SHM, although in a few cases the velocity distribution is a minor perturbation from the SHM. In the left panel, the closed curves represent the DAMA/LIBRA (dark red; \cite{Bernabei:2013xsa}), CoGeNT (red and light red; \cite{Aalseth:2012if}), and CRESST (magenta; \cite{Angloher:2011uu}) signals.  The green line shows the current XENON100 upper limit \cite{Aprile:2013doa},   black crosses the combined CDMS II and Edelweiss limits \cite{ahmed2011} and the blue lines the limits from COUPP \cite{Behnke:2012ys}, KIMS \cite{Kim:2012rza}, and SIMPLE \cite{Felizardo:2011uw}.  The right panel displays projected sensitivities for Generation 2 (upper curves) and Generation 3 (lower curves) experiments. The light green curves are for XENON, dark green for LUX and LZ, light red for  1000 kg Argon in DEAP/CLEAN and blue two different models for the SuperCDMS experiment (including Silicon).  Plots made using DMTools, \texttt{dmtools.brown.edu}, and references to projected sensitivities may be found on the DMTools website.}
\end{figure*}

The future goal of WIMP direct detection is to reach sensitivities at which the irreducible astrophysical neutrino background would become visible \cite{billard2013}.  We show the current status and future sensitivity limits for spin-independent WIMP-nucleon scattering in Fig. \ref{fig:sicurrent}.  The spin-dependent limits are similar in shape, but about six orders of magnitude weaker owing to the fact that they do not benefit from $A$-dependent enhancement.  Generation 3 experiments should hit the neutrino background for \mwimp$\sim 100$ GeV.  We refer the reader to Ref.~\cite{Baudis:2012ig} for a more detailed discussion of the experimental challenges and planned future experiments.

\section{Halo-independent comparisons between experiments}\label{sec:gvmin}
While in the next sections we consider WIMP parameter estimation using conventional likelihood and Bayesian inference methods, we will first discuss a different direct-detection analysis technique.  This method was developed when several experiments saw excess events above their expected backgrounds, in apparent contradiction with exclusion limits from other experiments. The CRESST~\footnote{CRESST is revising its background estimates c.f. Ref.~\cite{Kuzniak:2012zm}. } \cite{Angloher:2011uu}, CoGeNT \cite{Aalseth:2012if,collar2013b} and, most recently, CDMS Si \cite{Agnese:2013rvf} experiments see an excess in the time averaged event rate, while DAMA/LIBRA \cite{Bernabei:2013xsa} and CoGeNT \cite{Aalseth:2012if,collar2013b} have measured an annual modulation. Individually these data are compatible with light WIMPs, however there is no single point in the \mwimp-\sigmapsi \, plane which is consistent with all of these excesses and also the exclusion limits from other experiments. However comparing results in the  \mwimp-\sigmapsi \, plane requires a strong assumption to be made about the form of the WIMP velocity distribution, typically that it has the SHM form.

A particular issue in comparing the putative low-mass WIMP signals is that different experiments probe different parts of the WIMP velocity distribution, in particular its  highly uncertain high-velocity tail (see Sec. \ref{sec:theory:astrophysics}).  
The key insight of Drees \& Shan \cite{Drees:2007hr} and Fox et al. \cite{Fox:2010bz} was to think about the constraints from each experiment in terms of the velocity integral instead of \mwimp-\sigmapsi~space.  This insight allows results from different experiments to be compared in a way that is independent of the form of the velocity distribution.  This technique is called ``halo-independent'' modeling.

To describe the power of this technique, we follow the discussion of Fox et al., as their approach has been most widely adopted in the community.  As they do, we focus for simplicity on spin-independent elastic scattering.  The key point is that the scattering rate of Eqs. (\ref{eq:energyspectrum}) and (\ref{eq:sisigma}; assuming $f_n = f_p$) can be expressed as
\begin{eqnarray}\label{eq:drdq_vmin}
        \frac{{\rm d} R_i}{{\rm d} Q} = X( m_\chi, m_T^i) \rho_\chi \sigma_{\mathrm{p}}^{\mathrm{SI}}A^2 F^2(Q,A) g(v_{\mathrm{min}}),
\end{eqnarray}
for an experiment $i$ with target nucleus $T = (A,Z)$.  Here,
\begin{eqnarray}
        g(v_{\mathrm{min}}) = \int_{v_{\mathrm{min}}} {\rm d} v v f(\mathbf{v},t),
\end{eqnarray}
where $f(\mathbf{v},t)$ is the three-dimensional velocity distribution, is the velocity integral. (This expression differs from the alternative notation that is sometimes used, in which $f(v)$ is taken to be the one-dimensional speed distribution.)  Eq. (\ref{eq:drdq_vmin}) separates the velocity-dependent part of the event rate from the rest of the physics.  Note that for fixed \mwimp, $g(v_{\mathrm{min}})$ has no further dependence on the WIMP particle properties.  Fox et al. proposed that  Eq. (\ref{eq:drdq_vmin}) be inverted, and constraints or limits on the WIMP interaction are reported in terms of
\begin{eqnarray}
       \rho_\chi \sigma_{\mathrm{p}}^{\mathrm{SI}} g(v_{\mathrm{min}}) = \frac{{\rm d} R_i/{\rm d} Q}{X(m_\chi,m_T^i)A^2 F^2(Q,A)} \,,
\end{eqnarray}  
for specific choices of the  WIMP mass, which determines the relationship between nuclear recoil energy $Q$ and \vmin~for each target nucleus.  Plotting constraints/limits on this expression as a function of \vmin, allows one to see which experiments probe the same part of the WIMP velocity distribution and whether or not they are compatible.

The strength of this scheme is that it allows a direct comparison between experiments for a fixed interaction model, and independent of the WIMP velocity distribution.  It is an extremely useful consistency check, and has served a great service to the community as such. However its main drawback is that the comparison has to be done for fixed WIMP mass (but see Ref. \cite{Drees:2008bv}), and therefore it is not as useful for parameter estimation.

This scheme has been generalized to incorporate multiple isotopes in one experiment \cite{Gondolo:2012rs}, energy-dependent experimental response functions \cite{DelNobile:2013cta}, annual modulation \cite{Frandsen:2011gi,Gondolo:2012rs,HerreroGarcia:2012fu,DelNobile:2013cta}, inelastic dark matter \cite{Bozorgnia:2013hsa,Miao:2013sqa}, and other effective operators \cite{Shan:2011ka,DelNobile:2013cva}.  The current consensus is that it is impossible to reconcile all experiments in the spin-independent framework, regardless of the WIMP velocity distribution, unless the experiments are woefully miscalibrated at small nuclear recoils (see also Refs. \cite{Fox:2011px,Schwetz:2011xm}).

\section{This work: Bayesian inference and parameter estimation}\label{sec:simulation}
\subsection{Mock Experiments}

In order to demonstrate the power of an ensemble of direct-detection experiments to unveil WIMP physics, we generate mock data sets using several idealized experiments, and use Bayesian inference techniques to reconstruct WIMP particle and velocity-distribution parameters.  The idealized experiments are representative of `Generation 2' dark matter detectors.  We choose benchmark points in WIMP physics parameter space that lie below current `Generation 1' sensitivities. These experiments and benchmark points are meant to show what information about WIMPs may be gleaned in the early discovery days.  Of course, if WIMPs are conclusively discovered in Generation 2 experiments, the statistics will improve dramatically with Generation 3 experiments.  If WIMPs are not conclusively discovered in Generation 2 detectors, our results are still relevant for the early days of discovery for Generation 3 experiments, only that the cross sections that will be probed are smaller than quoted here.

Each experiment is characterized by the target material used, the range of nuclear recoil energies to which the experiment is sensitive and the total exposure (which takes into account the detector efficiency). For the sake of simplicity, and so that we are not wed to experimental realities that may change with time, we assume a step-function detector efficiency.  In other words, we assume that experiments are completely insensitive to nuclear recoils above and below the analysis window, and that the sensitivity is constant within the analysis window.  Note that in general, the sensitivity is energy-dependent (see, for example, Fig. 1 in Ref. \cite{Aprile:2012nq}).  The experimental parameters used in this work are summarized in Table~\ref{tab:Experiments}. 

For each experiment, we divide the energy range into bins and generate Asimov data for each bin \cite{cowan2013}. This entails setting the observed number of events equal to the expected number of events for the benchmark models in each bin. In this way, we can analyse how the underlying physics of different benchmarks and experiments will affect parameter reconstructions, without having to worry about the influence of Poisson fluctations in the data.  In reality, however, Poisson noise will be important for WIMP parameter estimation \cite{Strege:2012kv}.  This is especially true for the benchmark models for which only a few dozen (or fewer) events are expected in each experiment.  Thus, the credible intervals we show in the next three sections are representative of WIMP parameter uncertainties, but in reality they could look quite different.

We assume that all experiments have perfect energy resolution and no background contamination. Clearly, for realistic experiments these assumptions would not hold, especially for most low-threshold experiments (these have nontrivial backgrounds). However, finite energy bins mimic finite energy resolution and the parameter estimation will not be significantly changed if the energy resolution is better than our $1.0$ keV energy bin width.
It is important to emphasise that the reconstructions we present here should be taken as a best-case scenario in which experimental uncertainties are negligible.

\begin{table*}[t]
  \setlength{\extrarowheight}{3pt}
  \setlength{\tabcolsep}{12pt}
  \begin{center}
	\begin{tabular}{c|cm{2cm}m{2cm}m{2cm}m{2cm}m{2cm}}
	Experiment & Target & Target Mass (amu) & Energy Range (keV) & Exposure (ton-yr) & Energy bin width (keV) \\
	\hline\hline
	Xenon & Xe & 131  & 7-45 \cite{aprile2010} & 2.0 & 1.0 \\
	Argon & Ar & 40 & 30-100 \cite{grandi2005, DarkSide} & 2.0 & 1.0 \\
	Germanium & Ge & 73 & 8-100 \cite{bauer2013} & 0.2 & 1.0 \\
	Silicon & Si & 28 & 7-100 \cite{Agnese:2013rvf} & 0.2 & 1.0 \\
	COUPP-500 & CF$_3$I & 12,19,127 & 10-200 \cite{COUPP500} & 1.0 & 190.0 \\
	\end{tabular}
  \end{center}
\caption{Parameters used for the mock experiments in this work. All experiments are assumed to have perfect resolution and zero background.}
\label{tab:Experiments}
\end{table*}

\subsection{Bayesian Inference}
In order to infer WIMP physics from our mock data sets, we employ Bayesian statistics.  Bayes' theorem is 
\begin{eqnarray}
  \mathcal{P}(\{p\},H|\{d\}) = \frac{\mathcal{L}(\{d\}|\{p\},H) \mathcal{P}(\{p\},H)}{\mathcal{P}(\{d\}|H)}.
\end{eqnarray}
Here, $\mathcal{L}(\{d\}|\{p\},H)$ is the well-known likelihood function that a data set $\{d\}$ is collected given parameters $\{p\}$ for a hypothesis $H$ for the underlying model. $\mathcal{P}(\{p\},H|\{d\})$ is the posterior probability for the model parameters $\{p\}$ given the data $\{d\}$, and is the distribution we report in all following sections.  $\mathcal{P}(\{p\},H)$ is the prior probability for $\{p\}$, which we will define separately in each following section.  Finally, $\mathcal{P}(\{d\}|H)$ is the probability of obtaining the data set $\{d\}$ under the hypothesis for the underlying model.  This probability is often called the ``evidence'' for the model, and while it plays a critical role in Bayesian model selection (i.e., in deciding which hypotheses are a better fit to the data set), for our purposes it is simply a proportionality constant.

Since we are working with Asimov data, we use a binned likelihood function instead of the unbinned likelihood function that is a less lossy choice for relatively small data sets \cite{Peter:2011eu}.  This means that our likelihood function is the product of Poisson probabilities for finding $N_i^o$ observed events in bin $i$ given the theoretically expected $N_i^e$ events for parameter values $\{p\}$,
\begin{equation}
  \mathcal{L}(\{N_i^o\}|\{N_i^e\}) = \prod_i^{N_{\mathrm{bins}}}\frac{(N_i^e)^{N_i^o}}{N_i^o!}e^{N_i^e}.
\end{equation}

We explore the posterior probability of each ensemble of mock data sets using the \textsc{MultiNest} nested sampling code \cite{Feroz:2007kg,Feroz:2008xx,Feroz:2013hea}.
  For most of the benchmark points we explore in later sections, we set the parameters \texttt{efr=0.4}, \texttt{tol}$=10^{-5}$, and use $10^4$ live points.  We will comment when we set these parameters to other values.  We achieve efficient convergence with these parameters, and find robust values of the Bayesian evidence.  As the authors of \textsc{MultiNest} have noted, \textsc{MultiNest} converges faster when the dimensionality of the space used for mode separation is smaller than the number of dimensions if the number of dimensions is large.

In the next three sections, we show many examples of marginalized 1- and 2-dimensional posteriors.  These are probability distribution functions of the one or two parameters of interest.  They may be obtained (as here, in the case of a 2D posterior) by integrating the posterior over the remainder of the theoretical parameters,
\begin{eqnarray}
  \mathcal{P}({p_1,p_2}|\{d\}) = \int \mathcal{P}(\{p\}|\{d\})\prod_{i=3}^N dp_i .
\end{eqnarray}

\section{Non-directional experiments and WIMP particle properties}\label{sec:WIMPphys}
In this section, we illustrate which WIMP particle properties can be recovered from direct-detection experiments.  We use only one fiducial velocity distribution: a Maxwell-Boltzmann distribution with a hard cut-off at the Galactic escape velocity.  This is not to say that differences in velocity distributions cannot be important, as we will see in the following two sections, but we make this choice so as to focus on WIMP particle properties in this section.

\subsection{Spin-dependent and -independent interactions}\label{sec:WIMPphys:sd}
In this section, we consider the capability of Generation-2 experiments to distinguish between spin-independent and -dependent scattering, assuming WIMP particle-physics benchmark parameters that are just beyond current sensitivities.  We follow the general arc of Pato \cite{Pato:2011de}, but show how constraints on WIMP physics parameters depend both on the types of experiments available and on assumptions about the specific parameters in the WIMP velocity distribution.

\begin{table}[t]
  \setlength{\extrarowheight}{3pt}
  \setlength{\tabcolsep}{3pt}
  \begin{center}
    \begin{tabular}{c|cc}
      Experiment &Isotope &Mass Fraction \\
      \hline \hline
      Xenon & $^{129}$Xe &0.26 \\
      &$^{131}$Xe &0.21 \\
      &$^{132}$Xe &0.53 \\
      \hline
      Argon &$^{40}$Ar &1 \\
      \hline
      Germanium &$^{70}$Ge &0.21 \\
      &$^{72}$Ge &0.28 \\
      &$^{73}$Ge &0.08\\
      &$^{74}$Ge &0.43\\
      \hline
      Silicon &$^{28}$Si &0.95\\
       &$^{29}$Si &0.05 \\
      \hline
      COUPP-500 &$^{12}$C &0.06 \\
      &$^{19}$F &0.29\\
      &$^{127}$I &0.65
    \end{tabular}
  \end{center}
  \caption{\label{tab:sd:isotopes}Isotopic abundances in experiments for Sec.~\ref{sec:WIMPphys:sd}.}
\end{table}

\begin{table*}[t]
  \setlength{\extrarowheight}{3pt}
  \setlength{\tabcolsep}{3pt}
  \begin{center}
    \begin{tabular}{c|ccccccc}
     Benchmark &\mwimp (GeV) &$\rho_\chi$\sigmapsi/$m_p^2$ &$\rho_\chi$\sigmapsd/$m_p^2$ $\,(10^{-46} \hbox{GeV}^{-1} \,\hbox{cm}^{-1})$ &$a_n/a_p$ &\vlag (km/s) &$\sigma_v$ (km/s) &\vesc (km/s) \\
     \hline\hline
     1 &10 &3 &$3\times 10^6$ &-1 &220 &155 &544 \\
     2 &50 &3 &$3\times 10^5$ &-1 &220 &155 &544 \\
     3 &50 &3 &$3\times 10^3$ &-1 &220 &155 &544 \\
     4 &50 &0.03 &$3\times 10^5$ &-1 &220 &155 &544\\
     5 &50 &0.03 &$3\times 10^5$ &0 &220 &155 &544 \\
    \end{tabular}
  \end{center}
\caption{\label{tab:sd:benchmarks}Benchmark points for Sec.~\ref{sec:WIMPphys:sd}.  For the cross-section-related parameters, $\rho_\chi\sigma_{\mathrm{p}}/m_p^2 = 3 \times 10^{-46}\hbox{ GeV}^{-1}\,\hbox{cm}^{-1}$ corresponds to 1 zb ($10^{-45}$ cm$^2$) if $\rho_\chi = 0.3\hbox{ GeV cm}^{-3}$.}
\end{table*}

Because the sensitivity of an experiment to spin-dependent interactions depends sensitively on the isotopic abundances within the target volume, in this section we deviate from the rest of this work (and from Pato \cite{Pato:2011de}) by including realistic isotopic abundances in our model experiments.  The isotopes and their mass fractions within each experiment are given in Table \ref{tab:sd:isotopes}.  It is particularly important to have a realistic mass fraction of the experiments for isotopes sensitive to spin-dependent interactions, those with odd atomic numbers.  For the most part, the abundances match nature, which is what is used for most experiments.  There are a couple of exceptions---we lump the natural abundances of any Xenon isotope above $A=132$ in with $^{132}$Xe; and $^{76}$Ge with $^{74}$Ge.  While $^{76}$Ge is important for neutrinoless double-beta decay experiments, the difference in isotope mass is small for our purposes, so we model these heavy, even-nucleon isotopes of Germanium together.  Another difference from the rest of this work is that we treat COUPP-500 as having two energy bins.  This is meant to mimic a possible strategy for COUPP-500, to increase the energy threshold of the experiment once several tens of events have been found at a low threshold, in order to find the e-folding scale for the energy spectrum.  We model COUPP-500 as having two energy bins: 10-30 keV, and 30-50 keV.  One key point in this discussion is that COUPP-500 is the experiment with by far the best sensitivity to spin-dependent interactions, although it is the experiment with the worst energy sensitivity.

\begin{figure*}[t]
  \includegraphics[width=0.32\textwidth]{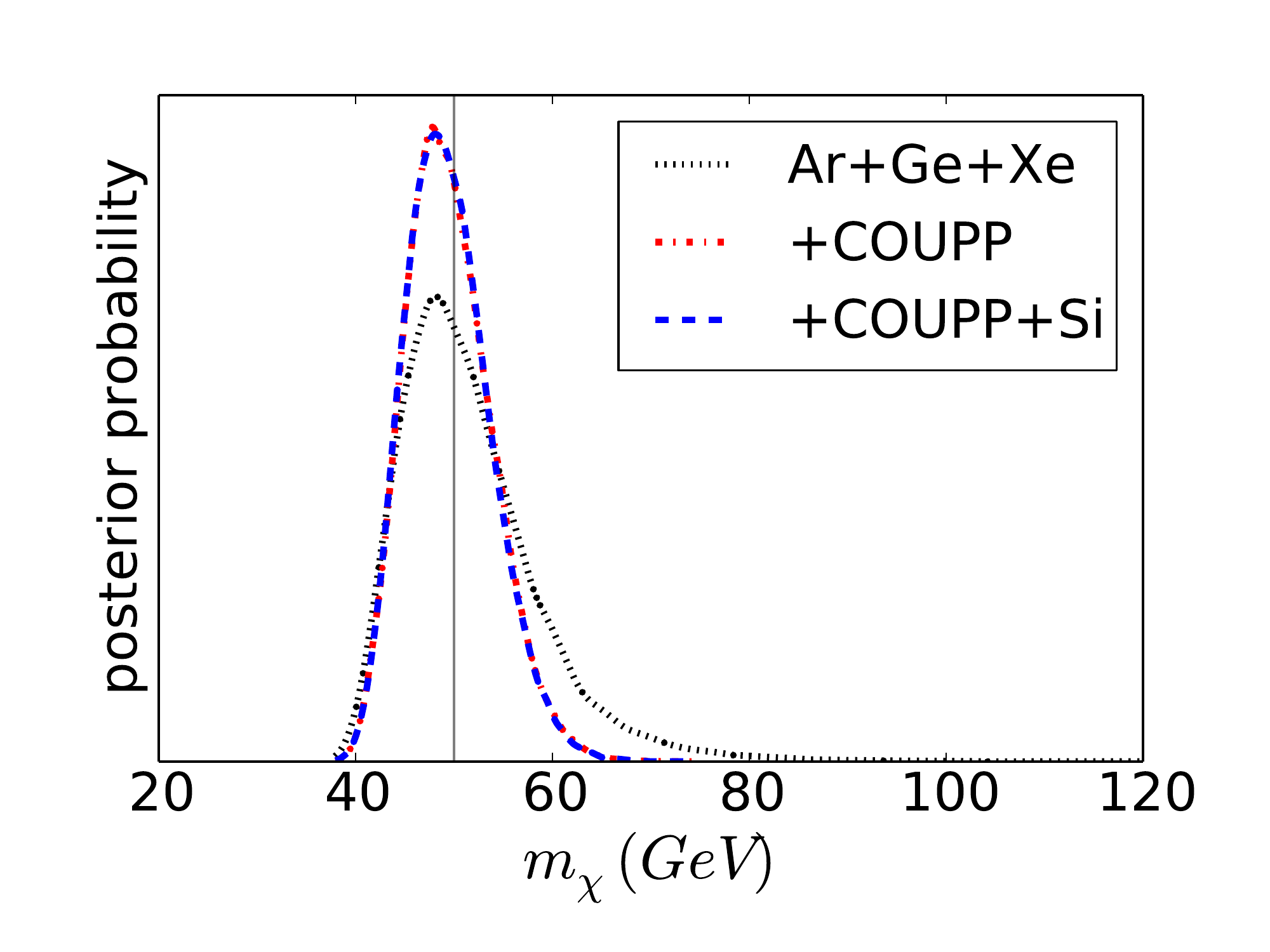}  \includegraphics[width=0.32\textwidth]{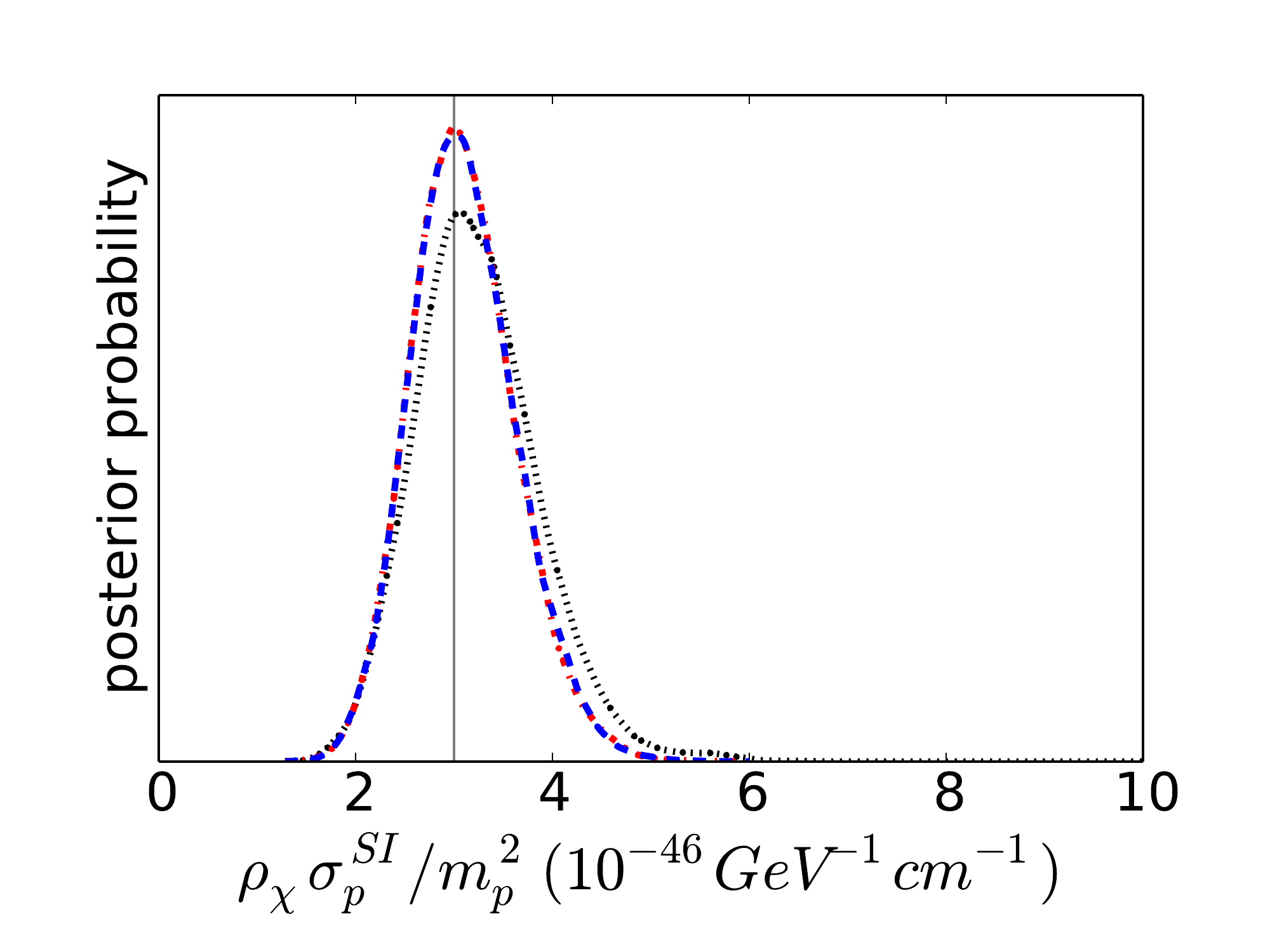} \includegraphics[width=0.32\textwidth]{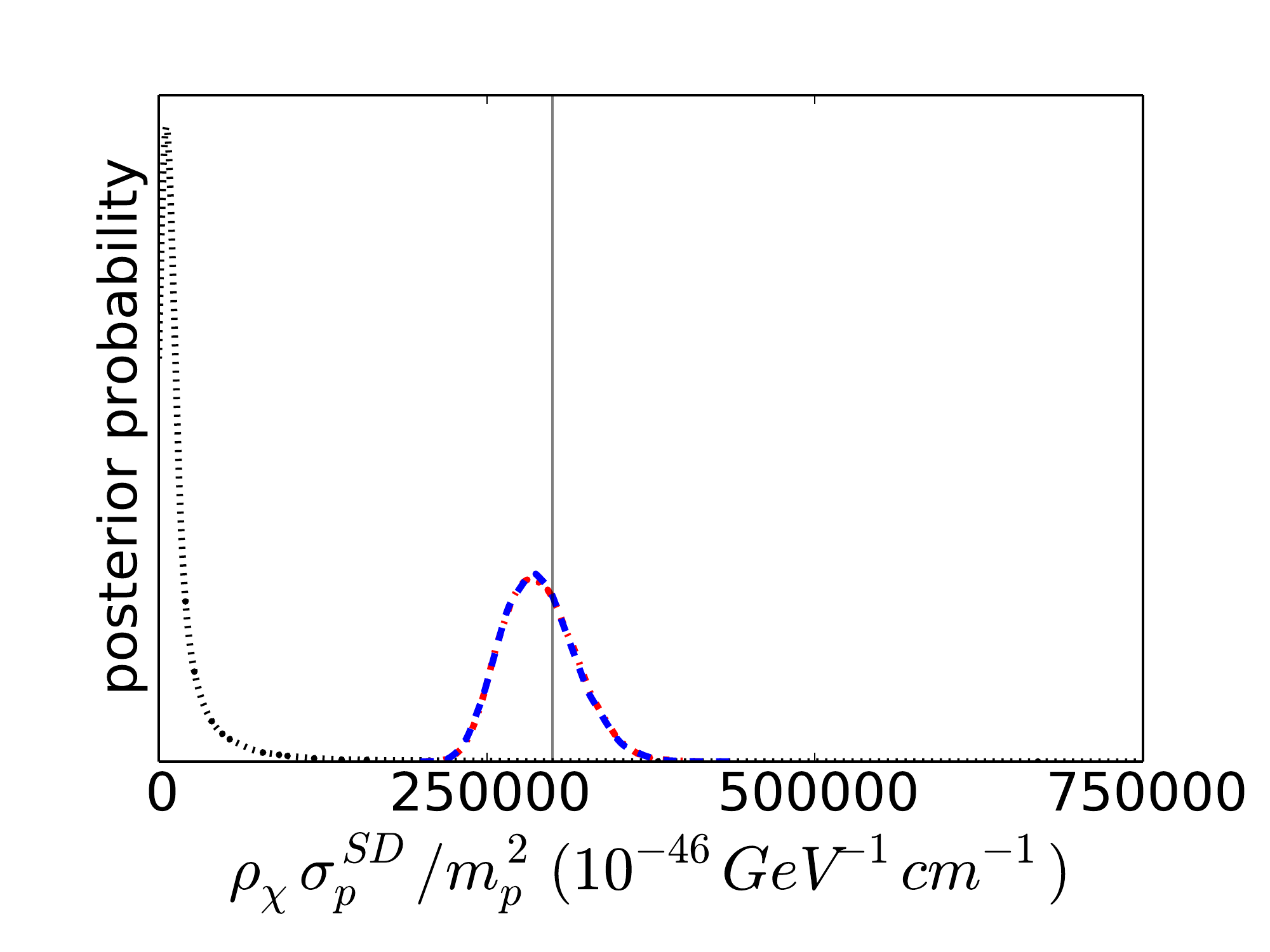} \\
  \includegraphics[width=0.32\textwidth]{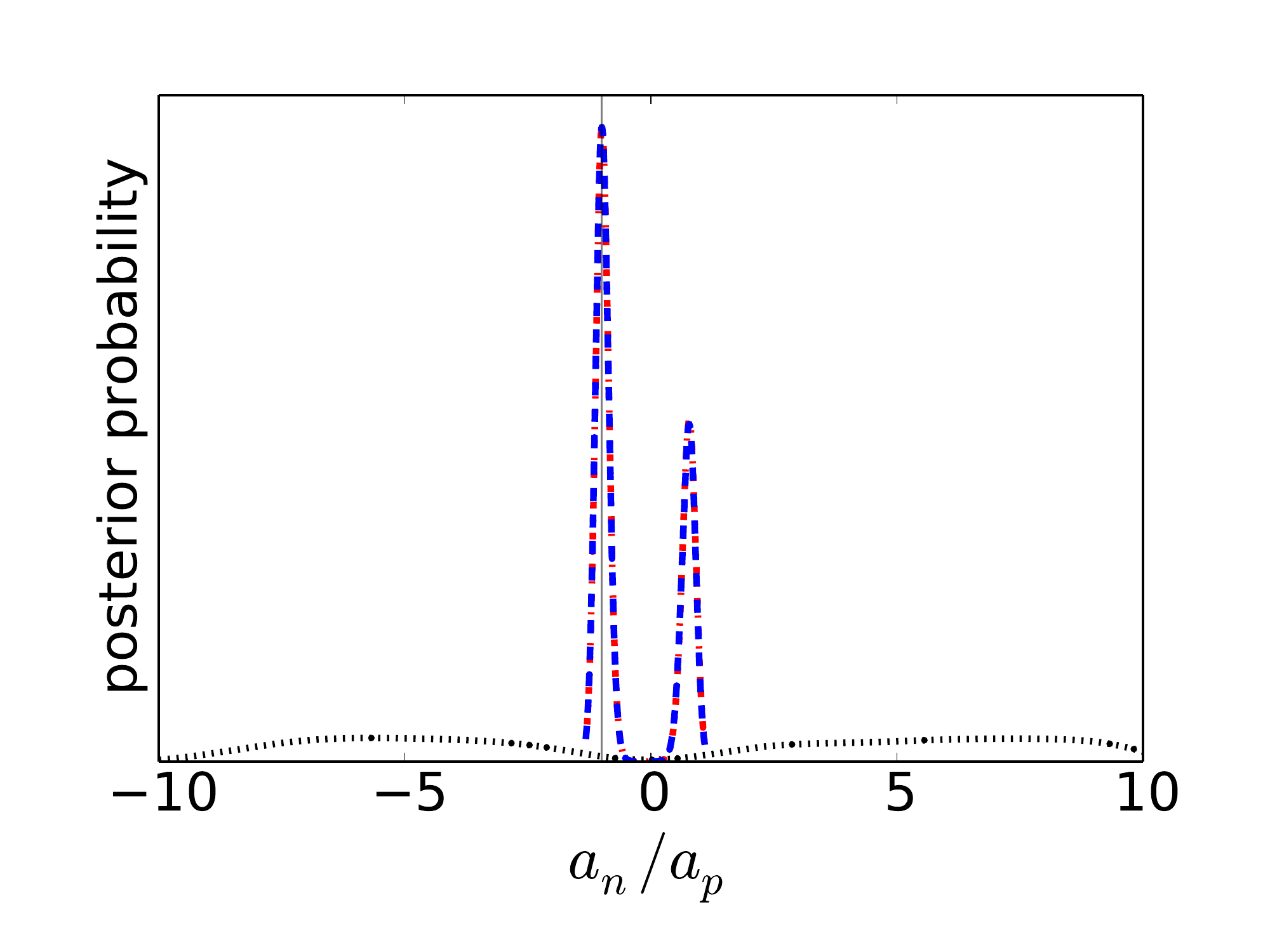}  \includegraphics[width=0.32\textwidth]{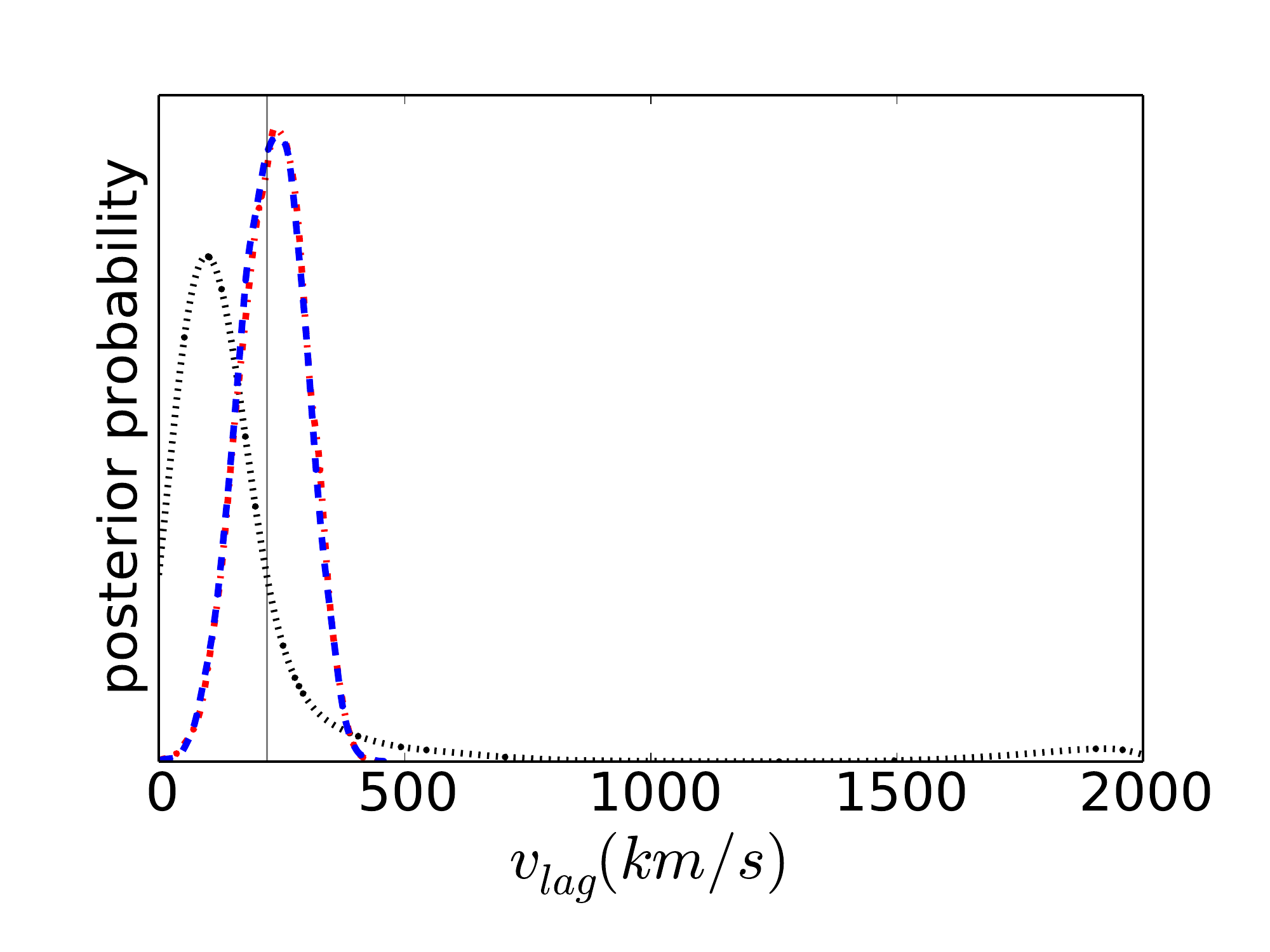} \includegraphics[width=0.32\textwidth]{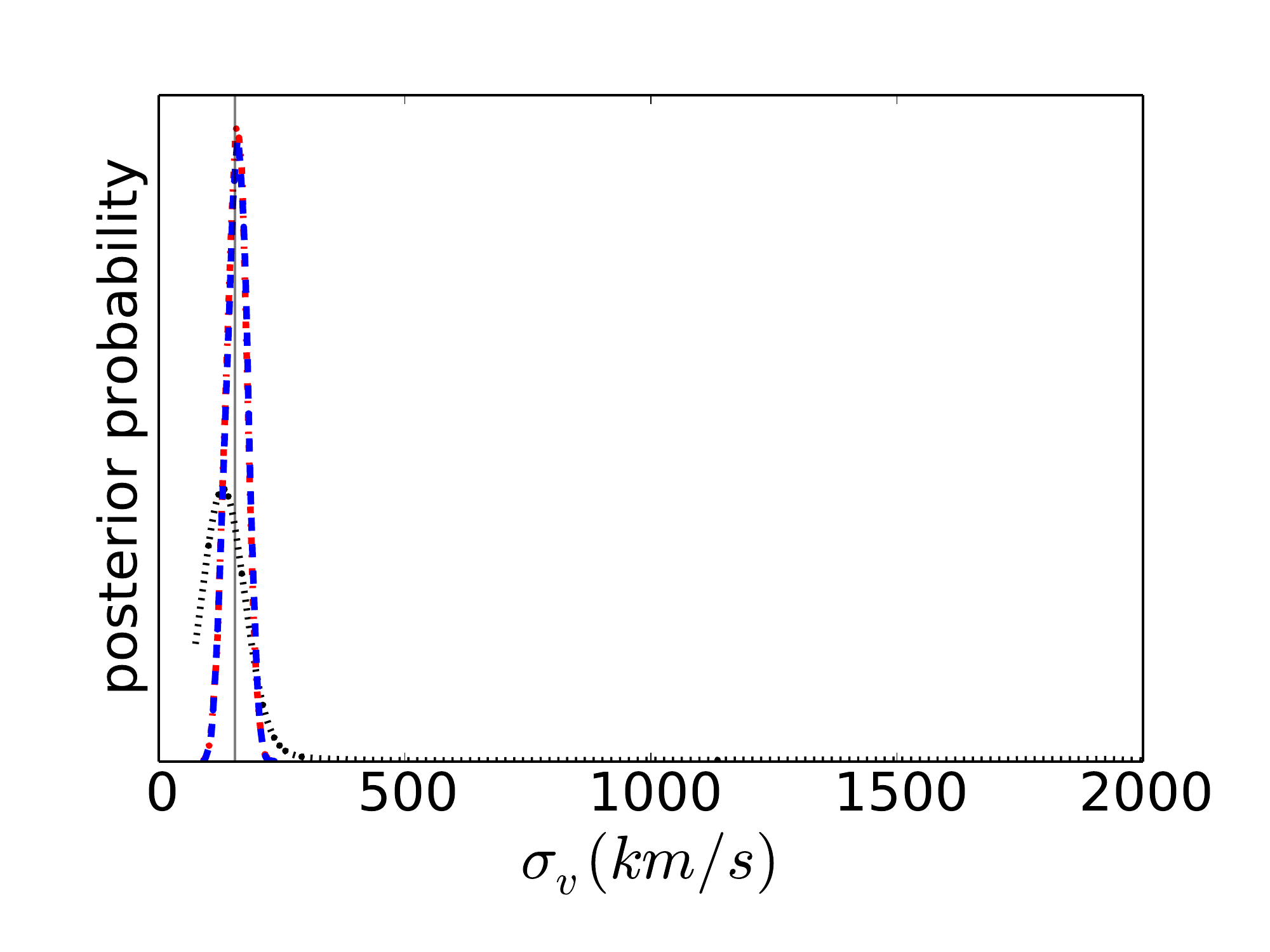}
  \caption{\label{fig:2trip}{\bf Marginalized one-dimensional posteriors for Benchmark 2 as a function of experiment ensemble.}  For the top row and the left-hand panel on the bottom row, we show posteriors with the strong velocity priors.  For the right two bottom panels, we show the Maxwell-Bolzmann velocity parameters inferred when weak velocity priors are used.  The vertical lines show the benchmark parameter values.}
\end{figure*}

\begin{figure*}
  \includegraphics[width=0.32\textwidth]{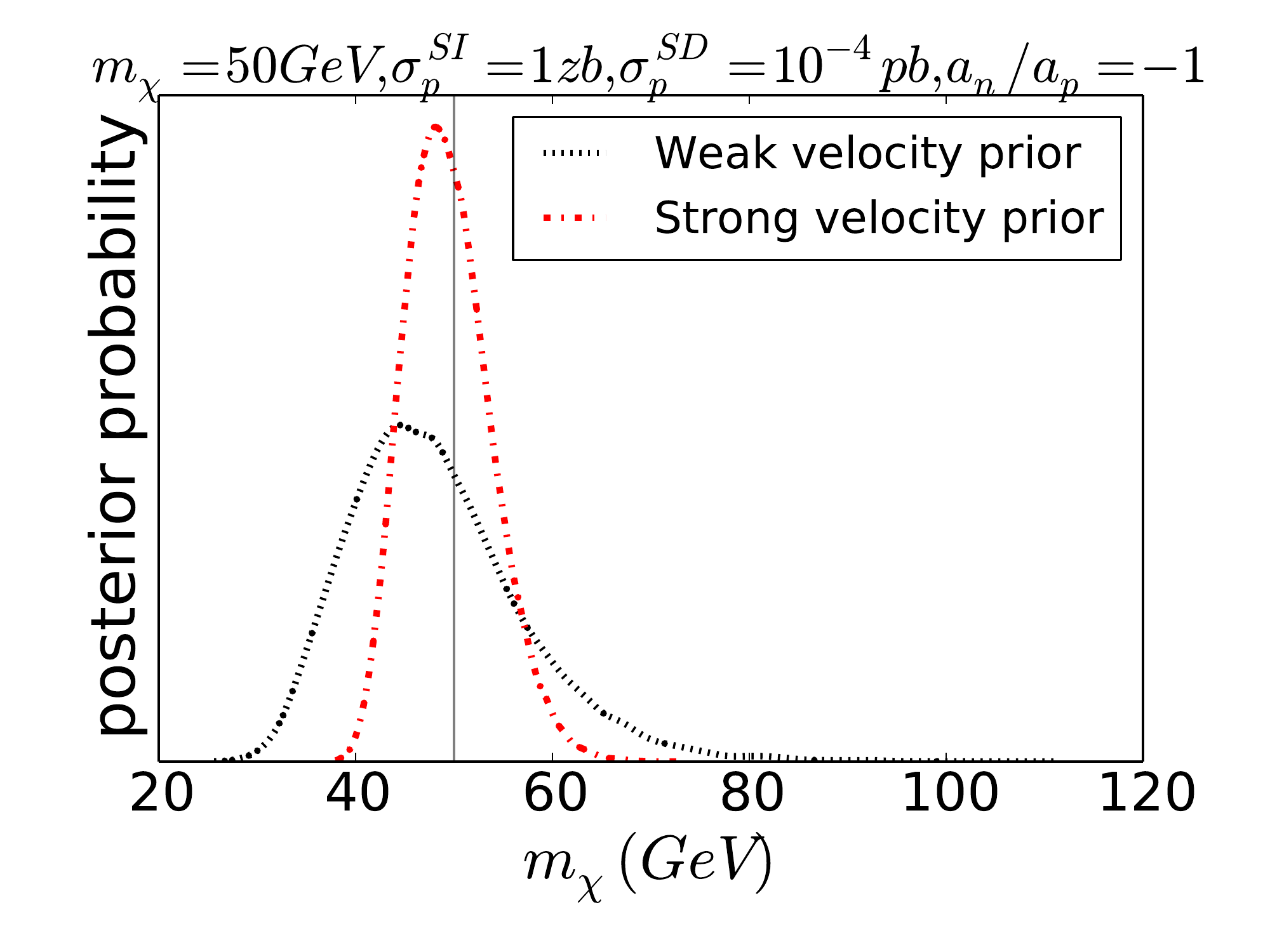} \includegraphics[width=0.32\textwidth]{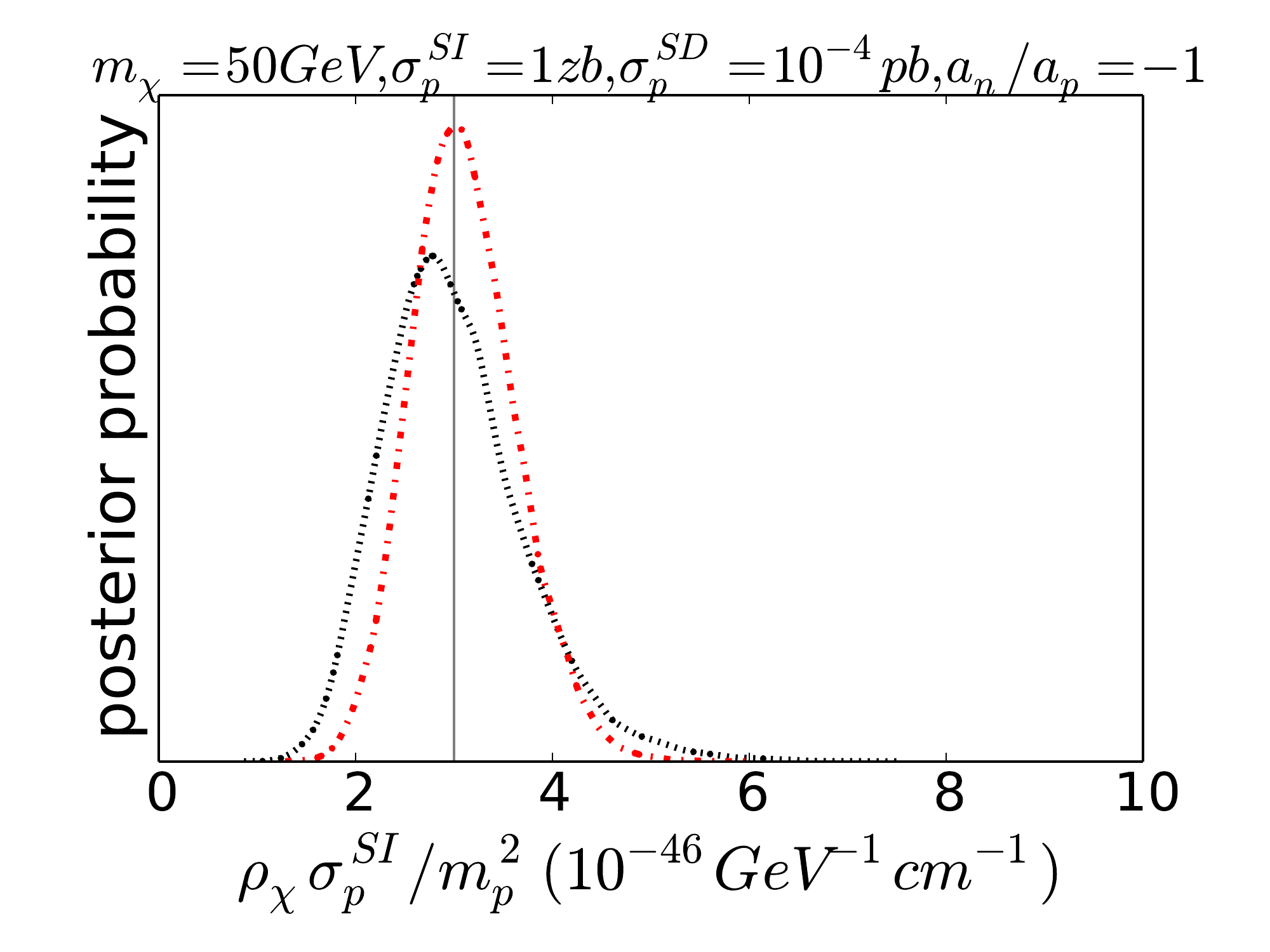} \\
  \includegraphics[width=0.32\textwidth]{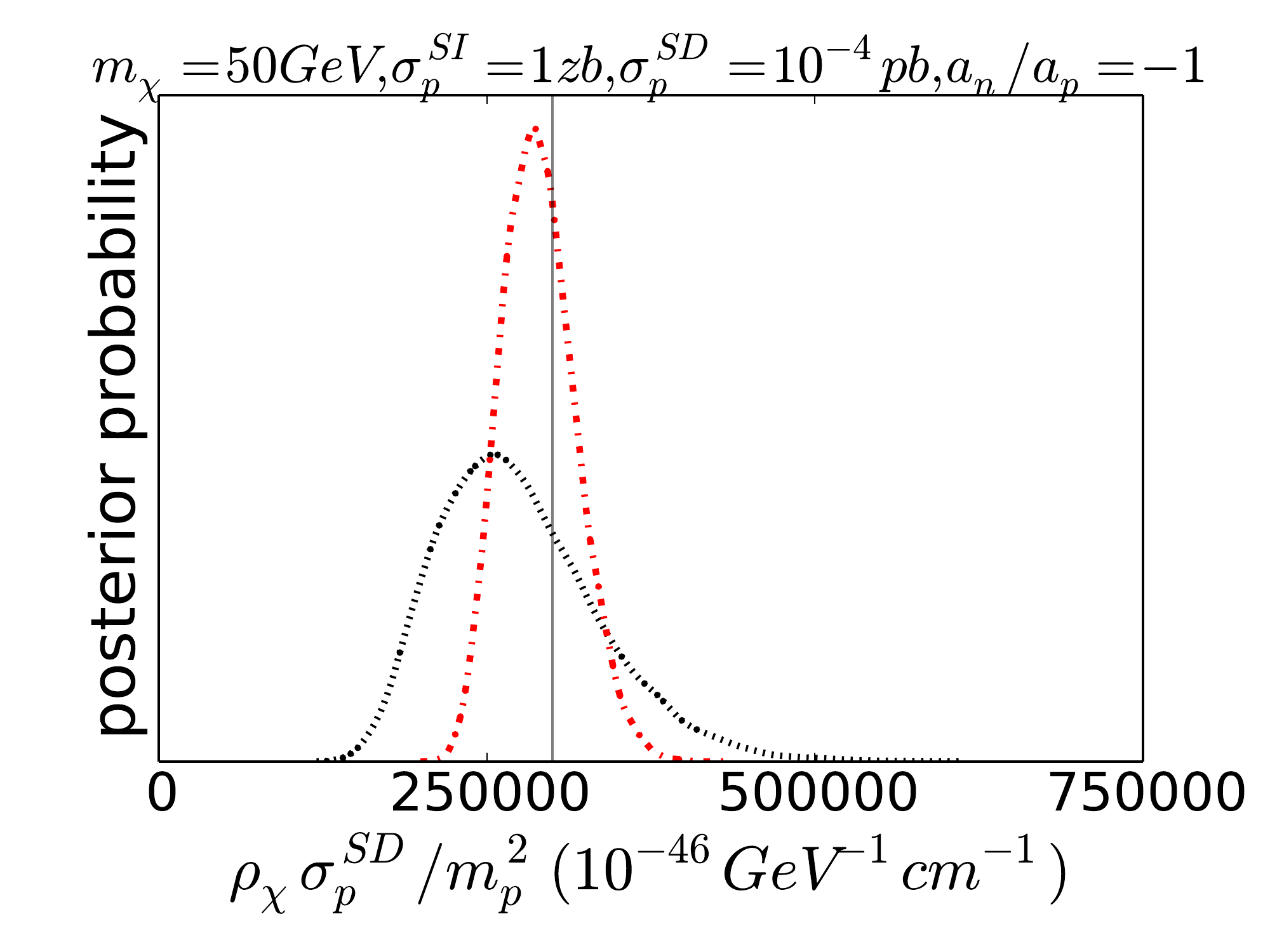} \includegraphics[width=0.32\textwidth]{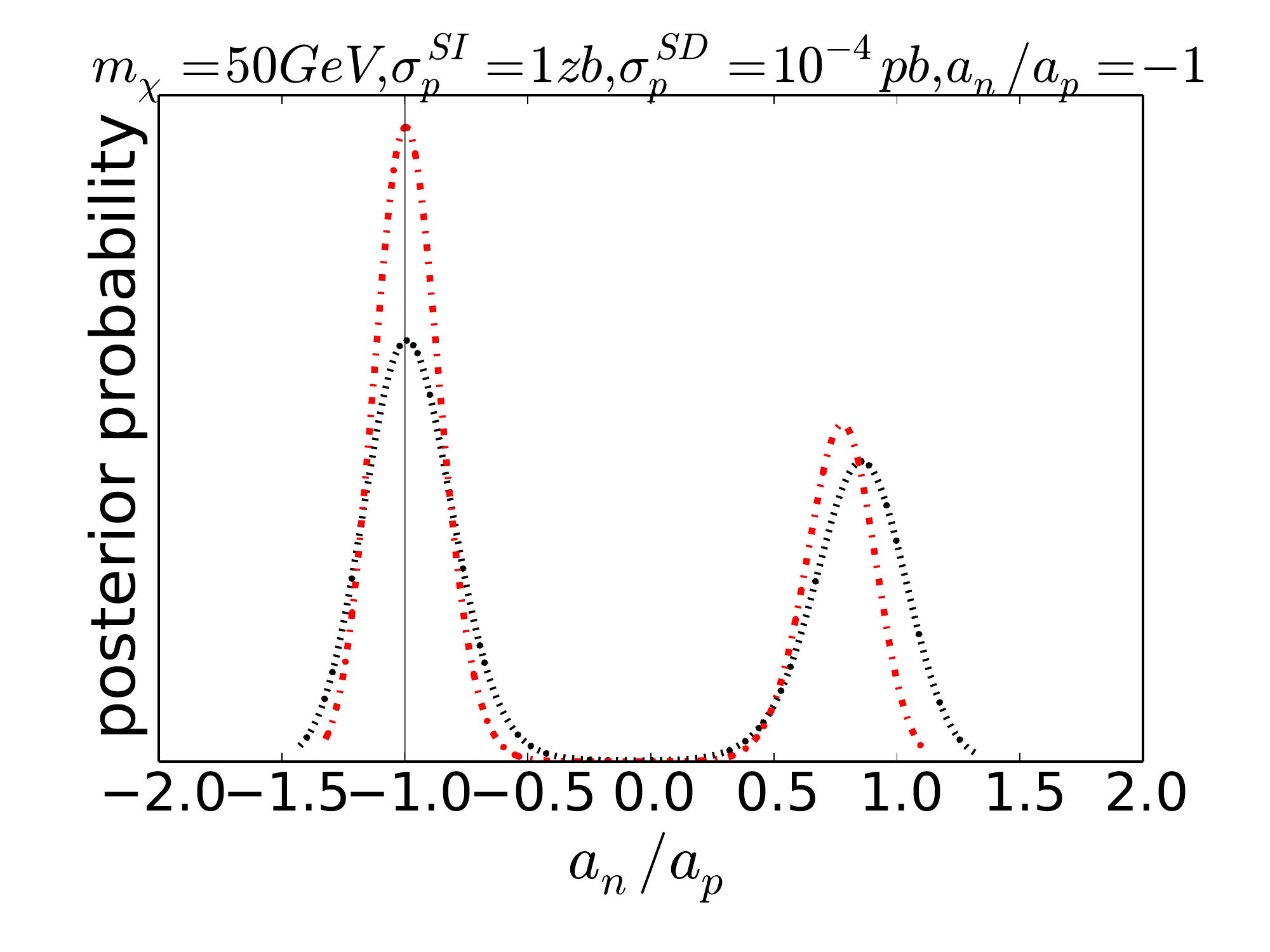}
  \caption{\label{fig:2doub}{\bf Marginalized one-dimensional posteriors for Benchmark 2 with weak or strong velocity priors.}  In this case, we use all five experiments for constraints.}
\end{figure*}

In Table \ref{tab:sd:benchmarks}, we show the benchmark WIMP points we use for the spin-dependent and spin-independent analysis.  Most of our benchmarks (2-5) center on a WIMP with mass \mwimp$= 50 $ GeV, which is the mass to which most experiments have the best sensitivity (see Fig. \ref{fig:sicurrent}).  We show one example of a low-mass WIMP (10 GeV; Benchmark 1).  We also investigated high-mass WIMPs, but the trends were not significantly different than for \mwimp$=50$~GeV.  The cross sections are chosen such that they lie just below current exclusion limits (assuming a SHM velocity distribution) for the given WIMP mass.  For spin-independent scattering, our fiducial cross sections are 1 zb ($10^{-45}\hbox{ cm}^2$) for \mwimp$=50$~GeV.  The fiducial spin-independent cross section is two orders of magnitude higher for Benchmark 1 because of the reduced experimental sensitivity for low-mass WIMPs.  We note that experiments with extremely low thresholds (e.g., CDMSlite \cite{Agnese:2013lua}, CoGeNT \cite{Aalseth:2012if}, DAMIC \cite{barreto2012}) can contribute significantly to constraints in this region of parameter space.  Recall that we are enforcing isospin symmetry in both our benchmark models and our fits.  For our analysis, we constrain the quantity $\rho_\chi$\sigmapsi$/m_p^2$ (or with \sigmapsd) instead of \sigmapsi, since these parameters are degenerate with each other.

For spin-dependent scattering, we choose a spin-dependent WIMP-nucleon cross section $=10^{-39}\hbox{ cm}^2$ ($10^{-3}$ pb) for Benchmark 1, and one order of magnitude smaller for the higher-mass benchmarks.  Primarily we consider the case in which $a_n/a_p = -1$, an MSSM-inspired choice.  In Benchmark 5, we consider the extreme case in which there is no coupling to neutrons, only to protons: $a_n/a_p = 0$.  We consider cases in which we expect a similar number of events resulting from spin-dependent and spin-independent scattering (Benchmarks 1 and 2), a case in which spin-independent scattering dominates (Benchmark 3), and in which spin-dependent scattering dominates (Benchmarks 4 \& 5).

We sample the posterior distribution for the WIMP parameters (the 7 parameters listed in Table \ref{tab:sd:benchmarks}) using the prior probabilities given in Table \ref{tab:sd:priors}.  There are two different sets of priors on the velocity parameters in order to evaluate both the effects of the prior volume on the posterior as well as the possibility of reconstructing the WIMP velocity distribution from small-ish numbers of events.  The ``weak priors'' are broad, flat priors about the fiducial values.  These prior ranges are quite wide, but especially for Benchmark 1, span a wide range of shapes in the high-speed tail of the WIMP speed distribution.  The ``strong priors'' are flat within a $\sim 10\%$ range about the fiducial velocity-distribution parameters.

\begin{table}[t]
  \setlength{\extrarowheight}{3pt}
  \setlength{\tabcolsep}{3pt}
  \begin{center}
        \begin{tabular}{r|cc}
        Parameter & Prior type & Prior Range \\
        \hline\hline
        \mwimp (GeV) &log-flat & [0.1-$10^4$]\\
        \hline
        $\rho_\chi$\sigmapsi$/m_p^2\,(10^{-46} \hbox{GeV}^{-1} \,\hbox{cm}^{-1})$  &log-flat & [0.007-$10^5$]\\
        \hline
        $\rho_\chi$\sigmapsd$/m_p^2\,(10^{-46} \hbox{GeV}^{-1} \,\hbox{cm}^{-1})$  &log-flat & [0.1-$10^9$]\\
        \hline
        $a_n/a_p$ &flat & [-10,10] \\
        \hline
        \vlag (km/s) Weak prior &flat &[0-2000] \\
          Strong prior &flat &[200-240] \\
        \hline
        $\sigma_v$ (km/s) Weak prior &flat &[0-2000] \\
          Strong prior &flat &[140-170] \\
          \hline
        \vesc (km/s) Weak prior &flat &[490-600] \\
        Strong prior &flat &[540-550] \\
        
        \end{tabular}
  \end{center}
\caption{\label{tab:sd:priors}Priors for the spin-dependent analysis in Sec.~\ref{sec:WIMPphys:sd}.}
\end{table}

We discuss Benchmark 2 first.  In Fig.~\ref{fig:2trip}, we show marginalized one-dimensional posteriors for the WIMP parameters as a function of which experiments are included in the analysis.  Each linetype corresponds to a different ensemble of experiments used in the analysis; the black dotted lines include only Argon, Germanium, and Xenon, which have somewhat limited spin-dependent sensitivity.  Those data are mostly dominated by Xenon, and these isotopes are primarily sensitive to \sigmansd, not \sigmapsd.  The red dot-dashed line includes COUPP, and the blue dashed lines add Silicon to the mix.  The vertical solid lines show the benchmark parameter values.  The main conclusion of these plots, and in fact of much of this subsection on spin-dependent scattering, is that COUPP-500 is crucial to characterizing WIMPs that have significant spin-dependent interactions with nuclei, even if it has poor energy sensitivity.  Not only does it have good spin-dependent sensitivity, but it is complementary to Germanium, Xenon, and Silicon experiments in that it is primarily sensitive to \sigmapsd, not \sigmansd.  All but the bottom center and bottom right panels use the strong velocity priors.  The bottom center and bottom right panels show the WIMP velocity parameters inferred from the experimental ensembles.  Clearly, the data are strong enough to overcome the prior to make strong statements about \vlag~and $\sigma_v$, especially if COUPP-500 is included.  As we show in Sec.~\ref{sec:astrophysics}, velocity-distribution constraints improve with the number of experiments with different target nuclei, that see at least a few tens of events.  This leads to improved contraints for the WIMP particle properties.  However, the data are not strong enough to overcome the prior on \vesc; the posterior is almost completely flat over the prior range for this parameter.

\begin{figure}
  \includegraphics[width=0.32\textwidth]{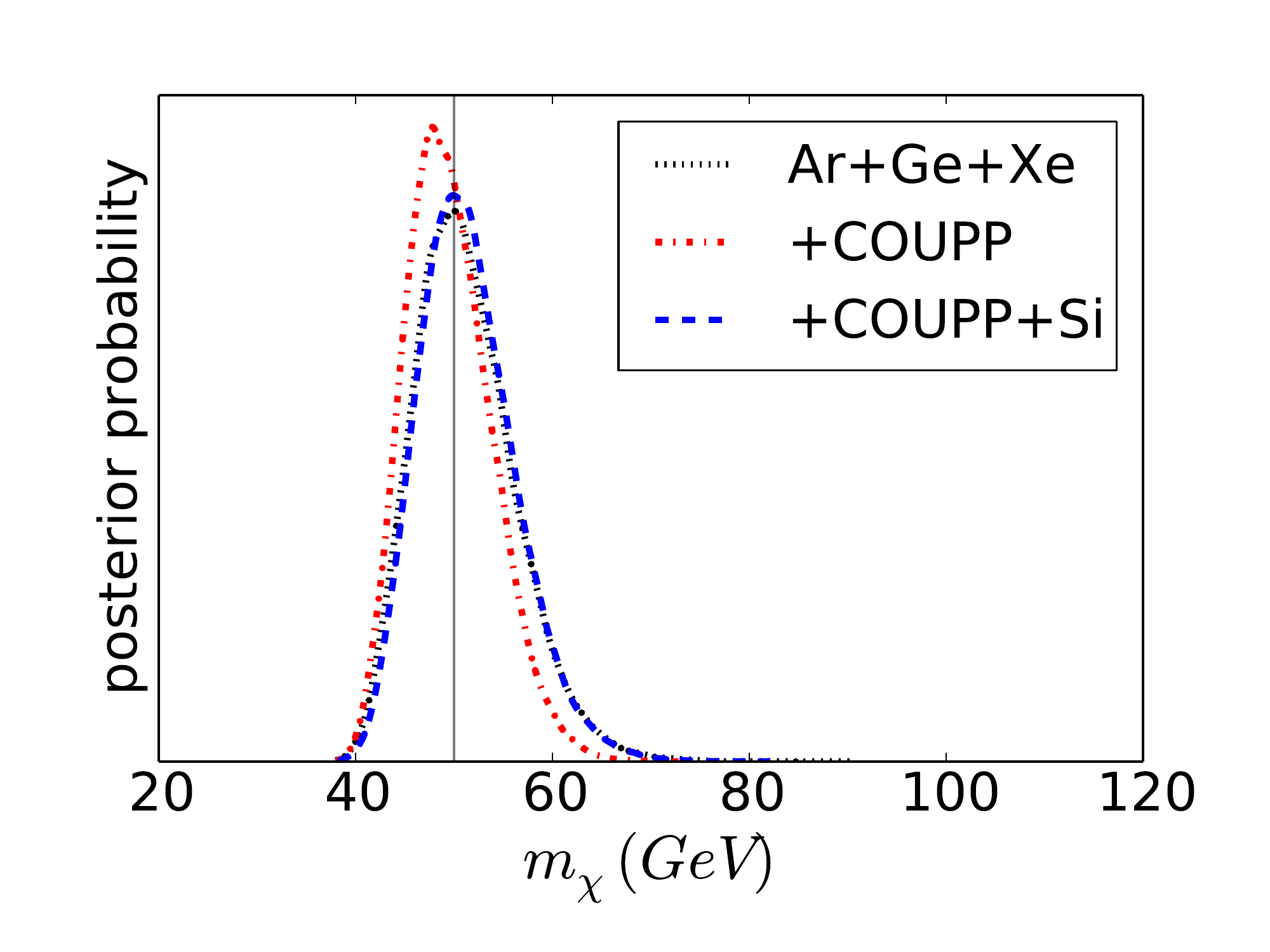} \\
  \includegraphics[width=0.32\textwidth]{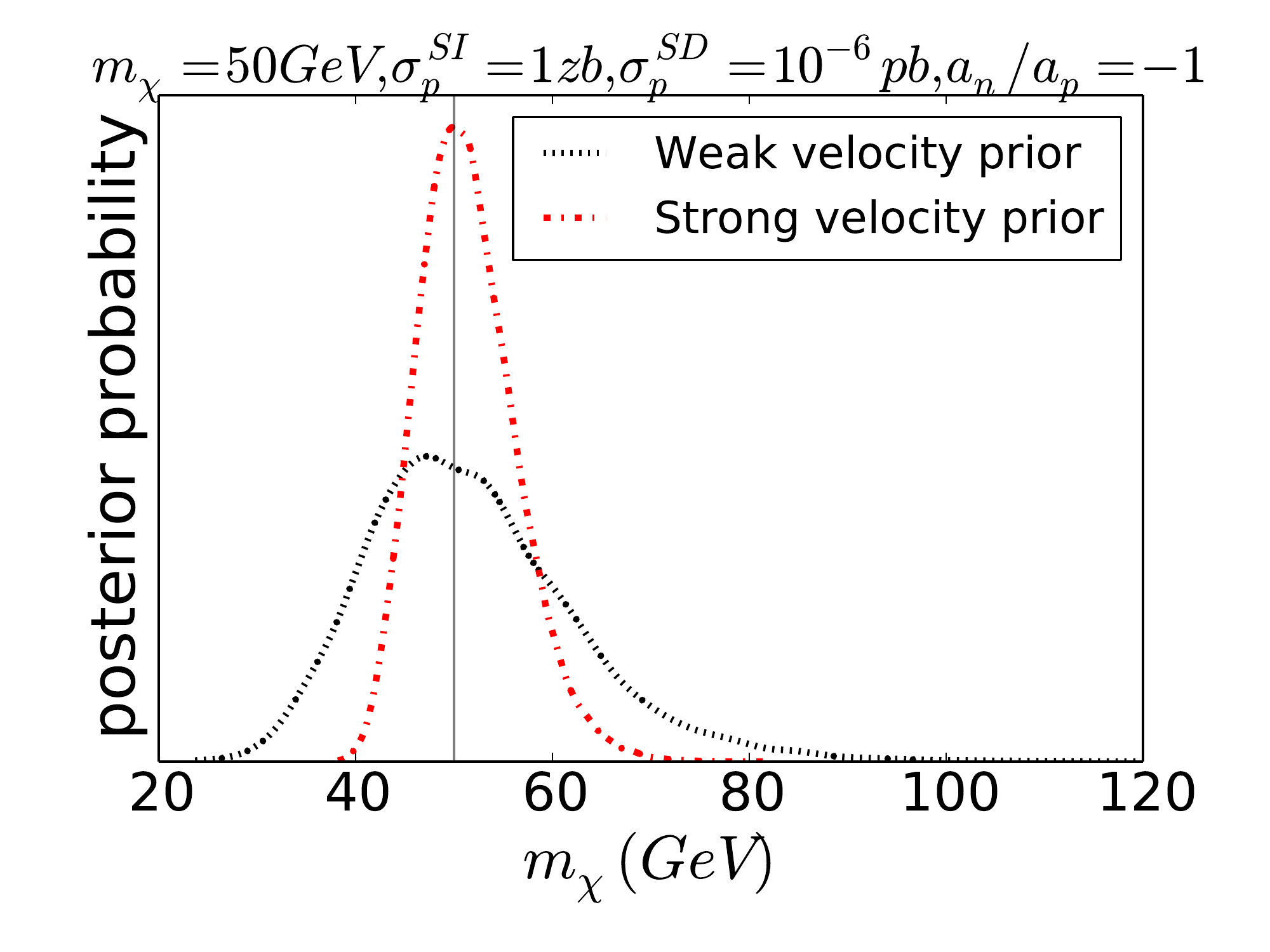} 
  \caption{\label{fig:3tot}{\bf Marginalized one-dimensional posteriors for Benchmark 3 for \mwimp.}  In the top panel, we show the \mwimp~posteriors under the assumption of strong velocity priors.  Each line corresponds to a different ensemble of experiments.  In the bottom panel, we show the effects of the velocity priors using data from all five mock experiments.}
\end{figure}

We show the dependence of the posterior on the choice of velocity prior for Benchmark 2 in Fig. \ref{fig:2doub}.  The WIMP mass and spin-dependent cross section show the largest improvement in the width of the posteriors with the strong velocity priors relative to the weak ones, reducing the width of the posterior by approximately a factor of two.  However, there is only modest improvement in either the spin-independent cross section or the neutron-to-proton coupling ratio $a_n/a_p$.  Similar to Pato \cite{Pato:2011de}, we find that it is difficult to distinguish negative from positive values of $a_n/a_p$ with our particular ensemble of mock experiments.  This situation is likely only to improve if experiments become more highly enriched in odd-spin isotopes.

In Benchmarks 3 and 4, we either crank down the spin-dependent scattering (3) or the spin-independent scattering (4) while leaving other parameters unchanged from Benchmark 2.  Those cranked-down cross sections are low enough that they produce negligible numbers of events in the experiments relative to the dominant interaction type.  For Benchmark 3, similar conclusions to Benchmark 2 hold.  We find similar constraints on the velocity parameters as with Benchmark 2.  We correctly find only an upper limit on $\rho_\chi$\sigmapsd$/m_p^2$, and cannot constrain $a_n/a_p$.  In Fig. \ref{fig:3tot}, we show constraints on \mwimp~as a function of experiment, and as a function of velocity prior.  For $\rho_\chi$\sigmapsi$/m_p^2$, the trends with ensemble of experiment and velocity prior look similar to those for \mwimp~shown in Fig. \ref{fig:3tot}.

\begin{figure*}
  \includegraphics[width=0.32\textwidth]{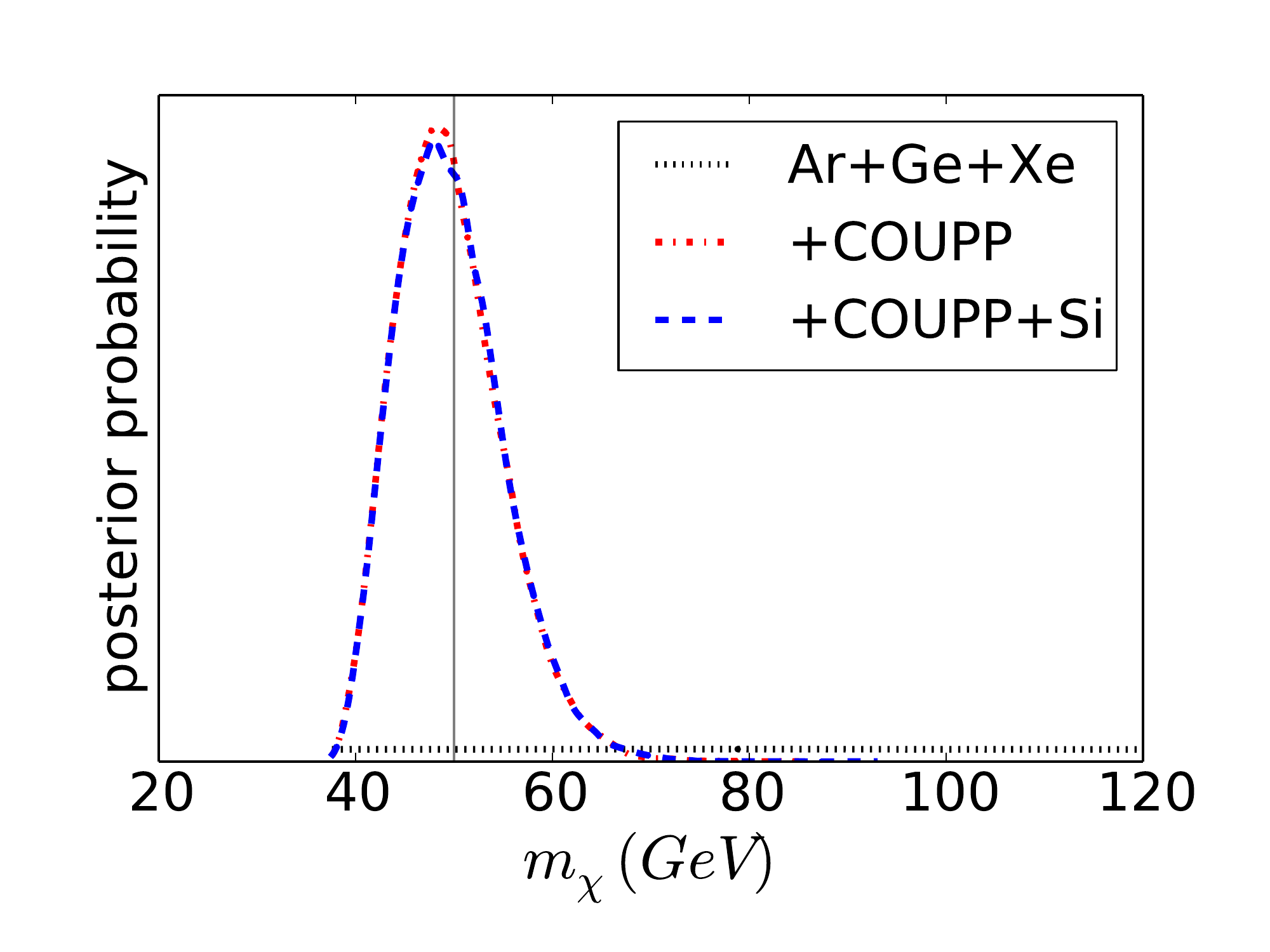}  \includegraphics[width=0.32\textwidth]{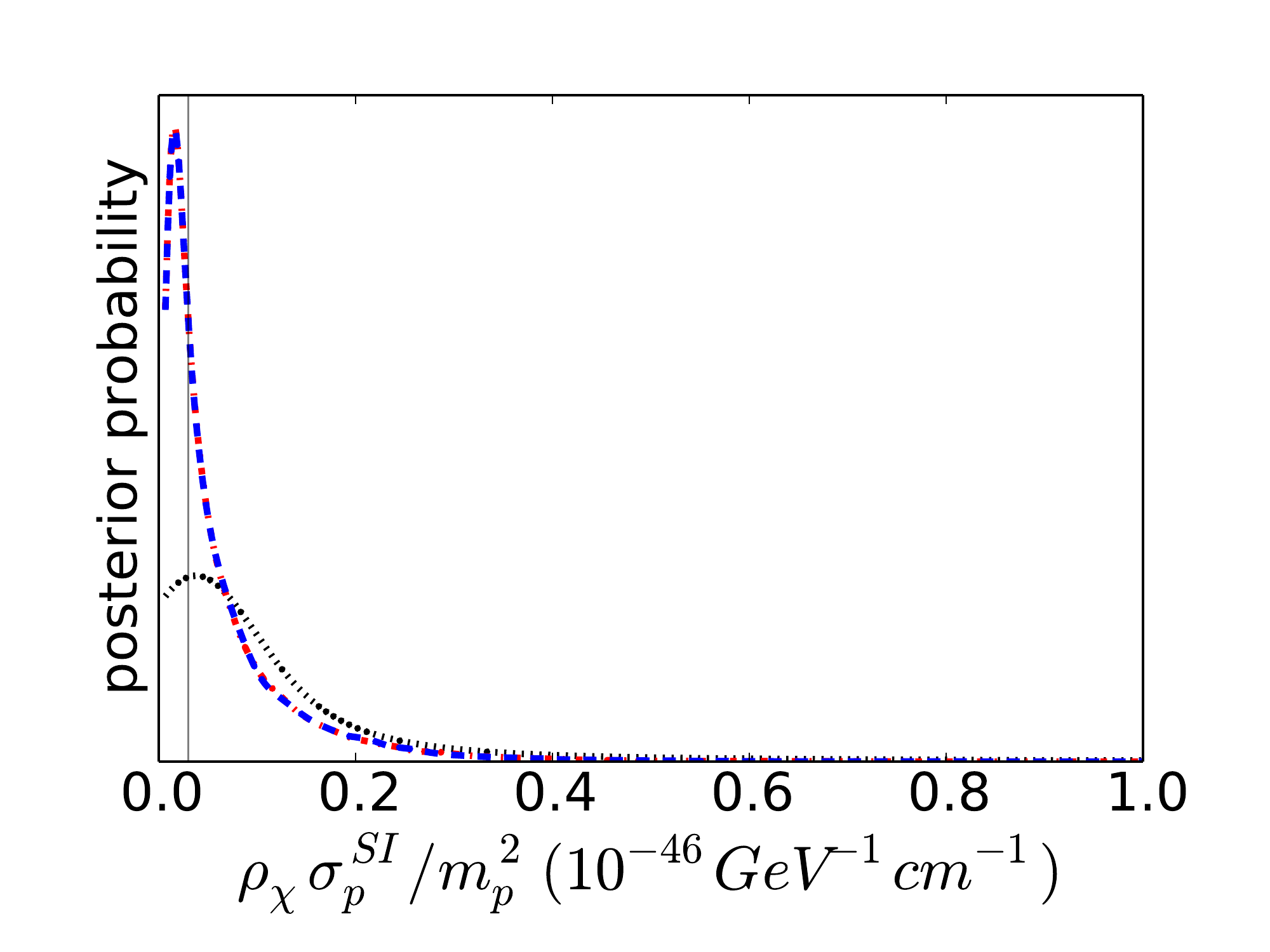} \includegraphics[width=0.32\textwidth]{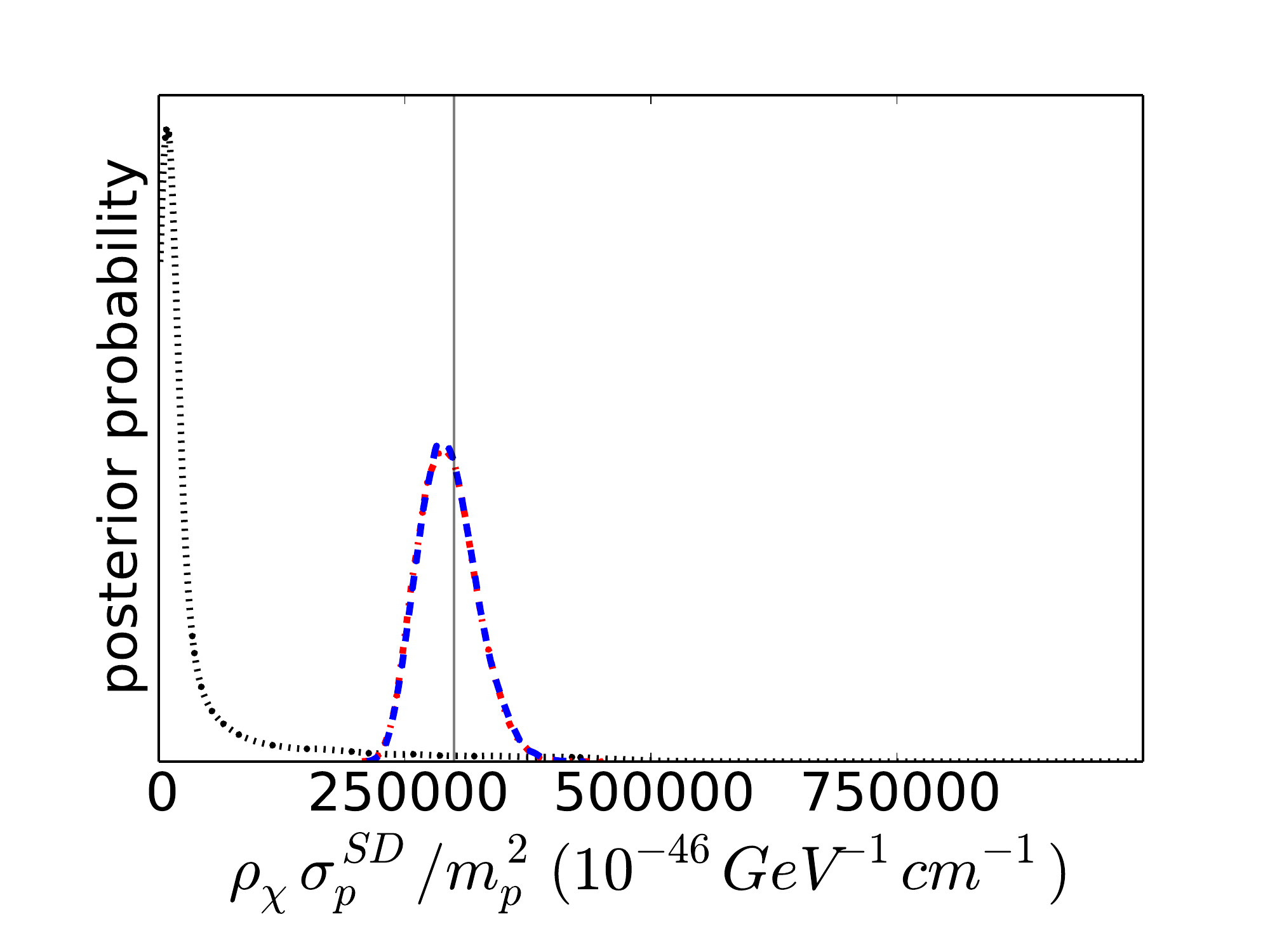} \\
  \includegraphics[width=0.32\textwidth]{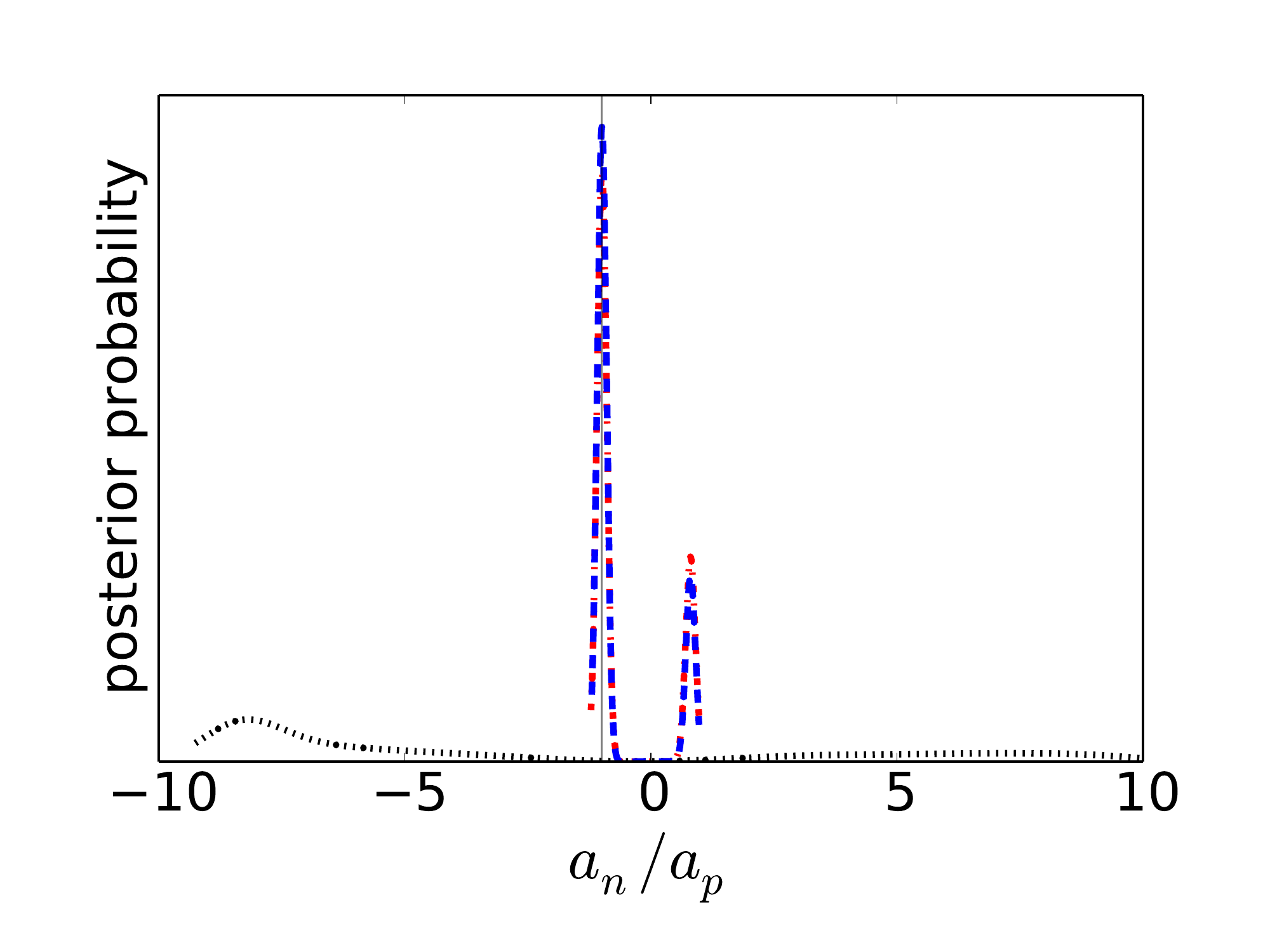}  \includegraphics[width=0.32\textwidth]{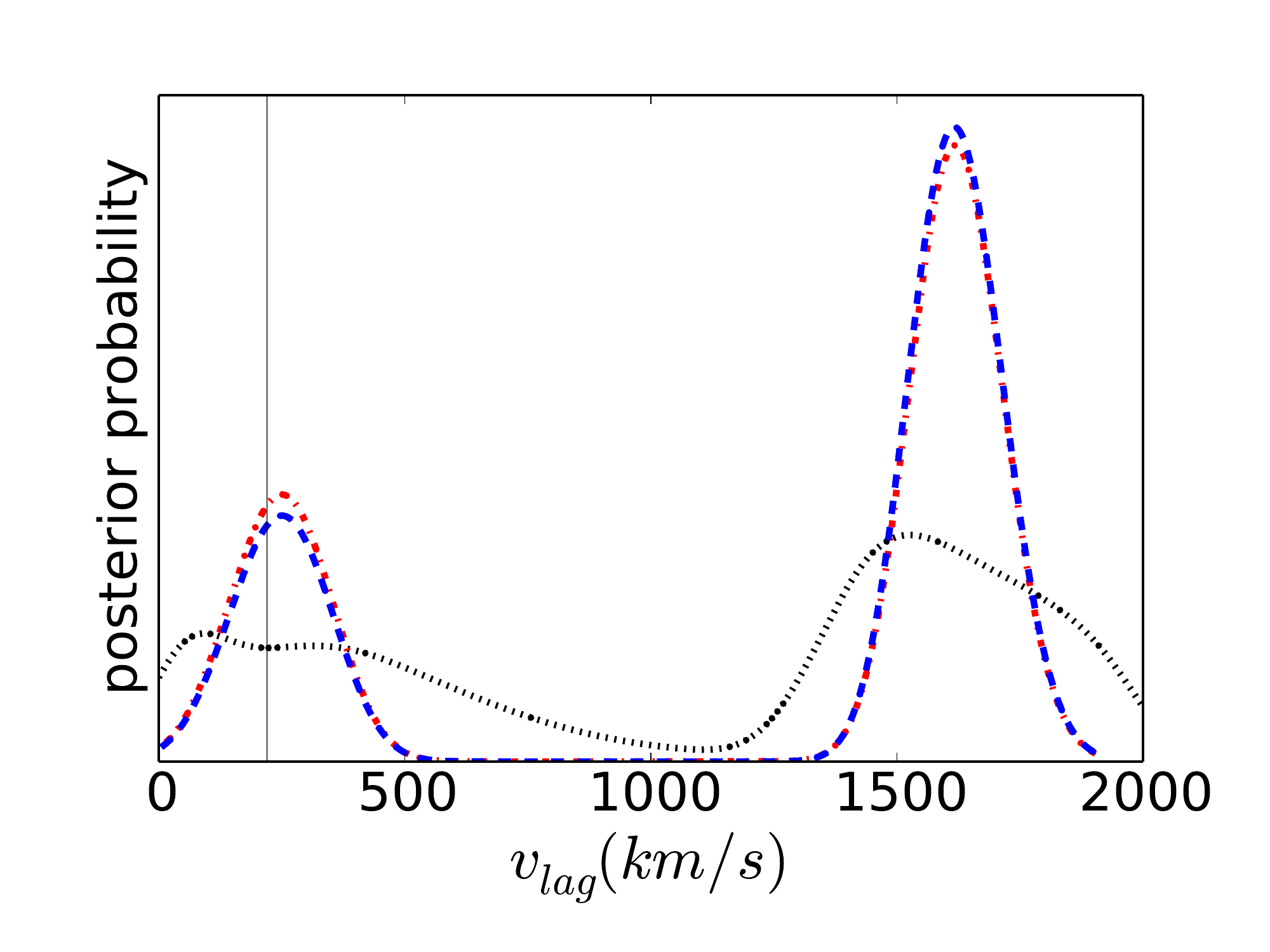} \includegraphics[width=0.32\textwidth]{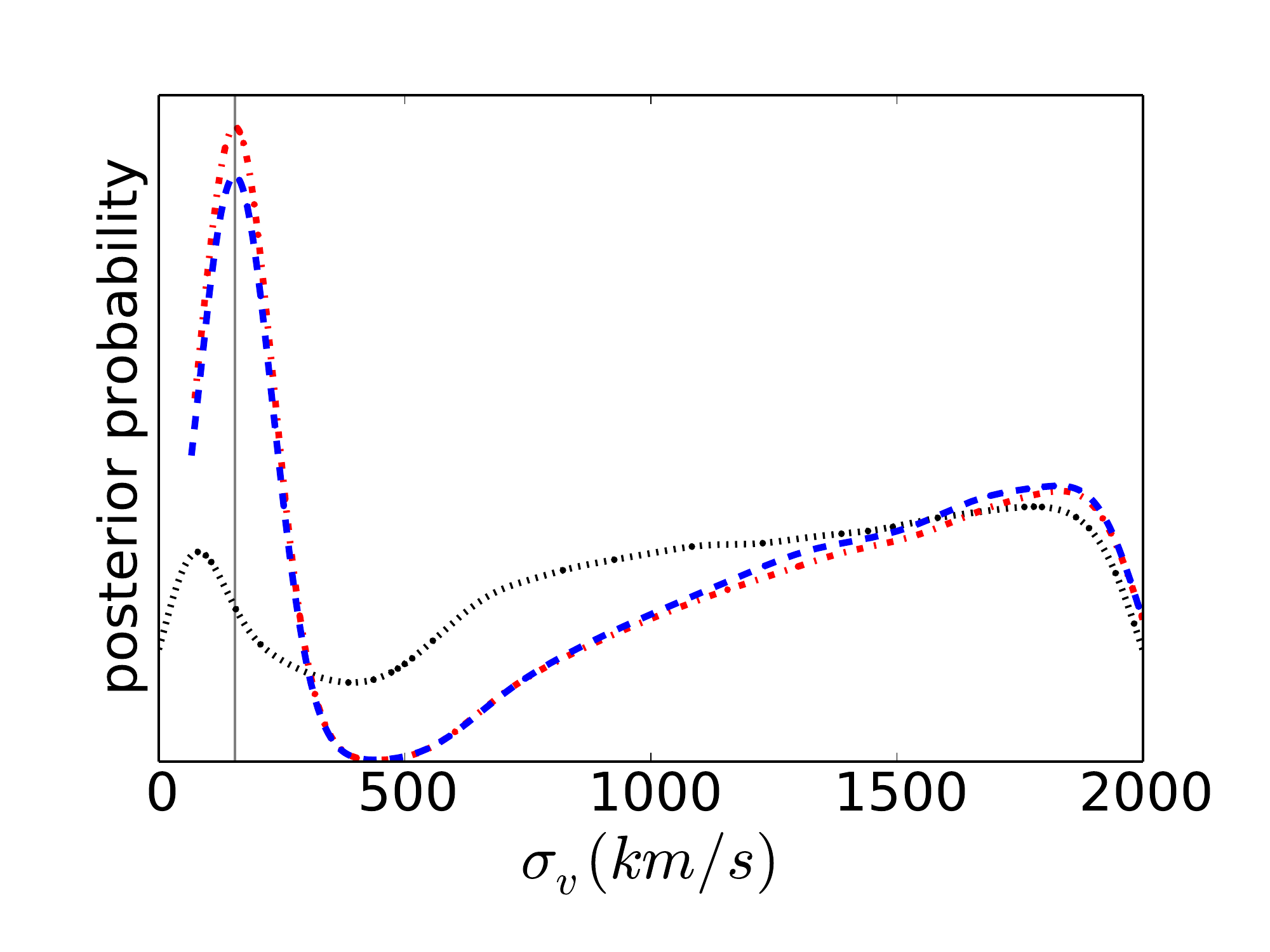}
  \caption{\label{fig:4trip}{\bf Marginalized one-dimensional posteriors for Benchmark 4 as a function of experiment ensemble.} The meanings of the lines are the same as in Fig. \ref{fig:2trip}.  Again, for all parameters but the velocity-related parameters, strong velocity priors are used.  For the velocity parameters, we use weak velocity priors.}
\end{figure*}

\begin{figure*}
  \includegraphics[width=0.32\textwidth]{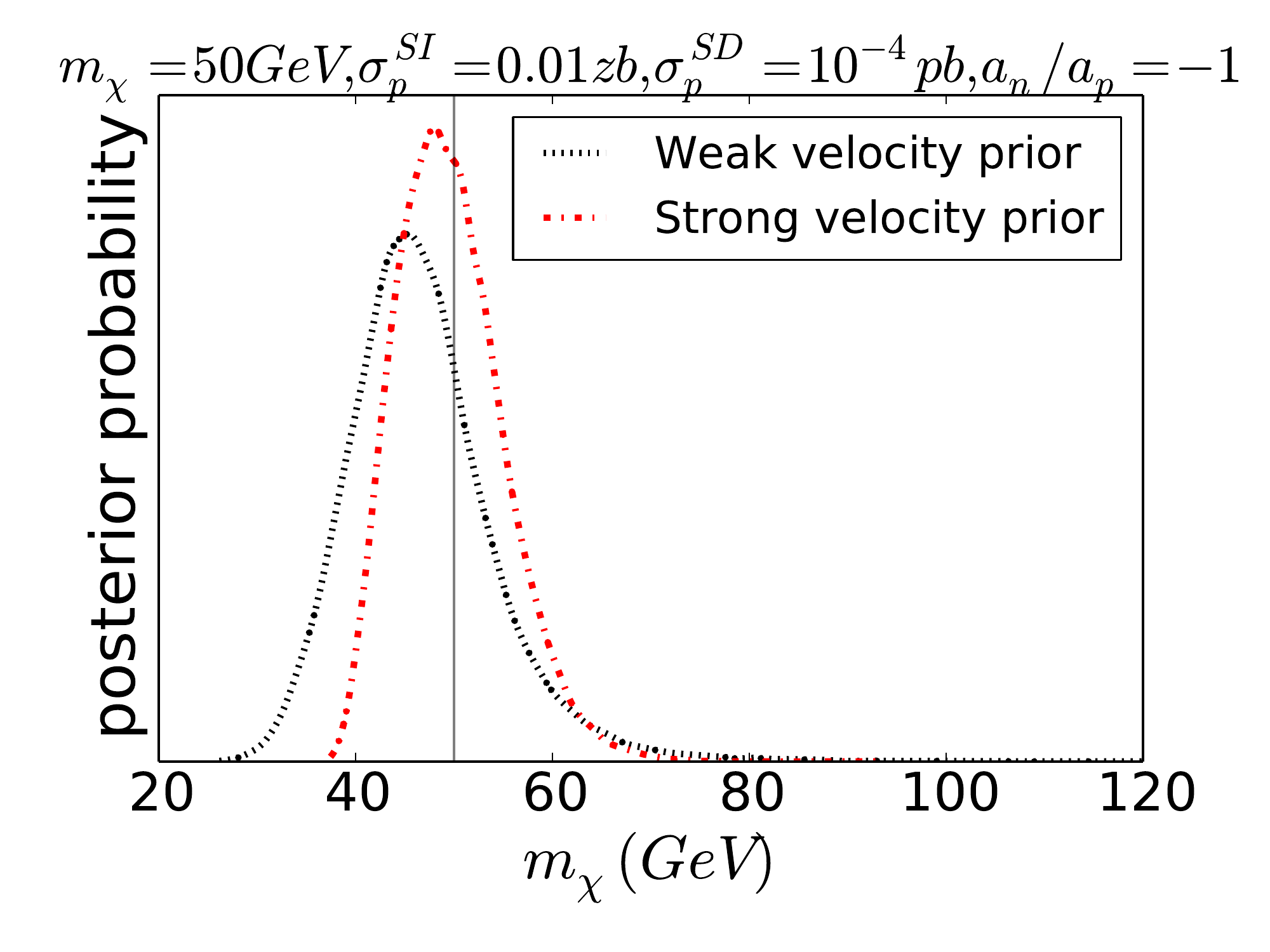} \includegraphics[width=0.32\textwidth]{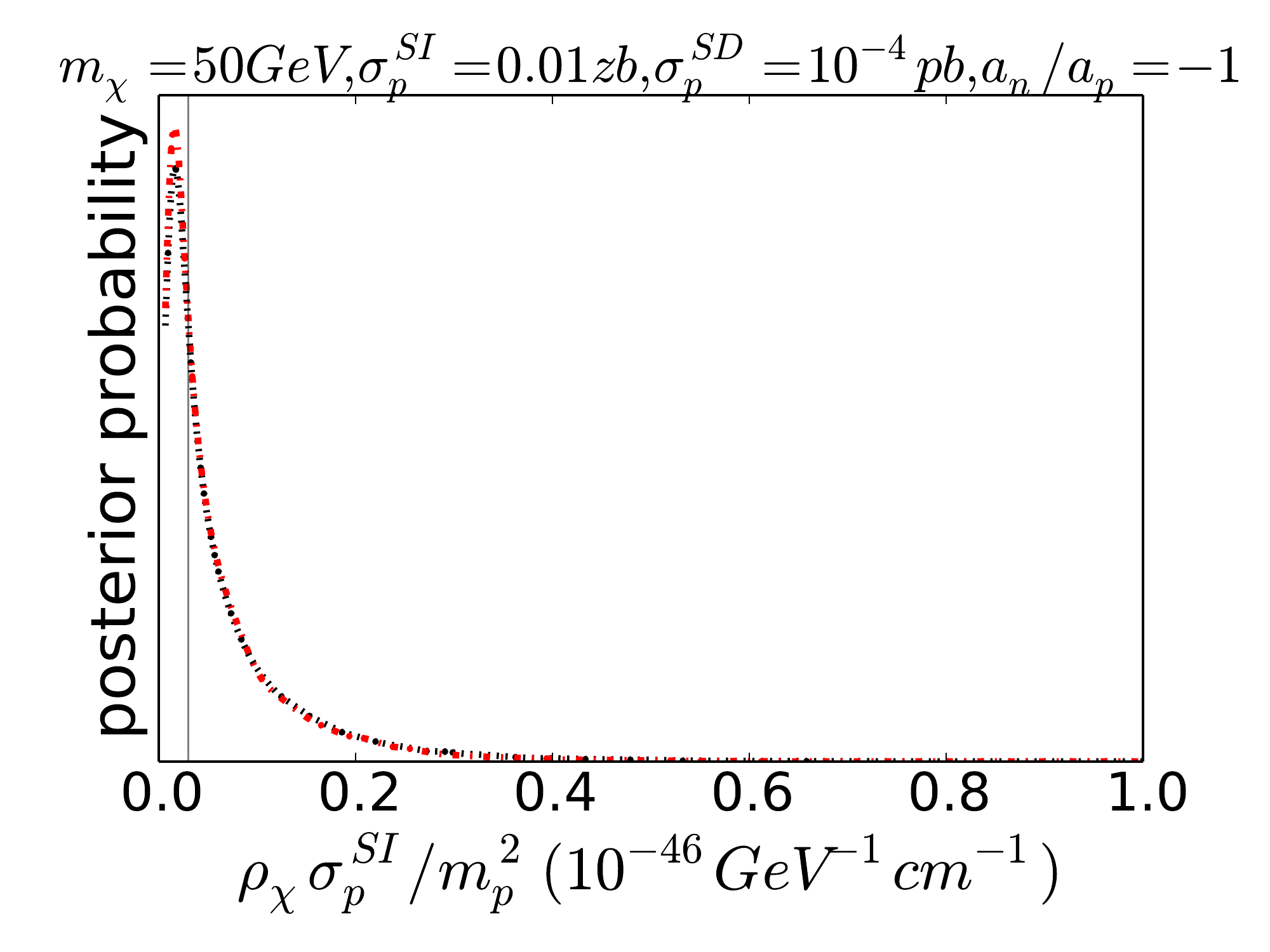} \\
  \includegraphics[width=0.32\textwidth]{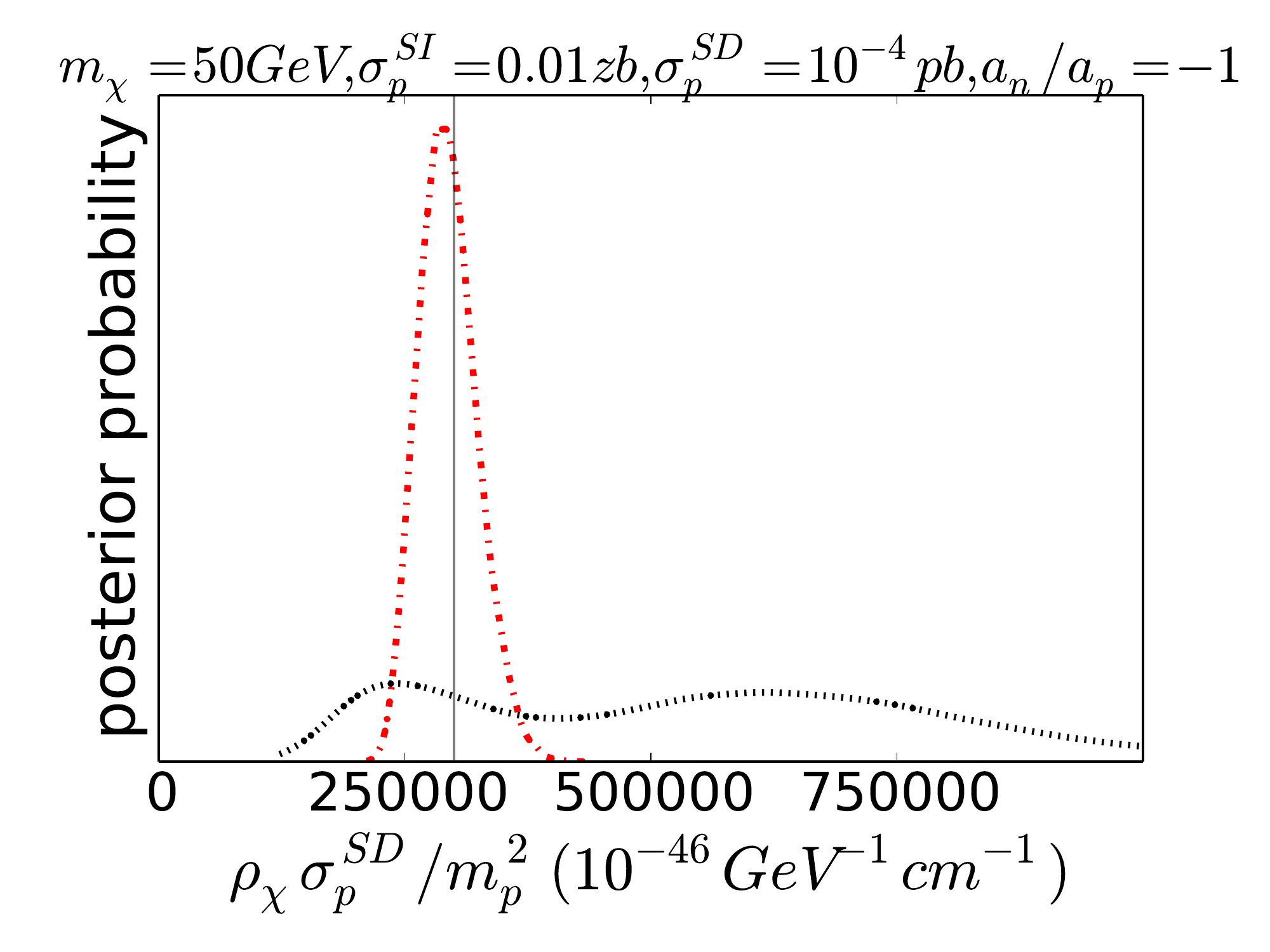} \includegraphics[width=0.32\textwidth]{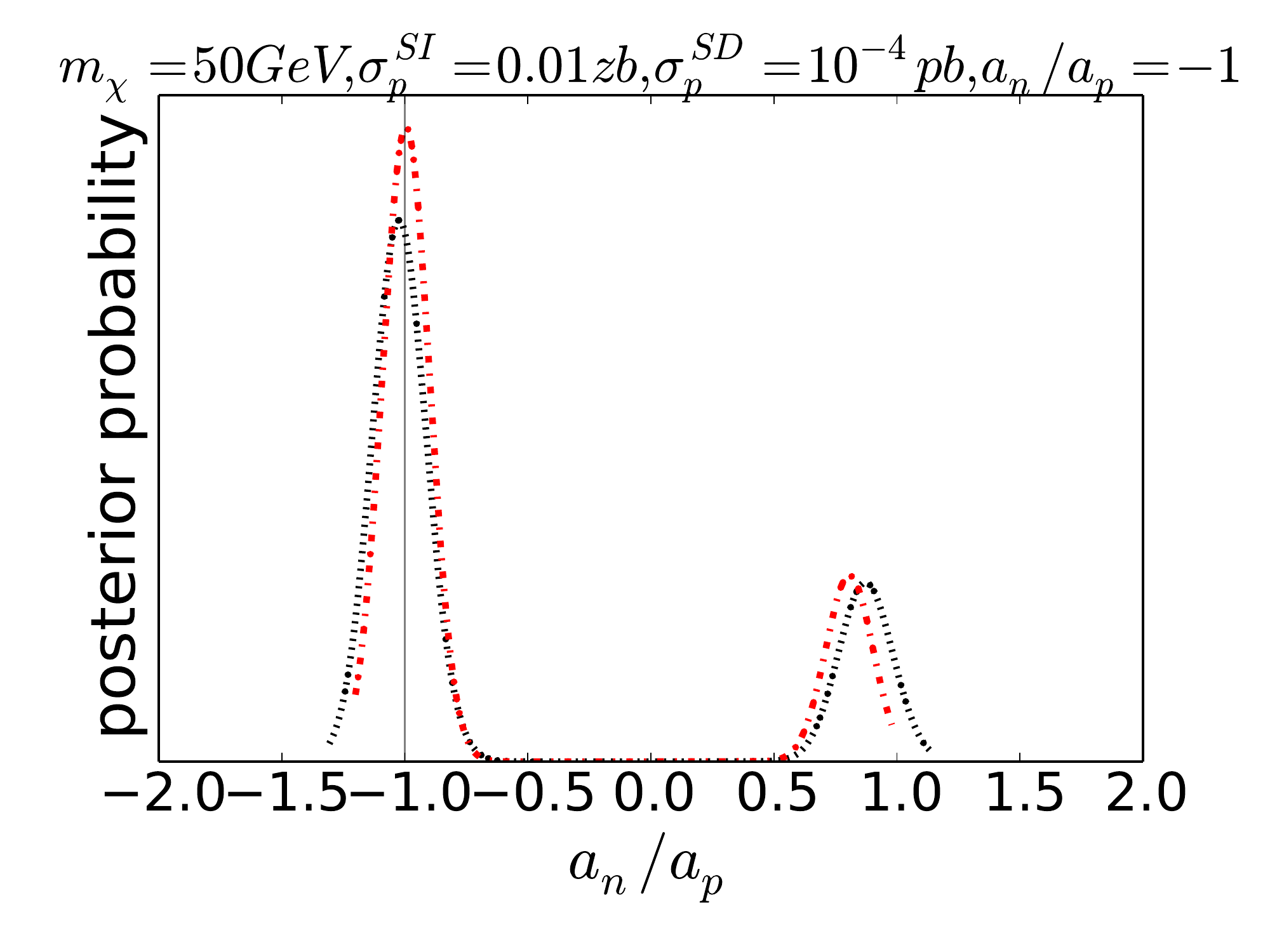}
  \caption{\label{fig:4doub}{\bf Marginalized one-dimensional posteriors for Benchmark 4 with weak or strong velocity priors.}  In this case, we use all five experiments for constraints.}
\end{figure*}

The situation is different for Benchmark 4, in which case we dial down the spin-independent cross section, such that only spin-dependent interactions dominate the recoil spectrum.  We illustrate our results in Figures \ref{fig:4trip} and \ref{fig:4doub}.  The lines on the plots have the same meanings as in Figures \ref{fig:2trip} and \ref{fig:2doub}.  In this case, the experiments do not meaningfully constrain the velocity distribution under the assumption of weak velocity priors, so the contraints on the WIMP physics parameters in Fig. \ref{fig:4trip} are more prior dominated.  While our priors for the velocity parameters are perhaps overly broad, they illustrate how important it is to get the model for the WIMP velocity distribution right in order to untangle data that are highly spin-dependent-interaction dominated.  This arises because Xenon and COUPP-500 are the experiments with the most events, but COUPP-500 does not have the energy resolution for this combination of experiments to yield better velocity-distribution constraints.  As we show in Sec.~\ref{sec:astrophysics}, one typically needs to get a number of events in at least three experiments with good energy resolution to get a good constraint on the WIMP mass.  In our case, because of the small event totals in several experiments, constraints are quite poor with only the Xenon, Argon, and Germanium experiments---for example, we only get a lower limit on the WIMP mass.

However, with the addition of the COUPP-500 and Silicon experiments, the constraints on \mwimp~are still reasonable (uncertainties of $\sim 20\%$) even with the weak velocity prior, although the posteriors are biased as a result of the large velocity prior volume.  Thus, even with the types of Generation-2 experiments planned, there ought to be reasonable sensitivity to WIMP masses.  Moreover, constraints on $a_n/a_p$ are similar to those achieved for Benchmark 2.

\begin{figure}[t]
  \includegraphics[width=0.32\textwidth]{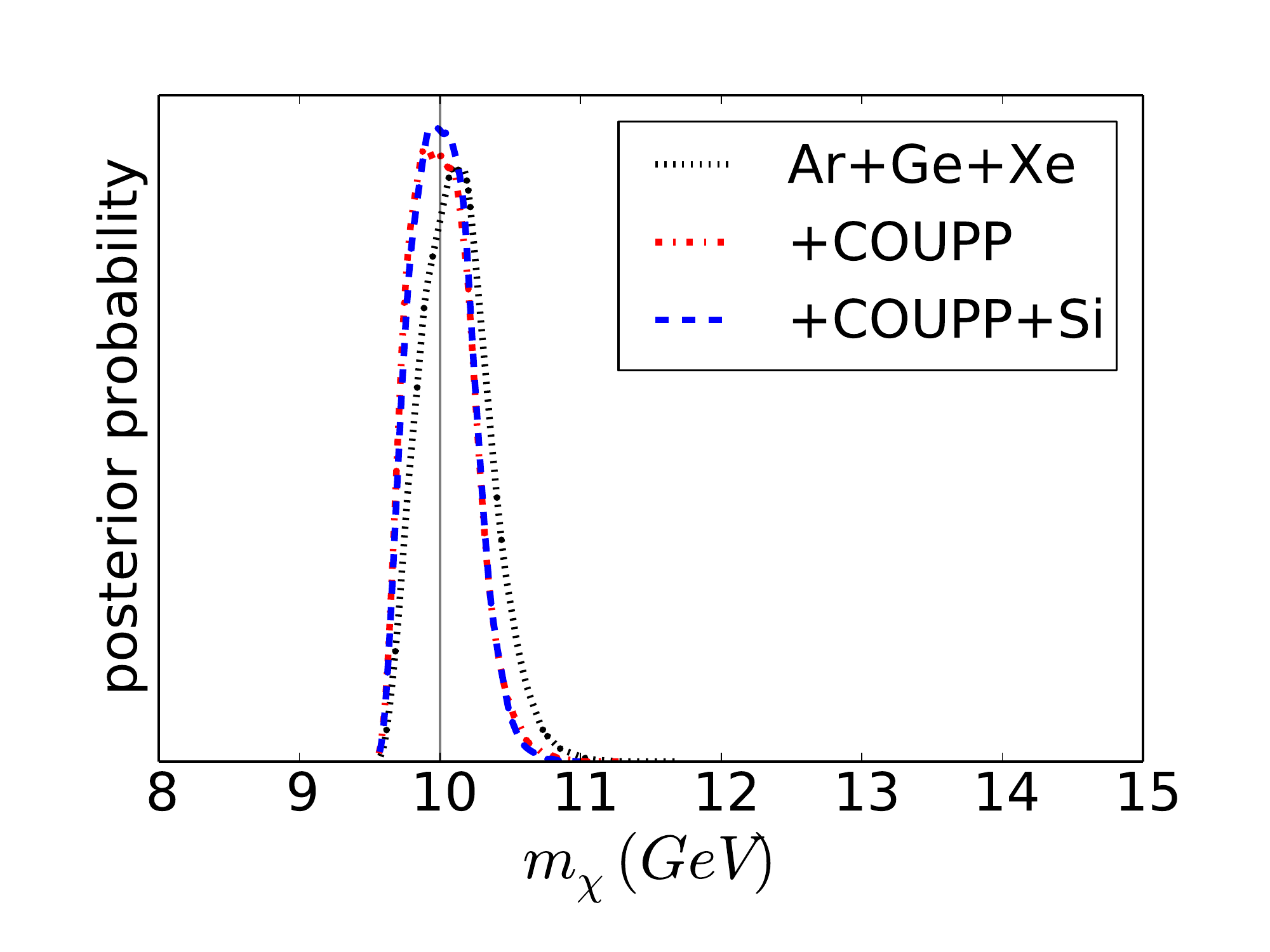} \\
  \includegraphics[width=0.32\textwidth]{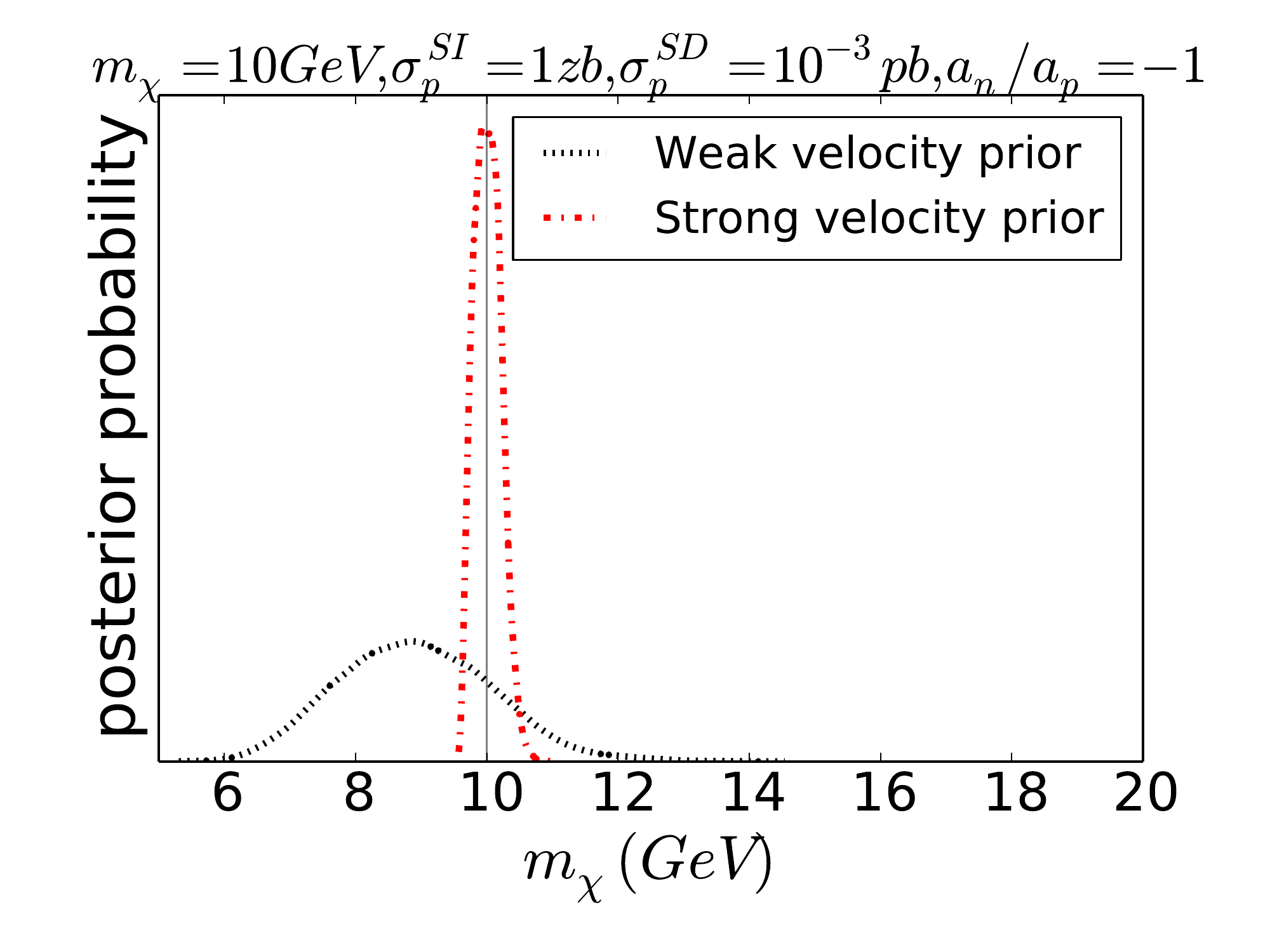}
  \caption{\label{fig:1mass}{\bf WIMP mass estimated with Benchmark 1 data.}  In the top panel, we show the marginalized WIMP mass posterior with three different sets of experiments.  We assume strong velocity priors.  We see that COUPP and Silicon narrow the posterior distribution.  However, the real driver of the narrowness of the posterior distribution is the velocity prior (bottom panel, all five experiments).}
\end{figure}

Next, we consider Benchmark 5, the case in which there is \emph{no} spin-dependent coupling to neutrons, only to protons.  In this case, we have similarly dialed down the spin-independent interactions to a negligible level.  The key trends we find for Benchmark 4 hold here: The big three experiments, Argon, Germanium, and Xenon do not say much; Germanium and Xenon are primarily sensitive to WIMP-neutron interactions.  COUPP saves the day with its sensitivity to WIMP-proton interactions, but only with strong velocity priors.  There is little sensitivity to the WIMP velocity distribution because of the poor energy resolution of COUPP, but the WIMP physics parameters (cross section, WIMP mass, $a_n/a_p$) are somewhat better constrained with the strong velocity prior.  Interestingly, $a_n/a_p$ is constrained to be in the range $-0.3$ to 0.1 even with weak velocity priors, and the spin-indepedent cross section is constrained to be vanishingly small.  However, constraints on the WIMP mass and $\rho_\chi$\sigmapsd$/m^2_\chi$ are extremely poor (i.e., highly prior-dominated) without a velocity prior.  Again, this is because of the (lack of) energy resolution in COUPP.

Finally, we explore constraints as a function of WIMP mass.  In Fig. \ref{fig:1mass}, we show constraints on \mwimp~for a benchmark \mwimp$=10$ GeV (Benchmark 1).  We show constraints as a function of experimental ensemble with strong velocity priors in the top panel, and the difference between strong and weak velocity priors on the five-experiment ensemble in the bottom panel.  The velocity prior dramatically affects the width of the posterior distribution for the WIMP mass.   We find that this is because, again, the velocity distribution is poorly constrained by the data, which translates into poor constraints on the WIMP mass and cross sections.  However, the uncertainties on $a_n/a_p$ are similar to those achieved by the Benchmark 2 experiments, and are relatively insensitive to the velocity prior.

The following are the key points from this section:
\begin{itemize}
  \item The parameter that is most robustly constrained, regardless of velocity distribution, is the absolute magnitude of the ratio of spin-dependent couplings to neutrons and protons, $|a_n/a_p|$.  This is robustly found only when COUPP-500, the only experiment we consider with strong couplings to protons, is considered.  Unfortunately, determining the sign of $a_n/a_p$ was not possible with any of our benchmarks.  Pato \cite{Pato:2011de} suggests that the sign may only be determined with several ton-years of exposure with several different odd-spin isotopes.
  \item Because COUPP is the only experiment we consider with strong spin-dependent WIMP-proton couplings, it greatly aids WIMP physics parameter estimation.  However, on account of its non-existent energy resolution, it does not help much with WIMP velocity distribution reconstruction if spin-dependent WIMP-proton couplings are large compared to other couplings.
  \item We achieve the best constraints on \mwimp, $\rho_\chi$\sigmapsi$/m_p^2$, $\rho_\chi$\sigmapsd$/m_p^2$, and the velocity parameters for moderate-mass ($\sim 50-200$ GeV) WIMPs with comparable spin-independent and spin-dependent interaction rates.  This is in part a consequence of our choice in the ensemble of experiments, and in part is a consequence of the kinematics of scatter.  It is important to have a diverse set of experiments, with different target isotopes, and with some energy resolution.  The poorest constraints came for Benchmark 1, our low-mass WIMP benchmark.  Unsurprisingly, experiments with low energy thresholds will be critical to unveiling the properties of low-mass WIMPs.
  \item The escape velocity from the Galaxy \vesc~is never constrained by the data, at least for our benchmark points.
\end{itemize}

\subsection{Effective operators}\label{sec:WIMPphys:effective}
\begin{table*}[htbp]
\begin{center}
    \begin{tabular}{ c | c  c  c  c  c  c  c  p{10cm} }
    \hline\hline
    Experiment &  Target material [A]& Target mass [kg]&Exposure [yr] &  Efficiency & Energy window [keV]  \\ \hline
    Xenon & Xe [131] & 1000 & 2 & 1/6 & 7-45  \\ \hline
    Germanium & Ge [73] & 100 & 2 & 1/3 & 8-100  \\ \hline
    C4 & Ge [73] & 100 & 1 & 1 & 0.001-3 \\ \hline
    \end{tabular}
    \caption{Key experimental parameters for benchmark experiments considered in Sec.~\ref{sec:WIMPphys:effective}.}\label{tab:experiments_operators}
\end{center}
\end{table*}
\begin{table}[htbp]
\begin{center}
    \begin{tabular}{ c | c  c  c  c  p{10cm} }
    \hline \hline
    Parameter & $h_1$ [GeV$^{-2}$] & $h_2$ [GeV$^{-3}$]&  $\ell_1$ [-] & $\ell_2$ [GeV$^{-1}$] \\ \hline
    Normalization & $10^{-9}$ & $10^{-9}$ & $10^{-9}$ & $10^{-8}$  \\ \hline
    Prior range & 0.1--100 & 0.1--100 & 0.01--20 & 0.01--20\\ \hline
    Fiducial value & 5 & 5 & 5 & 1  \\ \hline
    \end{tabular}
\caption{Normalizations, prior ranges, and fiducial values for the four cross-section parameters used to generate Asimov data analysed in Figs.~\ref{fig:singlepar_h1}, \ref{fig:singlepar_h2}, \ref{fig:singlepar_l1}, 
 and \ref{fig:2par} of Sec.~\ref{sec:WIMPphys:effective}.}\label{tab:priors_operators}
\end{center}
\end{table}
\begin{table}[htbp]
\begin{center}
    \begin{tabular}{  c | c  c  c  c  c  c  c  p{10cm} }
    \hline\hline
    Experiment &  $h_1$=5 & $h_2$=5& $\ell_1$=5& $\ell_2$=1\\ \hline
    Xenon & 129 & 142 & 12 & 37  \\ \hline
    Germanium & 17 & 17 & 3 & 6\\ \hline
    C4 & 4 & 0 & 80482 & 86 \\ \hline
    \end{tabular}
    \caption{Total number of events expected for each of the three benchmark experiments in Asimov data discussed in Sec.~\ref{sec:WIMPphys:effective}.}\label{tab:nevents_operators}
\end{center}
\end{table}
\begin{figure*}
\begin{center}
\includegraphics[height=3.3cm,keepaspectratio=true]{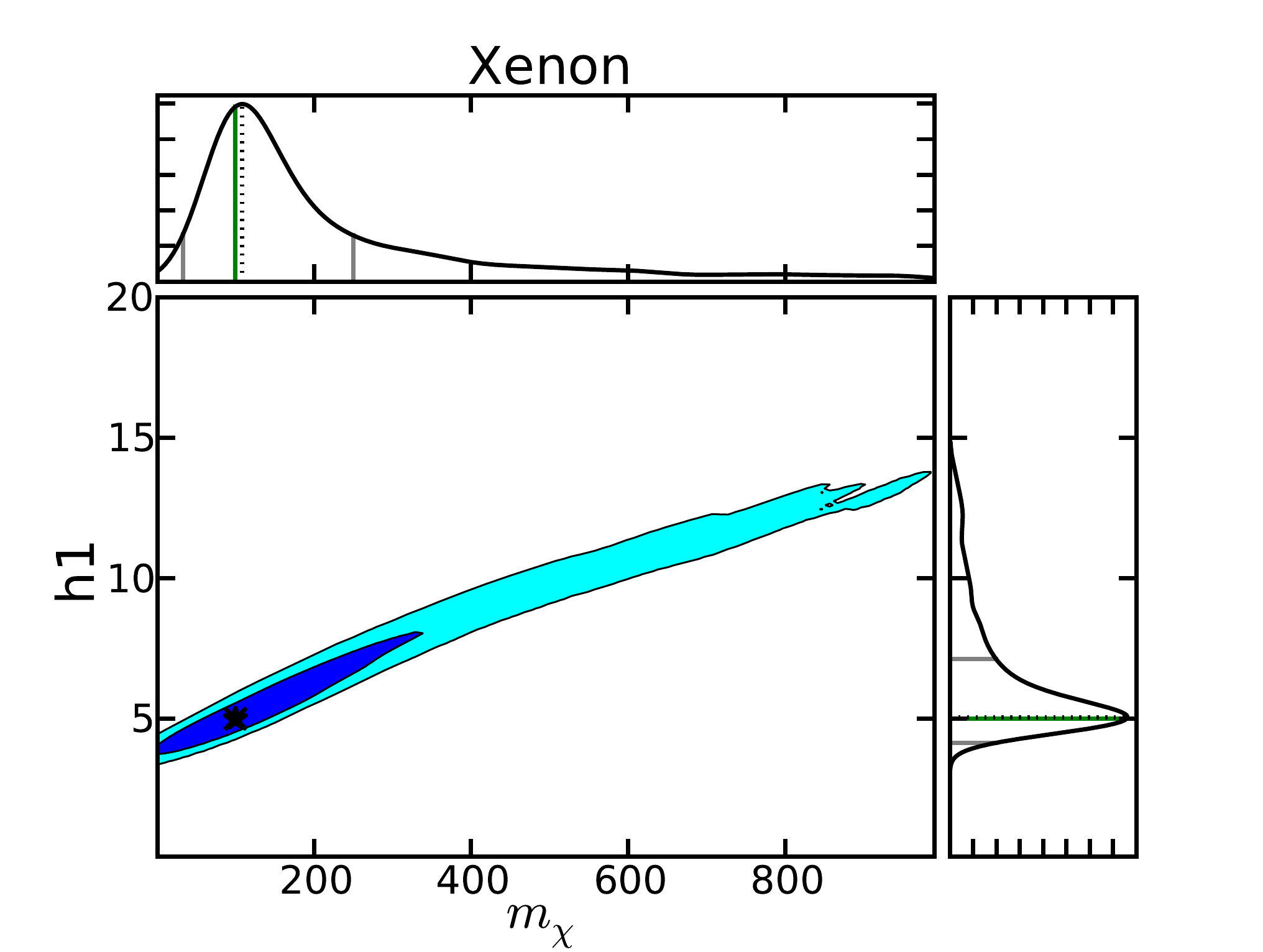}
\includegraphics[height=3.3cm,keepaspectratio=true]{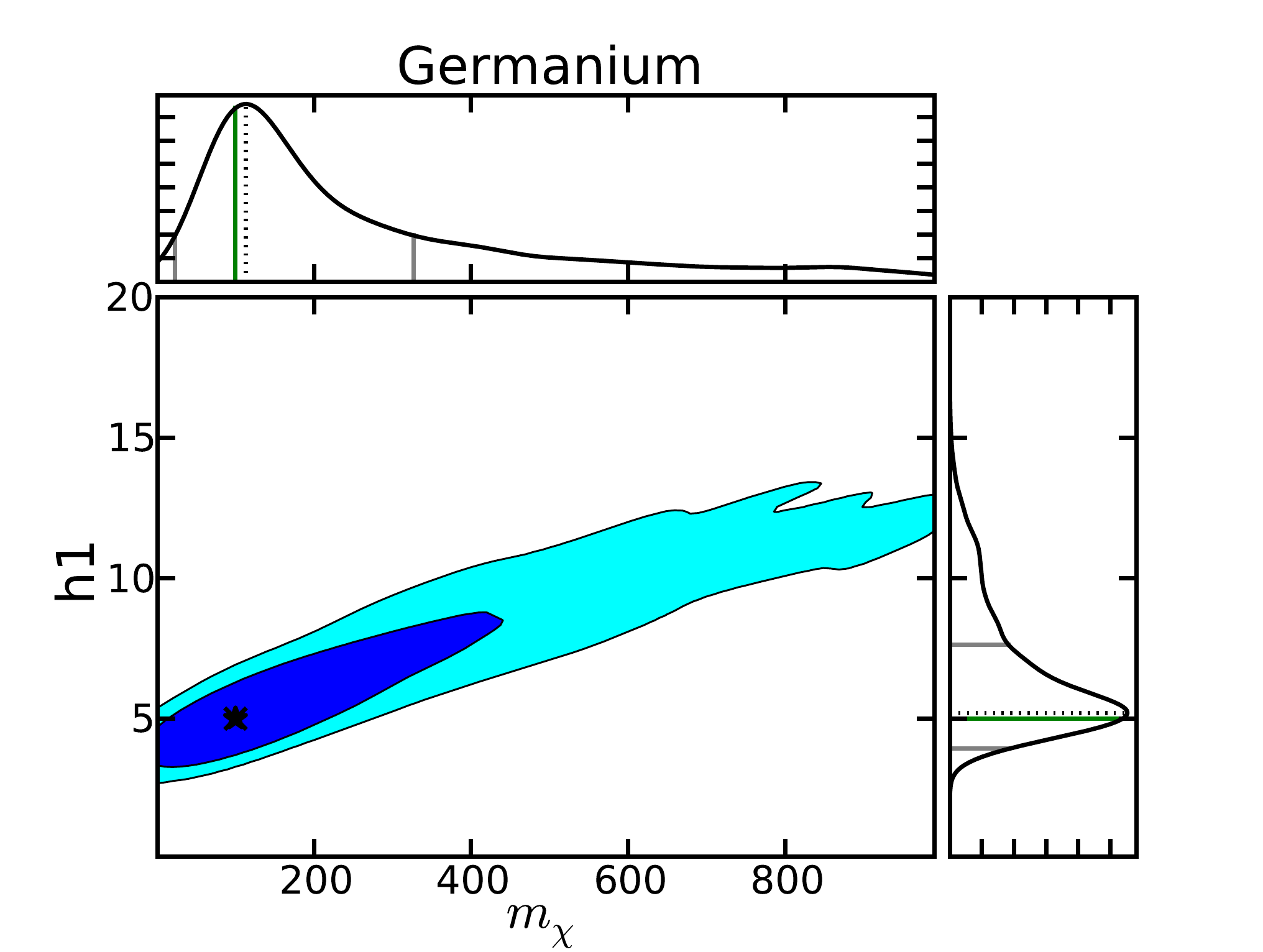}
\includegraphics[height=3.3cm,keepaspectratio=true]{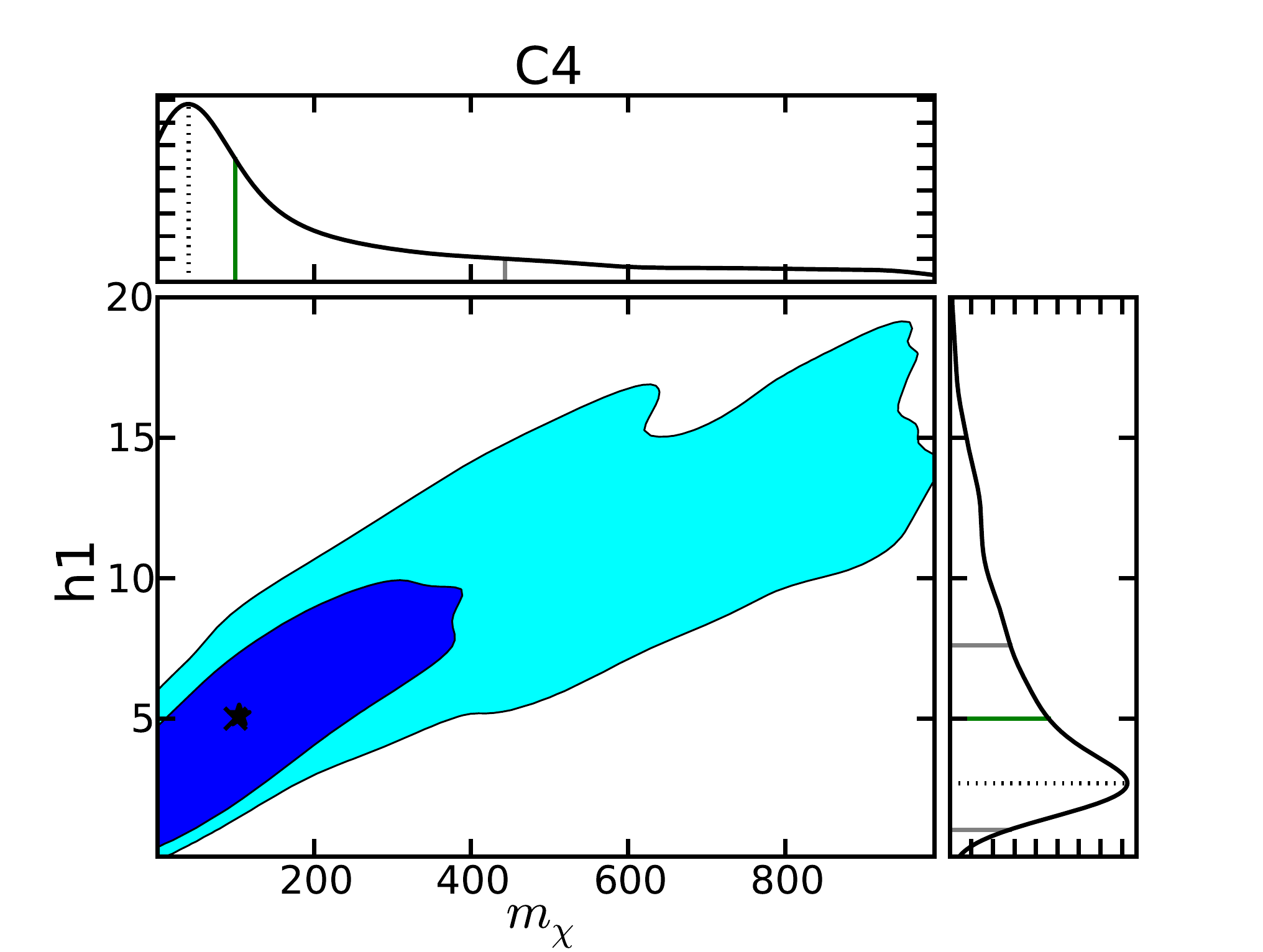}
\includegraphics[height=3.3cm,keepaspectratio=true]{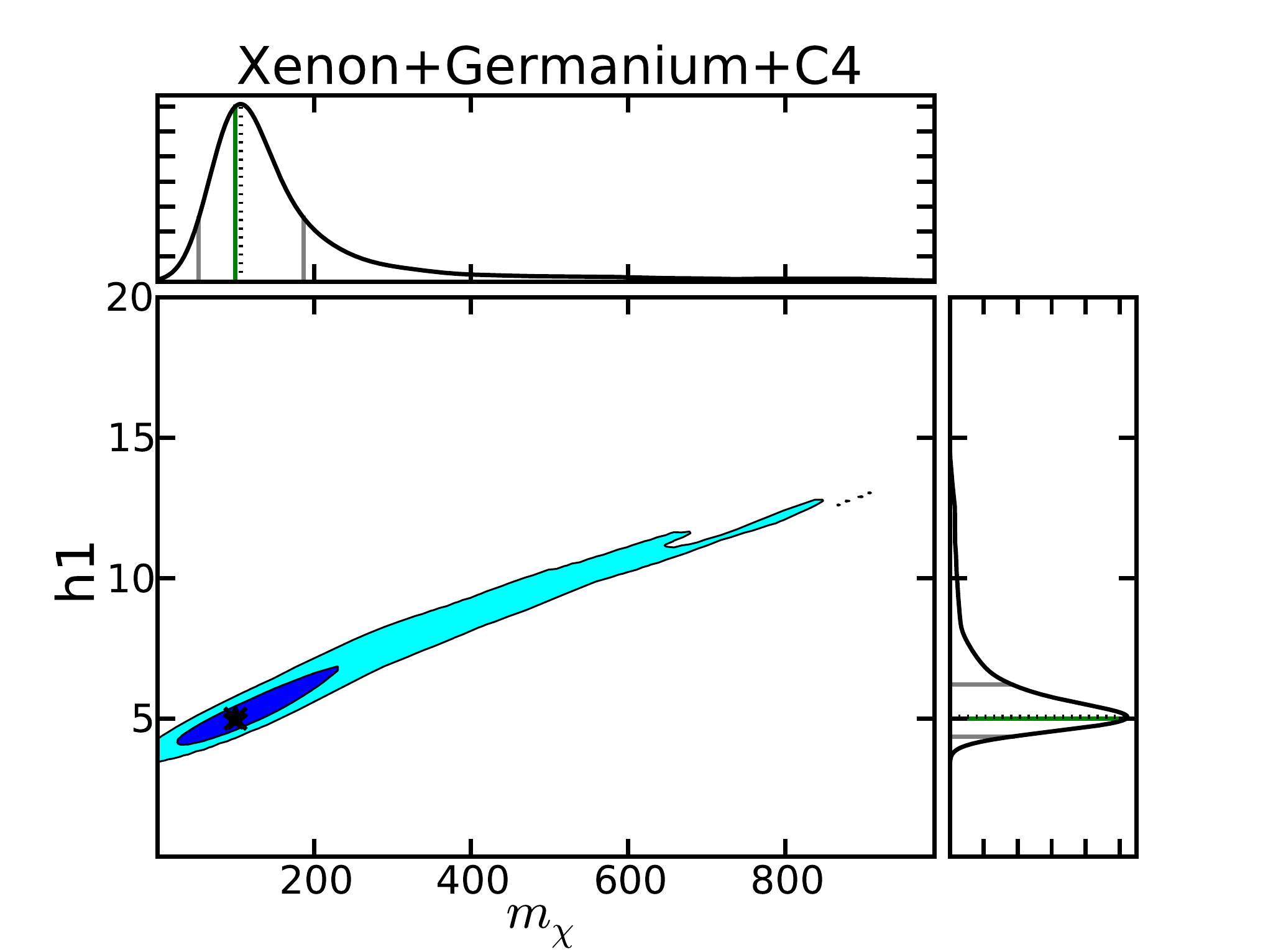}
\caption{{\bf Posterior probability distributions with the $68\%$ and $99\%$ confidence-level intervals, and corresponding marginal distributions, for the $h_1$ effective operator.} The posteriors are recovered from a simulation generated for three different experiments only with the $h_1$ operator. Free parameters are dark-matter particle mass and the cross-section parameter $h_1$ (notice that the normalization and units of $h_1$ are given in Table \ref{tab:priors_operators}). The input values for the simulation are marked with a cross, and the maximum-posterior values are marked with a star. In marginalized distributions, the green solid line marks the input value, the dotted black marks the maximum likelihood value, and the grey lines mark the $68\%$ confidence interval. The precision of the parameter estimation improves by a factor of a few when data from different experiments are combined, as shown in the right-most panel.\label{fig:singlepar_h1}}
\end{center}
\end{figure*}
\begin{figure*}
\begin{center}
\includegraphics[height=3.3cm,keepaspectratio=true]{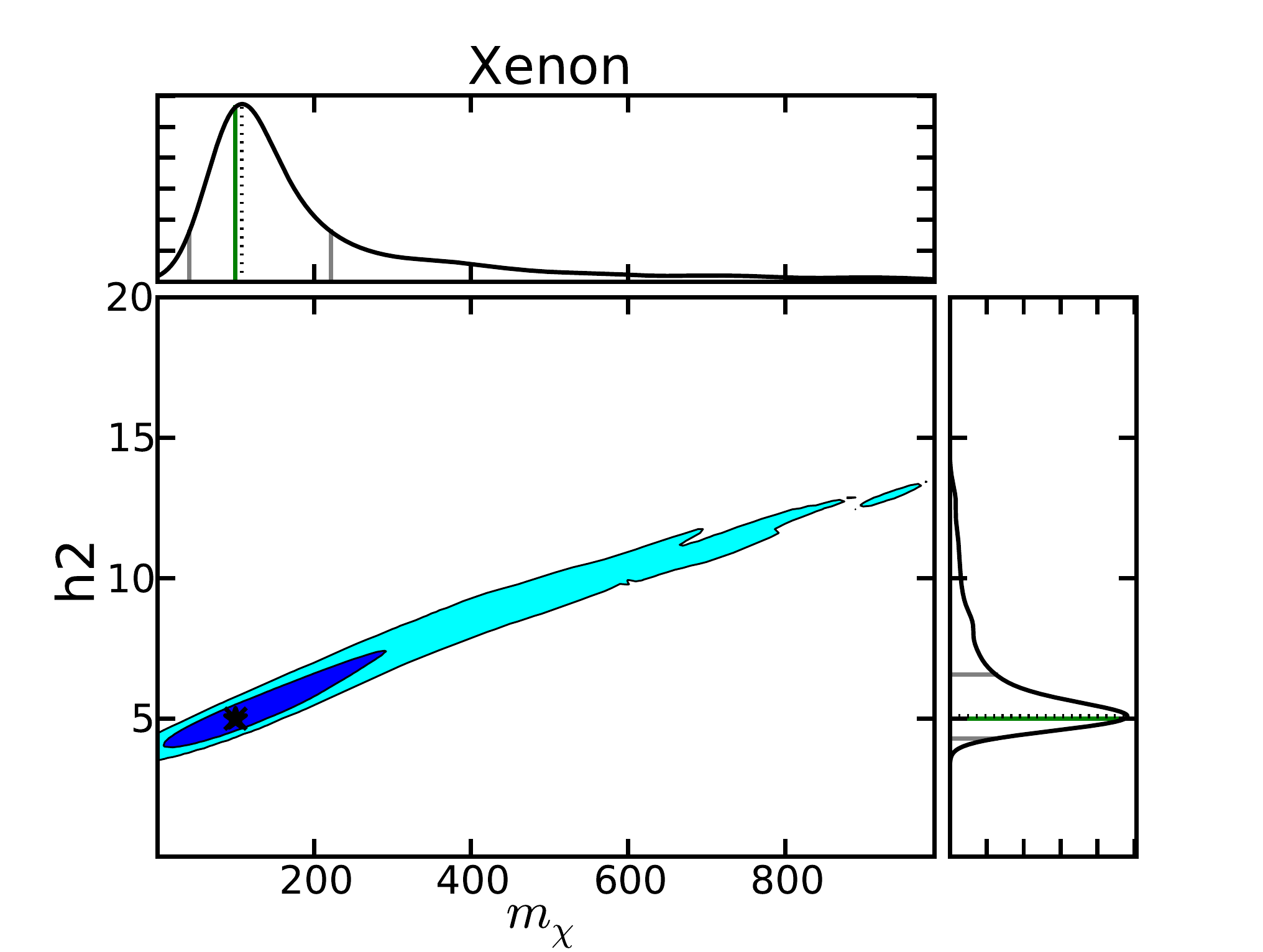}
\includegraphics[height=3.3cm,keepaspectratio=true]{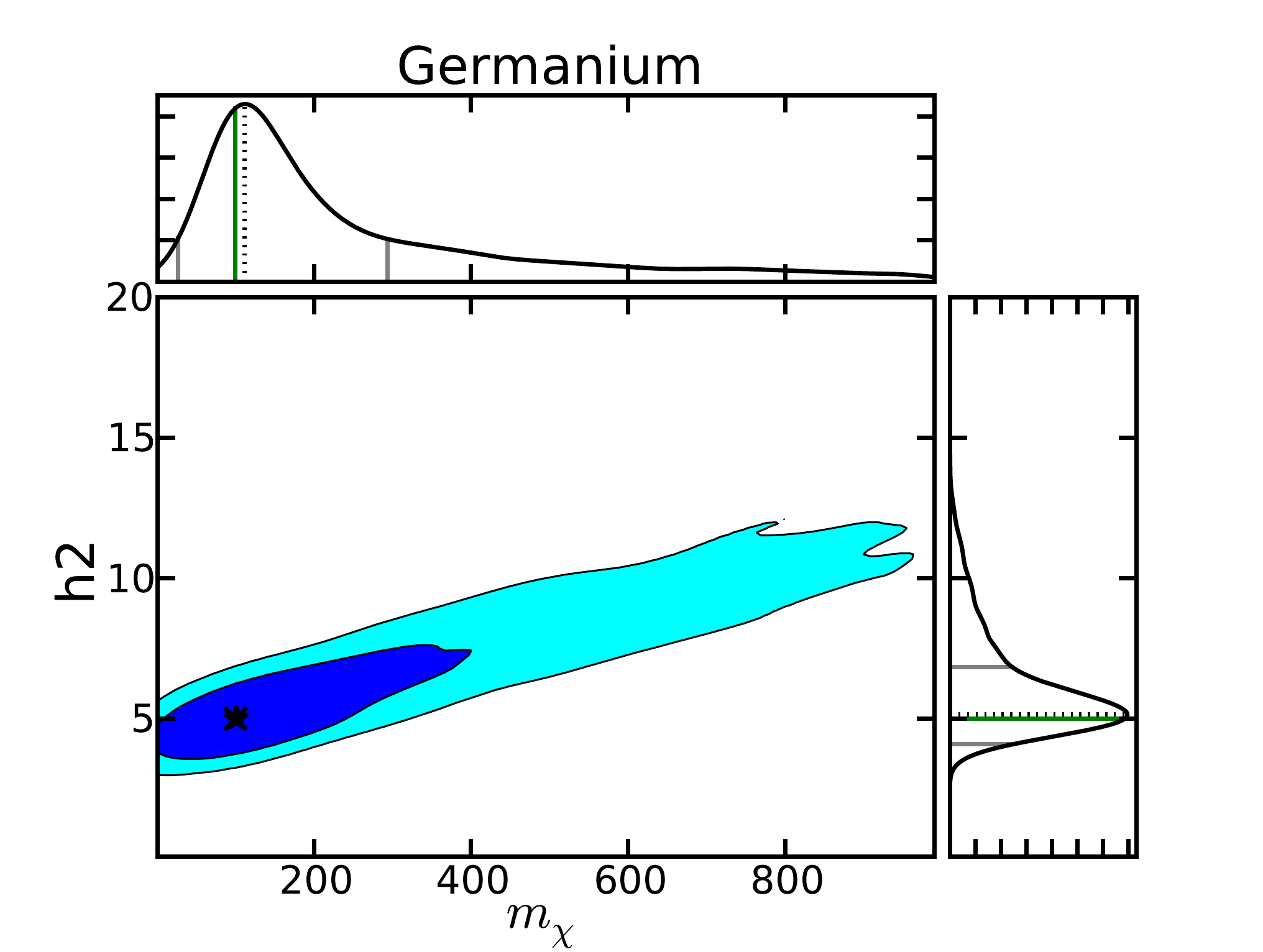}
\includegraphics[height=3.3cm,keepaspectratio=true]{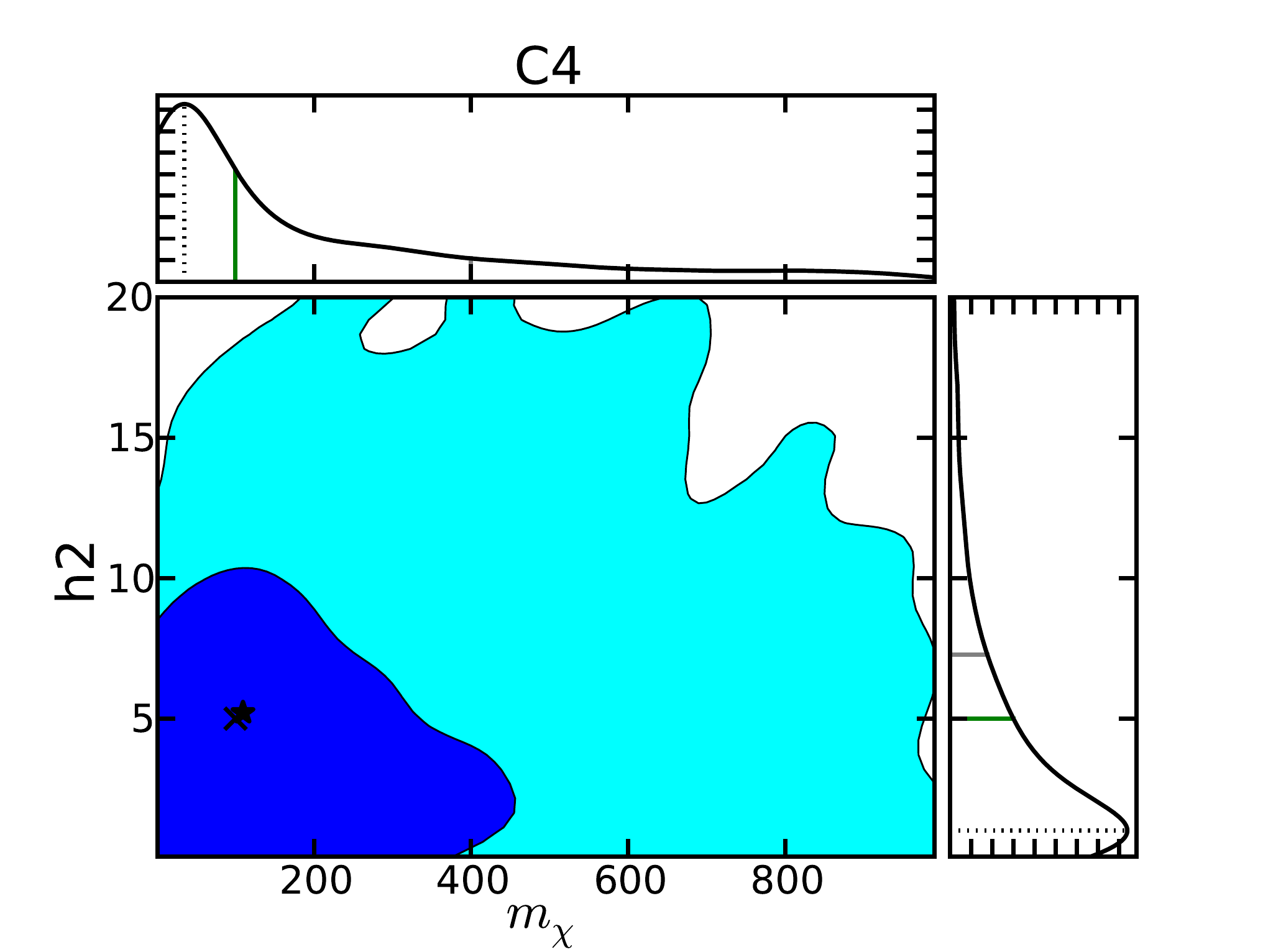}
\includegraphics[height=3.3cm,keepaspectratio=true]{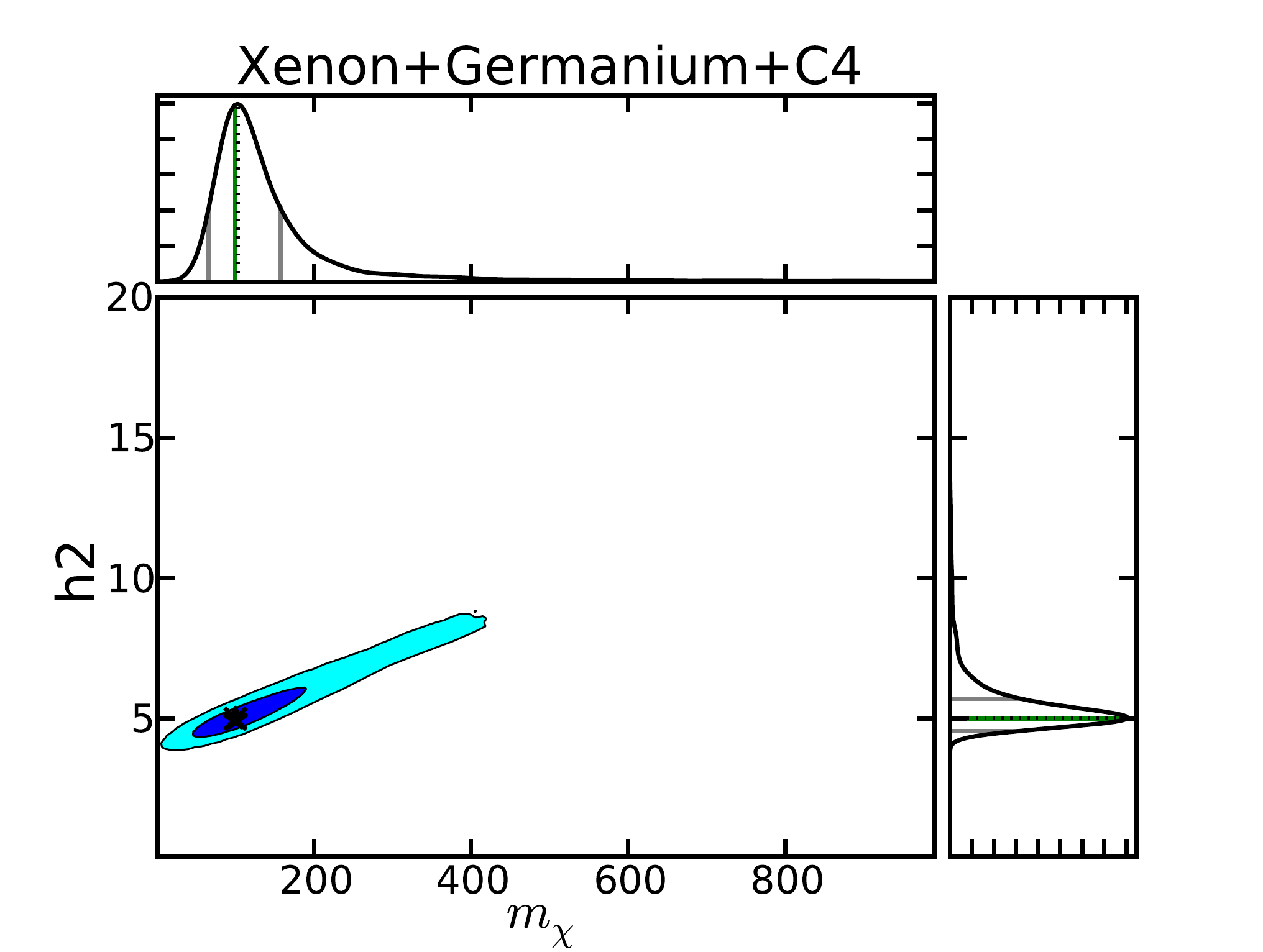}
\caption{{\bf Same as Fig.~\ref{fig:singlepar_h1}, but with $h_2$.} \label{fig:singlepar_h2}}
\end{center}
\end{figure*}
\begin{figure*}
\begin{center}
\includegraphics[height=3.3cm,keepaspectratio=true]{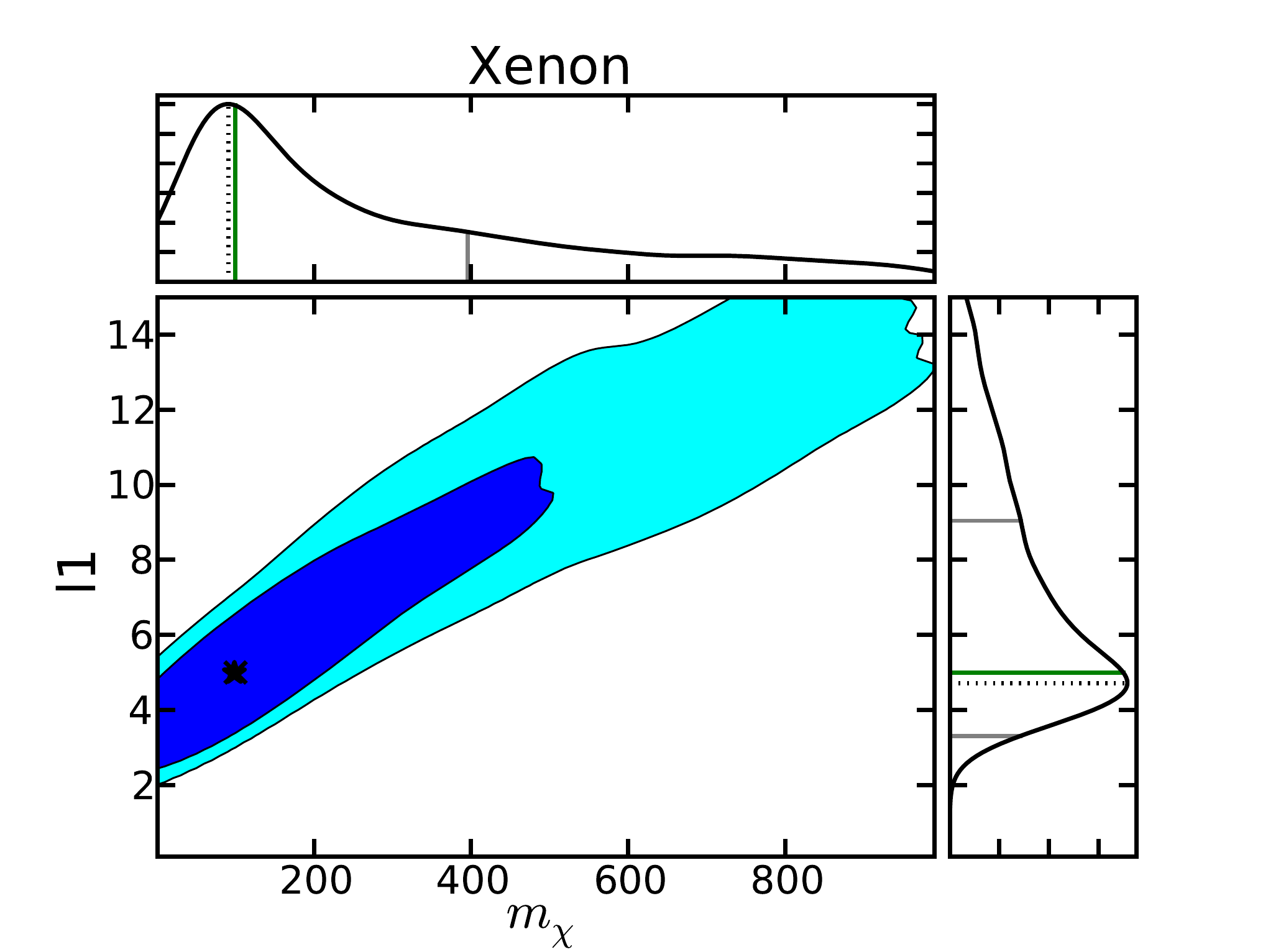}
\includegraphics[height=3.3cm,keepaspectratio=true]{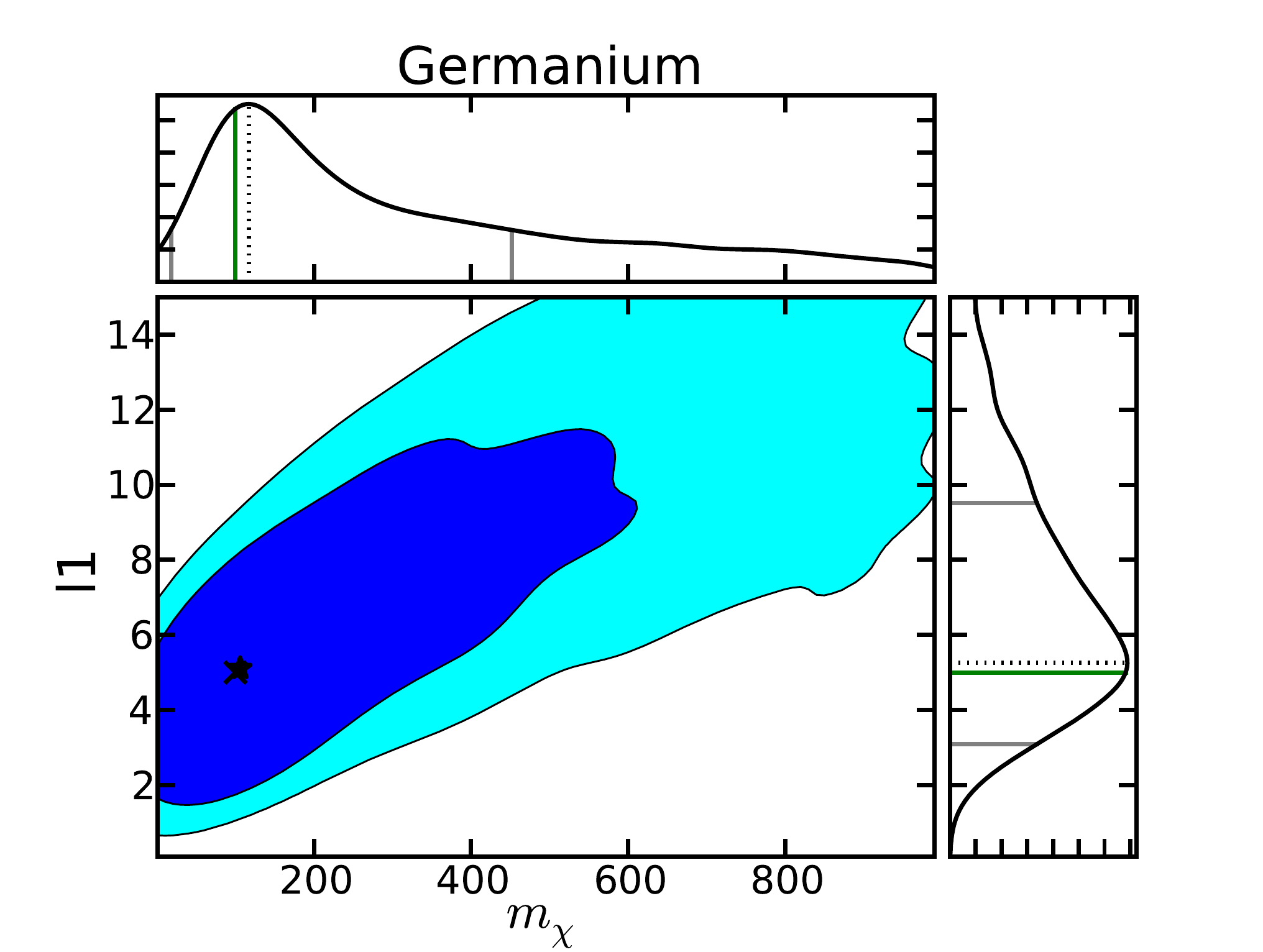}
\includegraphics[height=3.3cm,keepaspectratio=true]{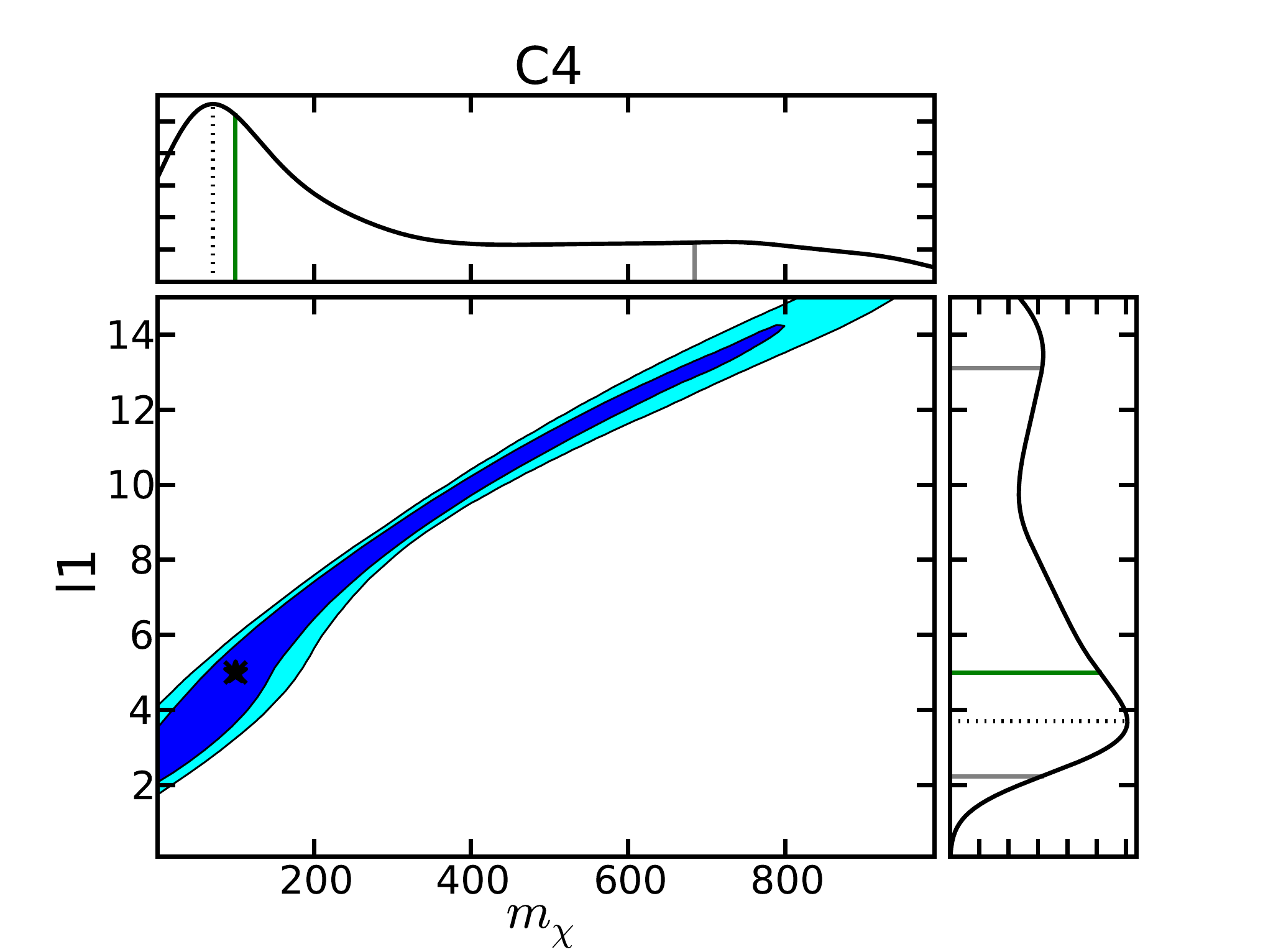}
\includegraphics[height=3.3cm,keepaspectratio=true]{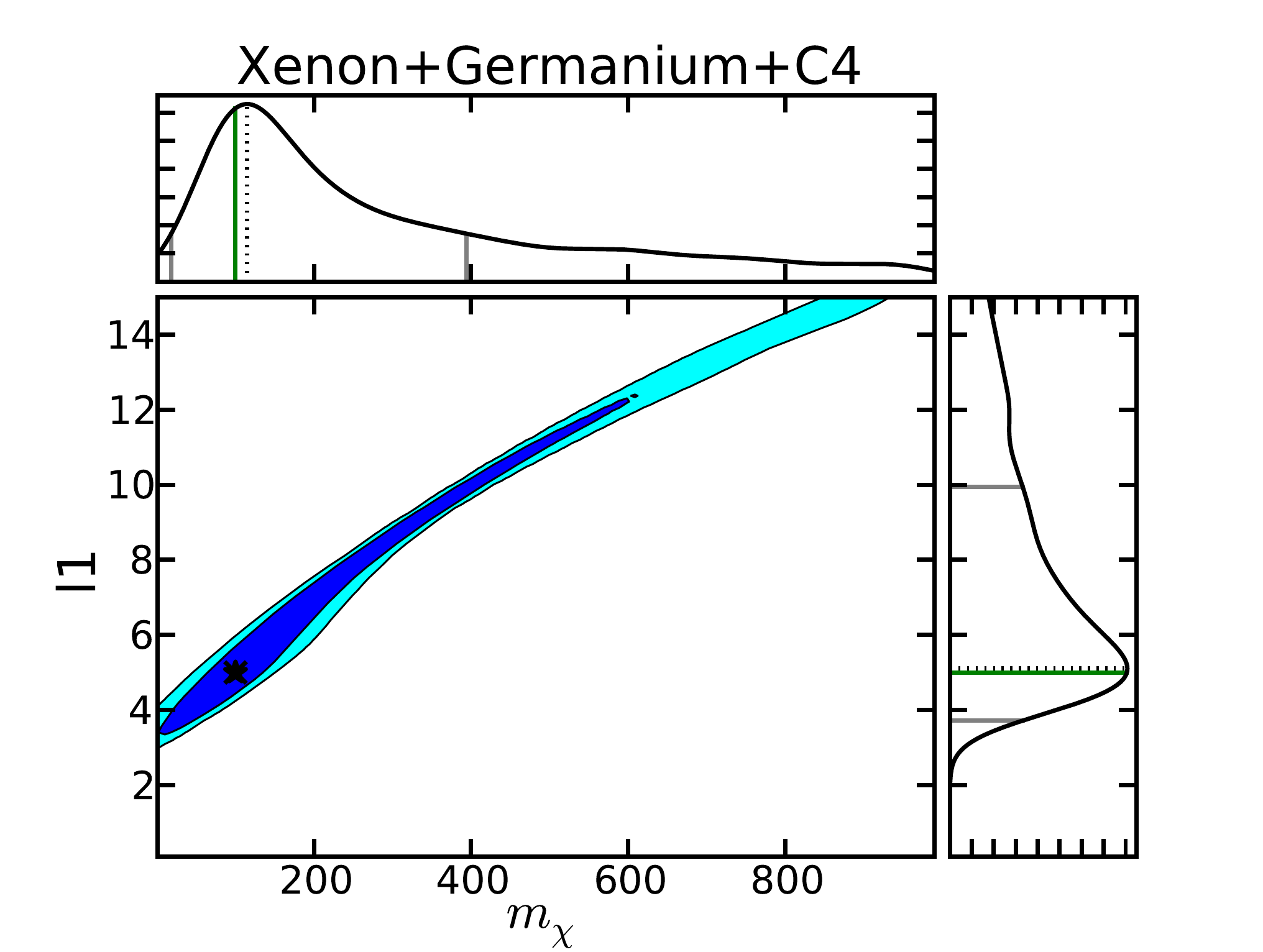}
\caption{{\bf Same as Fig.~\ref{fig:singlepar_h1}, but with $\ell_1$.}  \label{fig:singlepar_l1}}
\end{center}
\end{figure*}
In this section, we address the question of how well the different underlying physical scenarios discussed in Sec.~\ref{sec:theory:effective} can be reconstructed with direct detection. 
For the most part, this section summarizes points of Ref.~\cite{gluscevic2013}, but we also wish to point the reader to Ref.~\cite{McDermott:2011hx} for a similar discussion. 

To evaluate theoretical limitations for Wilson-coefficient estimation with upcoming data, we consider three benchmark experiments described in Table \ref{tab:experiments_operators}, with energy-independent efficiency (within the experimental energy window), perfect energy resolution, and neglect backgrounds and Poisson noise.  We add an experiment to this section that is not included in other sections, namely an ultra-low-threshold and ultra-low-background germanium experiment inspired by C4 or CDMSlite \cite{collar2013b,Agnese:2013lua}.  An ultra-low threshold experiment is necessary in order to resolve the sharply falling energy spectra of operators $\ell_1$ and $\ell_2$ (Eq. (\ref{eq:Veff}) and Fig. \ref{fig:operators}).

We generate Asimov data using Eq.~(\ref{eq:dRdQ_operators}) with Wilson-coefficient values given by normalizations times the fiducial values listed in Table \ref{tab:priors_operators}. Each of our simulations was generated using only one of the scattering operators, while the rest of the coefficients are set to zero. The normalizations for the coefficients are chosen so as to produce about 1 event per year of exposure for experiment denoted as Xenon in Table \ref{tab:experiments_operators}
; for $h_1$, the fiducial value of $5\times10^{-9}\,$GeV$^{-2}$ at $m_\chi$=100~GeV corresponds to a cross-section of about \sigmapsi$\approx 10^{-45}$\,cm$^{-2}$. For the purposes of this section, we only simulate scenarios with $m_\chi$=100~GeV, and assume a Maxwellian velocity distribution, with the same fiducial astrophysical parameters as in Sec.~\ref{sec:WIMPphys:sd}.  
This velocity distribution is fixed in the analysis as well; we assume perfect knowledge of the WIMP velocity distribution.  We explore the dependence of the constraints on uncertainties in the WIMP velocity distribution in Ref. \cite{gluscevic2013}.  In this work, we analyze each data set separately, and also combine data from the three experiments. 

We are interested in the posterior probability distributions for the dark matter particle mass $m_\chi$ and a subset of the four Wilson coefficients $h_1$, $h_2$, $\ell_1$, and $\ell_2$. We allow the chosen coefficients to vary in the prior range given in Table \ref{tab:priors_operators} for the Wilson coefficients, and we explore mass prior range between 1 and 1000 GeV; we assume log-flat priors for all parameters.  
As described in Sec.~\ref{sec:simulation}, each of our likelihood functions is calculated per energy bin.  We choose a width of about $1$ keV for each bin (such that the number of bins depends on the span of the energy window), except for the case of ``C4'', where we use 20 bins.  We perform parameter estimation using \textsc{MultiNest} using \texttt{tol = 0.001} and 1000 live points.

As the first step, we perform parameter estimation for the Wilson coefficient corresponding to the operator that was used to generate the mock data at hand. This step explores how well the data can measure the value of a given Wilson coefficient, supposing we assume the right underlying effective operator. Illustrations of the recovered posteriors, including marginalized posteriors and $68\%$ and $99\%$ confidence-level intervals, are presented in Figs.~\ref{fig:singlepar_h1}, \ref{fig:singlepar_h2}, and \ref{fig:singlepar_l1}.
The constraints for $\ell_2$ show almost exactly the same morphology as for $\ell_1$, so we only show the results for $\ell_1$.  The input values for a given parameter are marked with a cross, and the recovered values (the values corresponding to the global peak of the posterior) are marked with a star. In corresponding marginalized distributions, the input values and the maximum-likelihood values are also denoted. We see that, when Xenon detects a dozen to a hundred events, the measurements tend to be quite accurate---the estimated parameter values are typically not biased. On the other hand, there remains a large degeneracy between the mass and the cross-section parameters, so the estimation has a limited precision---the 1$\sigma$ confidence-level intervals are relatively broad, even in the case where we assume the correct underlying operator. However, the precision of parameter estimation is significantly improved (in some cases by a factor of a few) when data sets from different experiments are combined. As we discuss throughout this paper, using different and complementary targets has a big impact on breaking degeneracies and improving precision in direct-detection data sets. In particular, due to a sharp fall-off of the recoil rate at high recoil energies in the case of some operators (such as those representing interactions with light mediators), experiments with very low energy thresholds (exemplified here with ``C4'') are complementary to those that probe broad energy windows, such as Xenon and Germanium.  

\begin{figure*}[t]
\begin{center}
\includegraphics[height=5cm,keepaspectratio=true]{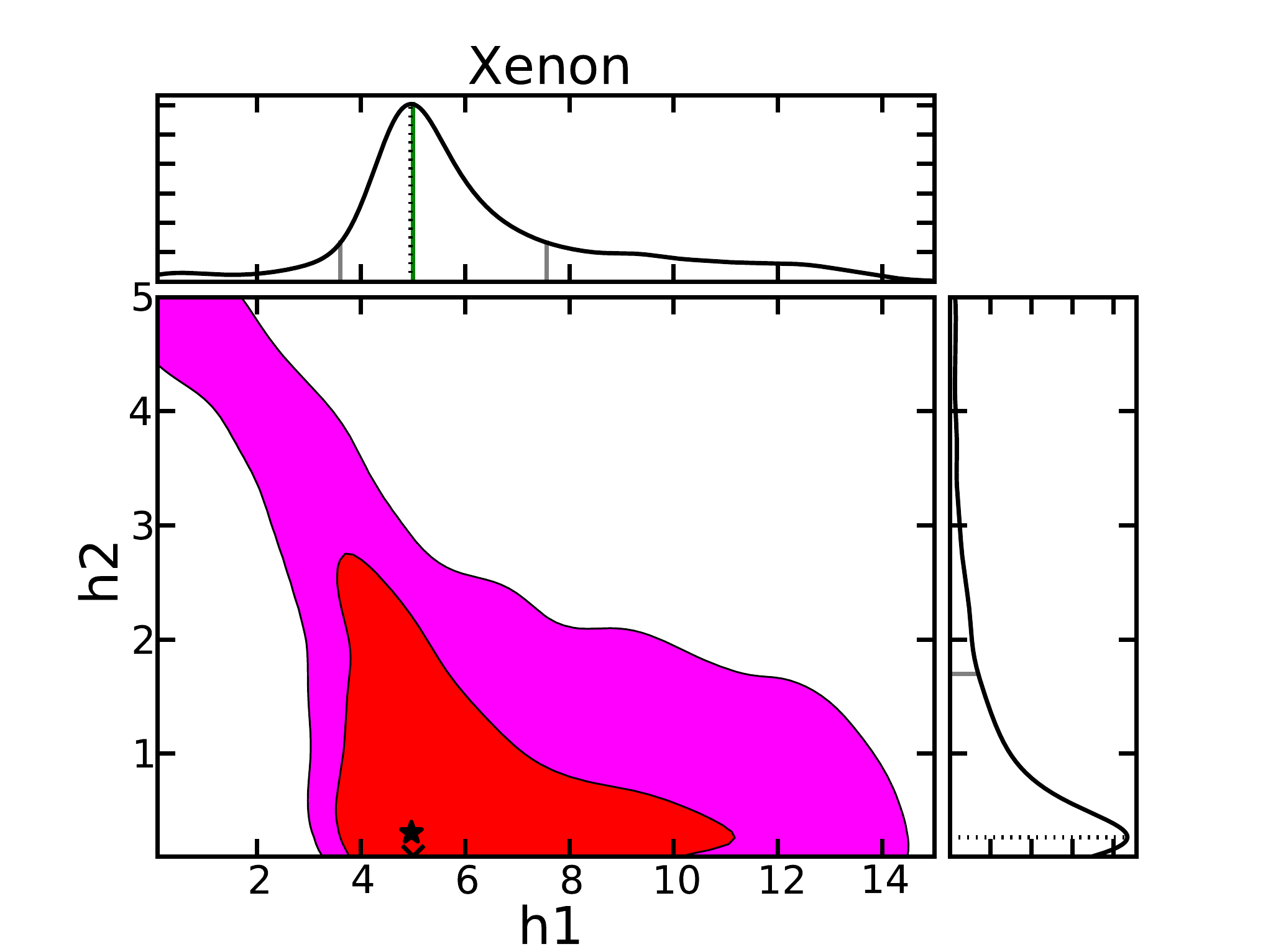}
\includegraphics[height=5cm,keepaspectratio=true]{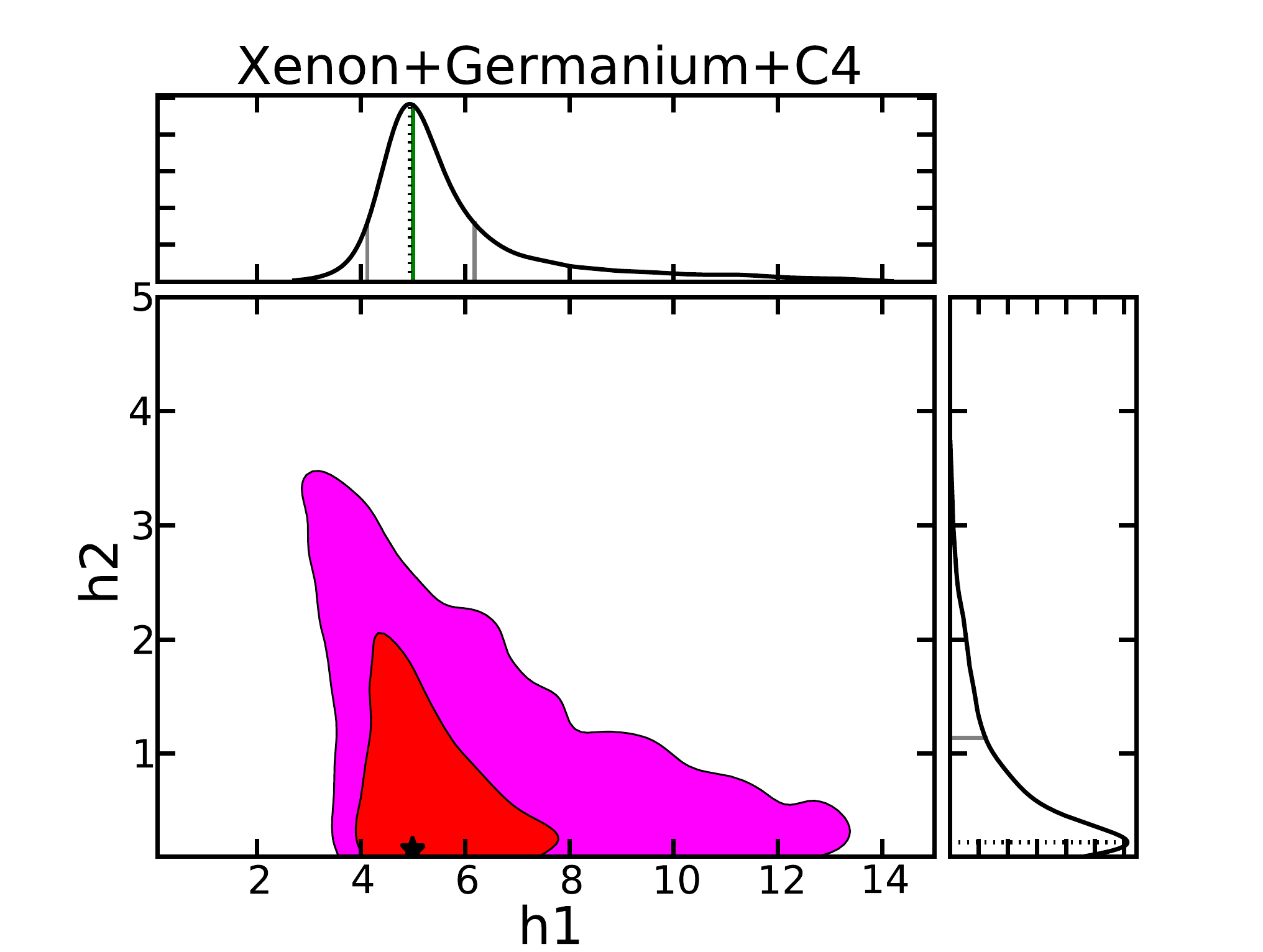}
\includegraphics[height=5cm,keepaspectratio=true]{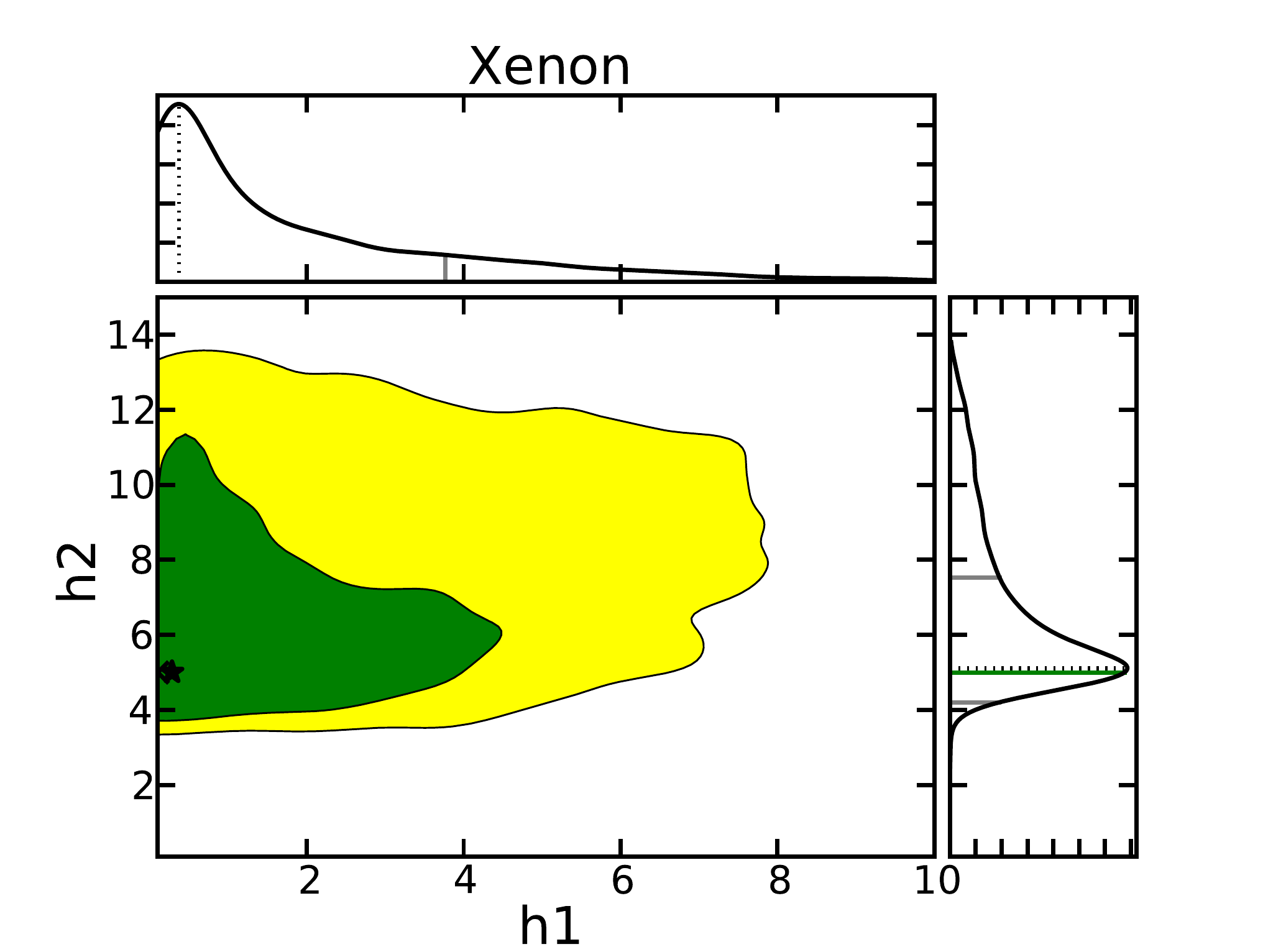}
\includegraphics[height=5cm,keepaspectratio=true]{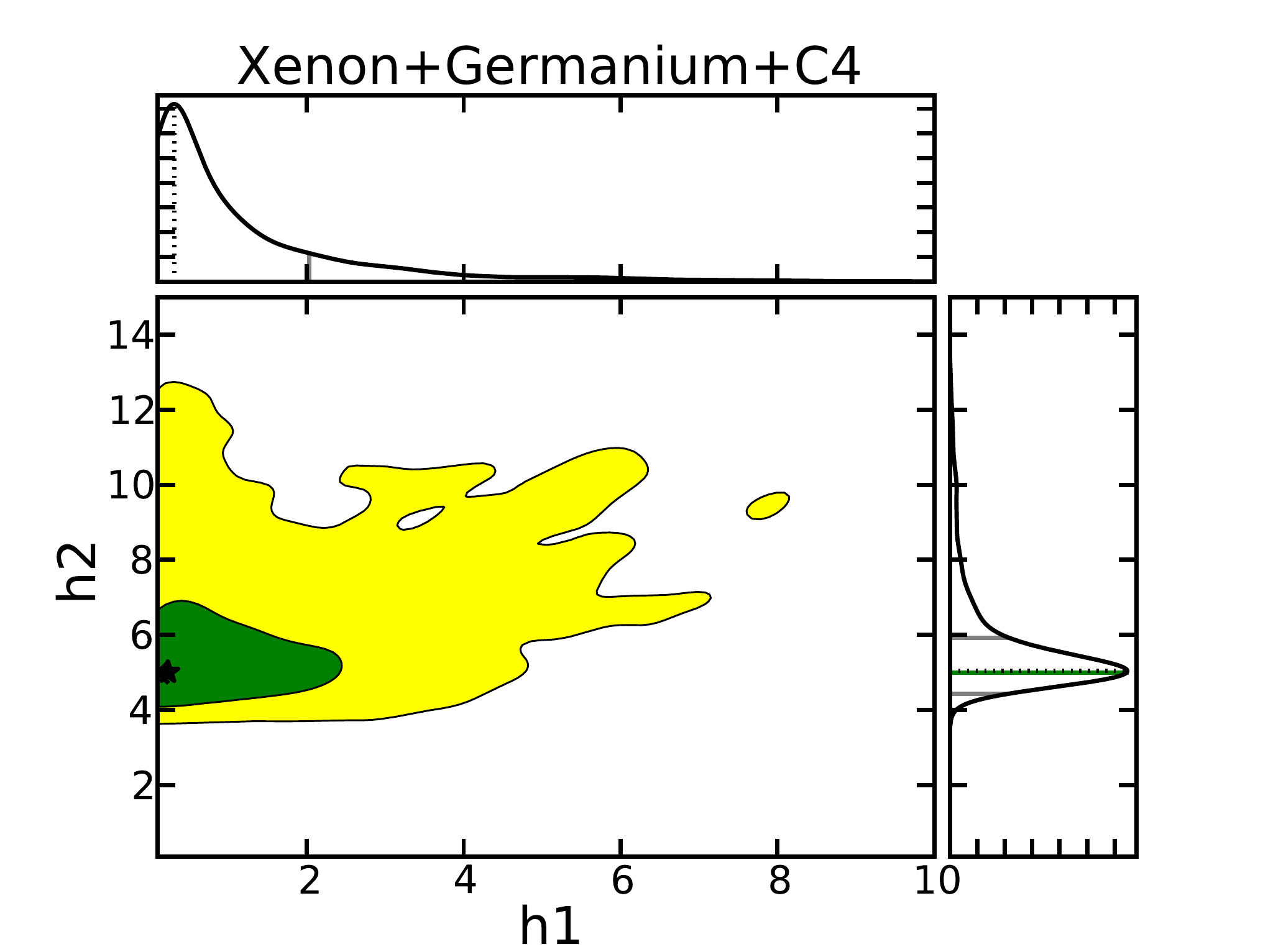}
\caption{{\bf A simulation containing only $h_1$ (two upper panels), and only $h_2$ (two lower panels), setting all other operators to zero, is analyzed by letting the mass, $h_1$, and $h_2$ to vary as free parameters.} In both cases, the data are able to pick out the dominant parameter, estimating the other parameter to be zero.  A cross denotes the input value, and the star represents the ``best-fit'' value, i.e. maximum of the posterior distribution.  The precision of parameter estimation crucially depends on the addition of data from an experiment with ultra-low energy threshold. \label{fig:2par}}
\end{center}
\end{figure*}

As the second step, we use the same mock data sets and perform parameter estimation, including two of the Wilson-coefficients as free parameters. This step tests if the data are good enough to distinguish different underlying scenarios, if the analysis is performed more agnostically, with no assumptions about which operator dominates the recoil spectrum. The results of this exercise are shown in Fig.~\ref{fig:2par}, for the case where the underlying theory only has $h_1$ (two upper panels), and only $h_2$ (two lower panels). In both cases, the estimation is again relatively accurate---the data are able to pick out which of the parameters dominates the recoil spectrum---but the combination of the broad-energy-window data with the low-recoil-energy data from ``C4'' is crucial in breaking the degeneracy between different operators, and shrinking the confidence intervals by a large factor.

\subsection{Inelastic and exothermic dark matter}\label{sec:WIMPphys:inelastic}
In this section, we consider the possibility of uncovering inelastic WIMP scattering (in the iDM and exoDM frameworks; see Sec.~\ref{sec:theory}) in direct-detection experiments.  In this case, we model experiments according to Table \ref{tab:Experiments}.  We illustrate experimental capabilities to uncover inelastic scattering using the three benchmark points detailed in Table \ref{tab:inelastic:benchmarks}.  In each case, we are explicitly setting to zero any spin-dependent or effective-operator interactions, and are considering only spin-independent interactions, for both event simulation and analysis.  We choose two benchmark points (2 \& 3) with a moderately heavy WIMP of \mwimp$=50$ GeV.  We also checked higher WIMP masses (specifically, \mwimp$=200$ GeV).  However, the results were similar to the \mwimp$=50$ GeV case, but for the well-known fact that it is typically only possible to set lower bounds for the WIMP mass and cross section because of the strong degeneracy between \mwimp~and \sigmapsi.  This degeneracy results from the fact that energy spectra are essentially indistinguishable for \mwimp$\gg m_{\mathrm{T}}$ unless the splitting between states in the WIMP multiplet $\delta$ is large (see Eq. (\ref{eq:vmin-inelastic})).

For the \mwimp$=50$ GeV iDM benchmarks, we choose two different values of $\delta: 0\hbox{ keV}$ and $40\hbox{ keV}$.  The former is a purely elastic case, but we are interested to see how well one can constrain an elastic interaction in the case in which one is agnostic about the number of states in the WIMP multiplet.  The second case is one with a mild splitting.  Of course, larger splittings are typically easier to distinguish because they can significantly alter the shape and magnitude of the energy spectrum for at least one experiment.  In this particular case (Benchmark 3), Fig. \ref{fig:idmspectrum} shows that the purely exponential part of the recoil spectrum lies above threshold for Xenon and Argon, but not Germanium and Silicon.  However, in the case of Silicon, few events are expected regardless of the magnitude of $\delta$ for our choice of spin-independent cross section, so it is not influential in parameter estimation. 

\begin{table*}[t]
  \setlength{\extrarowheight}{3pt}
  \setlength{\tabcolsep}{3pt}
  \begin{center}
    \begin{tabular}{c|cccccc}
     Benchmark &\mwimp (GeV) &$\rho_\chi$\sigmapsi/$m_p^2$ $\,(10^{-46} \hbox{GeV}^{-1} \,\hbox{cm}^{-1})$ &$\delta$ (keV) &\vlag (km/s) &$\sigma_v$ (km/s) &\vesc (km/s) \\
     \hline\hline
     1 &5 &300 &-50 &220 &155 &544 \\
     2 &50 &3 &0 &220 &155 &544 \\
     3 &50 &3 &40 &220 &155 &544\\
    \end{tabular}
  \end{center}
\caption{\label{tab:inelastic:benchmarks}Benchmark points for Sec.~\ref{sec:WIMPphys:inelastic}.  For the cross-section-related parameters, $\rho_\chi$\sigmapsi$/m_p^2 = 3 \times 10^{-46}\hbox{ GeV}^{-1}\,\hbox{cm}^{-1}$ corresponds to 1 zb ($10^{-45}$ cm$^2$) if $\rho_\chi = 0.3\hbox{ GeV cm}^{-3}$.}

\end{table*}

We also choose one exoDM benchmark (1).  In this case, we consider a low-mass WIMP (5 GeV).  This benchmark point lies on the low end of the CDMS-Silicon allowed region for exoDM \cite{Frandsen:2013cna}.

\begin{figure*}
    \includegraphics[width=0.32\textwidth]{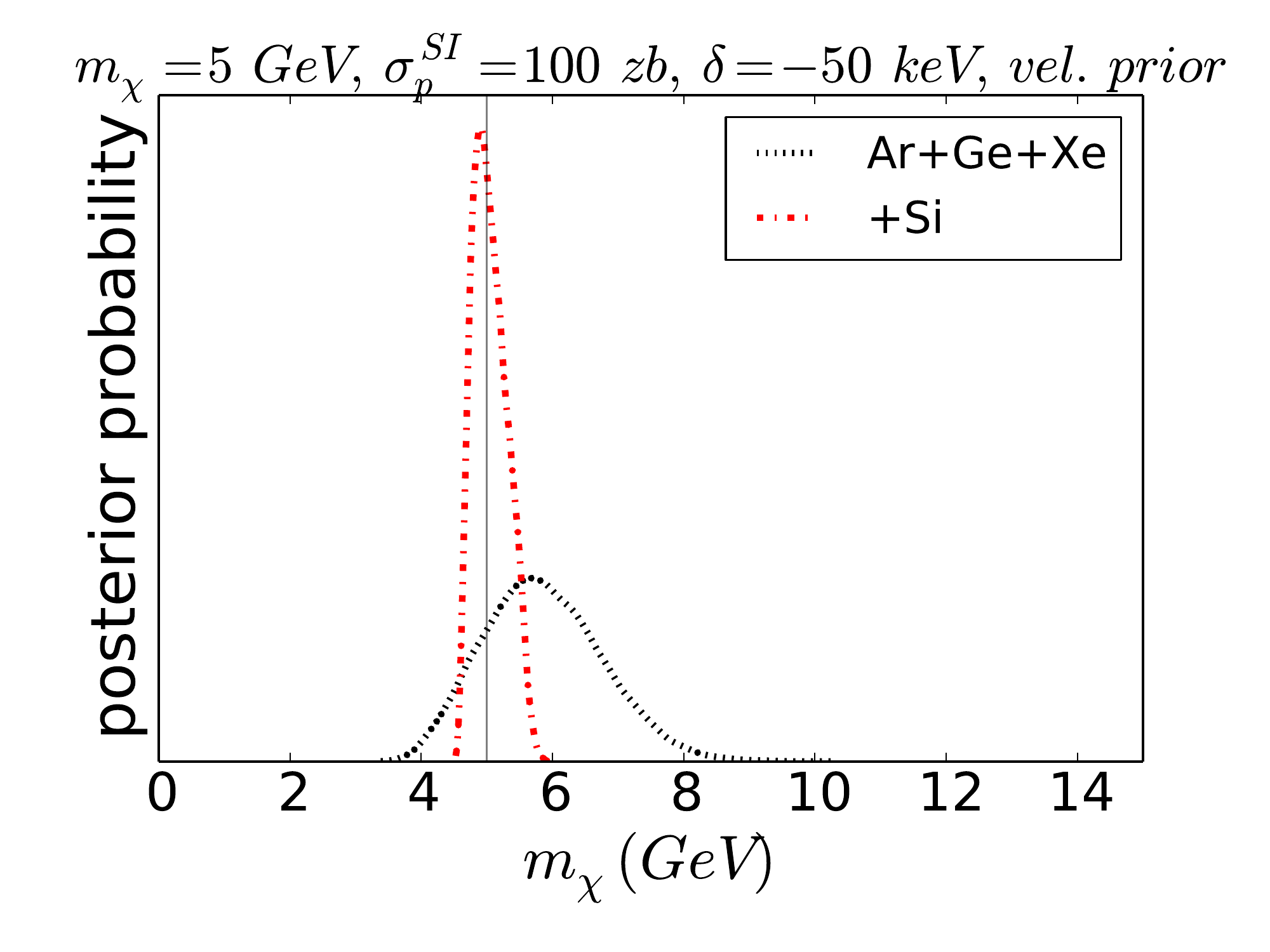} \includegraphics[width=0.32\textwidth]{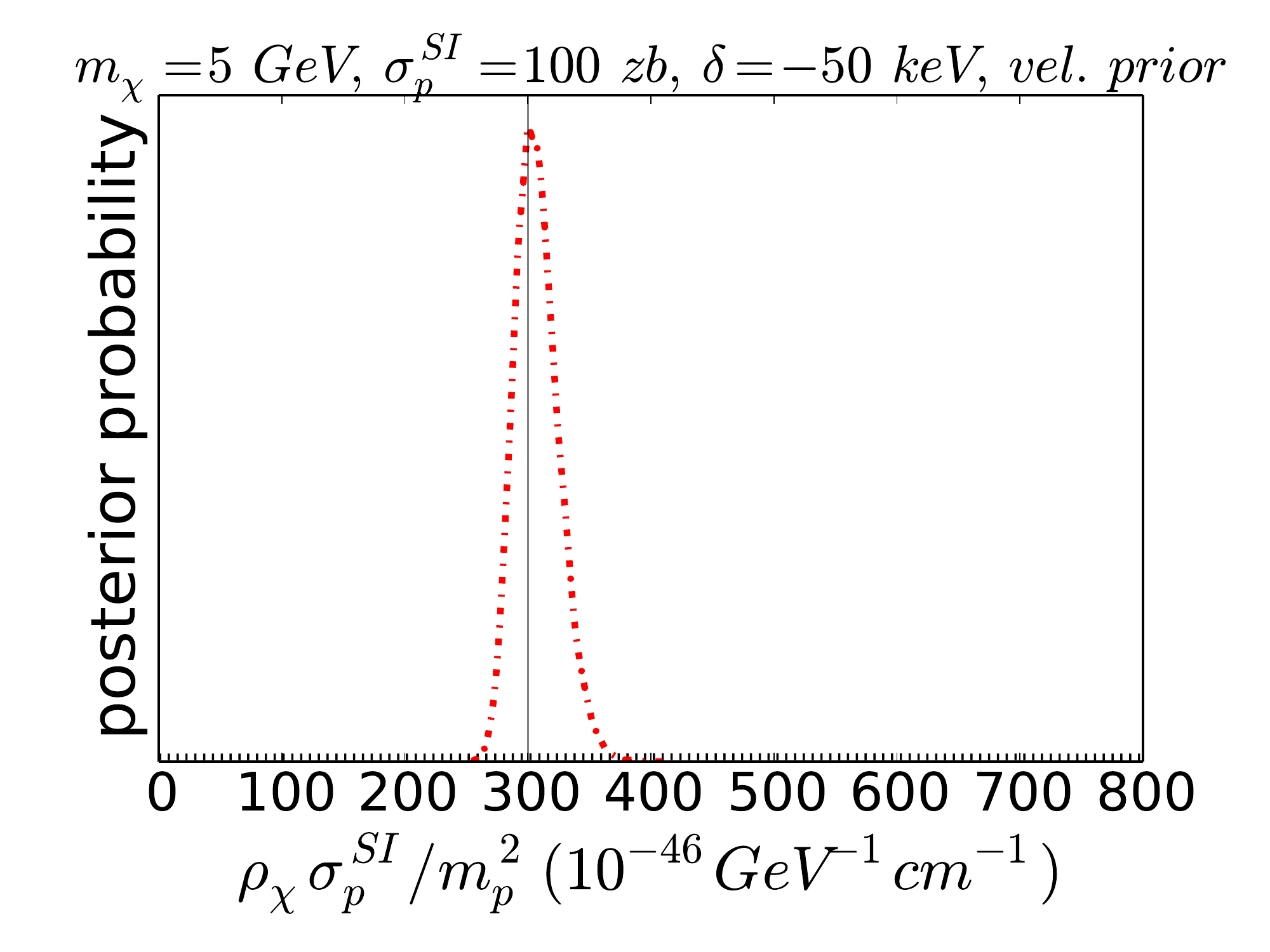} \includegraphics[width=0.32\textwidth]{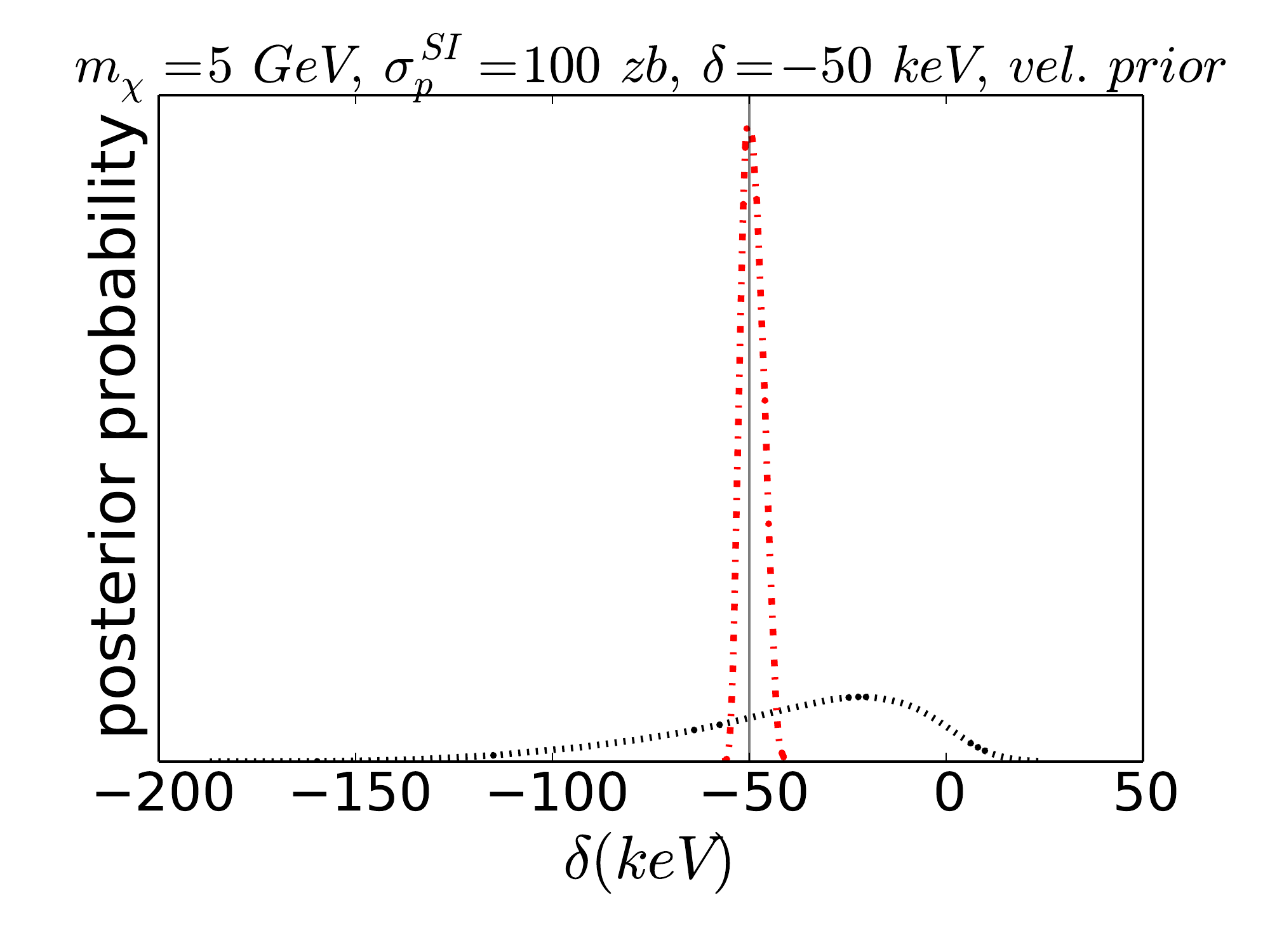} \\
  \includegraphics[width=0.32\textwidth]{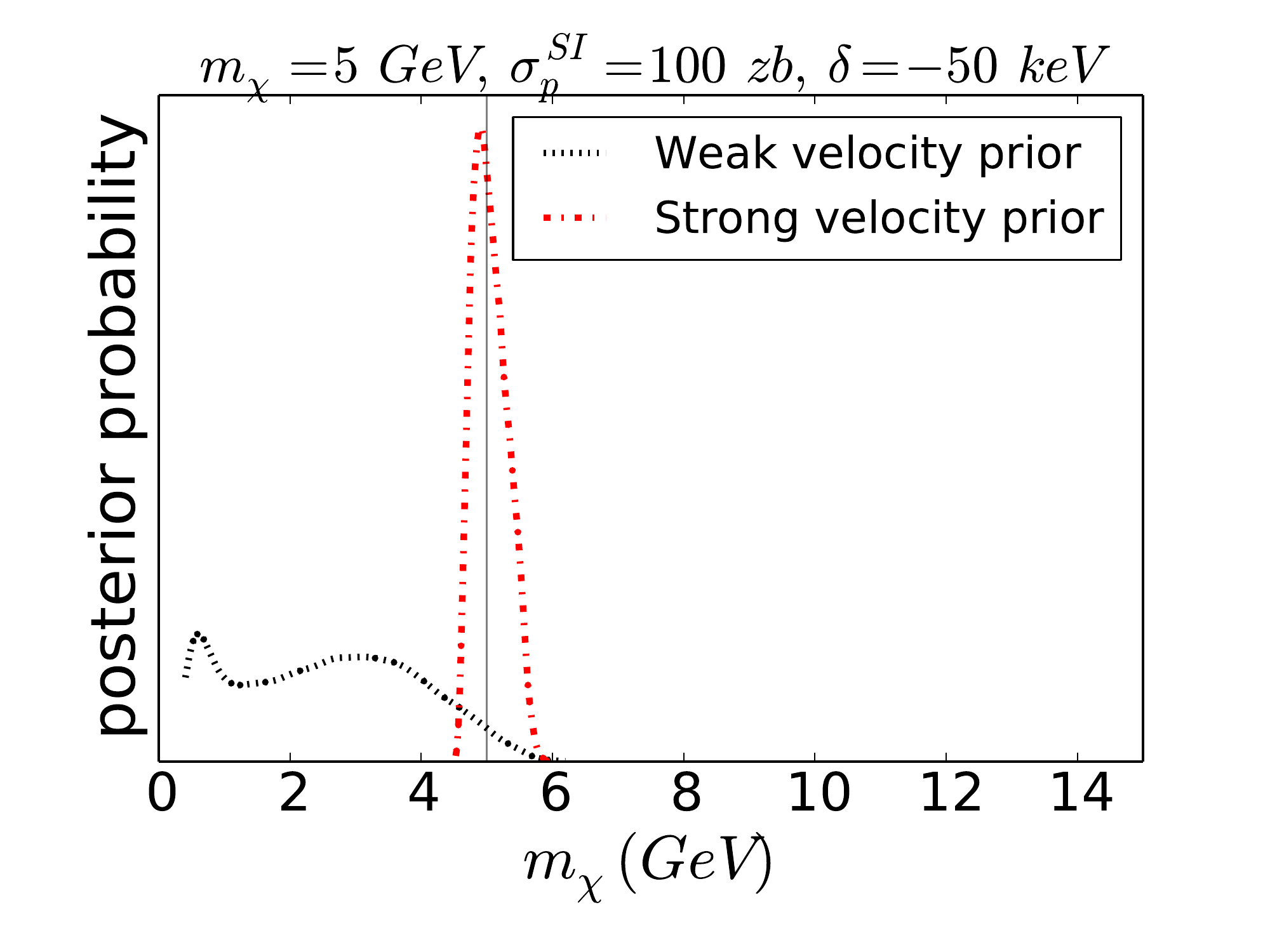} \includegraphics[width=0.32\textwidth]{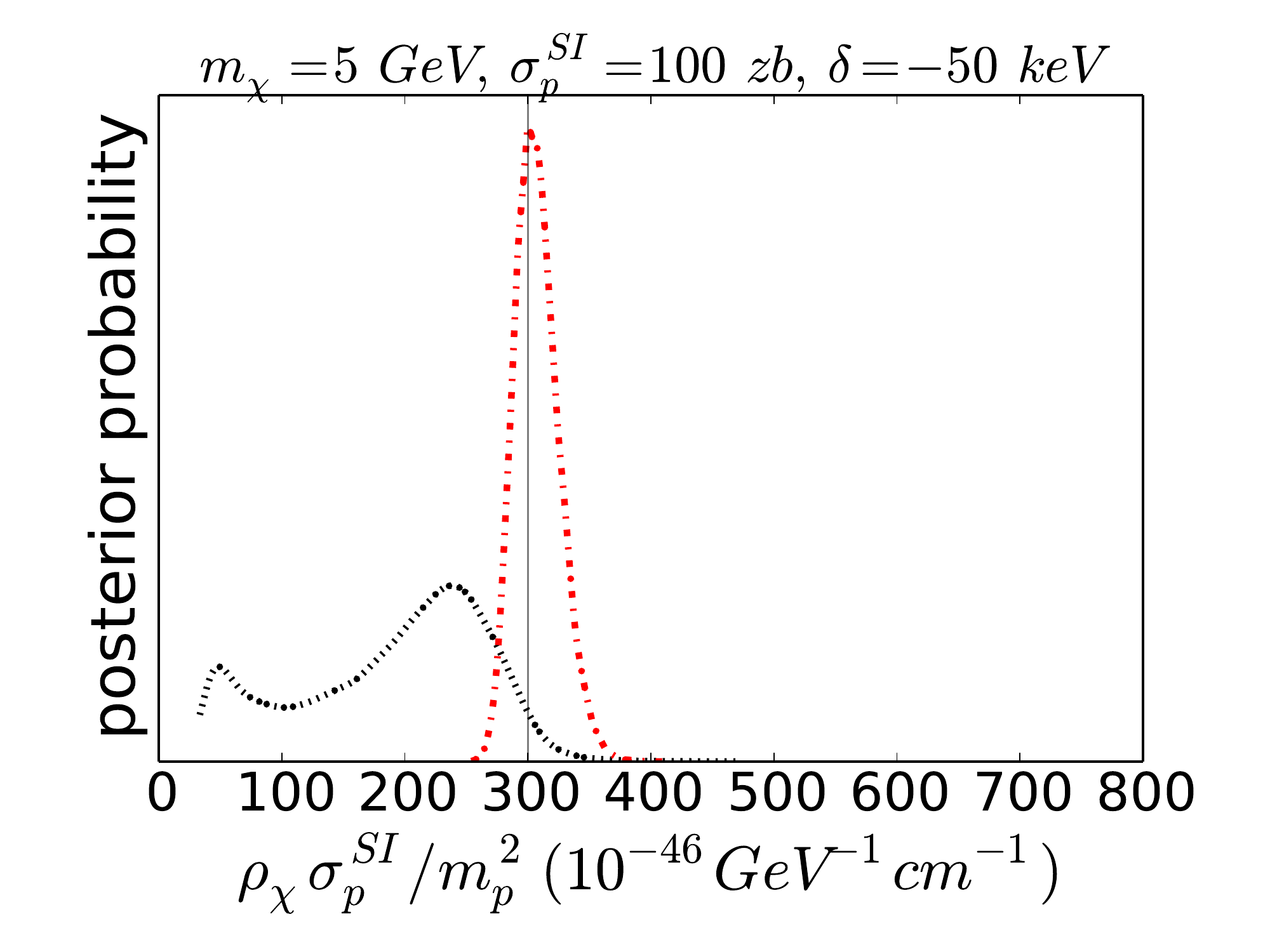} \includegraphics[width=0.32\textwidth]{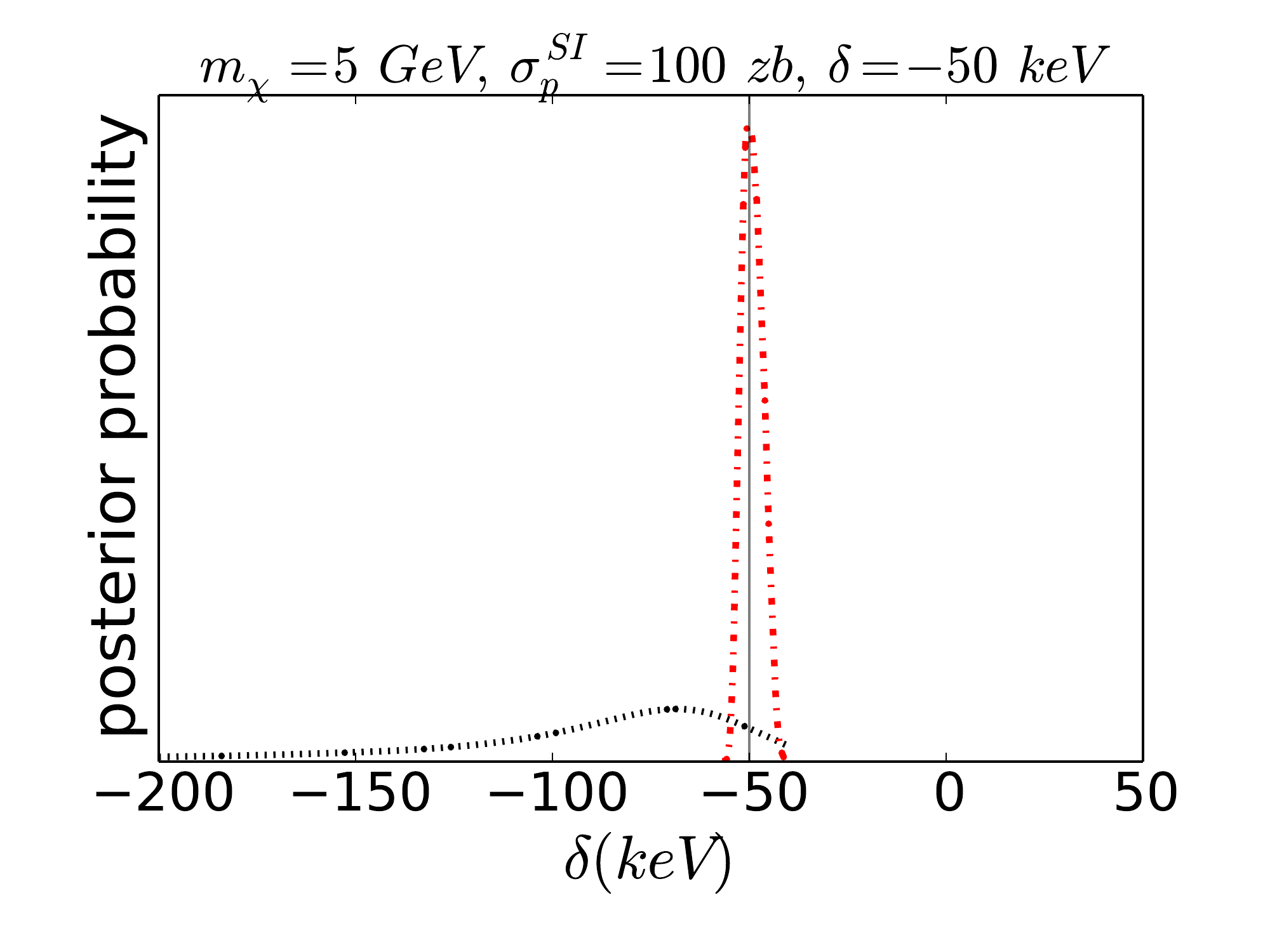}
\caption{\label{fig:exodm5}{\bf Marginalized one-dimensional posterior probabilities for Benchmark 1.}  In the top row, we show the difference in posterior probabilities for a combination of three experiments (Argon, Germanium, and Xenon), and with four experiments (including Silicon).  In this row, we use the strong velocity priors.  In the bottom row, we show posteriors using the four experiments, but with either strong or weak velocity priors.  The experiments do not strongly constrain the velocity distribution, and its large prior volume weights the posterior away from the fiducial parameter values.}
\end{figure*}

For our parameter estimation, we use the priors and parameter ranges given in Table \ref{tab:inelastic:priors}.  We consider both strong and weak velocity priors, as in Sec.~\ref{sec:WIMPphys:sd}.

\begin{table}[t]
  \setlength{\extrarowheight}{3pt}
  \setlength{\tabcolsep}{3pt}
  \begin{center}
        \begin{tabular}{r|cc}
        Parameter & Prior type & Prior Range \\
        \hline\hline
        \mwimp (GeV) &log-flat & [0.1-$10^4$]\\
        \hline
        $\rho_\chi$\sigmapsi$/m_p^2\,(10^{-46} \hbox{GeV}^{-1} \,\hbox{cm}^{-1})$  &log-flat & [0.007-$10^5$]\\
        \hline
        $\delta$ (keV) &flat & [-1000,1000] \\
        \hline
        \vlag (km/s) Weak prior &flat &[0-2000] \\
          Strong prior &flat &[200-240] \\
        \hline
        $\sigma_v$ (km/s) Weak prior &flat &[0-2000] \\
          Strong prior &flat &[140-170] \\
          \hline
        \vesc (km/s) Weak prior &flat &[490-600] \\
        Strong prior &flat &[540-550] \\
        
        \end{tabular}
  \end{center}
\caption{\label{tab:inelastic:priors}Priors for the inelastic and exothermic dark-matter analysis.}
\end{table}

We consider the exoDM case, Benchmark 1, first.  In Fig. \ref{fig:exodm5}, we show one-dimensional marginalized posteriors for \mwimp, $\rho_\chi$\sigmapsi$/m_p^2$, and $\delta$.  In the top row, we show posteriors for the three standard big experiments---Argon, Germanium, and Xenon---alone, and then with the addition of the Silicon experiment.  In the former case, essentially zero events are expected for the Argon and Xenon experiments, but of order 10 events for Germanium and a few hundred for Silicon.  For the top row of Fig. \ref{fig:exodm5}, we apply the strong velocity priors.  We find, as expected, that the addition of the Silicon experiment drastically improves parameter estimation, especially for the cross section (middle panel).  

\begin{figure}
  \includegraphics[width=0.32\textwidth]{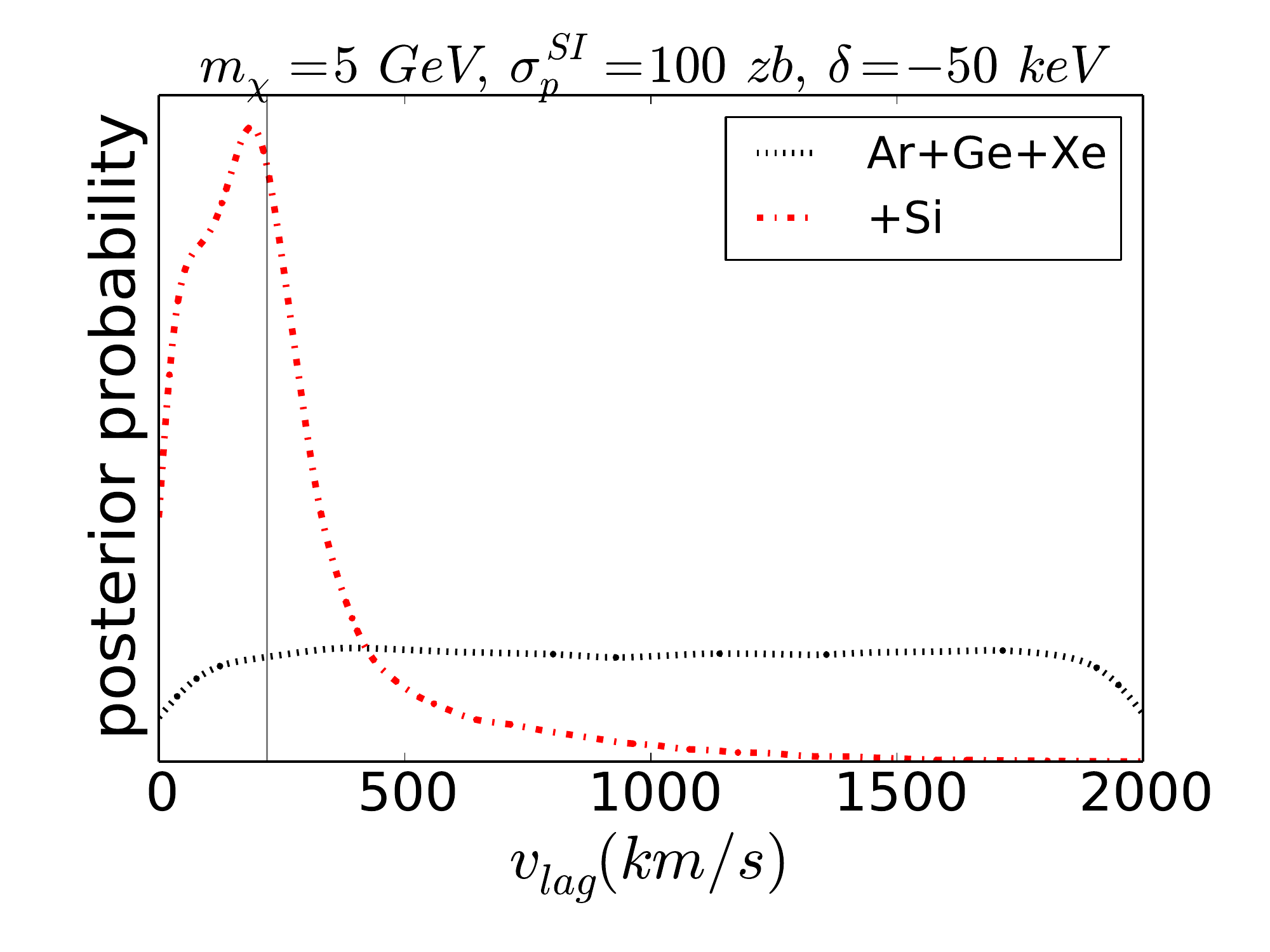} \\
  \includegraphics[width=0.32\textwidth]{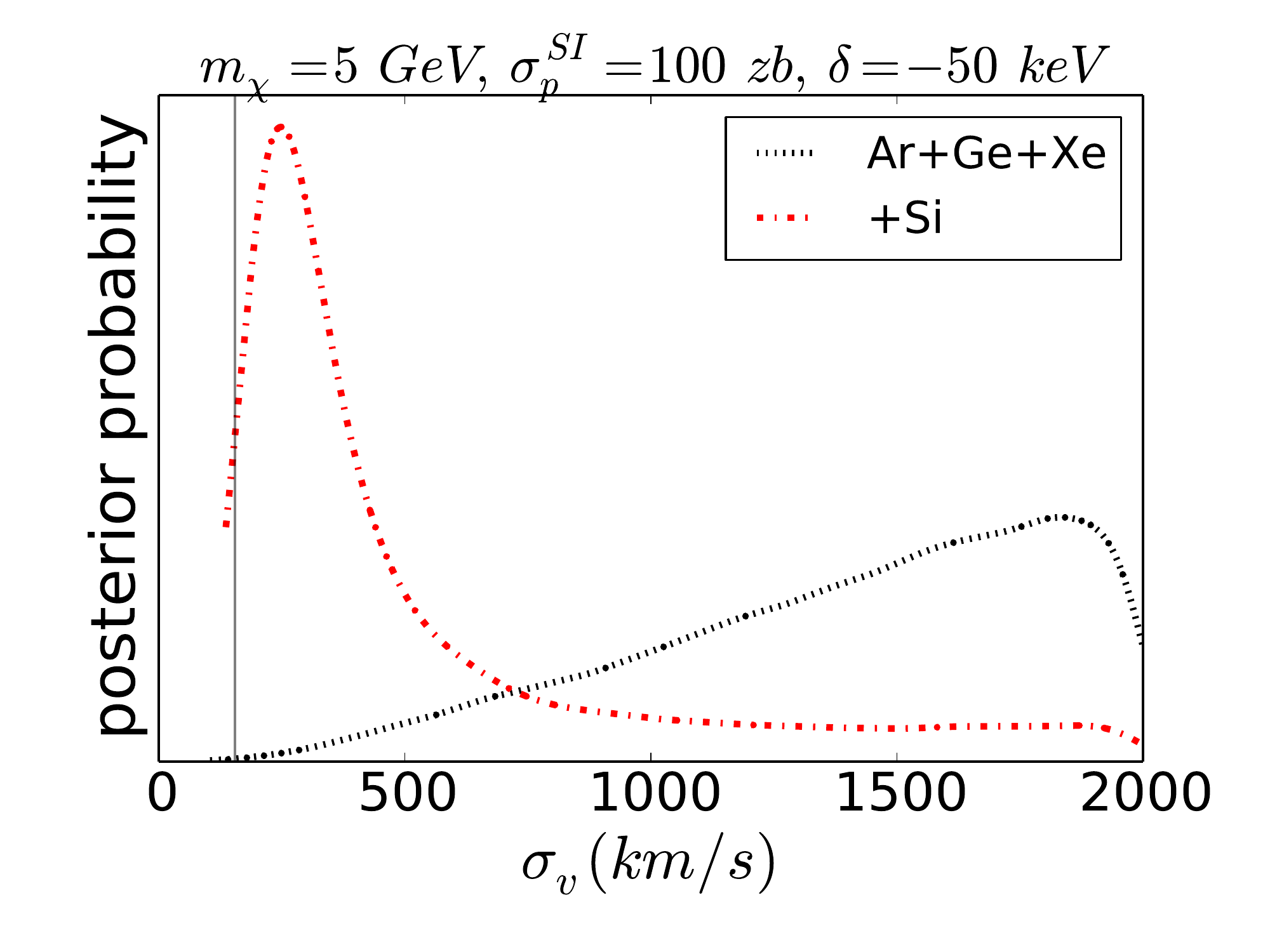} 
  \caption{\label{fig:exodm_vel}{\bf Velocity parameters for Benchmark 1 with weak velocity priors.}  The line types correspond to different ensembles of experiments.}
\end{figure}

In the second row, we show the difference in the posteriors assuming either weak or strong velocity priors, including information from all four experiments.  We find that the posteriors are offset from the fiducial values in the case in which we assume weak velocity priors, but are well-centered for the strong velocity priors.  This is because we have large prior volumes for the weak-velocity-prior case, and the data are not sufficient to significantly constrain the velocity distribution beyond the priors.  We illustrate this using Fig. \ref{fig:exodm_vel}.  Here we show the posteriors for \vlag~and $\sigma_v$~for the three- and four-experiment cases.  Because there are so few events for the big three experiments, there is essentially no constraint on the velocity parameters.  When Silicon is added, the constraints get better.  However, there is a long tail to large velocity parameters that leads to a long tail in \mwimp~and $\rho_\chi$\sigmapsi$/m_p^2$.  The down side of using Bayesian inference is that if both the prior volume is large and the data are not sufficiently constraining, parameter estimation is prior-dominated, not data-dominated.  This is what shifts the peaks of the posterior distribution from the benchmark values, even when we have essentially ``perfect'' data.  When the velocity priors are strong, and the prior volume is greatly reduced, the posteriors do center on the benchmarks.

We conclude that for exoDM, Silicon-based experiments and low energy thresholds for other experiments are necessary to characterize the WIMP properties, unless we have strong reasons to be confident about the shape of the WIMP velocity distribution.

\begin{figure*}
    \includegraphics[width=0.32\textwidth]{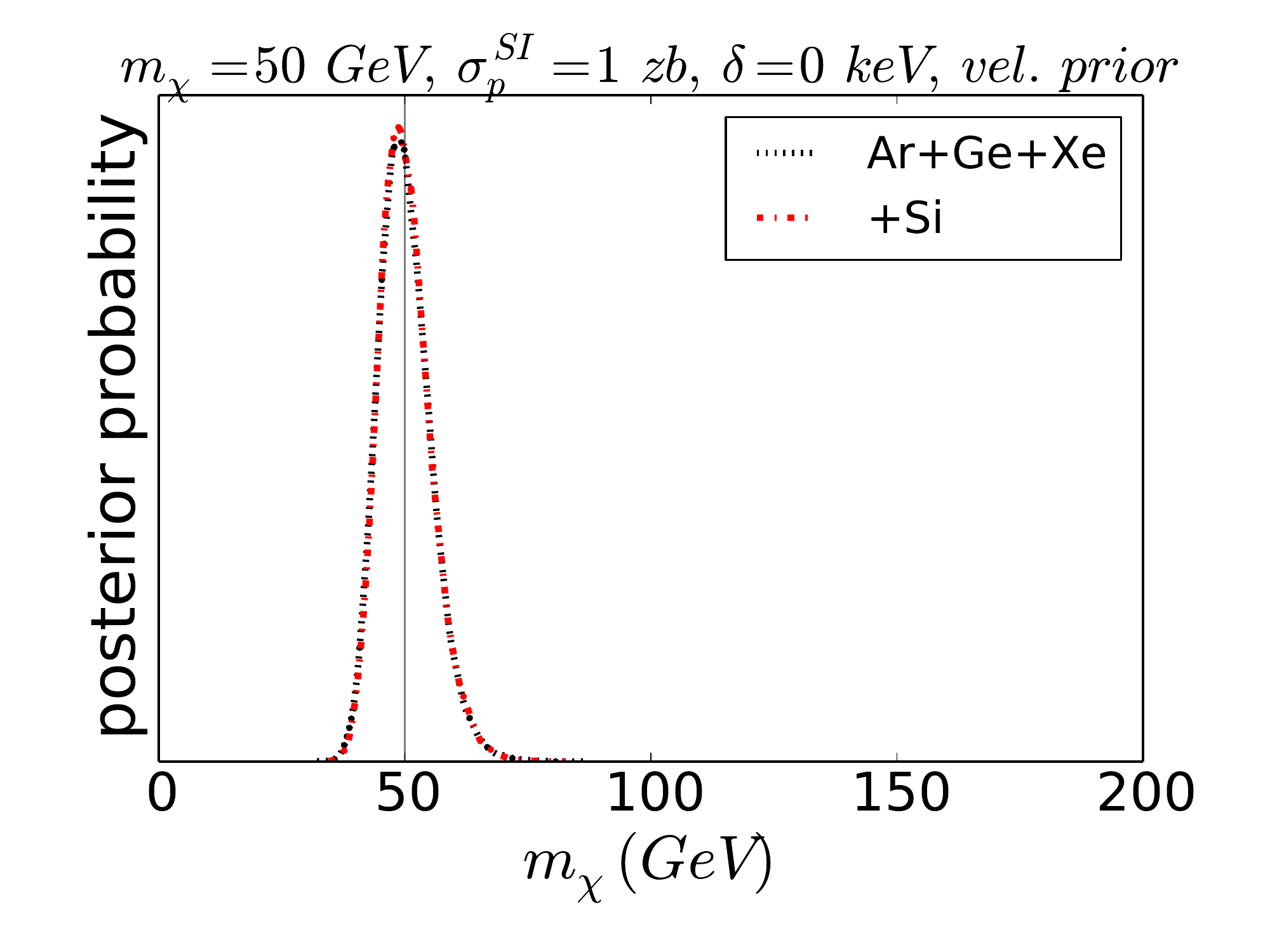} \includegraphics[width=0.32\textwidth]{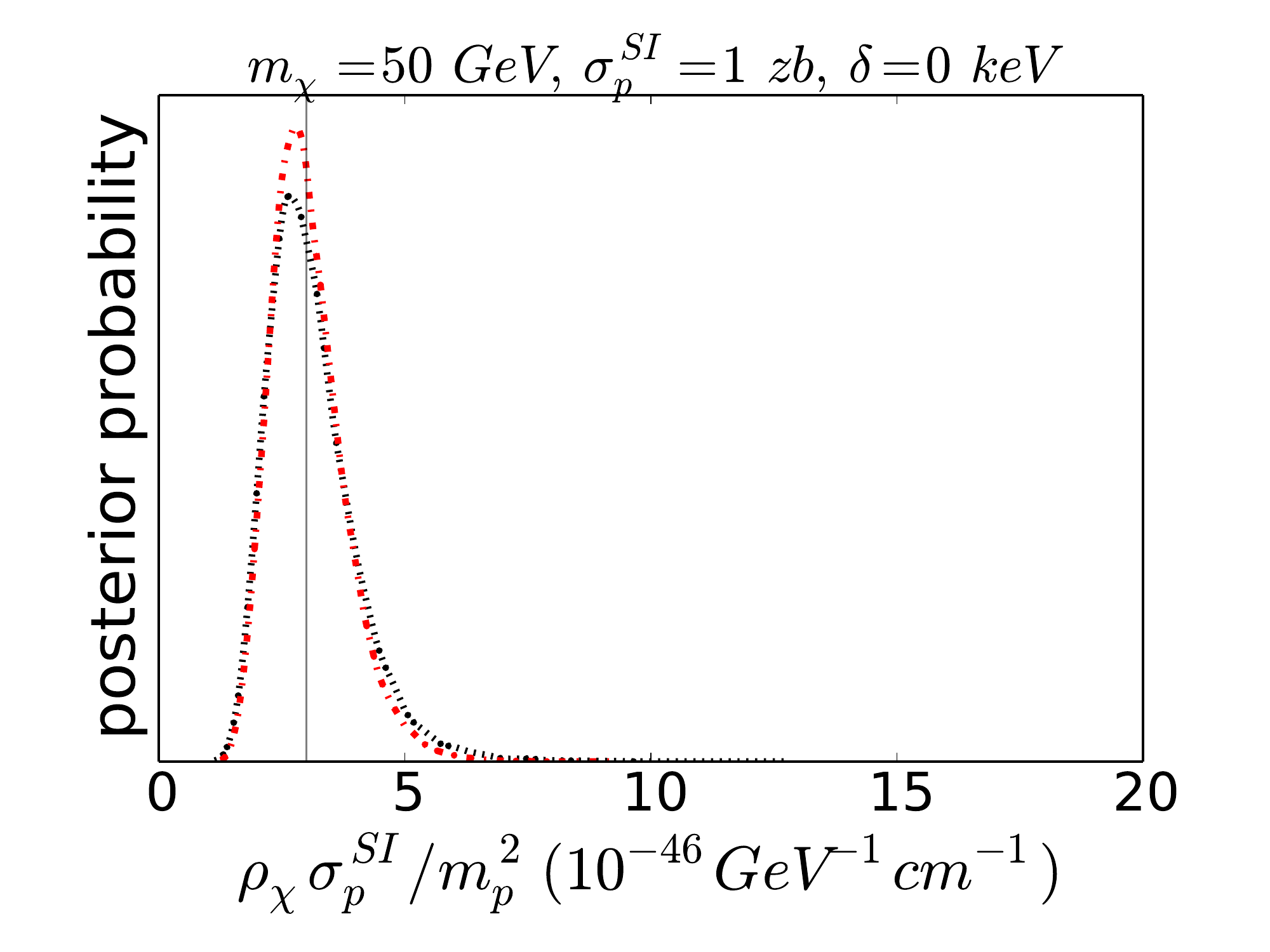} \includegraphics[width=0.32\textwidth]{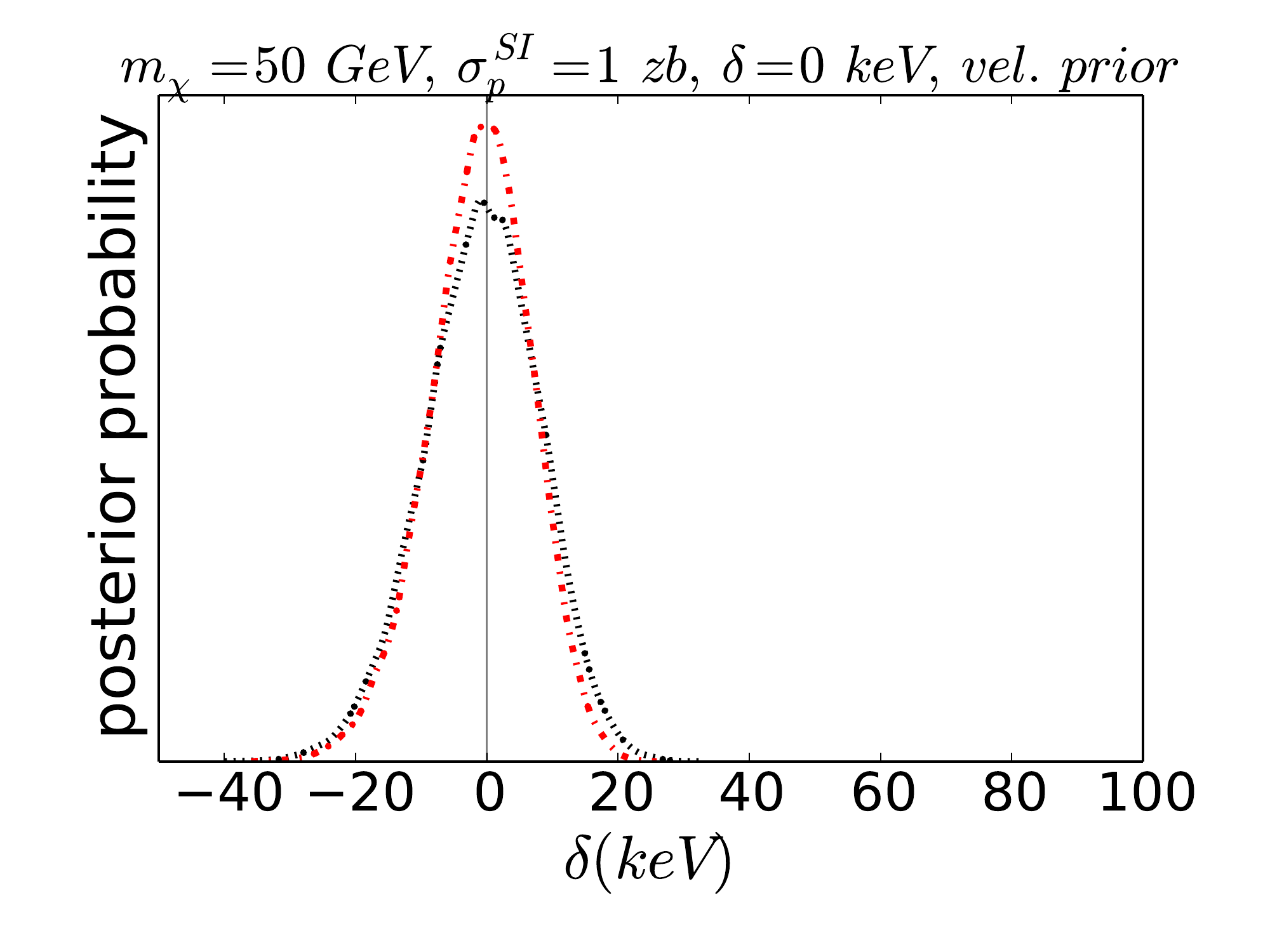} \\
  \includegraphics[width=0.32\textwidth]{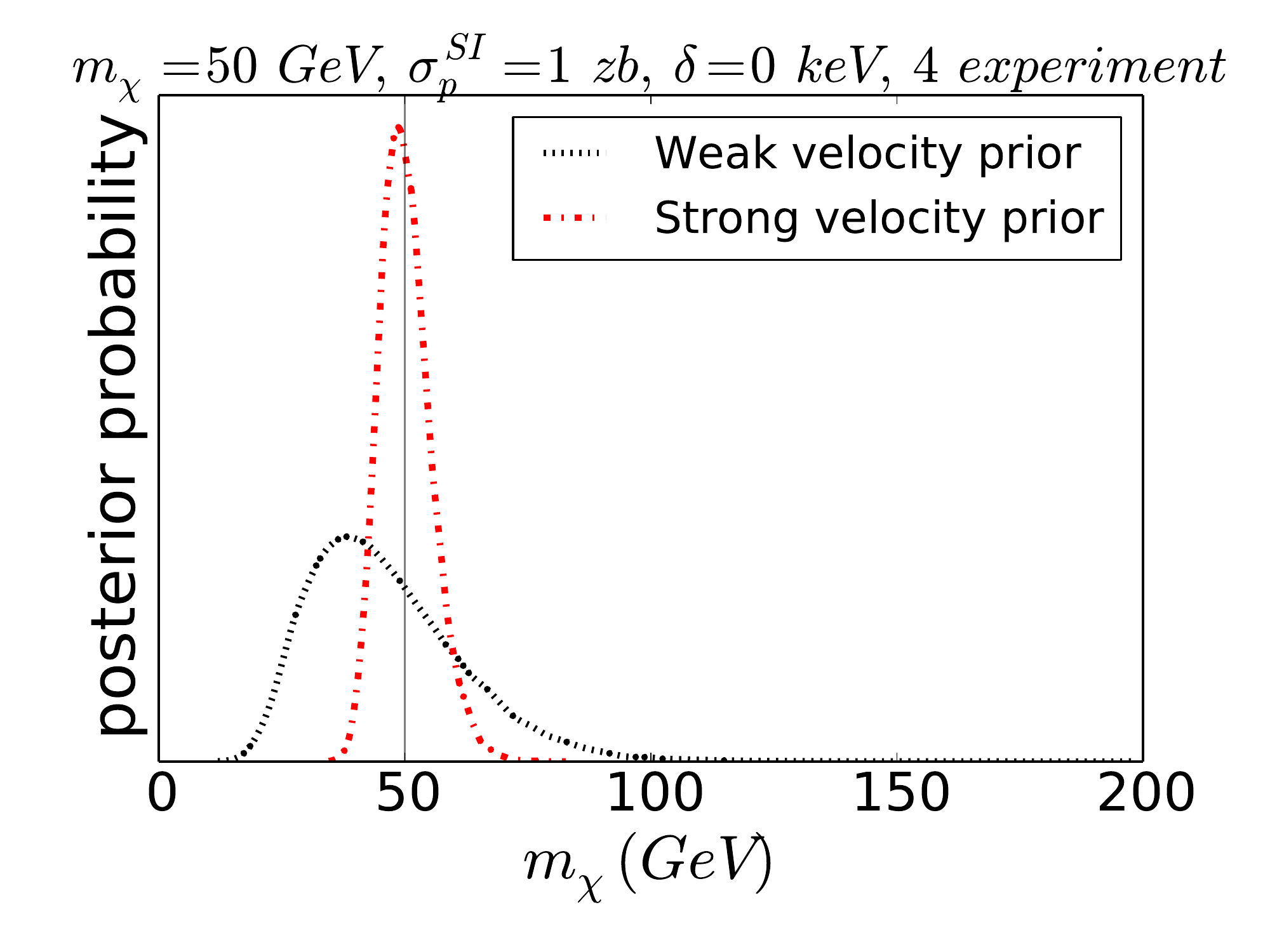} \includegraphics[width=0.32\textwidth]{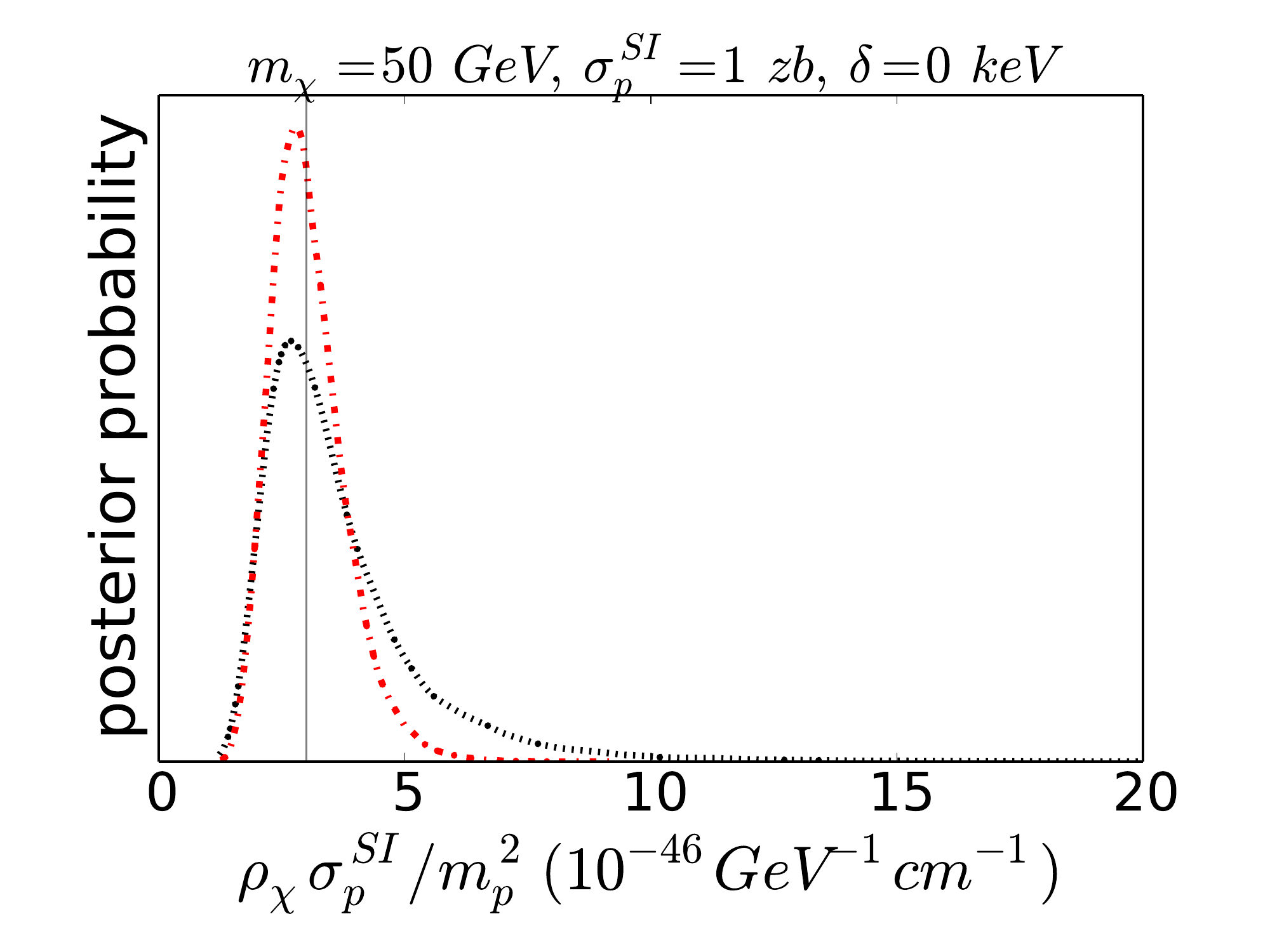} \includegraphics[width=0.32\textwidth]{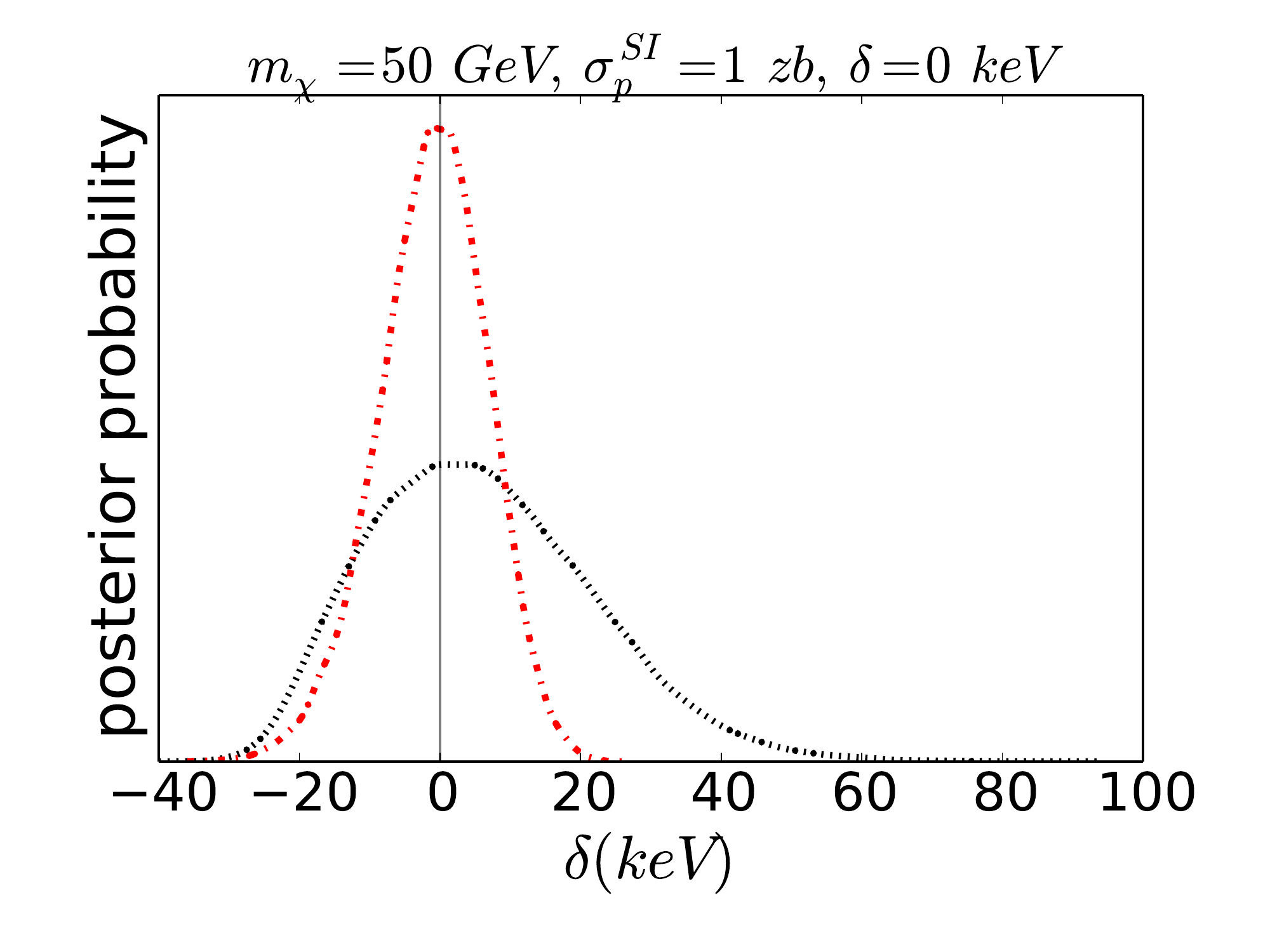}
\caption{\label{fig:inelastic_50_0}{\bf Marginalized one-dimensional posterior probabilities for Benchmark 2.}  Lines have the same meaning as for Fig. \ref{fig:exodm5}.  The relative narrowness of the posteriors for the parameters even without strong velocity priors indicates that the velocity constraints are better than for the exoDM case.}
\end{figure*}

Next, we consider Benchmark 2, the elastic-scattering case with \mwimp$=50$ GeV.  We show the one-dimensional posteriors for the particle-physics WIMP paramters in Fig. \ref{fig:inelastic_50_0}.  As with Fig. \ref{fig:exodm5}, we show the difference in posterior between using three and four experiments assuming strong velocity priors in the top row, and posteriors for the four-experiment case with weak and strong velocity priors in the bottom row.  Unlike Benchmark 1, Silicon adds little to the constraints.  This is because there are few events in the Silicon experiment relative to the other experiments.  The parameter for which the addition of the Silicon data is most important is $\delta$.  The three big experiments, Argon, Germanium, and Xenon, are quite constraining, even with weak velocity priors.  This is because the data are also sufficient to constrain the velocity distribution as well as the WIMP particle-physics parameters.  We conclude that for elastic WIMPs, we can constrain $\delta$ to $\pm (10-20)$ keV even with quite weak velocity priors.  We caution that the constraints are likely to worsen with real data, as Poisson noise can become important.  On the other hand, constraints should strengthen with larger data sets.

\begin{figure*}
    \includegraphics[width=0.32\textwidth]{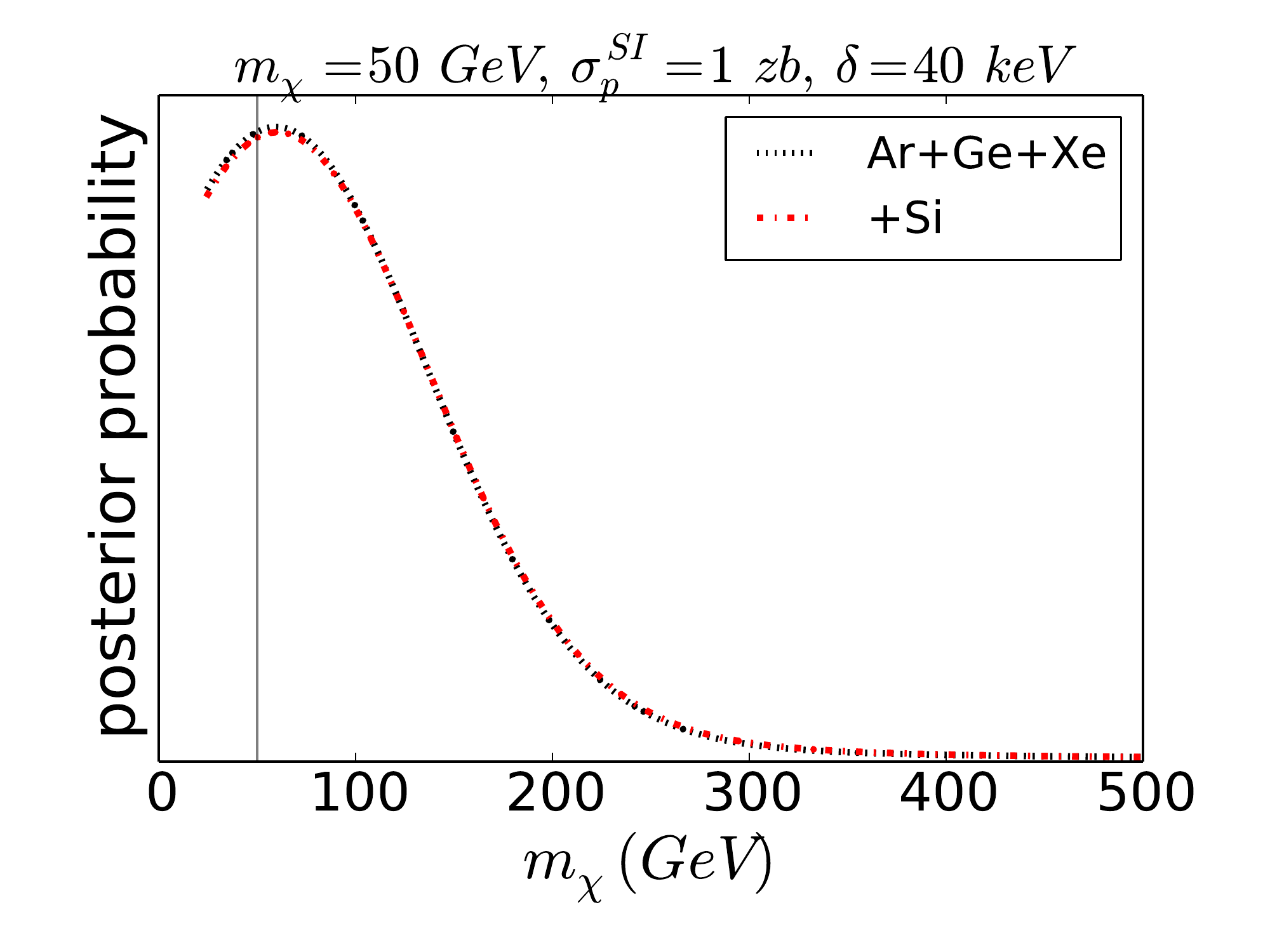} \includegraphics[width=0.32\textwidth]{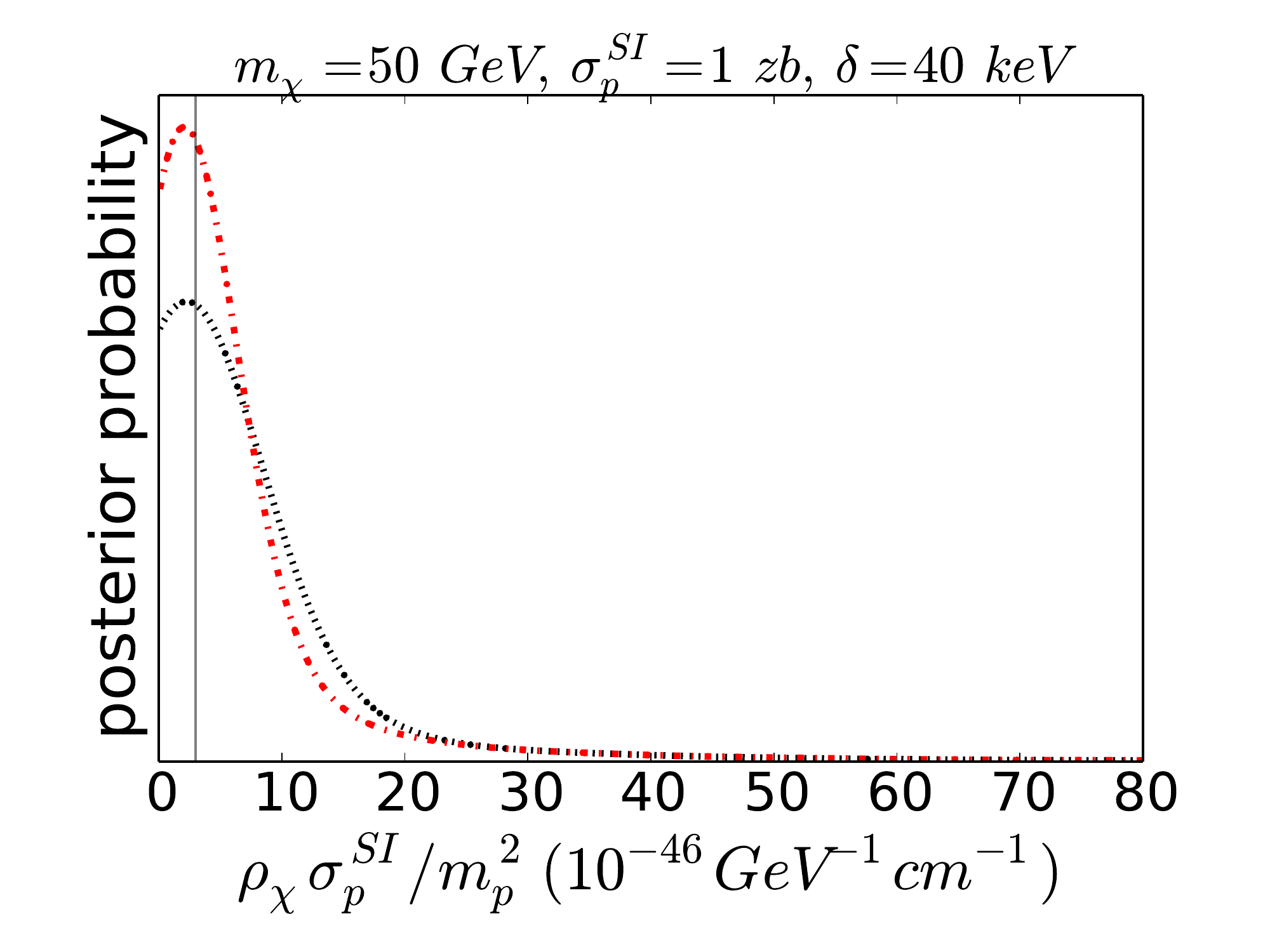} \includegraphics[width=0.32\textwidth]{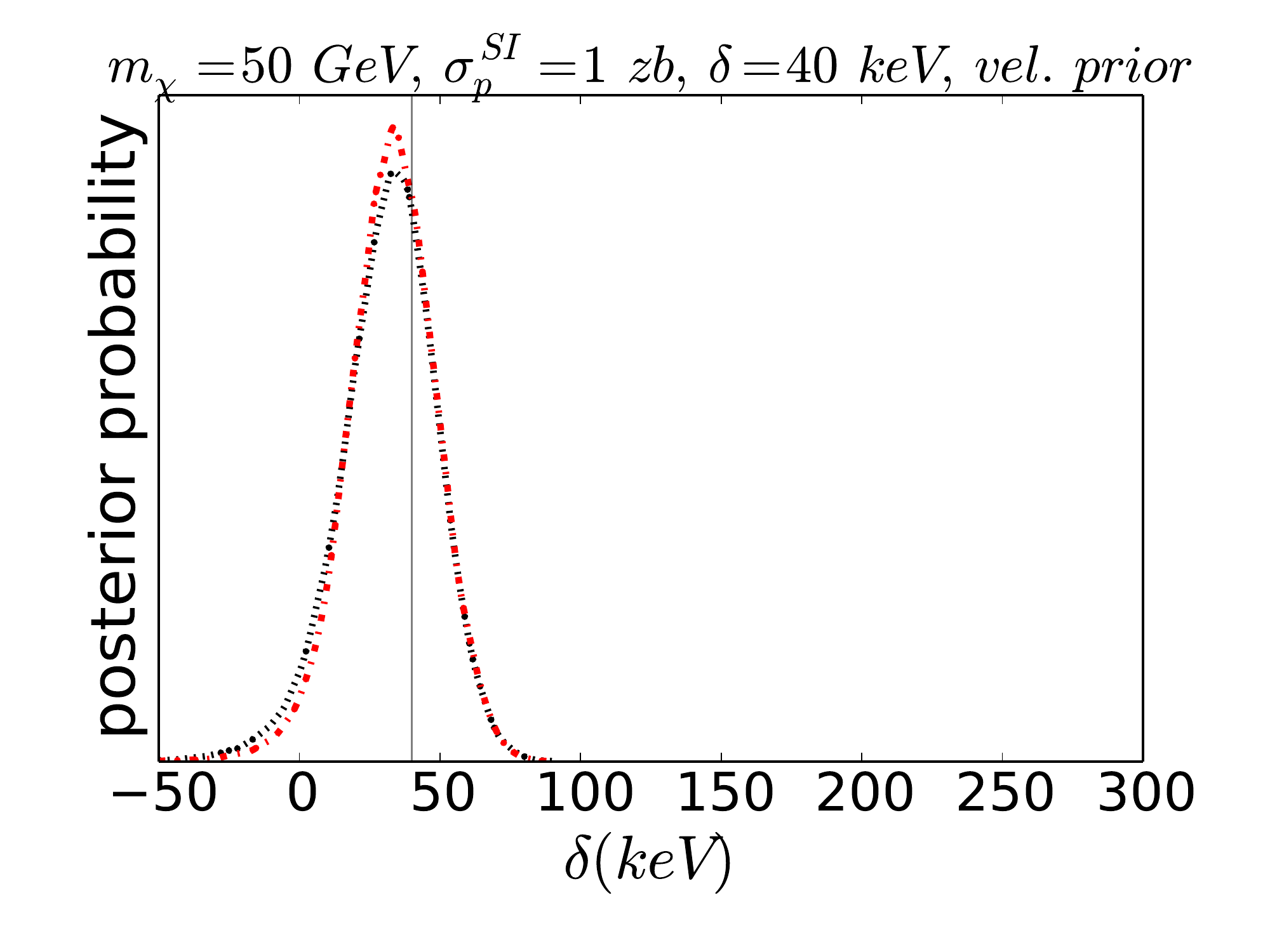} \\
  \includegraphics[width=0.32\textwidth]{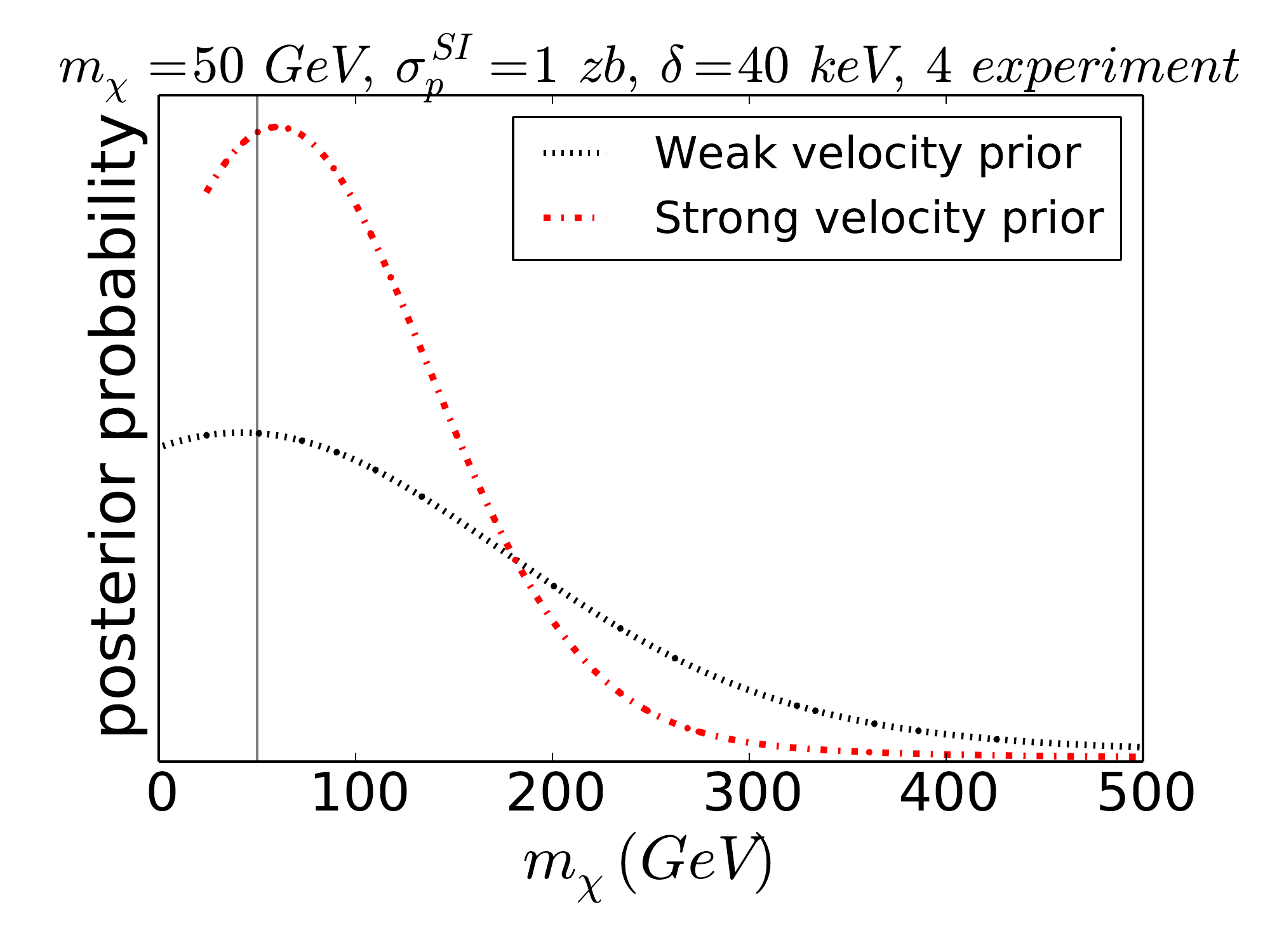} \includegraphics[width=0.32\textwidth]{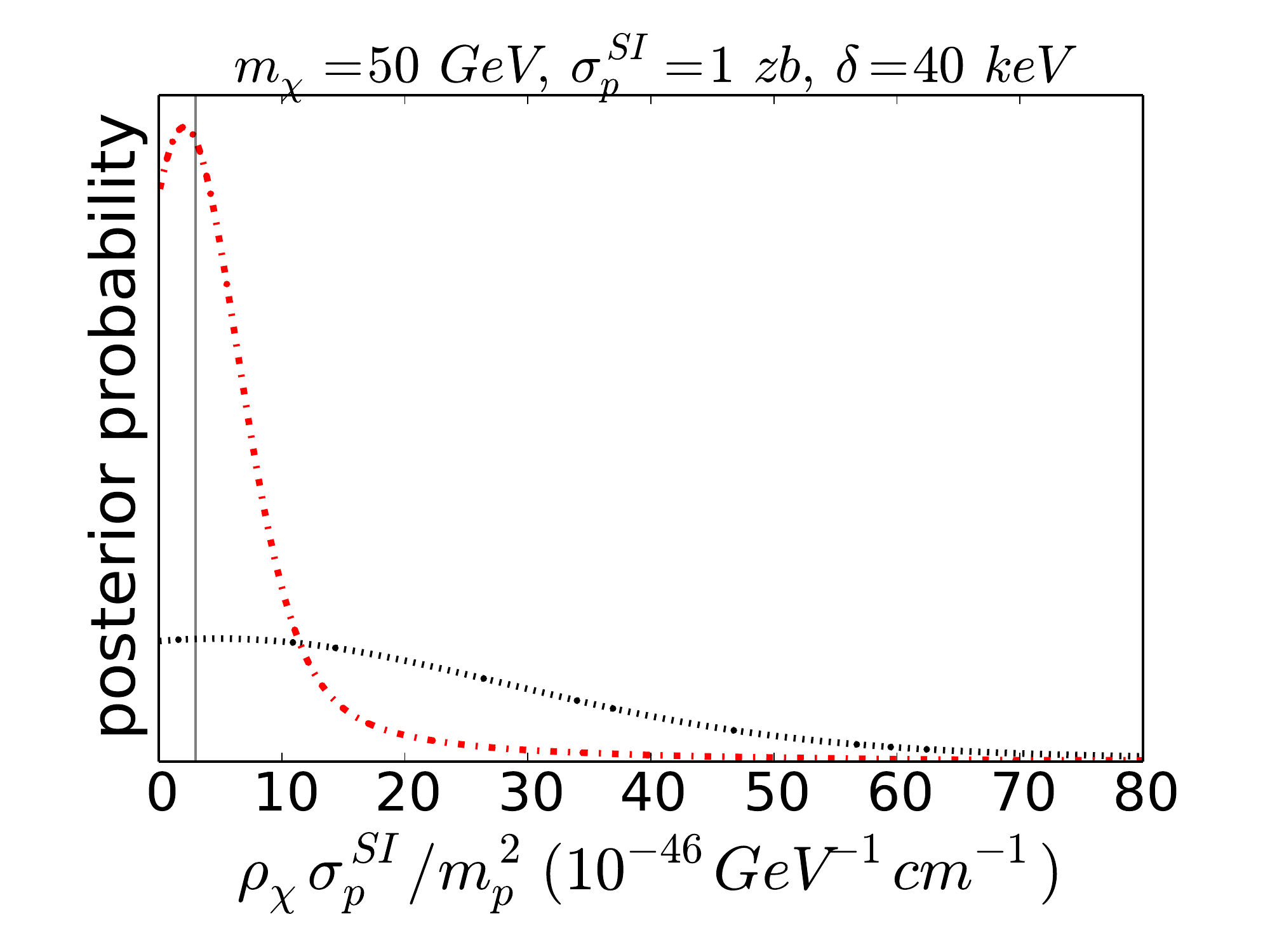} \includegraphics[width=0.32\textwidth]{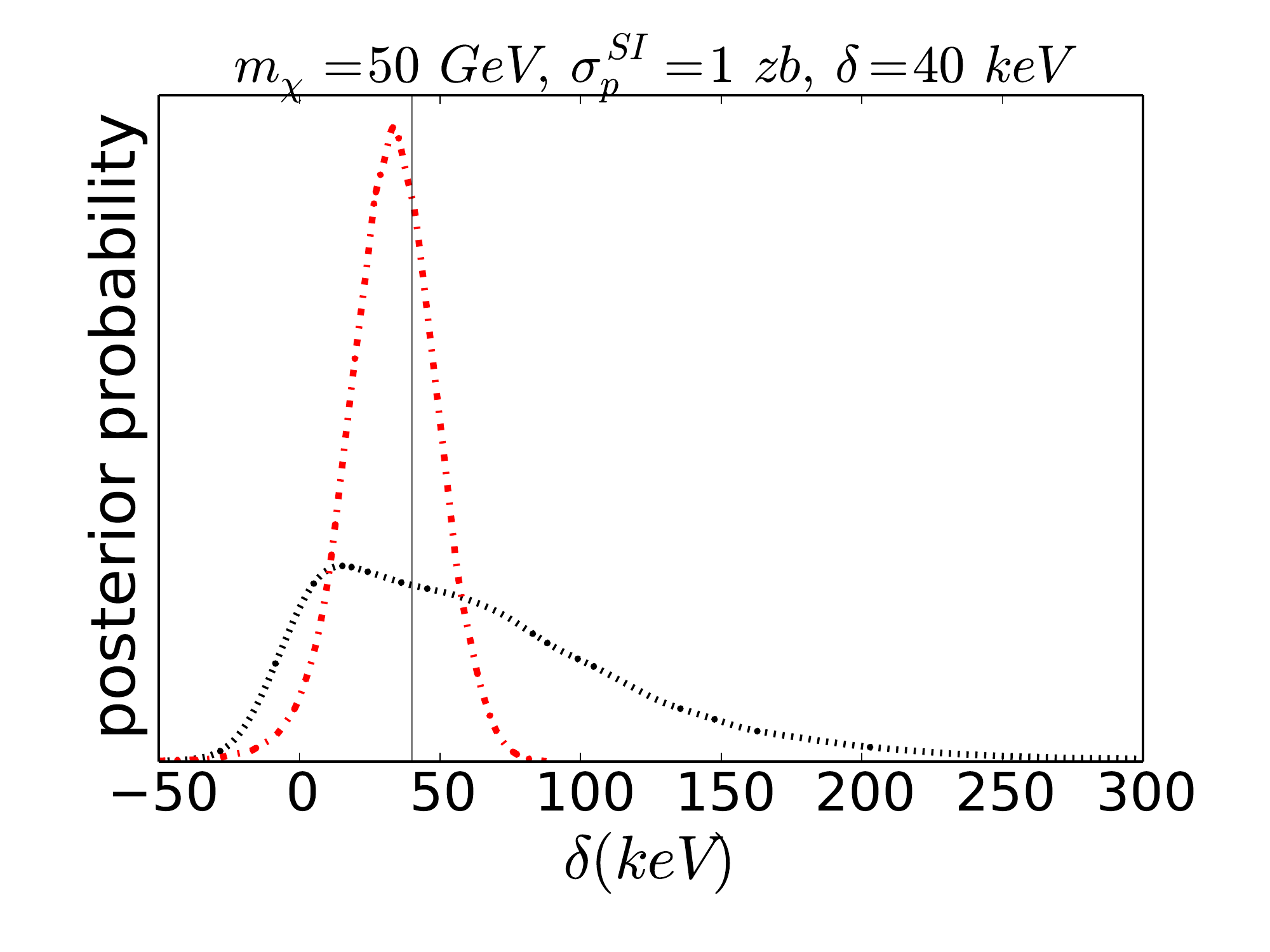}
\caption{\label{fig:inelastic_50_40}{\bf Marginalized one-dimensional posterior probabilities for Benchmark 3.}  The lines have the same meaning as in Fig. \ref{fig:exodm5}.}
\end{figure*}

Finally, we consider Benchmark 3, the one true iDM case we are considering in this section.  Unlike for Benchmark 2, the parameter constraints are overall worse across the board.  This is illustrated in Fig. \ref{fig:inelastic_50_40}.  The constraints on the WIMP mass are far weaker than for Benchmark 2, a fact that is not helped by the addition of the Silicon experiment, and which is velocity-prior-dominated.  With strong velocity priors, we can confidently say that WIMPs must scatter inelastically, but we cannot say much beyond that.  The key problem, which is also illustrated in Fig. \ref{fig:idmspectrum}, is that the thresholds of the experiments are high enough to not catch the low-recoil-energy rollover in the energy spectrum.  This means that there is a major degeneracy between the WIMP velocities and the energy state splitting $\delta$, which is shown in Eq. (\ref{eq:vmin-inelastic}).  This degeneracy encompasses the WIMP mass as well.  The Silicon experiment can in principle resolve the rollover, but sees so few events that it does not help much.  For iDM experiments, it is important that the energy thresholds be low enough to resolve the turnover in the energy spectrum.  In the purely elastic case, the energy spectrum is a falling exponetial for most conventional WIMP-nuclear operators---there is no break.  If a turnover in the energy spectrum is observed, it is evidence in favor of inelastic scattering.  It breaks the degeneracy between the WIMP velocity and $\delta$, and leads to better constraints all around.

In conclusion, we find that in order to characterize exoDM, it is important to have experiments with low energy thresholds and experiments with low atomic mass.  The low threshold is important such that experiments with heavy target nuclei will see some events.  For iDM, low thresholds and a variety of target nuclei are important in order to resolve the turnover in the energy spectrum.  While seeing the turnover in one experiment would be interesting (although possibly degenerate with the $h_2$ operator from Sec.~\ref{sec:WIMPphys:effective}), the pattern of recoil spectrum shapes in an ensemble of experiments would be a smoking gun for iDM. We could start setting interesting constraints on the WIMP particle parameters (and velocity distribution).  Importantly, this would greatly reduce the prior dependence of the velocity distribution parameter estimation.  In each figure in this section, we saw how much better parameter constraints were with the strong rather than weak velocity priors.  However, even for ``typical'' designs for Generation 2 experiments, it should be possible to distinguish $|\delta| > 20 $ keV from elastic dark matter for cross sections just below current sensitivities.

\section{Non-directional experiments and WIMP astrophysics}\label{sec:astrophysics}
\subsection{Overview}

In this section, we consider the prospects for reconstructing the particle physics properties of dark matter while accounting for astrophysical uncertainties in direct detection data, as well as the prospects for recovering the dark matter distribution function itself. Knowledge of the WIMP speed distribution is instrumental in calculating the expected event rate at direct detection experiments, influencing both the shape and normalization of the observed event spectra. However, poor assumptions about the speed distribution can lead to biased reconstructions of the dark matter mass and cross section (see e.g.\ Ref.~\cite{Strigari2009,Peter:2011eu}). It is therefore imperative that uncertainties in the WIMP distribution are accounted for, not least because this may allow us to measure any deviations from the standard assumptions. We briefly summarize previous attempts to accomodate these uncertainties, before illustrating how the dark matter properties can be reconstructed using a recently proposed method.

 As outlined in Sec.~\ref{sec:theory:astrophysics}, there are significant uncertainties in the parameters ($v_\textrm{LSR}$, $\sigma_v$, $v_\textrm{esc}$)  associated with the Standard Halo Model so a first step would be to incorporate these uncertainties into reconstructions. Strigari and Trotta \cite{Strigari2009} introduce a simple model of the Milky Way mass distribution, from which SHM velocity parameters can be derived. They then use projected stellar kinematics and direct detection data to fit both the model parameters and the dark matter properties. A more direct approach is to directly fit the SHM velocity parameters, incorporating their uncertainties into the fitting likelihood. This method has been considered by Peter \cite{Peter:2009ak}, and is typically used as a simple model of astrophysical uncertainties (especially in studies which focus on other aspects of direct detection, e.g.\ Ref.~\cite{arina2013}). These methods allow bias in the reconstructed WIMP parameters to be eliminated when the underlying speed distribution is indeed in the SHM form. However, as shown by Peter \cite{Peter:2011eu}, these methods fail when the distribution function differs from the standard Maxwellian case.

There have also been attempts to incorporate and fit more realistic distribution functions. Pato et al. \cite{pato2011} incorporate astrophysical uncertainties by using the distribution function of Lisanti et al. \cite{lisanti2010} and fitting the various shape parameters associated with it. In a more recent paper, Pato et al. \cite{pato2013} use projected direct detection data to fit a model of the Milky Way mass distribution, from which they derive a self-consistent distribution function using Eddington's formula. This means that the resulting speed distribution will be consistent with the underlying potentials of the galaxy's bulge, disk and dark matter, incorporating a broader range of shapes than the SHM alone. However, as the authors point out, velocity distributions from cosmological N-body simulations differ significantly from those expected from Eddington's formula. As with the Standard Halo Model, fitting a realistically-motivated distribution function is likely to result in biased reconstructions if the true distribution deviates significantly from the functional form used for fitting.

Methods which make no assumptions about the functional form of the speed distribution have had mixed success. Drees and Shan \cite{drees2008} developed a method for estimating the WIMP mass by calculating moments of the speed distribution. However, this method still introduces a bias into the reconstructed WIMP mass and performs more poorly for heavier WIMPs and when finite energy thresholds are considered. An empirical ansatz for the speed distribution has also been suggested, specifically dividing the WIMP speed into a series of bins, with the distribution being constant within each bin \cite{Peter:2011eu}. A similar method using bins in momentum-space has also been investigated \cite{Kavanagh:2012nr}. However, both of these still result in a significant bias in the reconstructed mass and cross section.

In the following, we consider the reconstruction prospects using the parametrization method presented in Ref.~\cite{Kavanagh:2013wba}. This involves parametrizing the logarithm of the WIMP speed distribution in the Earth frame as a polynomial in the WIMP speed $v$. This ensures that the resulting distribution is everywhere positive. We parametrize up to a conservative maximum speed $v_\textrm{max} = 1000 \textrm{ km s}^{-1}$ and use the precise form:

\begin{equation}
f(v) = v^2 \exp\left\{ -\sum_{k=0}^{N-1} a_k \tilde{P_k}\left(\frac{v}{v_\textrm{max}}\right)  \right\}\,,
\end{equation}
subject to the normalisation condition
\begin{equation}
\int_{0}^{v_\textrm{max}} f(v) \, \textrm{d}v = 1\,.
\end{equation}
We use a basis of shifted Legendre polynomials \(\tilde{P_k}\) of degree \(k = \{0,1,...,N-1\}\) for the parametrization. In this work, we use $N = 5$ basis polynomials. An in-depth analysis of the effect of varying the number of basis functions is left to a future paper. We attempt to fit the particle physics parameters, $m_\chi$ and \sigmapsi, and the speed distribution parameters $\left\{a_1,...,a_4\right\}$, with $a_0$ fixed by normalisation. This method has previously been illustrated using a single set of benchmark WIMP parameters \cite{Kavanagh:2013wba} and we aim to extend that analysis here.

\subsection{Benchmarks}

We generate mock data sets using three different sets of theoretical WIMP parameters, which are shown in Table~\ref{tab:astrophysics:benchmarks}. These benchmarks are chosen to span the range of sensitivity of direct detection experiments. WIMPs lighter than around 10 GeV typically have too little kinetic energy to excite nuclear recoils with energy of order a few keV. WIMPs which are significantly heavier than the mass of the detector nuclei lead to a degeneracy in the $m_\chi - \sigma_p$ plane which leads to a loss of sensitivity to the underlying WIMP mass. This degeneracy has been explored thoroughly in the literature (see, for example, Ref.~\cite{green2008,Peter:2011eu}), so we only consider an upper WIMP mass of 200 GeV. We have also chosen the cross section for each benchmark to be below the current exclusion limits from Xenon100 \cite{Aprile:2012nq}. We restrict the analysis to spin-independent interactions, in order to focus on tackling the astrophysical uncertainties of direct detection.

\begin{table}[t]
  \setlength{\extrarowheight}{3pt}
  \setlength{\tabcolsep}{3pt}
  \begin{center}
	\begin{tabular}{c|ccc}
	Benchmark & $m_\chi$ / GeV & \sigmapsi / $\textrm{cm}^2$ & $\sigma_p^{SD}$ / $\textrm{cm}^2$ \\
	\hline\hline
	A & 10 & $10^{-43}$ & 0 \\
	B & 50 & $10^{-45}$ & 0 \\
	C & 200 & $10^{-45}$ & 0 \\
	\end{tabular}
  \end{center}
\caption{Summary of particle physics benchmarks used in Section \ref{sec:astrophysics}.}
\label{tab:astrophysics:benchmarks}
\end{table}

\begin{figure}
  \includegraphics[width=0.49\textwidth]{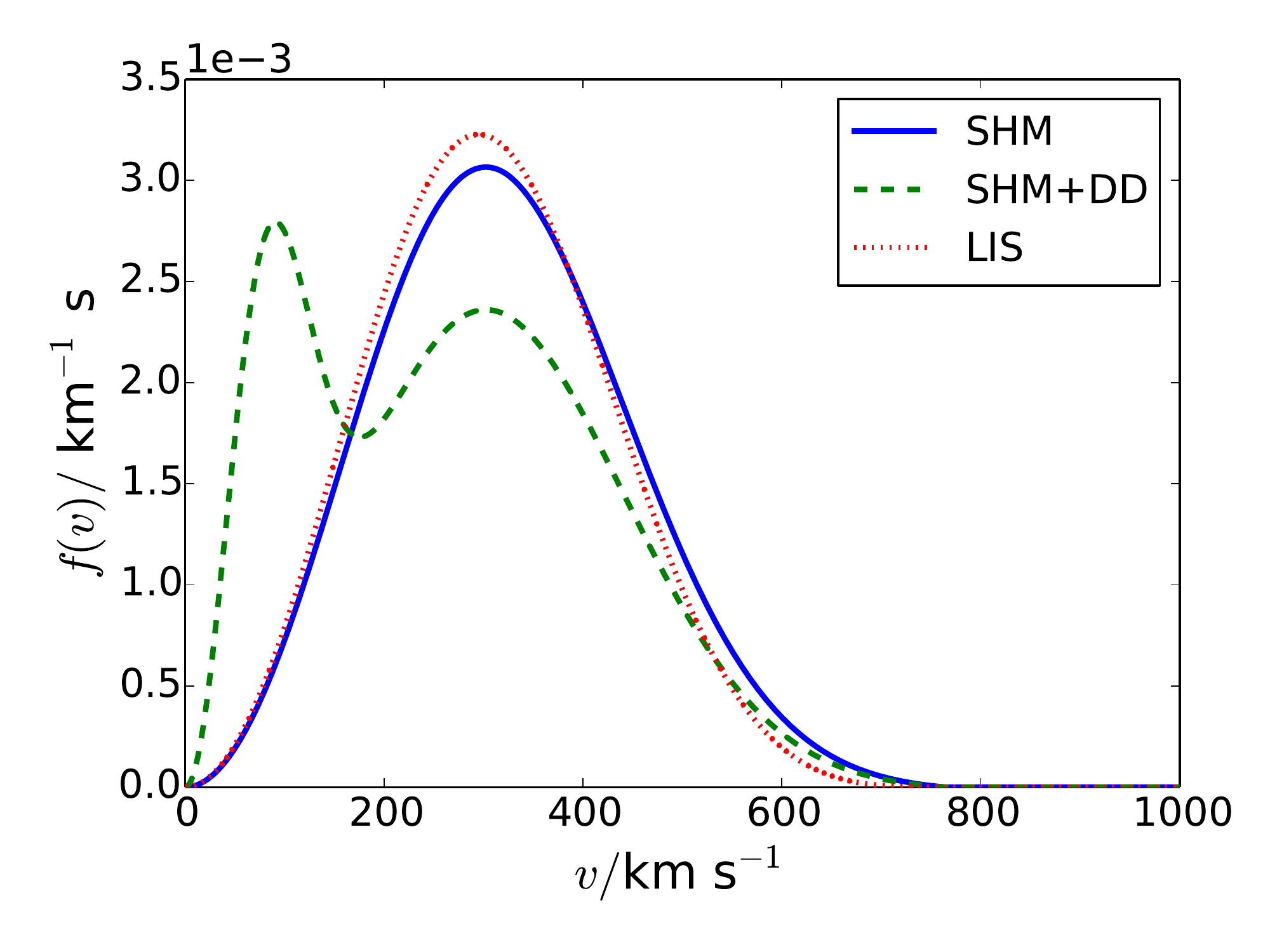}
  \caption{{\bf Benchmark speed distributions (in the Earth's reference frame) used in Sec.~\ref{sec:astrophysics}.}}
  \label{fig:astrophysics:speeddistributions}
\end{figure}

We consider three different underlying WIMP speed distributions. The first is a standard halo model, with the speed of the Local Standard of Rest (LSR) set to $220 \textrm{ km s}^{-1}$ and $\sigma_v = 155 \textrm{ km s}^{-1}$. The second is an SHM speed distribution with a 30\% overdensity contributed by a dark disk, which lags behind the Galactic distribution at $v_\textrm{lag} = 50 \textrm{ km s}^{-1}$ with $\sigma_v = 50 \textrm{ km s}^{-1}$. This is a relatively conservative dark disk scenario and is consistent with typical estimates of the dark disk contribution of between 25\% and 150\% \cite{Read:2008fh,Bruch:2008rx,Kuhlen:2013tra}. Finally, we also consider the functional form of Lisanti et.~al \cite{lisanti2010} which gives a good match to cosmological N-body simulations. In the Galactic frame, this has the form:

\begin{equation}\label{eq:lisanti}
f(\textbf{v}) = N \left[ \exp\left(\frac{v_\textrm{esc} - |\textbf{v}|^2}{kv_0^2}\right) -1\right]^k\, \Theta(v_\textrm{esc} - |\textbf{v}|) \,.
\end{equation}
In this work, we consider the specific case of $v_0 = 220 \textrm{ km s}^{-1}$ and spectral index $k = 2.0$. For all three distributions, we impose a hard cut off above the escape speed $v_\textrm{esc} = 544 \textrm{ km s}^{-1}$ in the Galactic frame. These benchmark distributions, which we will refer to as SHM, SHM+DD and LIS respectively, are illustrated in in Fig.~\ref{fig:astrophysics:speeddistributions}.

We assume that the local dark matter density $\rho_\chi$ is known exactly. As described in Sec.~\ref{sec:theory:astrophysics}, $\rho_\chi$ in fact has an uncertainty of around 50\%. However, the local dark matter density is entirely degenerate with the interaction cross section. We are therefore free to fix $\rho_\chi$ in our reconstructions, as long as we acknowledge that this 50\% uncertainty should now be associated with the reconstructed values of \sigmapsi. We note that we use only $N = 1000$ live points in \textsc{MultiNest} for this section, as the reconstructions are quite computationally intensive. Finally, we include in Table~\ref{tab:astrophysics:priors} a list of prior ranges used in the reconstructions.

\begin{table}[t]
  \setlength{\extrarowheight}{3pt}
  \setlength{\tabcolsep}{3pt}
  \begin{center}
	\begin{tabular}{c|cc}
	Parameter & Prior type & Prior Range \\
	\hline\hline
	$m_\chi$ / GeV & log-flat & [1,1000] \\
	$\sigma_p^{SI}$ / cm$^{2}$ & log-flat & [$10^{-46}$, $10^{-42}$] \\
	$\left\{a_k\right\}$ & linearly-flat & [-50, 50] \\
	\end{tabular}
  \end{center}
\caption{Summary of priors used in Section \ref{sec:astrophysics}.}
\label{tab:astrophysics:priors}
\end{table}

\subsection{Reconstructing $m_\chi$ and $\sigma_p^{SI}$}

\begin{figure}
  \includegraphics[width=0.49\textwidth]{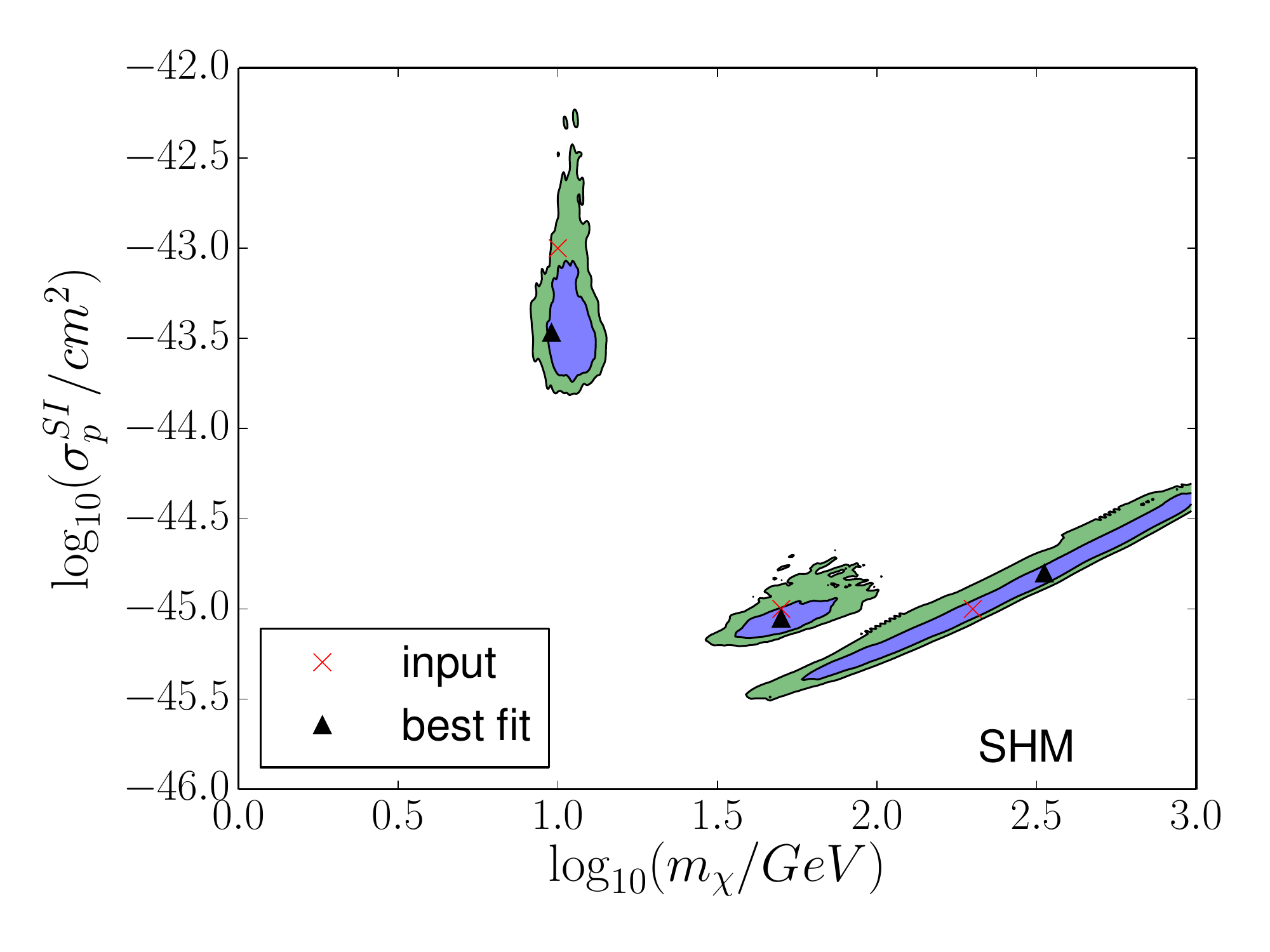}
  \includegraphics[width=0.49\textwidth]{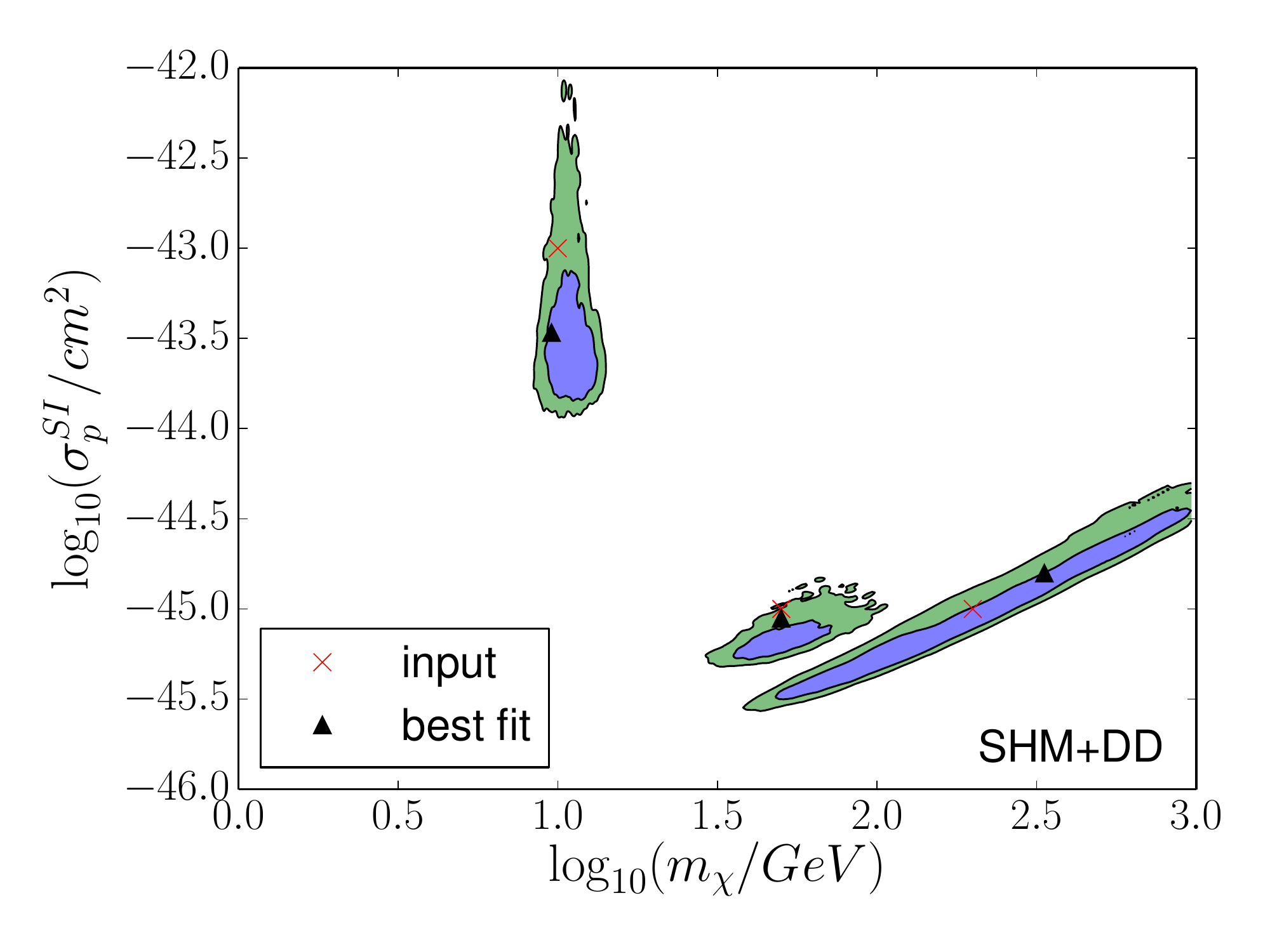}
  \includegraphics[width=0.49\textwidth]{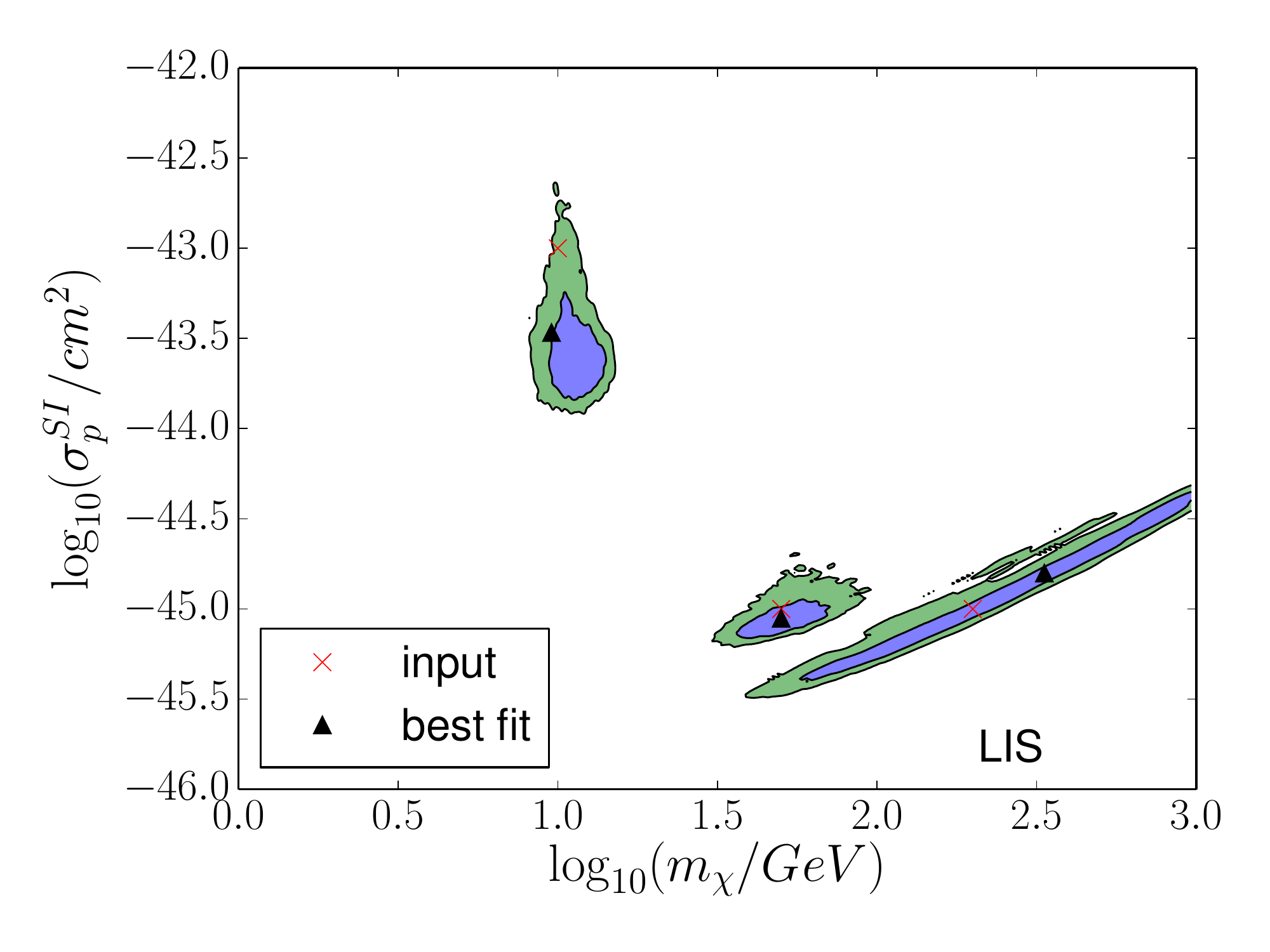}
  \caption{{\bf Marginalized posterior distributions obtained using the parametrization of Ref.~\cite{Kavanagh:2013wba}.} Inner and outer contours enclose the 68\% and 95\% credible regions respectively. The input values of the parameters are shown as red crosses, while the best-fit points are shown as black triangles. Data were generated using the SHM (upper), SHM+DD (middle) and LIS (lower) distributions.}
  \label{fig:astrophysics:allexperiments}
\end{figure}

We now present the results of parameter reconstructions for the particle physics parameters ($m_\chi$, $\sigma_p^{SI}$). Fig. \ref{fig:astrophysics:allexperiments} show the 68\% and 95\% credible contours in the $m_\chi$-$\sigma_p^{SI}$ plane for all 9 benchmarks obtained using all 5 experiments detailed in Table~\ref{tab:Experiments}. Also shown are the true values of the WIMP mass and cross section (crosses) and the best fit (i.e.\ maximum likelihood) values (triangles). In all 9 cases, the true values lie within the 95\% contours, indicating that there are no significant problems with the reconstructions, for a range of WIMP masses and distribution functions. For the 50 GeV WIMP, we also note that the best fit point for all three distributions is close to the true values. For the high mass case, there is significant degeneracy along a line \sigmapsi$\,\propto m_\chi$. This occurs in all experiments (regardless of whether astrophysical uncertainties are considered) and, as previously mentioned, is caused by a loss of sensitivity to the WIMP mass. If $m_\chi$ significantly exceeds the mass of the target nucleus, the recoil energy imparted becomes independent of $m_\chi$. Varying the WIMP mass then simply rescales the total number of WIMPs (for a given DM density), leading to the degeneracy along  \sigmapsi$\,\propto m_\chi$.

There also appears to be a bias in the reconstructed value of the cross section for both the 10 GeV and 50 GeV WIMPs. This occurs because for lower mass WIMPs, experiments probe only relatively high-speed WIMP speeds. There is therefore little information about what fraction of WIMPs lie at lower speeds, outside the sensitivity of the experiments. This problem is unavoidable in methods which make no assumptions about the speed distribution. Due to finite energy sensitivity windows, direct-detection experiments can only probe a finite range of speeds. This problem is worst for the case of 10 GeV WIMPs, for which only the high-$v$ tail of the distribution is probed, leading to a strong degeneracy in the cross section. We note, however, that this degeneracy in the cross section does not affect the reconstruction of the WIMP mass, with no significant bias observed along the $m_\chi$ direction.

Next we consider how the number of experiments impact these reconstructions, focusing on the reconstructed WIMP mass. Consider, for example, data from a single experiment. In attempting to fit particle- and astro-physics to this data, any change in the reconstructed WIMP mass can be exactly compensated by a change in the fitted WIMP speed distribution. This leads to a strong degeneracy and almost no constraints placed on the mass. Physically, a single experiment measures the energies of incoming WIMPs. Thus, a given nuclear recoil may be due to a heavy, slow-moving particle or a lighter, faster-moving particle. Incorporating data from different experiments allows this degeneracy to be broken, as a WIMP of a given mass and speed produces different recoil energies in different detectors.

\begin{figure*}
  \includegraphics[width=0.32\textwidth]{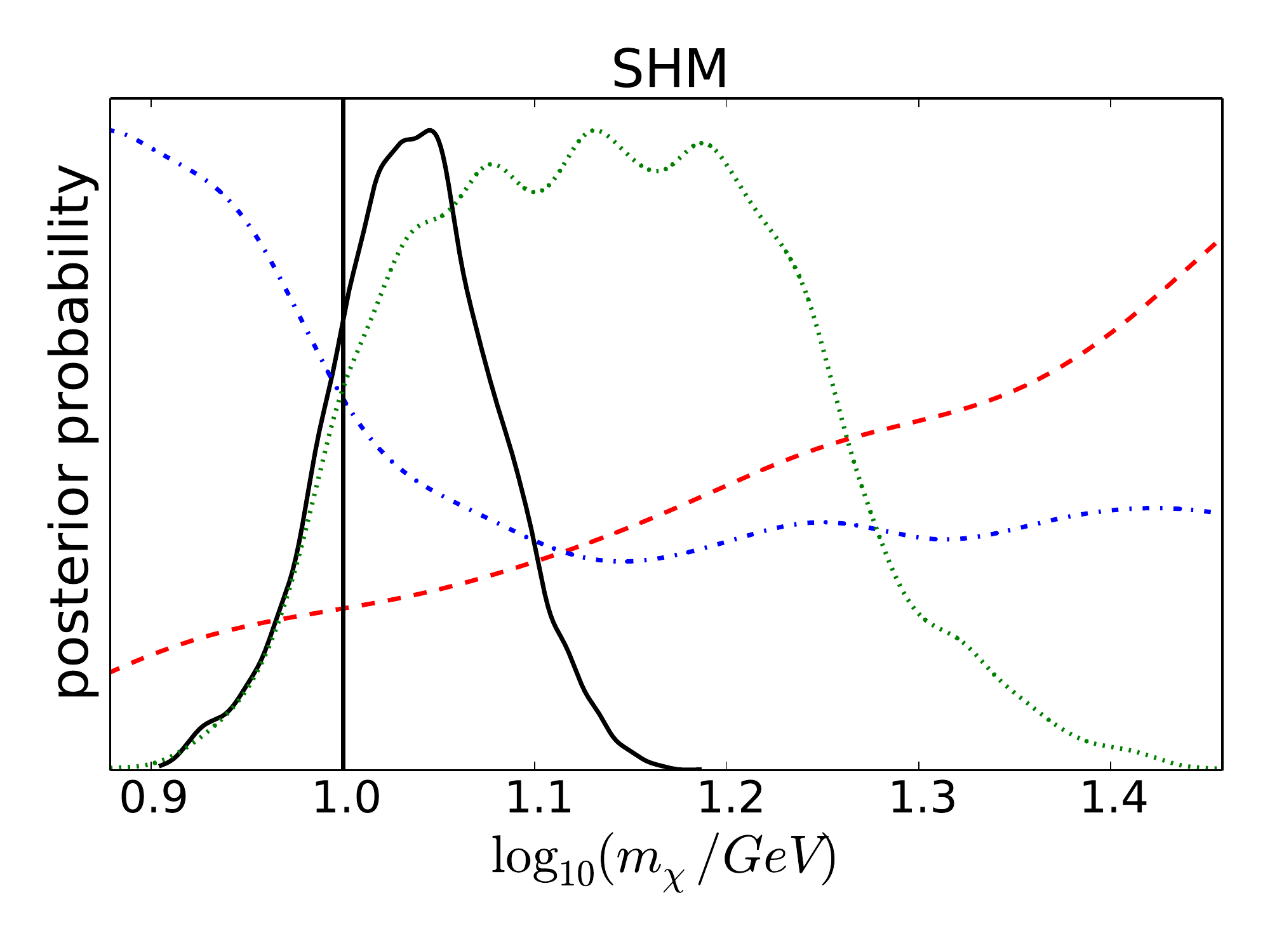}
  \includegraphics[width=0.32\textwidth]{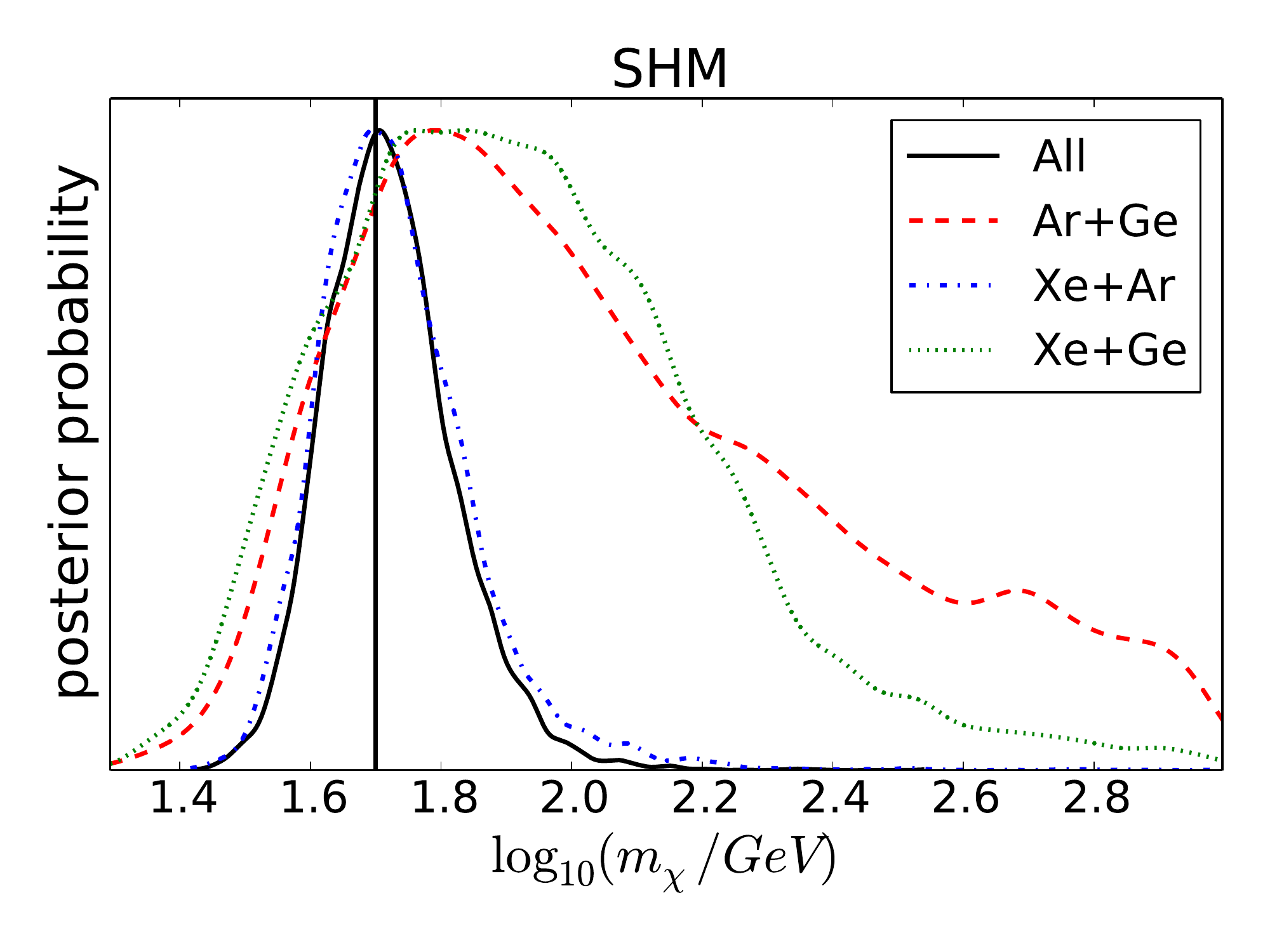}
  \includegraphics[width=0.32\textwidth]{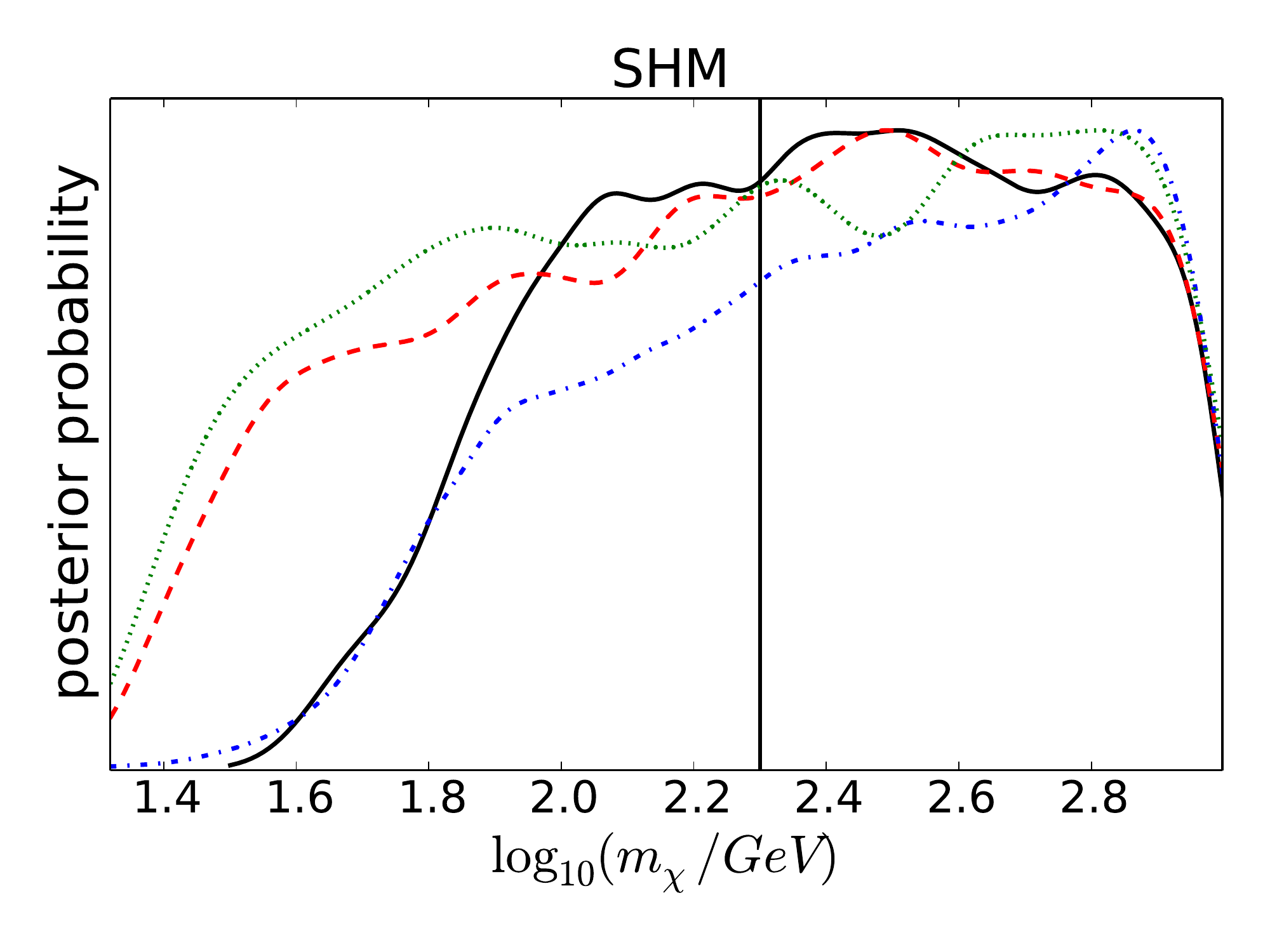}
  \includegraphics[width=0.32\textwidth]{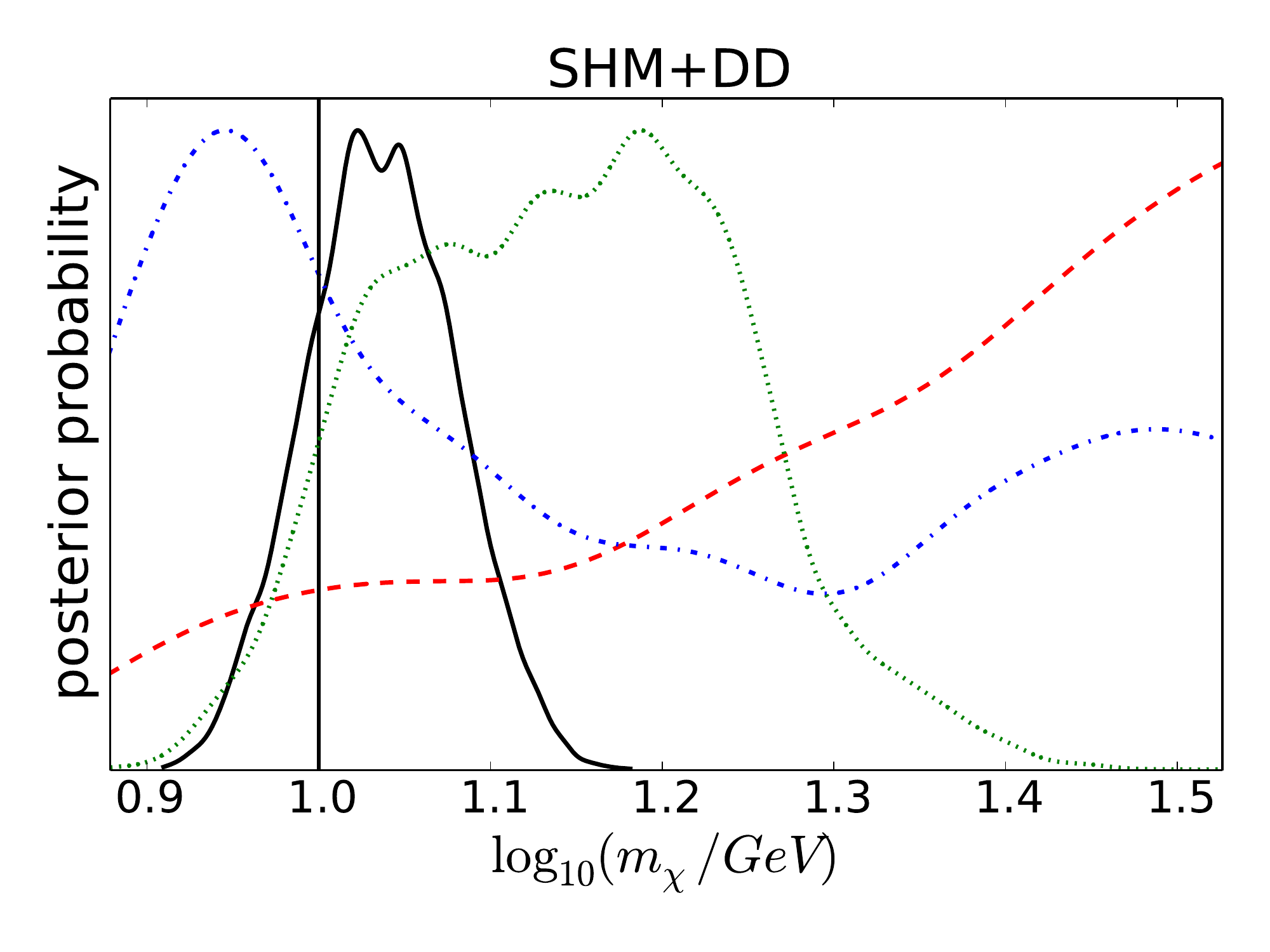}
  \includegraphics[width=0.32\textwidth]{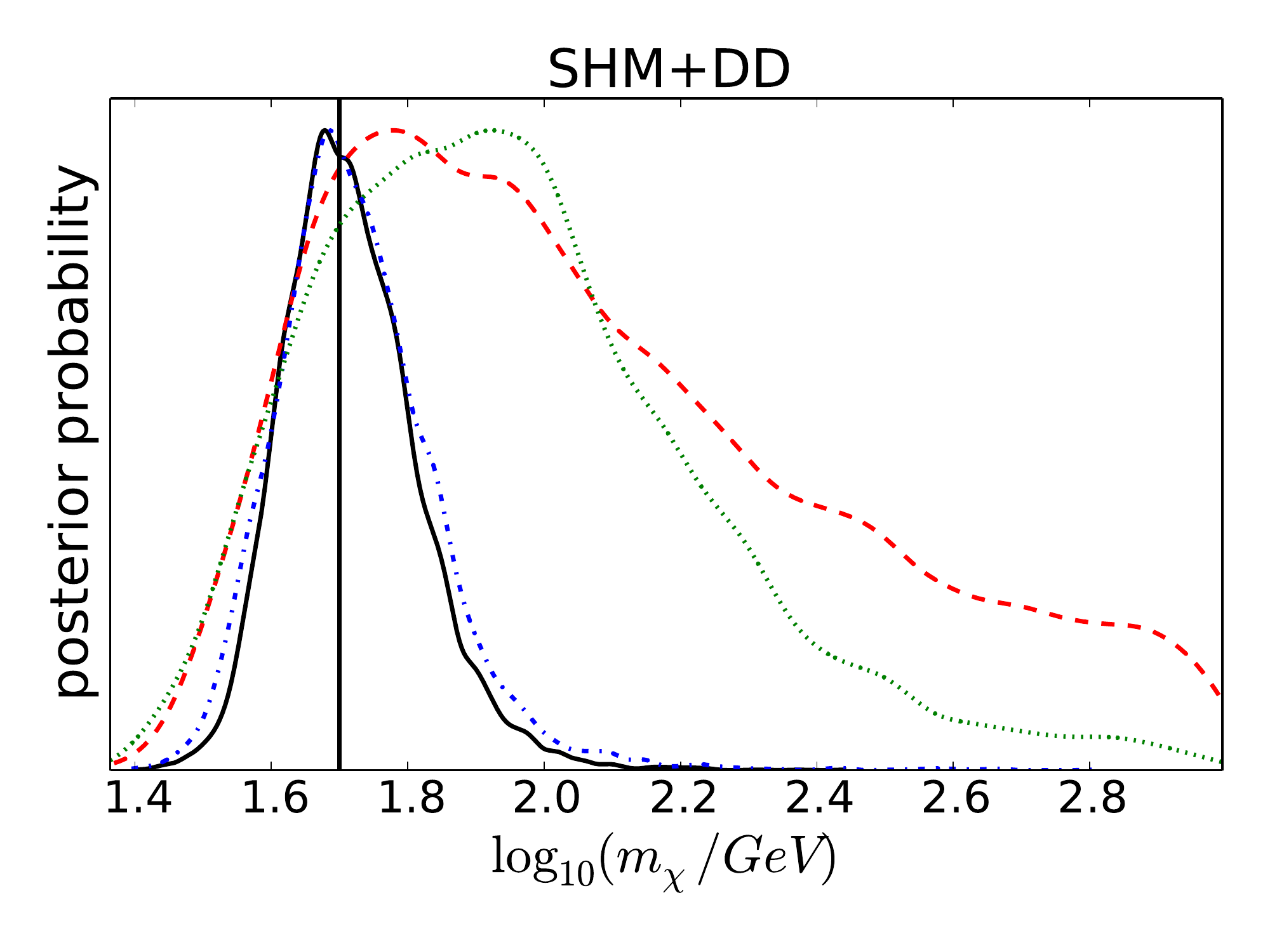}
  \includegraphics[width=0.32\textwidth]{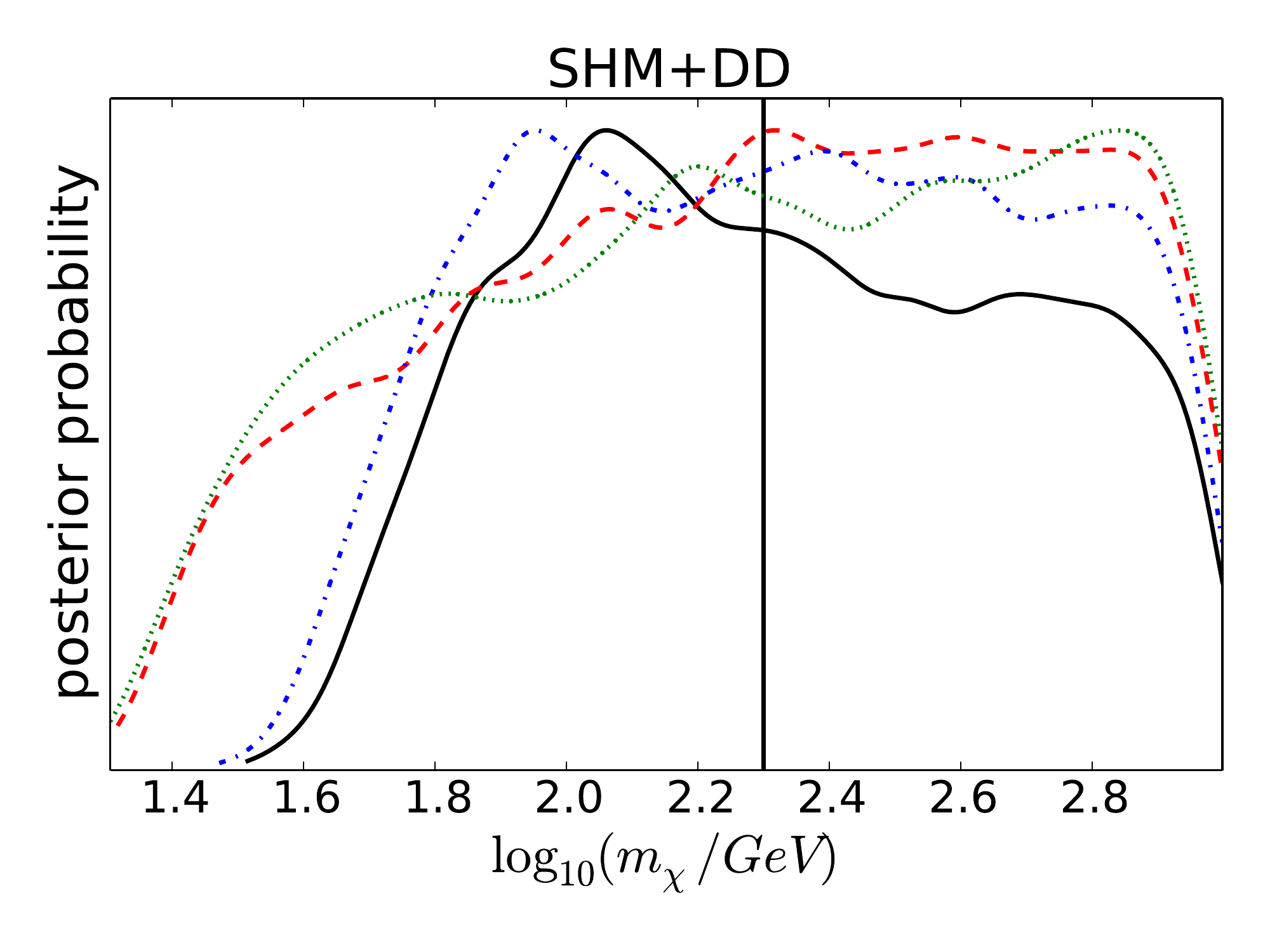}
  \includegraphics[width=0.32\textwidth]{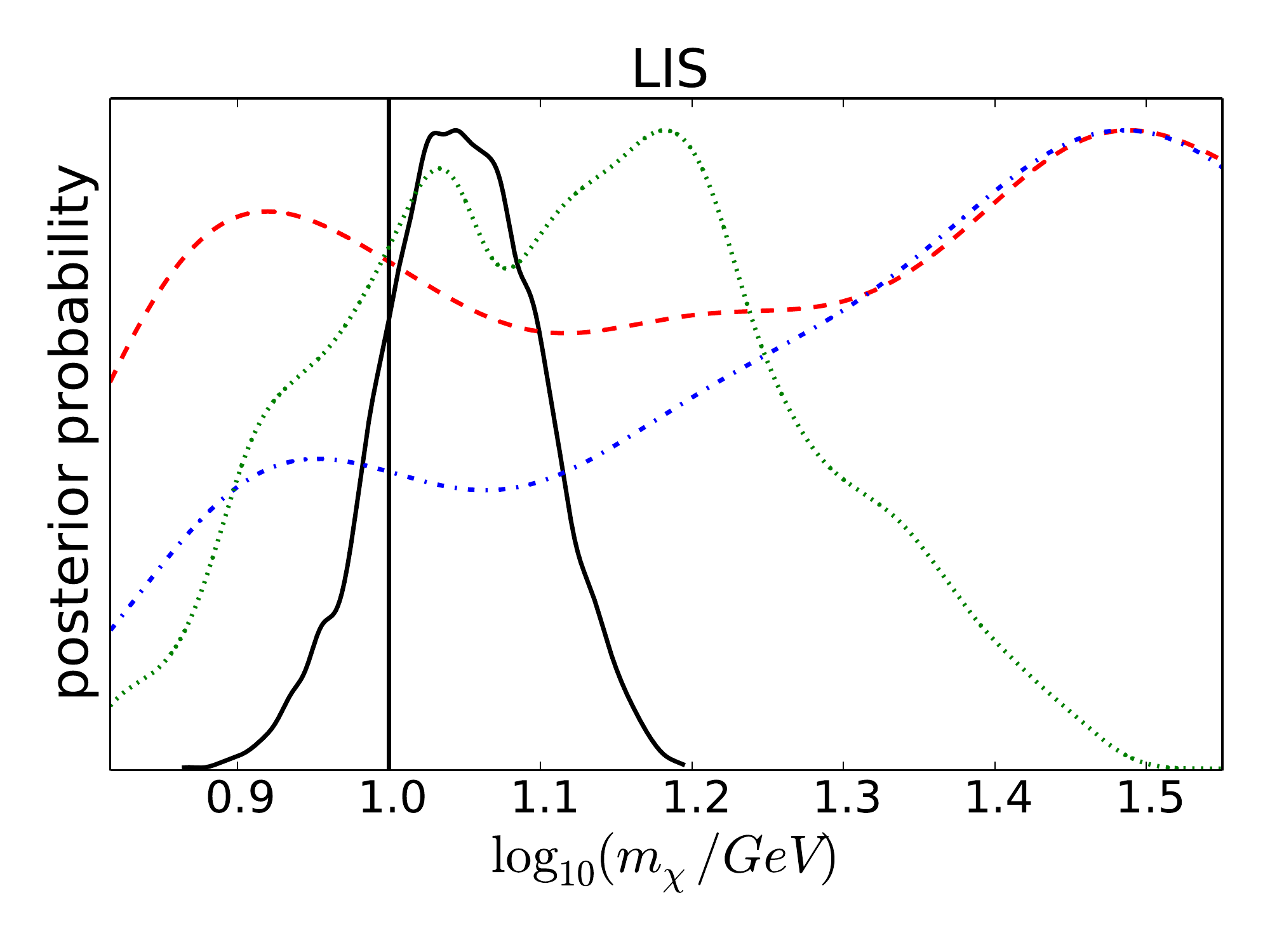}
  \includegraphics[width=0.32\textwidth]{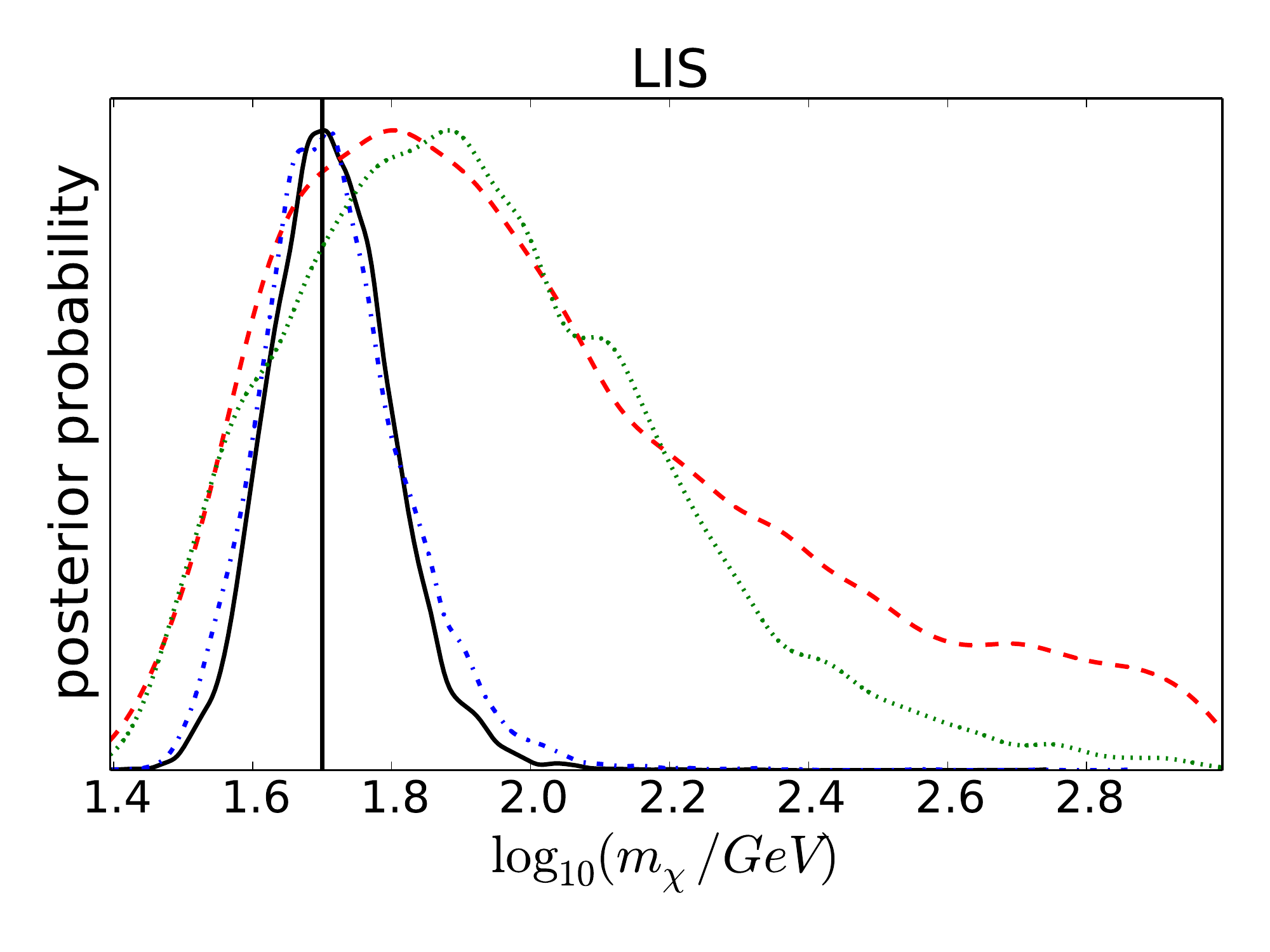}
  \includegraphics[width=0.32\textwidth]{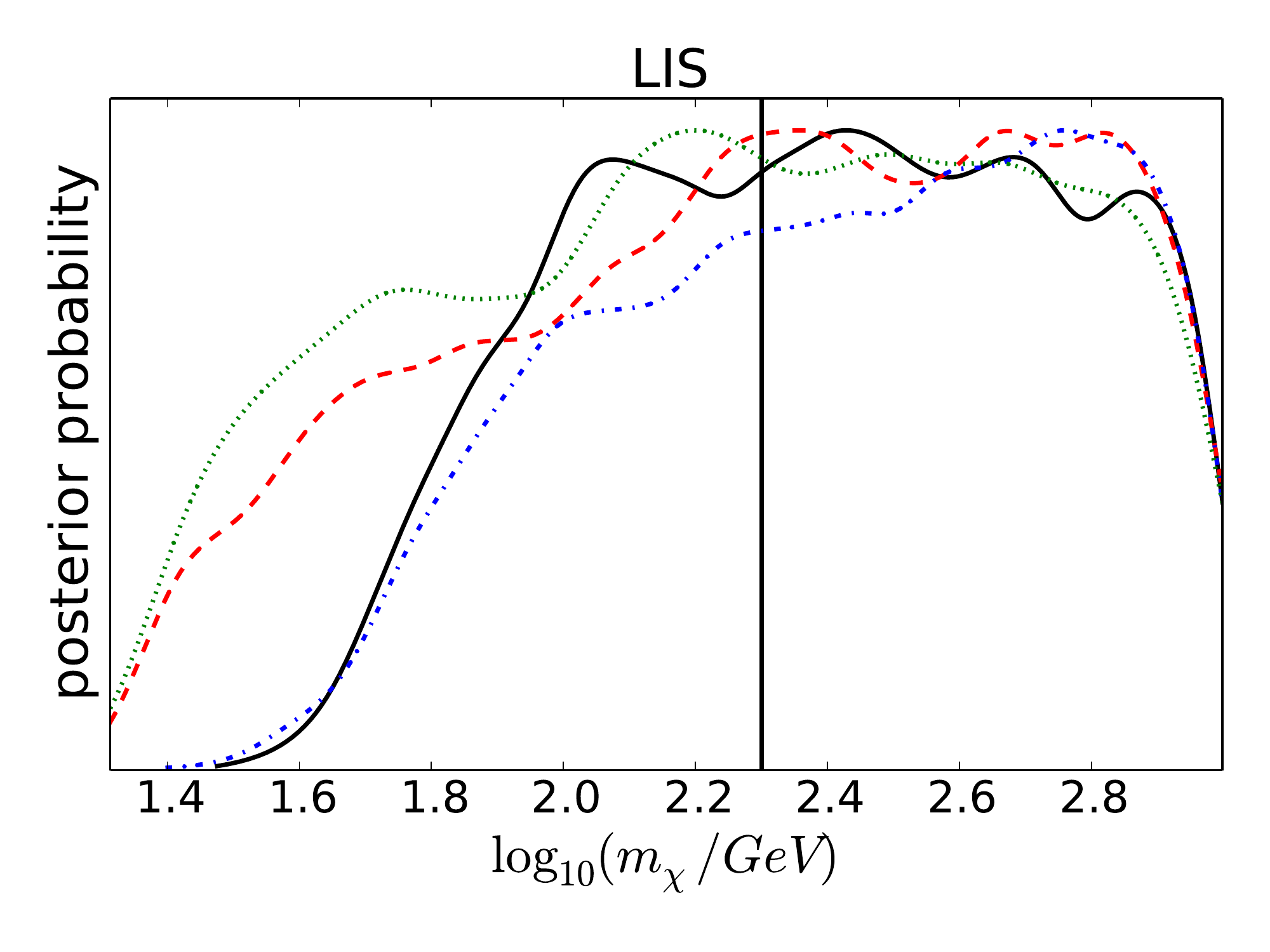}
  \caption{{\bf Marginalized 1-D posterior for the WIMP mass, obtained using the method of Ref.~\cite{Kavanagh:2013wba}, as a function of experiment ensemble.} Shown are the results using all 5 experiments presented in Table~\ref{tab:Experiments}, as well as using several different pairings of the Xe, Ar and Ge experiments. The vertical line marks the input mass.}
  \label{fig:astrophysics:MassRecon}
\end{figure*}

Fig.~\ref{fig:astrophysics:MassRecon} shows the marginalized posterior for the WIMP mass for all 9 benchmarks. We show the posterior obtained using data from all 5 hypothetical experiments, as well as the posterior obtained using data from several pairings of detectors: Ar+Ge, Xe+Ar and Xe+Ge. The reconstruction of the WIMP mass does not appear to differ significantly between different underlying distribution functions. This is a reflection of the fact that the parametrization used here is able to effectively marginalize over the astrophysical uncertainties. However, with fewer experiments, there is a larger space of distribution functions which can fit the data, leading to the increased uncertainty in $m_\chi$ which is observed.

In the case of a 200 GeV WIMP, the effect is least noticeable as the uncertainties on the WIMP mass will be large no matter which detectors we include. Because of the $m_\chi \propto\,$\sigmapsi~degeneracy described earlier, adding more data from experiments with a range of target masses adds little extra information about the WIMP mass. The effect is much more pronounced for 10 GeV and 50 GeV WIMPs, with the posterior probability distributions becoming significantly broader as we reduce the number of experiments used. As an example, consider the 68\% credible interval obtained for a 50 GeV WIMP with the SHM distribution function:
\begin{eqnarray}
\textrm{All experiments:} \qquad & m_\chi \in [41, 63] \textrm{ GeV} \\
\textrm{Xe \& Ge only:} \qquad & m_\chi \in [40, 138] \textrm{ GeV.}
\end{eqnarray}
The 68\% interval has widened by a factor of $\sim 4$ when we reduce the number of experiments to just two. There is less information to break the degeneracy between the WIMP mass and speed distribution and thus the constraints are weaker.

We note that a particular pairing of detectors may perform differently depending on the underlying WIMP parameters. For the 50 GeV case, Xenon and Argon experiments alone appear to be sufficient in recovering the correct WIMP mass with relatively high precision. In the 10 GeV case, however, this pairing produces very poor results. In this case, the thresholds for the Argon experiment are too low for light WIMPs to produce a significant signal in the detector. Because of the cut off in the distribution function at the escape speed, none of the low mass WIMPs have sufficient energy to scatter in the sensitivity window of the Argon experiment. Our ability to pinpoint the mass is therefore significantly reduced.

Using several experiments, we have therefore been able to reliably reconstruct the WIMP mass (though not the interaction cross section). As we have demonstrated, using only a small number of experiments can increase the uncertainties on the WIMP mass by a significant amount. However, different combinations of experiments will lead to smaller or larger uncertainties depending on the underlying WIMP mass being probed.

\subsection{Reconstructing $f(v)$}

We now consider how well $f(v)$ has been reconstructed using this method. In the figures below, we present the underlying speed distribution (solid line), as well as the best fit reconstructed distribution (dashed line). We also include an estimate of the uncertainty in the distribution function. This is obtained by marginalizing over the value of the distribution independently at each value of $v$ and calculating the range of the 68\% minimal credible interval at each value. This is shown as a shaded region. It should be noted that because $f(v)$ must be normalized, the uncertainties at different speeds are strongly correlated, so this shaded region should be taken as an illustrative uncertainty only. We have also rescaled the best fit distribution (and uncertainties) by a factor $\sigma_\textrm{rec}^{SI}/\sigma_\textrm{true}^{SI}$, the ratio of the reconstructed and true values of the SI cross section. This allows us to compare the reconstructed distribution with the underlying one, taking account of the degeneracy in \sigmapsi. This rescaling only has a significant effect for the $m_\chi = 10 \textrm{ GeV}$ case, where this degeneracy is most pronounced.

\begin{figure*}
  \includegraphics[width=0.32\textwidth]{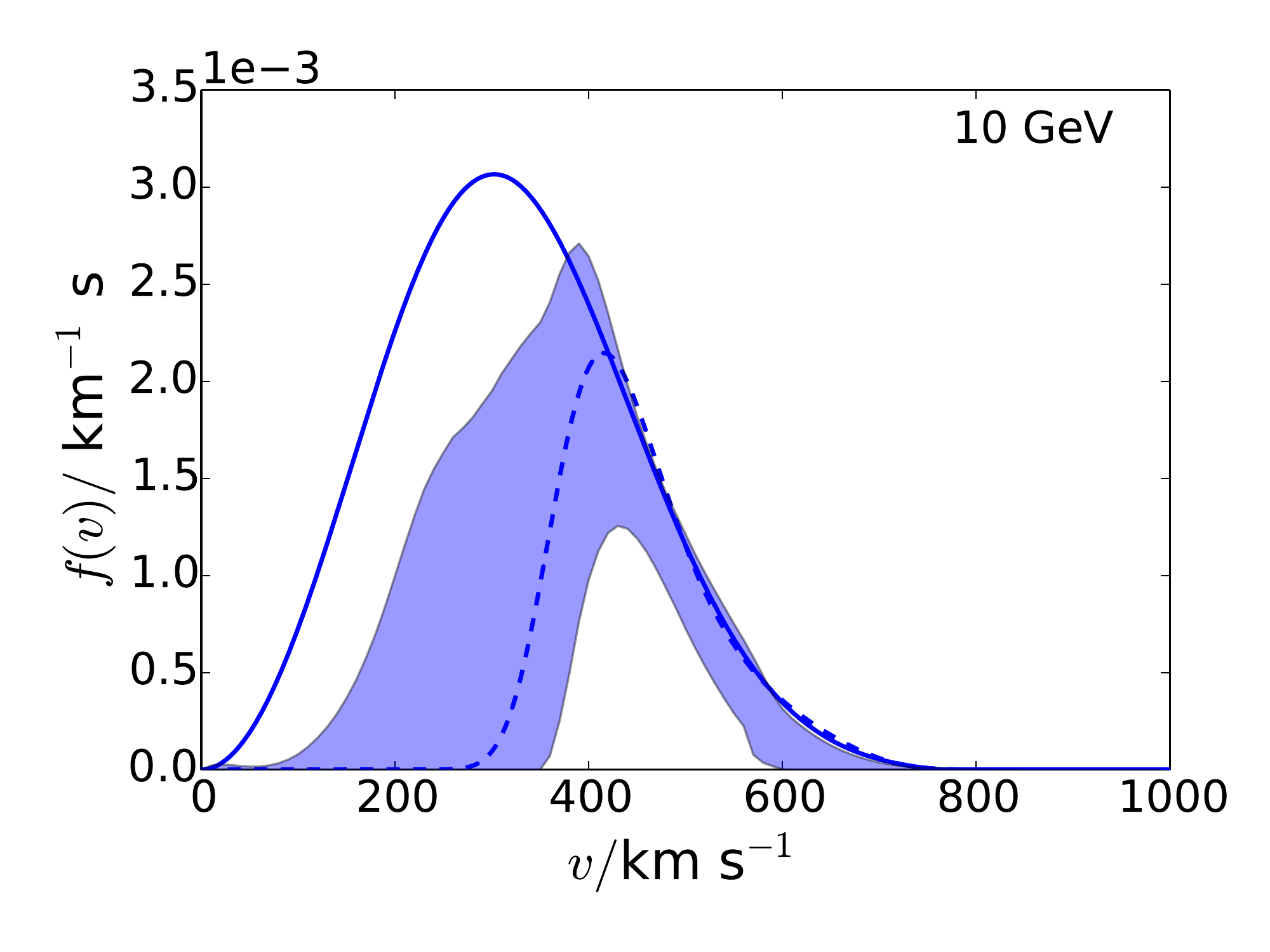}
  \includegraphics[width=0.32\textwidth]{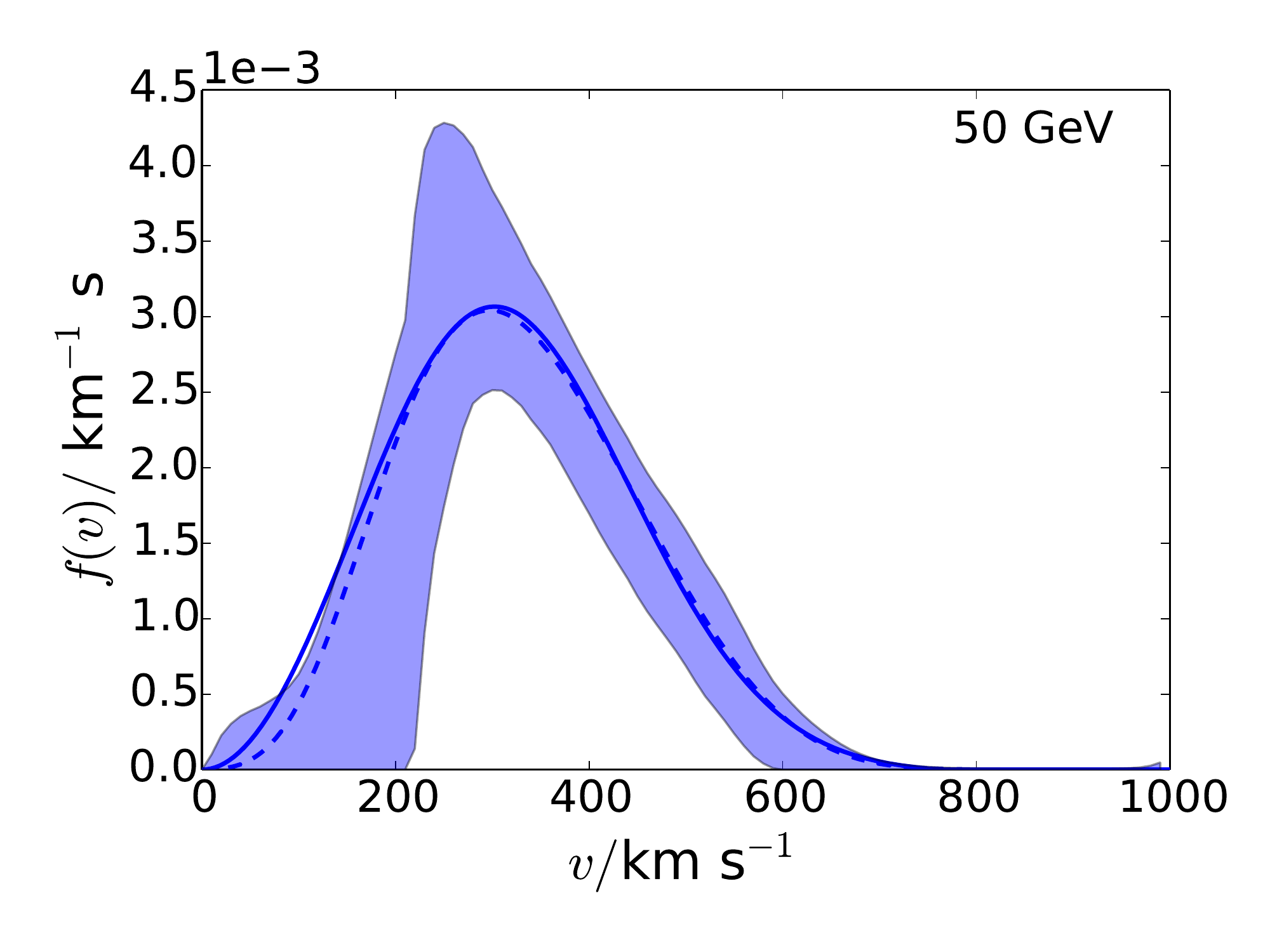}
  \includegraphics[width=0.32\textwidth]{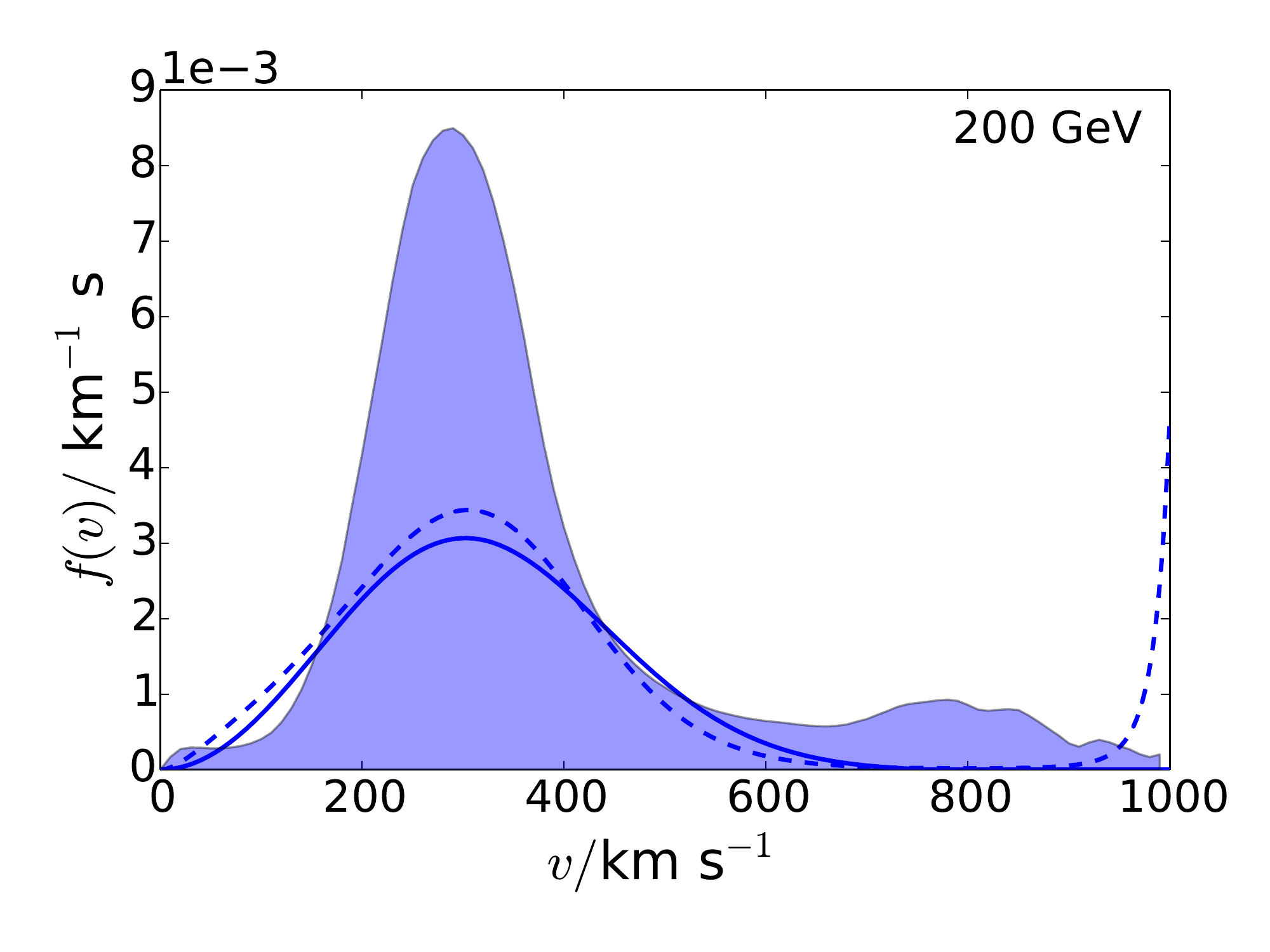}
  \includegraphics[width=0.32\textwidth]{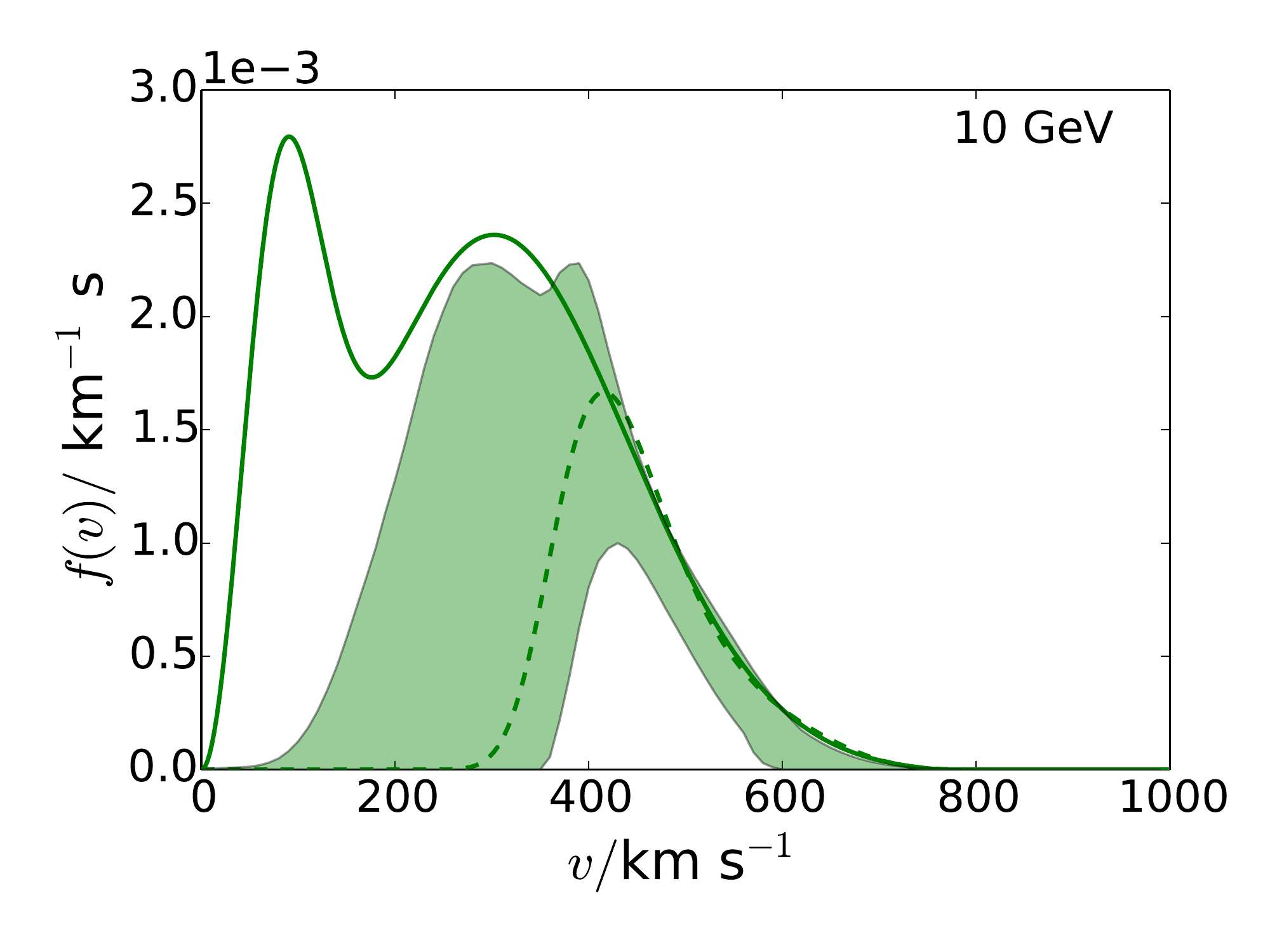}
  \includegraphics[width=0.32\textwidth]{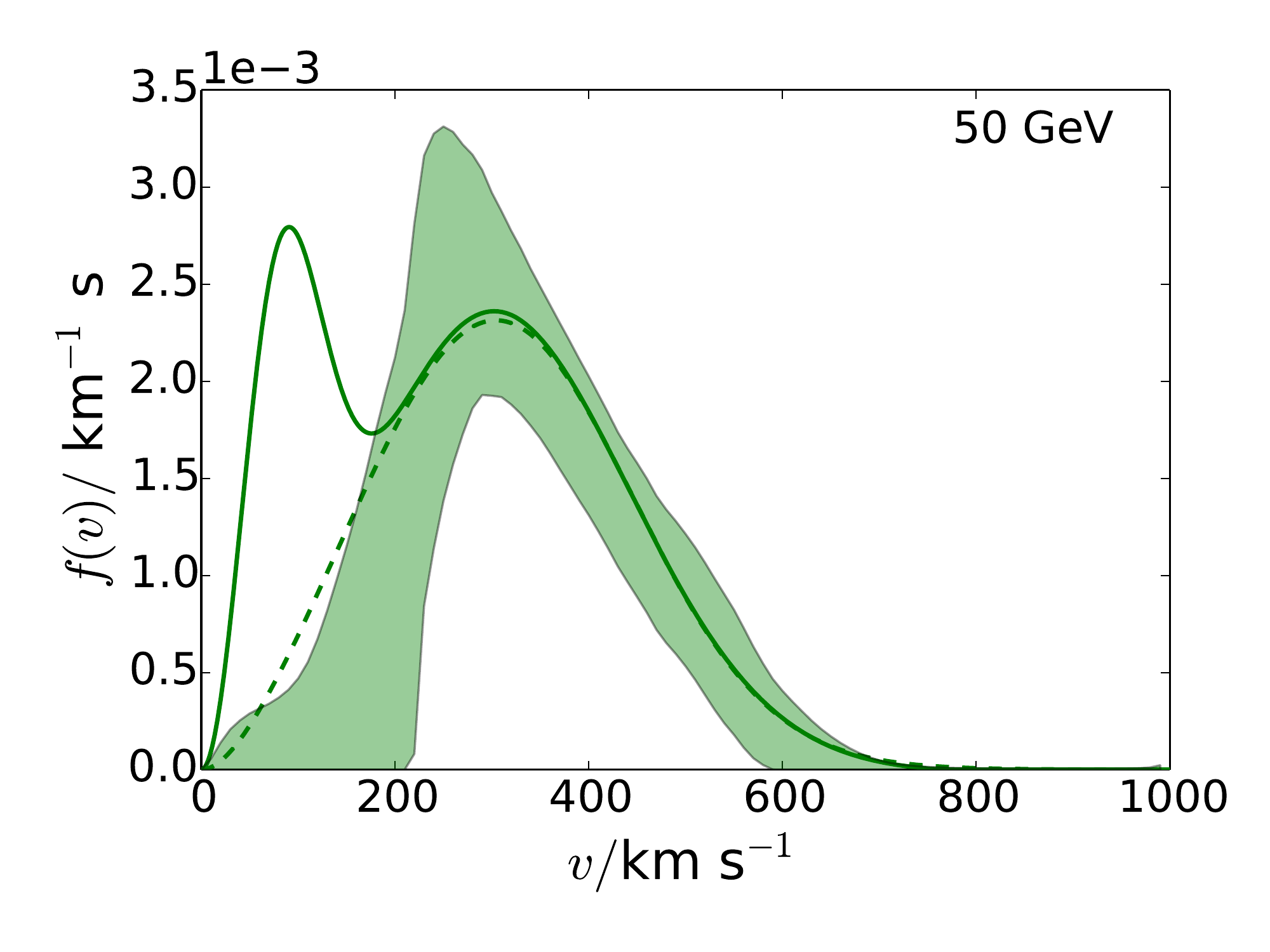}
  \includegraphics[width=0.32\textwidth]{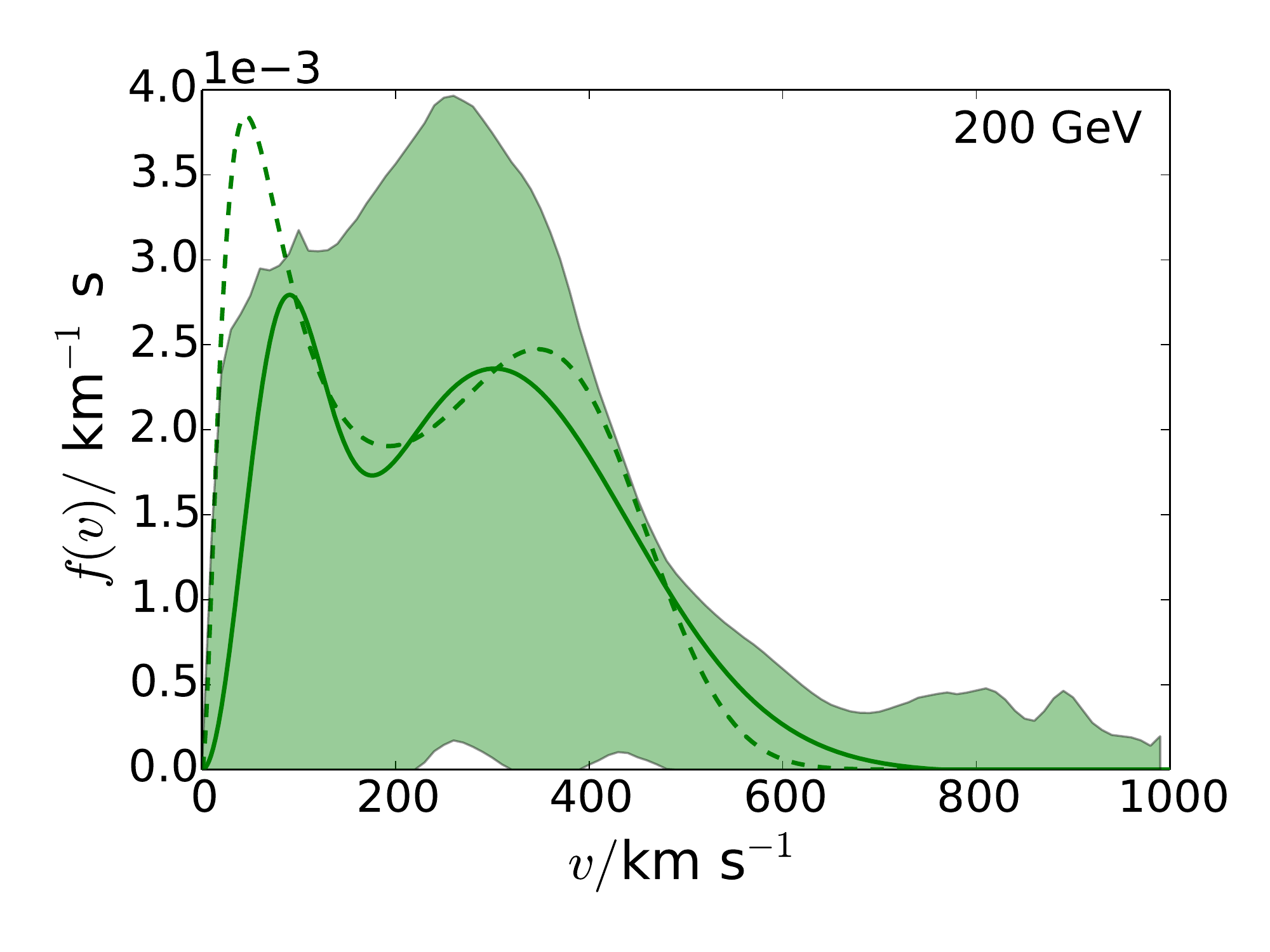}
  \includegraphics[width=0.32\textwidth]{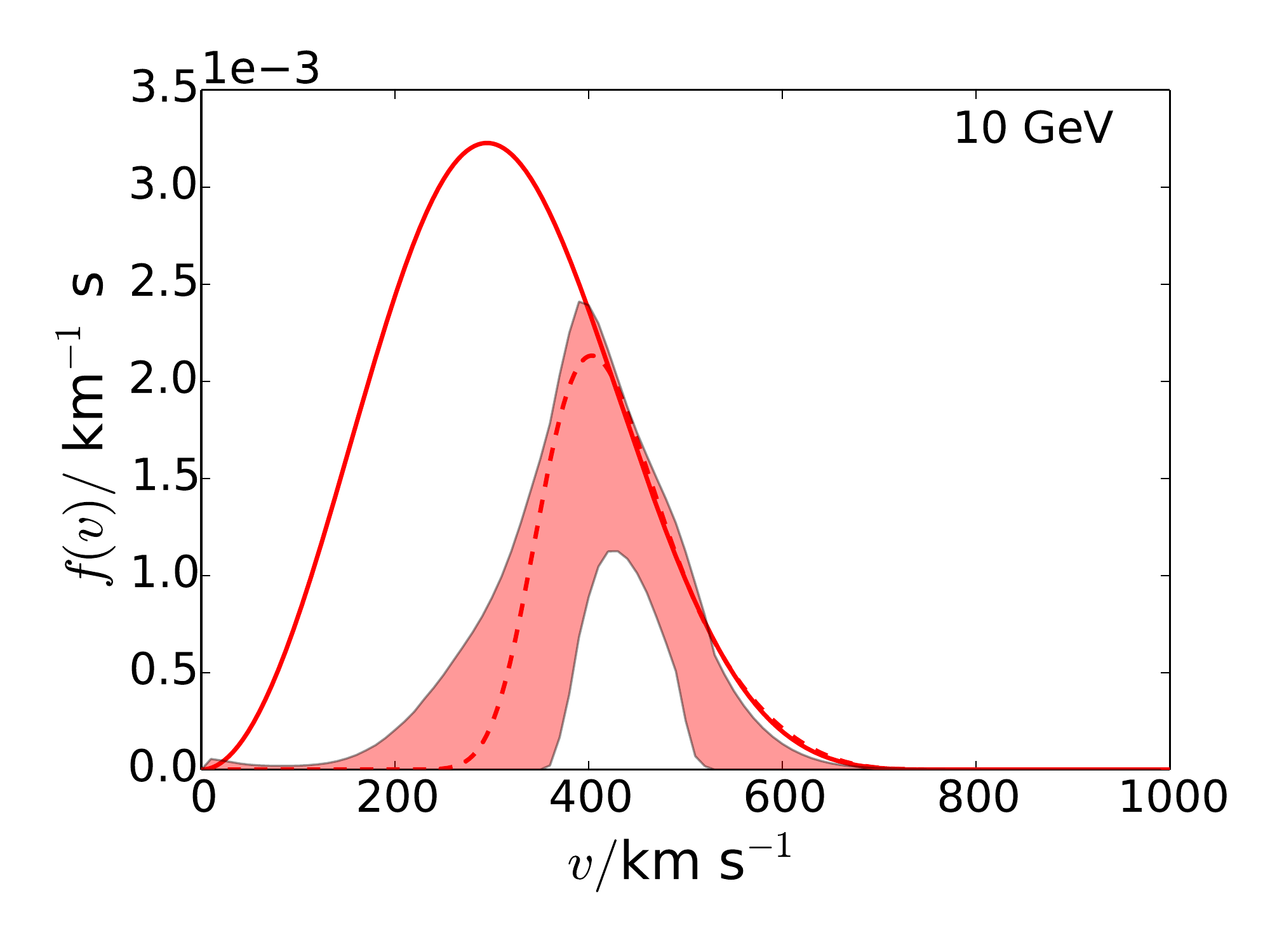}
  \includegraphics[width=0.32\textwidth]{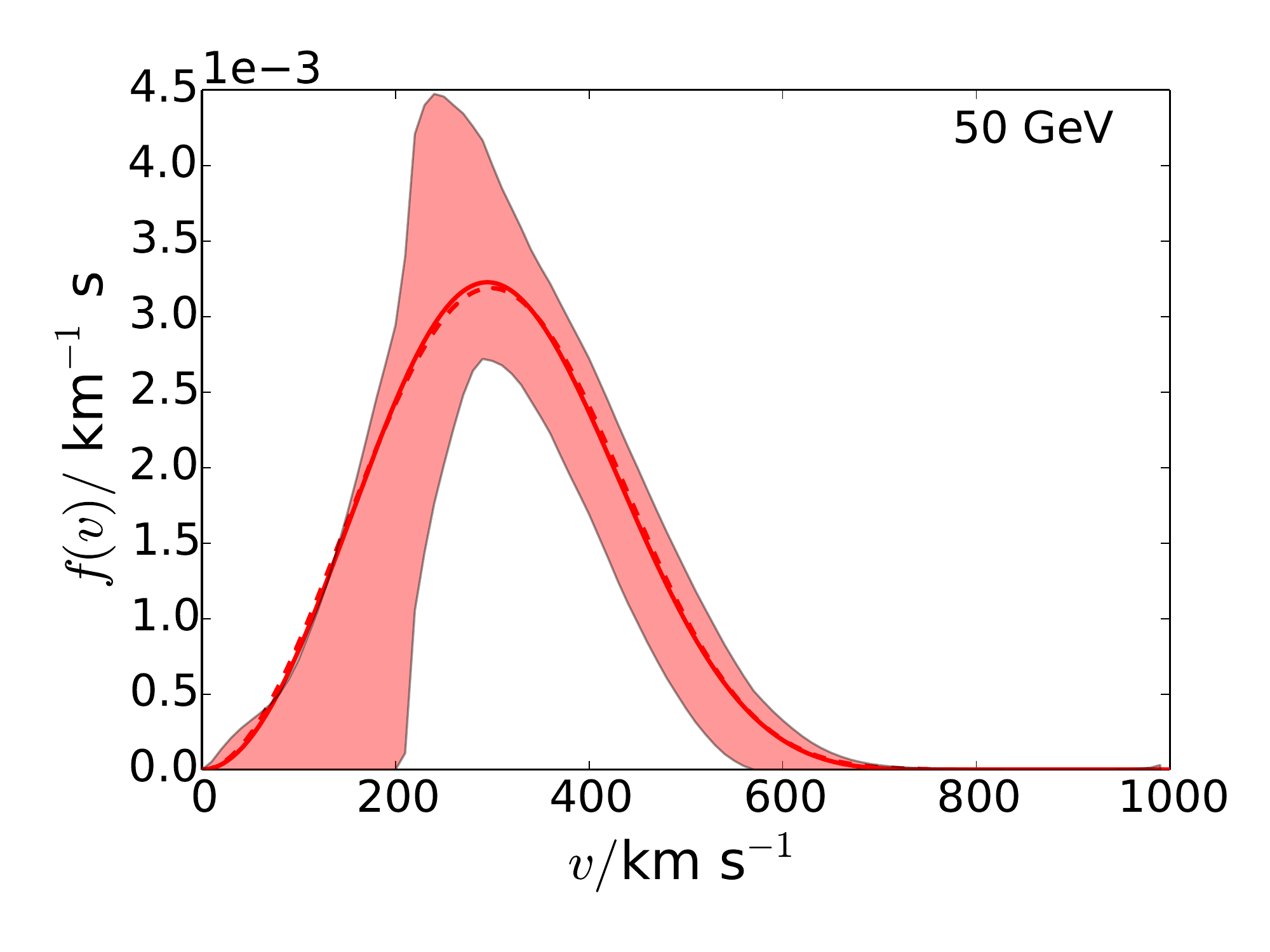}
  \includegraphics[width=0.32\textwidth]{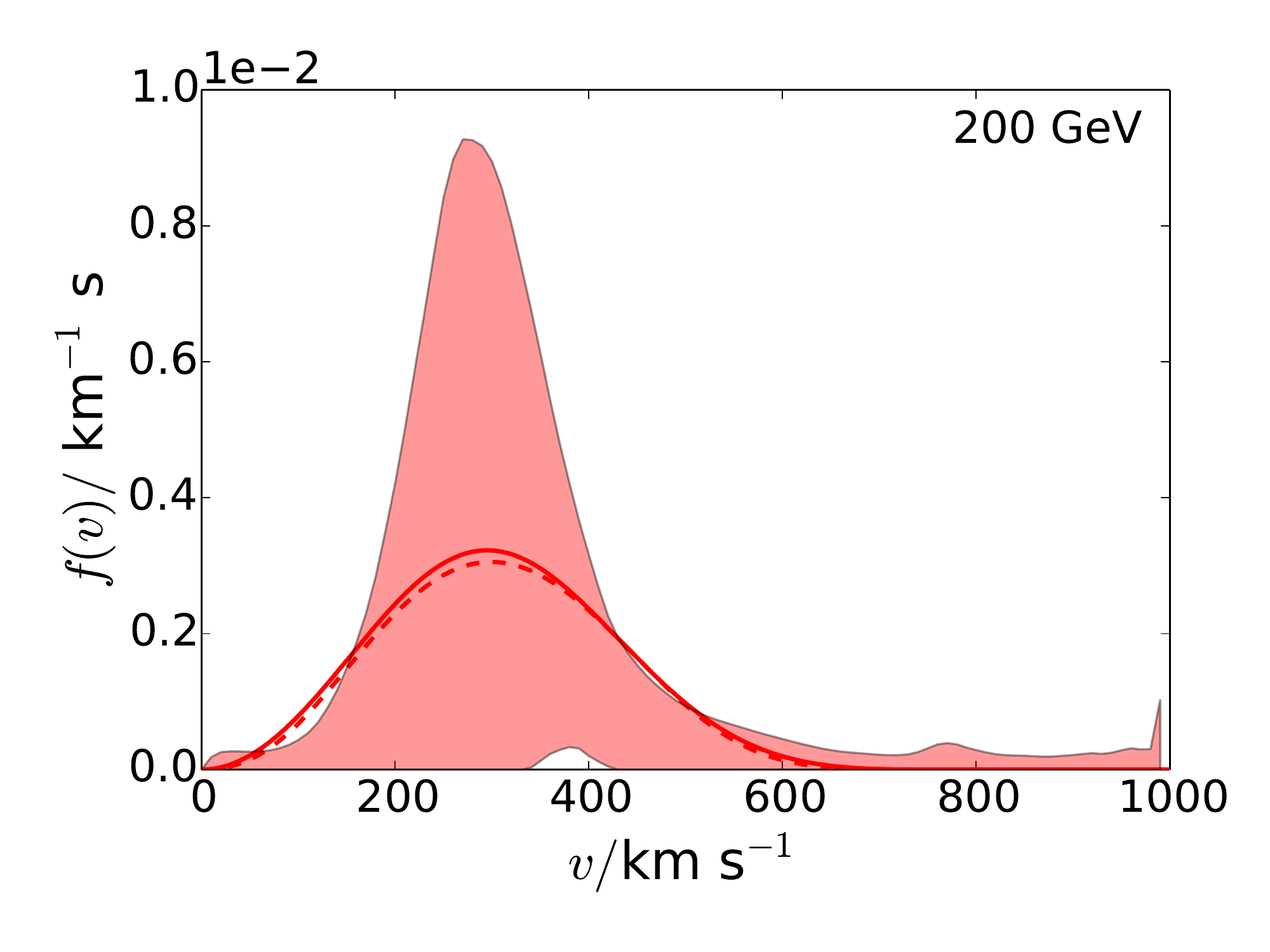}
  \caption{{\bf Reconstructed Earth-frame speed distributions for WIMP masses of 10 GeV (left column), 50 GeV (middle column) and 200 GeV (right column).} All 5 hypothetical experiments were used in the reconstruction. Three different underlying speed distributions were used, as described in the text: SHM (top row), SHM+DD (middle row), LIS (bottom row). The underlying distribution is shown as a solid line, the best fit is shown as a dashed line and the 68\% credible interval is shown as a shaded band. The reconstructed distribution has been rescaled by a factor $\sigma_\textrm{rec}^{SI}/\sigma_\textrm{true}^{SI}$, as described in the text.}
  \label{fig:astrophysics:speeds_allexperiments}
\end{figure*}

Fig.~\ref{fig:astrophysics:speeds_allexperiments} shows the reconstructed speed distribution for all 9 benchmarks, using all mock experiments. The reconstruction for a 10 GeV WIMP gives a close fit for all 3 speed distributions at speeds above $400 \textrm{ km s}^{-1}$. However, the reconstruction is insensitive to the structure of the distribution function below this speed. For a WIMP of mass 10 GeV, the lowest speed which can be probed by the experiments is $\sim 425 \textrm{ km s}^{-1}$, so we should not be surprised by this loss of sensitivity. We also note that for 10 GeV WIMPs the upper limit of the 68\% interval appears to trace the underlying distribution when we might expect the true distribution to lie reasonably centrally within this interval. However, this is a consequence of the strong correlations between values of $f$ at different speeds and the approximate nature of the uncertainties.

For a 50 GeV WIMP, the experiments are sensitive to a much larger range of speeds, from $v \sim 170 \textrm{ km s}^{-1}$ upwards. The reconstruction for this benchmark is therefore accurate over a much wider range. In the case of the SHM and LIS distributions, the best fit follows the underlying distribution closely over all speeds. However, in the case of the SHM+DD distribution, there is a significant deviation at low speeds. This does not impact the reconstruction of the WIMP mass, however, as the dark disk component lies mostly below the sensitivity of the experiments. For a 50 GeV WIMP, the SHM and SHM+DD distributions produce largely indistinguishable spectra and this is reflected in the reconstruction of the speed distribution.

WIMPs with higher masses will probe down to even lower speeds. For a WIMP of mass 200 GeV, the hypothetical experiments presented in this work would have a sensitivity down to $v \sim 80 \textrm{ km s}^{-1}$. This is why the experiments are sensitive to the dark disk component of the SHM+DD distribution in Fig.~\ref{fig:astrophysics:speeds_allexperiments}. We could attempt to increase the accuracy of the fit by increasing the number of basis functions in the parametrization but that is beyond the scope of the present work. Nonetheless, the double peaked structure of the underlying distribution function is recovered.

\begin{figure}
  \includegraphics[width=0.49\textwidth]{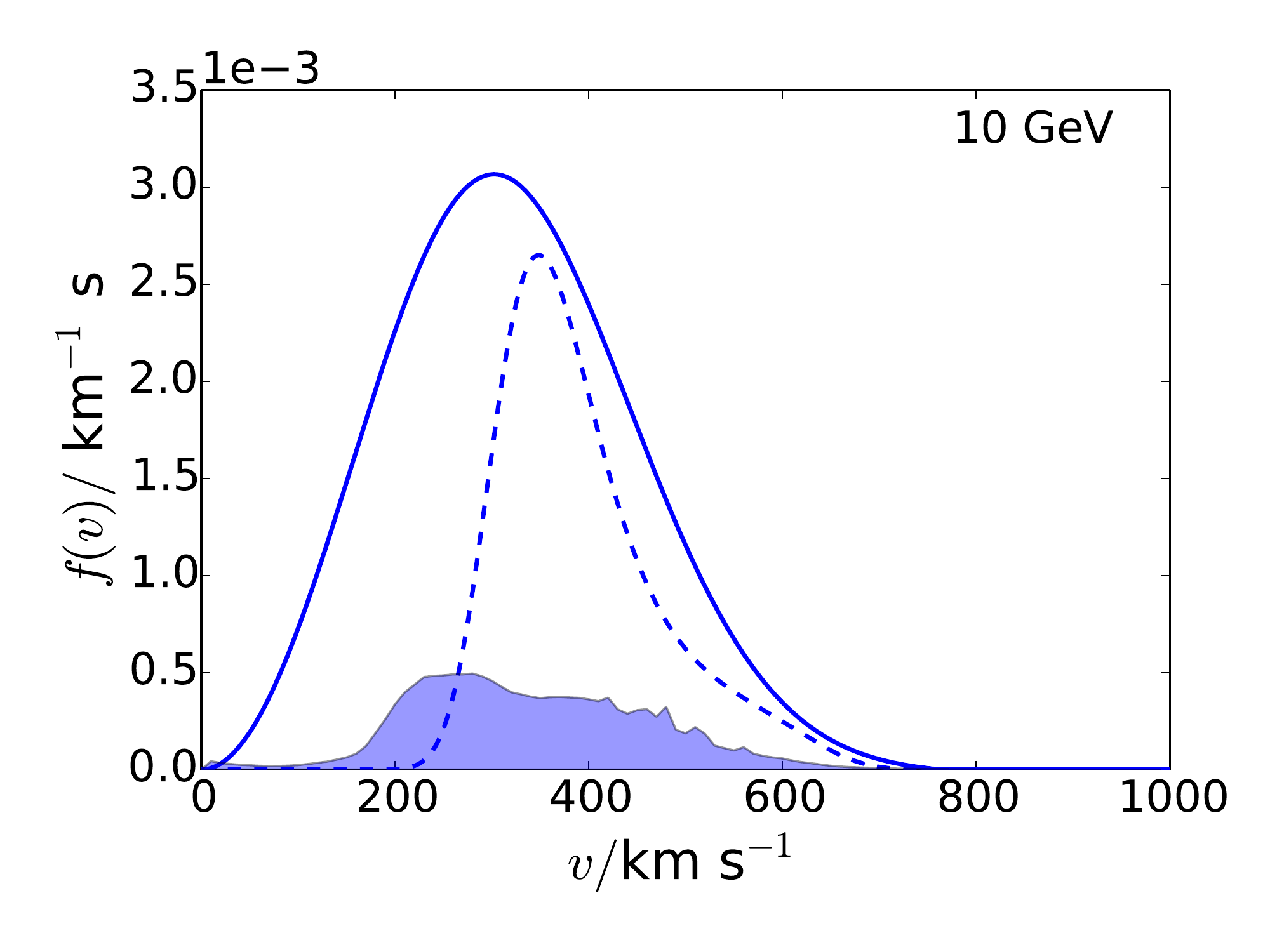}
  \includegraphics[width=0.49\textwidth]{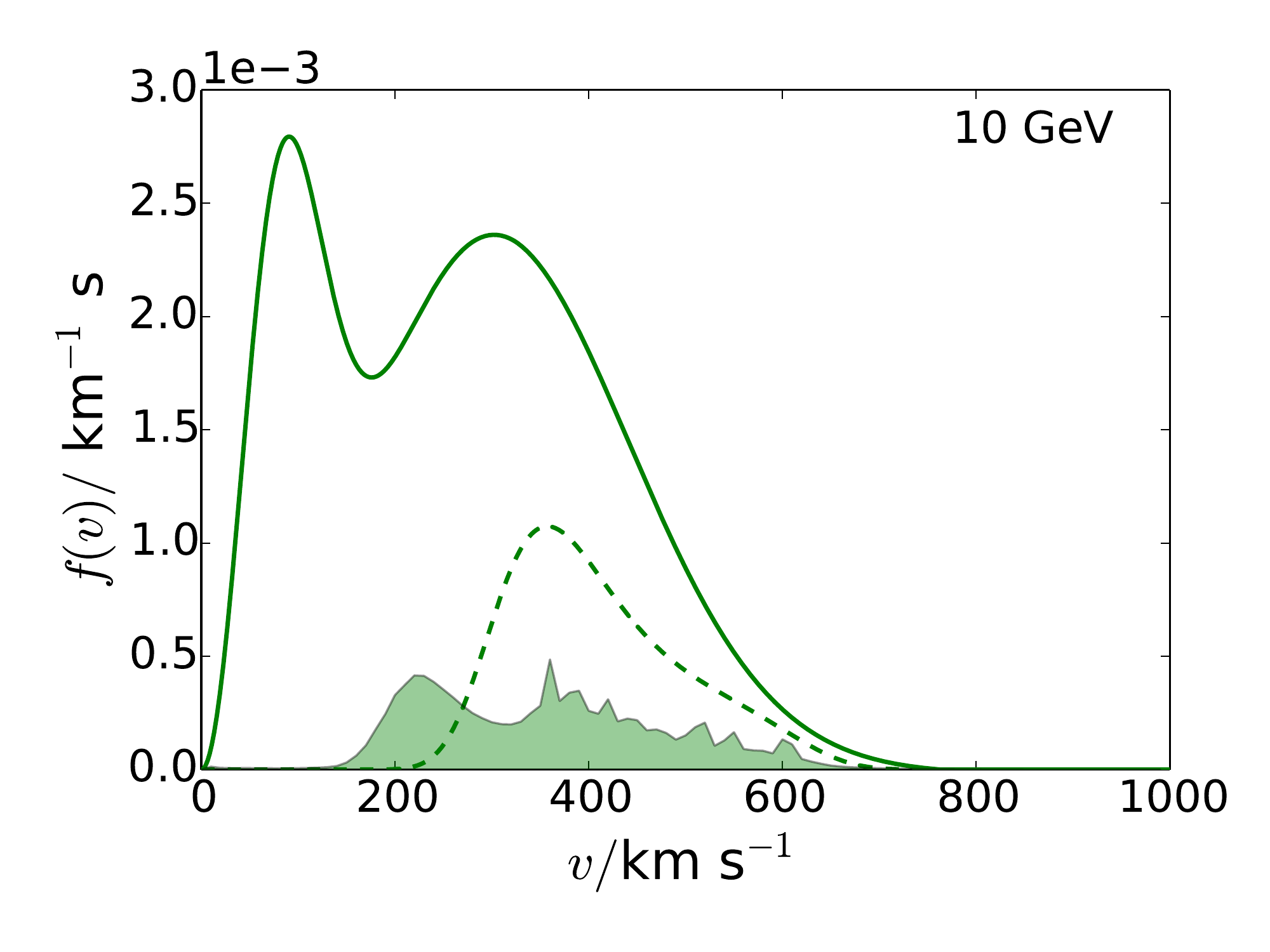}
  \includegraphics[width=0.49\textwidth]{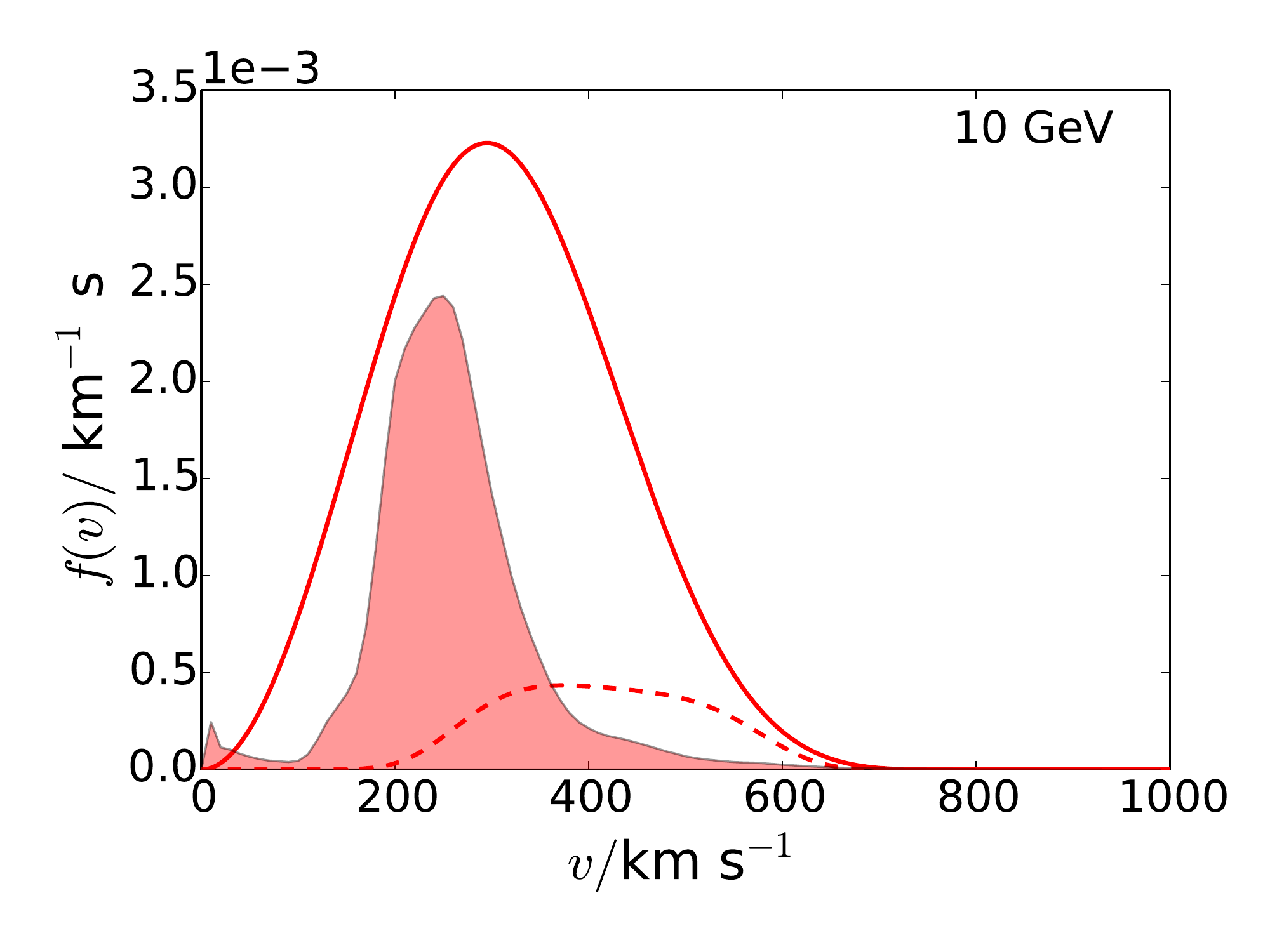}
  \caption{{\bf Reconstructed Earth-frame speed distributions for a 10 GeV WIMP.} The Xenon, Argon and Germanium experiments were used in the reconstruction. Three different underlying speed distributions were used, as described in the text: SHM (top row), SHM+DD (middle row), LIS (bottom row). The underlying distribution is shown as a solid line, the best fit is shown as a dashed line and the 68\% credible interval is shown as a shaded band. The reconstructed distribution has been rescaled by a factor $\sigma_\textrm{rec}^{SI}/\sigma_\textrm{true}^{SI}$, as described in the text.}
  \label{fig:astrophysics:speeds_XeArGe}
\end{figure}

We now compare with Fig.~\ref{fig:astrophysics:speeds_XeArGe}, in which we present reconstructions for a 10 GeV WIMP using Xenon, Argon and Germanium experiments (that is, we have discarded the data from both the Silicon and COUPP experiments). Here, we see a dramatic loss of accuracy in the reconstructed speed distribution. The Silicon and COUPP experiments consist of the lightest target nuclei and their removal from the dataset has increased the lowest accessible speed of the experiments up to $\sim 570 \textrm{ km s}^{-1}$. Moreover, for \mwimp$=10$ GeV, the Argon experiment observes no events (due to its high threshold energy). This results in a much broader range of allowed WIMP masses. This in turn results in a wide uncertainty in the range of speeds being probed by the experiments. This explains why the reconstruction appears to be shifted to lower speeds in some cases. This is not a failing of this particular parametrization, but is caused by the poor sensitivity of the Xe, Ar and Ge experiments to low speed WIMPs. The reconstructions for the heavier WIMPs are largely unaffected, as the range of accessible speeds for those benchmarks is dominated by these three remaining experiments.

\subsection{Discussion}

Using upcoming experiments, the prospects for reconstructing both the WIMP mass and distribution function are promising. We have demonstrated that by using a general parameterization of the speed distribution, we can reconstruct the WIMP mass without appealing to any assumptions about the astrophysics of dark matter. As we have explained, this process requires multiple experiments which use different target nuclei in order to yield useful information.  Using a wide range of experiments leads to a significant reduction in the uncertainties on the reconstructed WIMP mass. Simultaneously, we can extract the shape of the WIMP speed distribution, with the caveat that the results are only reliable over the range of speeds to which the experiments are sensitive.

However, any assumption-free method introduces an unavoidable degeneracy in the interaction cross section. Without knowing what fraction of WIMPs our experiments are probing, we cannot reliably extract the strength of WIMP-nucleon interactions. In order to tackle this problem, it is necessary to use experiments which probe as wide a range of WIMP speeds as possible. This can be achieved not only be widening the energy sensitivity windows of the experiments but also by using target nuclei with a range of masses, which naturally probe complementary regions of the WIMP speed distribution.

We have also shown how different combinations of particle physics and astrophysics are more reliably probed with certain sets of experiments. Low mass WIMPs require light targets which can probe the high-$v$ tail of the speed distribution. For a 50 GeV WIMP, a dark disk component was invisible to the experiments considered here, while a dark disk of 200 GeV WIMPs could be accurately reconstructed. However, the nature of dark matter and its speed distribution are \textit{a priori} unknown, so we cannot select the optimal experiments ahead of time. We must therefore use a range of different experimental targets to ensure sensitivity to a large range of WIMP masses and to ensure that features in the speed distribution can be reliably captured.

\section{Directional detection}\label{sec:directional}
Directional-detection experiments aim to measure the directions of WIMP-induced nuclear recoils in addition to their energies.  A primary strength of such experiments is their ability to distinguish signal events from background events.  The latter should be primarily isotropic; in contrast, the Earth's motion with respect to the Galactic rest frame leads to a strong angular dependence of the signal~\cite{Spergel:1987kx}.  For a smooth WIMP distribution, the event rate is strongly peaked in the direction opposite to the Sun's motion, towards the constellation Cygnus (or for low-energy recoils, in a ring around this direction~\cite{Bozorgnia:2011vc}).  Furthermore, the strength of the directional signal is expected to be relatively large for typical halo models; the maximum event rate in the peak direction may be roughly an order of magnitude larger than the minimum.  Comparing to annual modulation, where the signal is expected to be at the $\sim\!1-10\%$ level and may be mimicked by backgrounds that also modulate annually, the power of directional detection to confirm the dark-matter origin of a possible signal becomes clear.  With a detector capable of measuring the nuclear-recoil vectors in three dimensions and sufficient angular resolution, $\sim\!10$ events would be sufficient to reject isotropy~\cite{Copi:1999pw,Green:2006cb}.  With $\sim\!30-50$ events, the peak direction could be measured and the Galactic origin of the recoils confirmed~\cite{Billard:2009mf,Green:2010zm}.  However, the exposure that these experiments will require to detect the anisotropy of WIMP-induced recoils depends strongly on their ability to measure the senses ($+{\bf q}$ versus ${\bf -q}$) of low-energy recoils~\cite{Green:2006cb}. 

\begin{figure*}
  \includegraphics[trim=25 0 58 25,clip,height=0.28\textwidth]{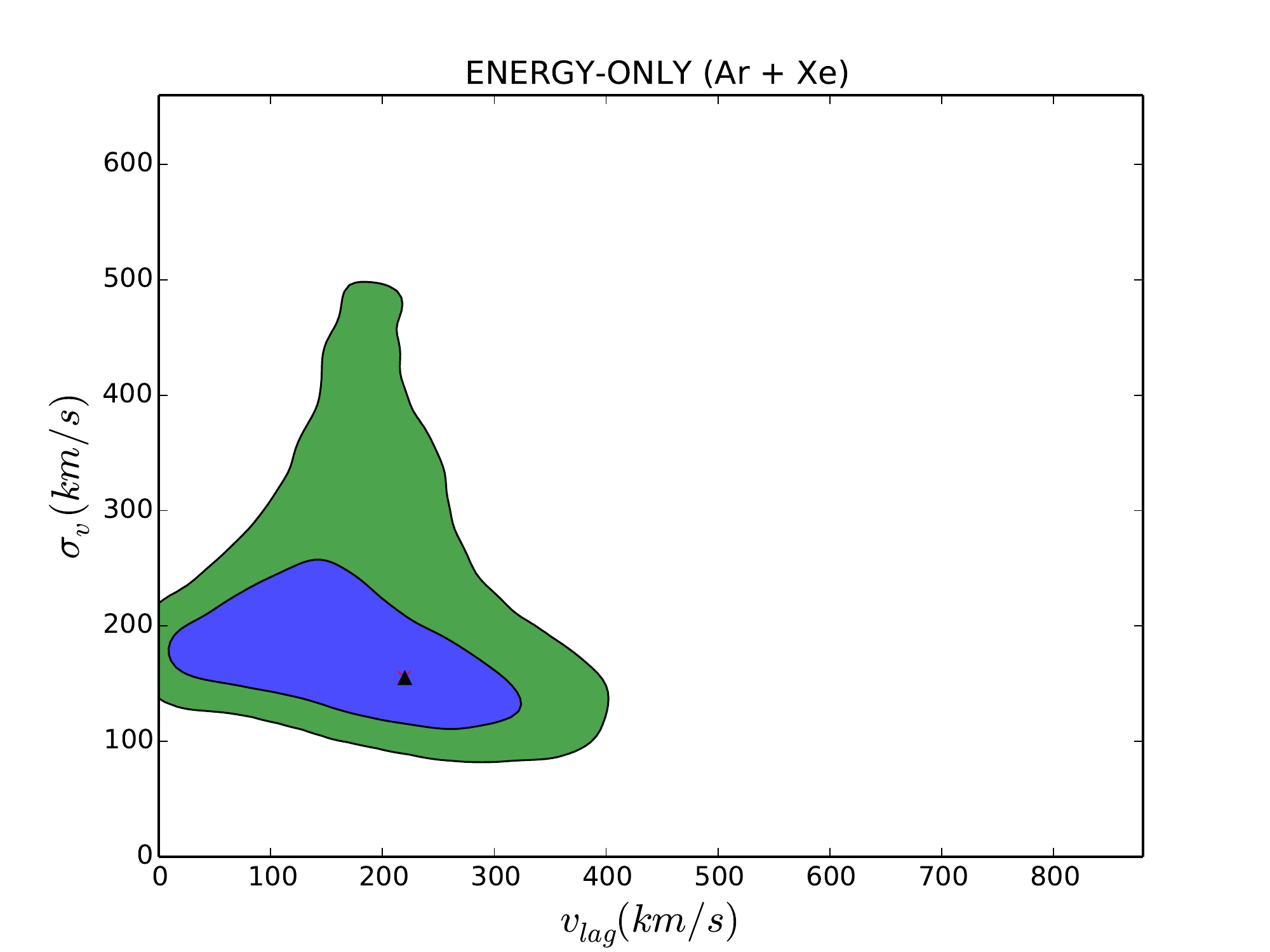}
  \includegraphics[trim=70 0 58 25,clip,height=0.28\textwidth]{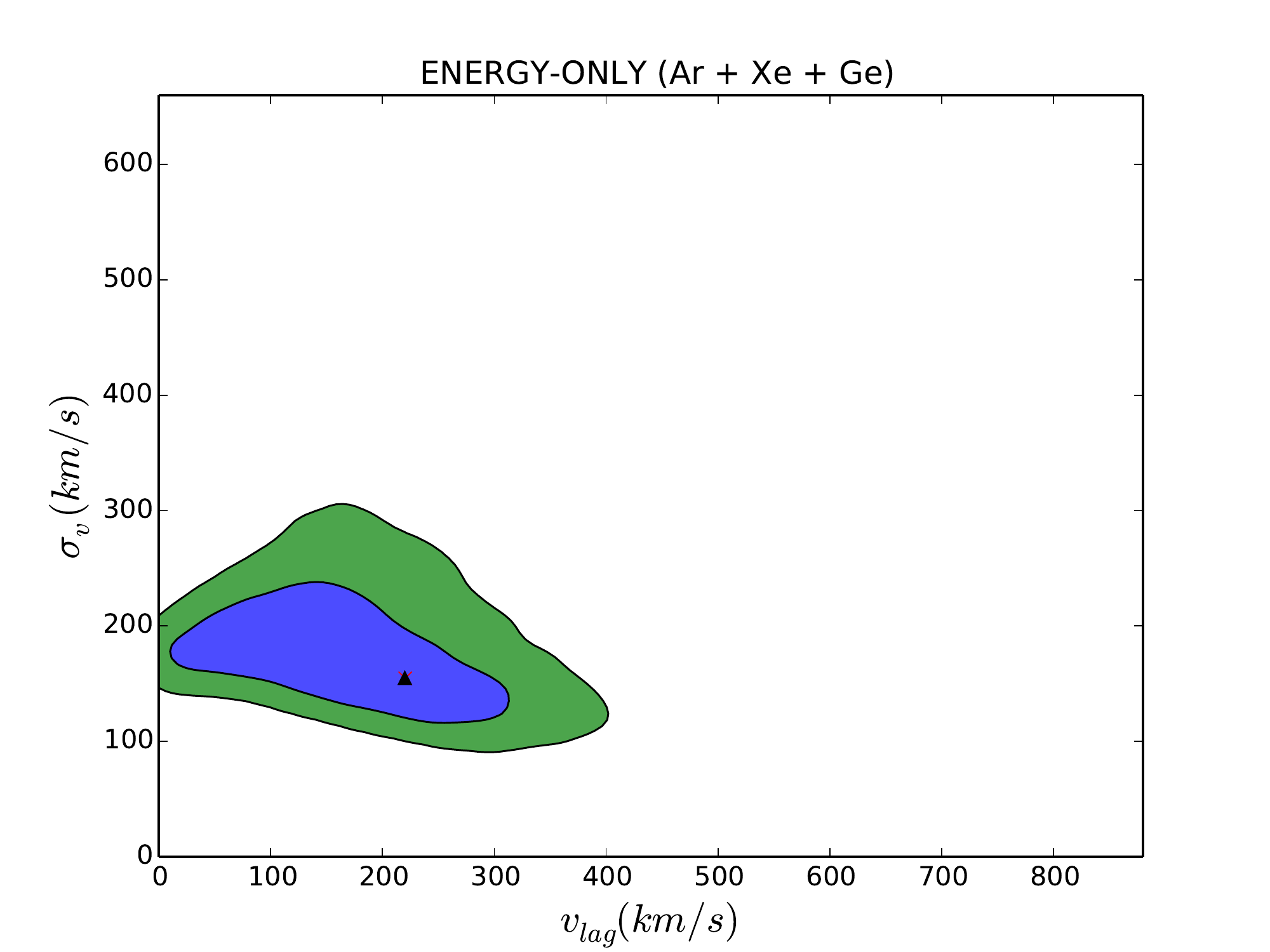}
 \includegraphics[trim=70 0 58 25,clip,height=0.28\textwidth]{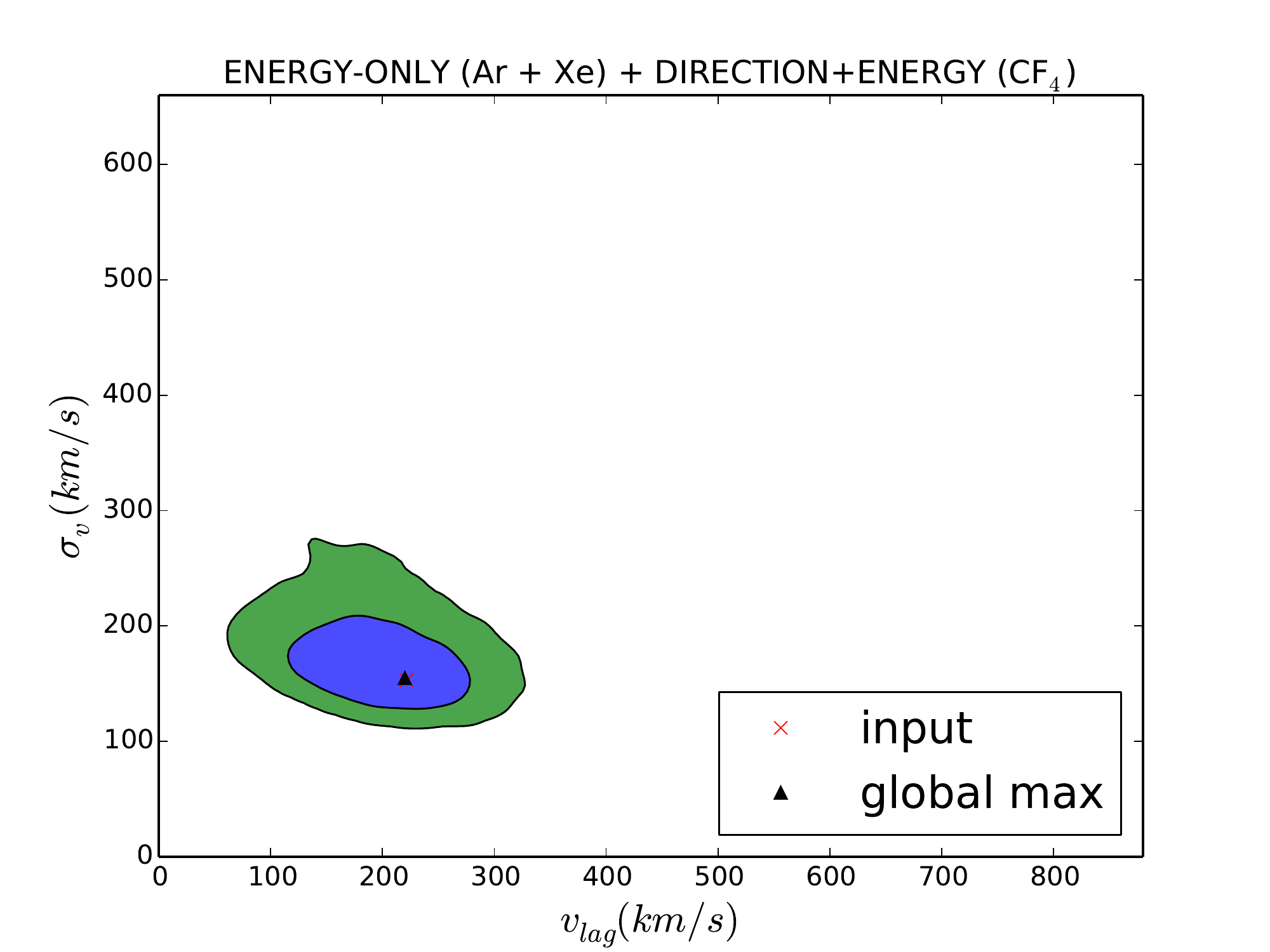}
  \caption{Posterior distributions for the $v_\mathrm{lag}$-$\sigma_v$ parameter space for the Asimov data sets of three ensembles of experiments.  Inner and outer regions enclose the 68\% and 95\% credible regions.  We assume the SHM/50-GeV WIMP (Benchmark 2) scenario with weak priors on $v_\mathrm{lag}$ and $\sigma_v$ (and fixed $v_\mathrm{esc}$) and flat priors on the WIMP mass. The left plot shows the contours for an ensemble consisting of non-directional Ar and Xe experiments observing $\sim\!25$ and $\sim\!350$ events, respectively.  The center plot shows the same for the same ensemble with an additional non-directional Ge experiment, while the right plot shows the same with an additional  MIMAC-like, directional CF$_4$ experiment; the exposures of these experiments have been tuned so that each observes $\sim\!20$ events (200 kg-year and 2000 kg-year, respectively).  We see that adding additional non-directional experiments to the ensemble does not significantly improve the 68\% contours in  $v_\mathrm{lag}$-$\sigma_v$, since these parameters are degenerate in the energy spectrum.  However, directional experiments can break such parameter degeneracies, resulting in tighter constraints (particularly on $v_\mathrm{lag}$, in this example).  Note also that other details of the SHM velocity distribution (such as the direction of the WIMP wind) can be recovered with a directional experiment to good accuracy \cite{Billard:2009mf}.}
  \label{fig:directional1}
\end{figure*}

Beyond simply providing background discrimination power, measurements of the directional event rate may also be crucial in determining the finer details of the WIMP velocity distribution.  The velocity distribution is related to the directional event rate as in Eq.  (\ref{eq:energythetaspectrum}).  Moreover, by noting the dependence in Eqs. (\ref{eq:sisdsigma}) and (\ref{eq:directionalcrosssection}) of the WIMP-nuclear cross section on the incoming WIMP velocity $\mathbf{v}$ and the resulting nuclear-recoil direction $\hat{\mathbf{q}}$, it can be shown that
\begin{equation}
\frac{d^2R(Q,\Omega_q,t)}{dQd\Omega_q} \propto \int d^3v\, \delta\left(\mathbf{v}\cdot\hat{\mathbf{q}} - v_{\mathrm{min}}(Q)\right) f(\mathbf{v},t).
\end{equation}
That is, the dependence of the directional event rate on the lab-frame velocity distribution $f(\bold{v},t)$ is determined by the integral on the right-hand side, known as the Radon transform of $f(\bold{v},t)$ \cite{Gondolo:2002np}.  Essentially, the content of this relation is simply that the rate of nuclear recoils with a given energy $Q$ and direction $\bold{q}$ is proportional to the number of WIMPs with velocities $\bold{v}$ that can induce such recoils -- i.e., that satisfy the non-relativistic scattering kinematics determined by $v_\mathrm{min}$.  Such velocities lie on the plane in velocity space defined by the 1-dimensional delta function in the Radon transform.

Interestingly enough, measuring the Radon transform of the velocity distribution provides information complementary to that provided by the energy spectrum.  This allows degeneracies in the energy spectrum to be broken.  As a simple illustration of this, we can compare the abilities of a directional and a non-directional experiment to constrain the parameters of the SHM velocity distribution.\footnote{We take a MIMAC-like experiment as our benchmark example of a directional detector.  Billard et al.~\cite{Billard:2011zj} have carried out a detailed study of the prospects of a ${\rm CF}_4$ detector, as proposed by the MIMAC collaboration. They consider an optimistic detector configuration with threshold $E_{\rm th} = 5 \, {\rm keV}$ and angular resolution $\sigma= 20^{\circ}$, zero background and $100 \%$ sense recognition and also a pessimistic configuration with 
$E_{\rm th} = 20 \, {\rm keV}$, $\sigma= 50^{\circ}$, 10 background events per ${\rm kg \, yr}$ and no sense recognition.  Our results stated here assume the optimistic configuration.}  For example, Lee and Peter~\cite{Lee:2012pf}  compared constraints that experiments with energy-only, direction-only, and direction+energy sensitivity can place on the halo lag speed (i.e., the relative speed of the dark-matter rest frame and the lab frame) and the halo velocity dispersion.  They showed that these two parameters are degenerate with only measurements of the energy spectrum; however, the information provided by measurements of the recoil direction is orthogonal, so that an experiment sensitive to both energy and direction can break this degeneracy (see \cite{Lee:2012pf} for more detail).  Note that this degeneracy remains even with an ensemble of non-directional experiments, as shown in Fig.~\ref{fig:directional1}.  Thus, combining our previous discussion with this simple consideration of the SHM velocity distribution, we can draw a key conclusion: an ensemble of multiple non-directional experiments may allow us to pin down the WIMP mass, cross section, and speed distribution, but directional experiments will offer the unique ability to fully constrain the WIMP \emph{velocity} distribution.

Of course, directional detectors can also constrain parameterized velocity distributions beyond the SHM approximation.  With ${\cal O}(1000)$ events, it may also be possible to detect and constrain the parameters of velocity substructures -- the direction and velocity of a stream, or the lag speed and velocity dispersion of a dark disk, for example~\cite{Lee:2012pf}.  It may also be possible to detect anisotropy in the velocity distribution.  For example,  Host and Hansen~\cite{Host:2007fq} showed that with $10^{4}$ events a ${}^{32}{\rm S}$ target with $100 \, {\rm keV} $ energy threshold could measure the velocity anisotropy parameter, $\beta = 1- \sigma_{\rm t}^2/\sigma_{\rm r}^2$ where $\sigma_{\rm r,t}$ are the radial and tangential velocity dispersions,  with a precision of $\sim\!0.03$.  

Furthermore, it may be possible to take a non-parametric approach -- analogous to the methods used in Sec.~\ref{sec:astrophysics}, but for the full 3-dimensional velocity distribution.  For example, Alves, Hedri and Wacker~\cite{Alves:2012ay} decomposed the velocity distribution in terms of the products of special functions of the integrals of motion (the relative energy, the z-component of the angular momentum and the magnitude of the angular momentum) and examined the precision with which the coefficients could be measured. They found that with ${\cal O}(1000)$ events, an ideal detector could discern the local velocity distribution found in the Via Lactea II simulation~\cite{Kuhlen:2008qj} 
from a Maxwellian distribution.

Directional experiments can also provide interesting constraints on the particle-physics side.  Importantly, the F and $^3$He target gases used in several of the current experiments offer sensitivity to spin-dependent interactions.  Since WIMP-nucleon spin-independent and spin-dependent cross sections are not necessarily correlated, directional experiments with relatively low exposures of $\sim\!0.1$ kg-year can rule out supersymmetric WIMP candidates that will remain out of reach of high-exposure, non-directional experiments only sensitive to spin-independent interactions~\cite{Mayet:2002ke,Moulin:2005sx}.  Furthermore, inelastic scattering produces a directional signal more strongly peaked in angular distribution than does elastic scattering.  Finkbeiner et al.~\cite{Finkbeiner:2009ug} examined inelastic-scattering models compatible with the DAMA claim of modulation, and showed that a gaseous Xe detector would require a relatively low exposure of $\sim\!1000$ kg-day to rule out or support the claim; measurements of the mass splitting $\delta$ might also be possible.

Finally, directional detectors may also provide a way around the neutrino background floor.  If if the WIMP-nucleon cross section indeed lies beneath the floor, discriminating neutrino-nucleon background events from the recoil spectrum or annual modulation alone would require large statistics and high exposures.  The ability of directional experiments to identify the predominantly Solar origin of background events would then be important~\cite{Monroe:2007xp,Gutlein:2010tq,billard2013}.

\section{Conclusion}\label{sec:conclusion}
In this work, we overviewed the power of ensembles of direct-detection experiments to characterize WIMPs.  We also reviewed the literature on the types of particle properties that WIMPs might have, and how they manifest themselves in experiments; summarized assumptions about WIMP astrophysics and their effects on WIMP direct detection; and showed the status of experiments.  We highlighted ``halo-independent'' methods for comparing experiments against each other for fixed interaction type and WIMP mass independently of the assumed WIMP velocity distribtuion.  We used Bayesian inference of mock Generation-2 experiments and data sets to examine what WIMP physics can be teased out of experiments in the early WIMP discovery days.  While we primarily focused on ``Generation 2'' experiments, we also highlighted what additional kinds of experiments may be useful to characterizing WIMPs in those early days.  It is vitally important to characterize WIMP physics using multiple types of direct-detection experiments in order to check for consistency among the classes of experiment (including colliders and indirect detection).  We must know if all types of experiment are seeing the same dark-matter WIMP.

In Sections \ref{sec:simulation} to \ref{sec:directional}, we explored the potential of ensembles of experiments to identify key pieces of WIMP physics.  Here, we summarize how the choice of experiments affects the ability to probe the different aspects of WIMP physics:
\begin{enumerate}
  \item In order to distinguish spin-independent scattering from spin-dependent WIMP-proton and WIMP-neutron scattering, we require a set of experiments with complementary sensitivity to each of these types of coupling.  In order to break degeneracies with the WIMP velocity distribution, it is also highly desirable that there be energy resolution on the order of $\sim$ keV.  With our fiducial set of Generation 2 experiments (liquid Argon and Xenon, cryogenic Germanium, and bubble-chamber CF$_3$I), spin-independent and spin-dependent scattering can be distinguished quite well.  For low-mass WIMPs, fits are aided somewhat with the addition of a Silicon experiment, but are still dominated by uncertainties in the WIMP velocity distribution.  WIMP masses can be estimated well if spin-independent or spin-dependent WIMP-neutron interactions dominate, given the proposed slate of Generation 2 experiments, but parameter estimate is significantly worse if spin-dependent WIMP-proton interactions dominate.  This is because the primary experiment to probe this interaction is COUPP.  Once WIMPs are discovered, we recommend that bubble-chamber experiments like COUPP explore a range of energy thresholds (and a range of target fluid).  This will yield a cumulative event rate, $R(>Q)$, which can be used to constrain the WIMP velocity distribution, and hence, the WIMP mass.  It would be useful to have additional experiments with energy resolution and spin-dependent WIMP-proton sensitivity.  If the spin-dependent WIMP-proton cross section is sufficiently large, directionally sensitive experiments can also be extremely helpful in constraining the WIMP velocity distribution, which should yield improved WIMP particle property parameter estimation.
    \item For inelastic dark matter, or to distinguish among non-relativistic effective operators, it is useful for the experimental ensemble to span a range of target nuclear masses, and for energy thresholds to be low.  We endorse the development of low-threshold experiments such as CoGeNT, CDMSlite, DAMIC, and MALBEK.  This is especially important in the case where long-range interactions between WIMPs and nuclei dominate over contact interactions (Sec. \ref{sec:WIMPphys}).  While a Silicon-based experiment does not add much to distinguishing between spin-independent and spin-dependent interactions, it is useful in identifying inelastic dark-matter models.
    \item In order to really characterize the WIMP mass, we strongly support the use of the method discussed in Kavanagh \& Green \cite{Kavanagh:2013wba} and Sec. \ref{sec:astrophysics}.  This method, modeling the logarithm of the WIMP speed distribution as a set of orthogonal polynomials, can constrain the shape of the WIMP speed distribution in the range in which experiments have sensitivity.  An accurate reconstruction of the shape of the WIMP velocity distribution is critical to unbiased estimates of the WIMP mass.  Depending on the WIMP mass, different sets of experiments dominate the posterior (Fig. \ref{fig:astrophysics:MassRecon}).  Therefore, since the WIMP mass is a priori unknown, having as wide a range of target nuclear masses as possible is desirable for WIMP searches.
    \item The WIMP velocity distribution is interesting in and of itself, as it is related to today's global properties of the Milky Way dark-matter halo, as well as its assembly history.  A combination of the Kavanagh \& Green method for the WIMP speed distribution reconstruction with the directionally sensitive methods discussed in Sec. \ref{sec:directional} can yield interesting constraints on the WIMP velocity distribution.  Different experiments probe different parts of the velocity distribution, and the velocity range depends strongly on the WIMP mass for fixed recoil energy.  For most current experiments, WIMP speeds of $\lesssim 100$ km s$^{-1}$ (relevant for dark disks) can only be probed in the WIMP mass is large (\mwimp$\gtrsim 200$ GeV).  Since the WIMP mass is unknown, we do not know \emph{a priori} down to what WIMP speed future experiments may probe. Low threshold experiments are critical to probing the WIMP speed distribution to low speeds, especially for low-mass WIMPs.
\end{enumerate}

In summary, we have clarified which pieces of WIMP physics can be explored by the currently proposed ensemble of Generation 2 experiments.  They can determine the relative strength of spin-independent and spin-dependent interactions; unveil non-relativistic effective operators; distinguish inelastic from elastic scattering for mass-splittings $|\delta| \gtrsim 20$ keV; estimate the WIMP mass to $\gtrsim 20\%$ without significant bias; and constrain the shape of the WIMP speed distribution in the range that experiments can kinematically probe.  What direct-detection experiments cannot do is estimate the absolute value of the cross sections without dubiously strong priors regarding the local WIMP mass density and the fraction of local WIMPs that experiments may kinematically probe.  However, we can reasonably estimate the ratios of cross sections and the shape of the velocity distribution.  With longer exposures for experiments with unpaired-nucleon target nuclei, we can estimate the sign of $a_n/a_p$ \cite{Pato:2011de}.  With experiments with lower thresholds, we can probe lower-speed WIMPs, a wider range of inelastic dark-matter models, and long-range WIMP-nuclear interactions.  We emphasize the fact that we use highly idealized experiments and data sets in this study.  In the future, it will be necessary to investigate real experimental configurations to provide accurate forecasts for WIMP parameter estimation using direct detection.

Our main conclusion is: we need a wide variety of direct-detection experiments in order to accurately infer WIMP properties (and their local phase-space density) once WIMPs are conclusively discovered.  There is currently pressure to reduce the number of future direct-detection experiments.  In our opinion, the selection of experiments must proceed with extreme caution so that we do not lose sensitivity to WIMP physics.  Different characteristics of WIMPs are probed best by different ensembles of experiments.  However, we do not know \emph{a priori} what properties WIMPs have.  Finding the right (and necessarily broad) ensemble of next-generation experiments is key to accurately and precisely estimating WIMP physics with those experiments in the early days of WIMP discovery.

\section*{Acknowledgments}
We thank Miguel Pato for helpful conversations. VG thanks Jiji Fan and Matthew Reece for enlightening conversations.
VG gratefully acknowleges support from the Friends of the Institute for Advanced
Study. AMG and BJK are both supported by STFC. AMG also acknowledges support from the Leverhulme Trust. BJK gratefully acknowledges access to the University of Nottingham High Performance Computing Facility.  This work was supported in part by an allocation of computing time from the Ohio Supercomputer Center to AHGP.


\end{document}